\documentclass[11pt,a4paper]{oxarticle}

\pdfoutput=1

\usepackage[letterpaper, body={6.5in, 9.25in}, top=1.0in, left=1.5in]{geometry}
\usepackage{amssymb} 

\usepackage{titlesec}

\titlespacing*{\subsubsection}
{0pt}{5pt plus 1pt minus 1pt}{5pt plus 1pt}

\titlespacing*{\subsection}
{0pt}{5pt plus 1pt minus 1pt}{0pt plus 1pt}

\usepackage[dvipsnames]{xcolor}
\usepackage{graphicx}
\usepackage{float}
\usepackage{amscd}
\usepackage{amssymb}
\usepackage{longtable}
\usepackage[bbgreekl]{mathbbol}
\usepackage[utf8]{inputenc}
\usepackage[sort&compress, comma, square, numbers]{natbib}
\usepackage{multirow}

\usepackage[bottom]{footmisc}
\usepackage{bm}
\usepackage{tikz}
\usepackage{tikz-cd}
\usepackage{braket}
\usepackage[pdfpagelabels=false]{hyperref}
\hypersetup{
	colorlinks,
	linkcolor={blue!50!black},
	citecolor={blue!50!black},
	urlcolor={blue!50!black}
}

\usepackage{tabularx}
\usepackage{caption}
\usepackage{cellspace}
\DeclareSymbolFontAlphabet{\mathbb}{AMSb}
\DeclareSymbolFontAlphabet{\mathbbl}{bbold}

\let\SS=\S 

\renewcommand{\S}{\Sigma}

\DeclareFontFamily{OT1}{pzc}{}
\DeclareFontShape{OT1}{pzc}{m}{it}{<-> s * [1.200] pzcmi7t}{}
\DeclareMathAlphabet{\mathpzc}{OT1}{pzc}{m}{it}

\newcommand{\cA}{\mathcal{A}}
\newcommand{\cB}{\mathcal{B}}
\newcommand{\cC}{\mathcal{C}}

\newcommand{\cE}{\mathcal{E}}

\newcommand{\cH}{\mathcal{H}}
\newcommand{\cI}{\mathcal{I}}
\newcommand{\cJ}{\mathcal{J}}

\newcommand{\cL}{\mathcal{L}}

\newcommand{\cN}{\mathcal{N}}
\newcommand{\cO}{\mathcal{O}}
\newcommand{\cP}{\mathcal{P}}

\newcommand{\cR}{\mathcal{R}}
\newcommand{\cS}{\mathcal{S}}

\newcommand{\cU}{\mathcal{U}}
\newcommand{\cV}{\mathcal{V}}

\DeclareFontFamily{U}{bbold}{}
\DeclareFontShape{U}{bbold}{m}{n}
{  <-5.5> s*[1.05] bbold5
	<5.5-6.5> s*[1.05] bbold6
	<6.5-7.5> s*[1.05] bbold7
	<7.5-8.5> s*[1.05] bbold8
	<8.5-9.5> s*[1.05] bbold9
	<9.5-11.5> s*[1.05] bbold10
	<11.5-16> s*[1.05] bbold12
	<16-> s*[1.05] bbold17
}{}

\newcommand{\IC}{\mathbb{C}}

\newcommand{\IE}{\mathbb{E}}
\newcommand{\IF}{\mathbb{F}}

\newcommand{\IH}{\mathbb{H}}

\newcommand{\IL}{\mathbb{L}}

\newcommand{\IP}{\mathbb{P}}
\newcommand{\IQ}{\mathbb{Q}}
\newcommand{\IR}{\mathbb{R}}
\newcommand{\IS}{\mathbb{S}}

\newcommand{\IU}{\mathbb{U}}

\newcommand{\IZ}{\mathbb{Z}}

\font\elevenrmfromseventeenrm = cmr17 at 11pt
\newcommand{\inbar}{\vrule height6.9pt depth-0.2pt width0.35pt}
\newcommand{\zero}{\hbox{{\elevenrmfromseventeenrm 0}\kern-3.5pt\inbar\kern1pt\inbar\kern2pt}}

\font\eightrmfromseventeenrm = cmr17 at 8pt
\newcommand{\ssinbar}{\vrule height5pt depth-0.1pt width0.3pt}
\newcommand{\sszero}{\hbox{{\eightrmfromseventeenrm 0}\kern-2.53pt\ssinbar\kern0.7pt\ssinbar\kern2pt}}

\newcommand{\mtB}{\text{B}}
\newcommand{\mtC}{\text{C}}

\newcommand{\mtI}{\text{I}}

\newcommand{\mtM}{\text{M}}
\newcommand{\mtN}{\text{N}}

\newcommand{\mtQ}{\text{Q}}

\newcommand{\mtS}{\text{S}}
\newcommand{\mtT}{\text{T}}
\newcommand{\mtU}{\text{U}}

\newcommand{\fa}{\mathfrak{a}}
\newcommand{\fb}{\mathfrak{b}}

\newcommand{\fm}{\mathfrak{m}}

\newcommand{\fp}{\mathfrak{p}}

\newcommand{\1}{\mathbf{1}}

\newcommand{\ib}{{\bar\imath}}

\font\eightrm=cmr8 at 8pt
\font\csc=cmcsc10

\newcommand{\defineas}{:=}
\newcommand{\asdefine}{=:}

\newcommand{\place}[3]{\vbox to0pt{\kern-\parskip\kern-7pt
		\kern-#2truein\hbox{\kern#1truein #3}
		\vss}\nointerlineskip}
\newcommand{\capt}[3]{\parbox{#1}{\renewcommand{\baselinestretch}{1.0}
		\caption{\label{#2}\small\it #3}}}

\newcommand{\smallfrac}[2]{\frac{\scriptstyle #1}{\scriptstyle #2}}

\newcommand{\+}{\phantom{-}}
\renewcommand{\=}{\;=\;}

\newcommand{\cref}[1]{Chapter \ref{#1}}

\DeclareFontFamily{U}{wncy}{}
\DeclareFontShape{U}{wncy}{m}{n}{<->wncyr10}{}
\DeclareSymbolFont{mcy}{U}{wncy}{m}{n}
\DeclareMathSymbol{\sha}{\mathord}{mcy}{"58}

\newcommand{\SO}{\text{SO}}
\newcommand{\SL}{\text{SL}}
\newcommand{\GL}{\text{GL}}
\newcommand{\Sp}{\text{Sp}}

\renewcommand{\Im}{\text{Im}}

\newcommand{\diag}{\text{diag}}

\newcommand{\ee}{\text{e}}
\newcommand{\me}{\text{e}}

\newcommand{\ii}{\text{i}}
\newcommand{\dd}{\text{d}}

\newcommand{\Gal}{\text{Gal}}

\newcommand{\Hg}{\text{Hg}}

\newcommand{\ish}[1]{\left| #1 \middle\rangle \!\right\rangle}

\newcommand{\wt}[1]{\widetilde{#1}}

\makeatletter
\g@addto@macro\bfseries{\boldmath}

\def\blindfootnote{\xdef\@thefnmark{}\@footnotetext}
\makeatother

\newcommand{\BP}[1]{\ket{B_#1}_\cP}
\newcommand{\BBP}[1]{{\vphantom{\ket{}}}_\cP \! \bra{B_#1}}

\newcommand{\HWR}{\text{HWR}_\cH(\cA)}
\newcommand{\HWRrep}{\text{HWR}(\cA)}
\newcommand{\HWRbar}{\text{HWR}_\cH(\overline{\cA})}
\newcommand{\dimHWR}{M}

\newcommand{\Rla}[1]{{{\bf r}(#1)}}
\newcommand{\Rl}{\Rla{l}}

\newcommand{\intscal}[1]{\ket{#1}_\cI}

\DeclareMathOperator{\lcm}{lcm}

\numberwithin{equation}{section}
\begin{document}
	%
	
	\proofmodefalse
	
	\thispagestyle{empty}  
	\begin{flushright} MITP--25--062\end{flushright}
	\begin{center}
		\null\vskip0.5in
		{\Huge Hodge Structures of Complex Multiplication Type from Rational Conformal Field Theories\\[12pt]}
		\vskip0.5cm
		{\csc Hans Jockers${}^{*1}$, Pyry Kuusela${}^{*\dagger2}$, and Maik Sarve${}^{*\flat3}$.\\[1cm]}
		\blindfootnote{\null\hskip-10pt \hfill ${}^1\,$jockers@uni-mainz.de \hfill ${}^2\,$pyry.r.kuusela@gmail.com \hfill 
			${}^3\,$maik.sarve@unimelb.edu.au\hfill \kern10pt}
		{\it
			${}^*$PRISMA+ Cluster of Excellence \& Mainz Institute for Theoretical Physics\\
			Johannes Gutenberg-Universit\"at Mainz\\
			55099 Mainz, Germany\\
		}
		\vskip20pt
	
		{\it
			${}^\dagger$Sorbonne Université, CNRS, Laboratoire de Physique Théorique et Hautes Energies,\\
			Campus Pierre et Marie Curie, 4 place Jussieu, F-75005, Paris, France\\	
		}
		\vskip20pt
		{\it
			${}^\flat$School of Mathematics and Statistics, University of Melbourne \\
Parkville, VIC 3010, Australia
		}

		\vspace*{1cm}
		{\bf Abstract}\\
		\end{center}
	\vskip-5pt
	\begin{minipage}{\textwidth}
		\baselineskip=15pt
		\noindent
		Under certain assumptions, we show that unitary rational $\mathcal{N}=(2,2)$ conformal field theories together with a certain generating set of Cardy boundary states in the associated boundary conformal field theories give rise to rational Hodge structures of complex multiplication type. We argue that these rational Hodge structures for such rational conformal field theories arising from infrared fixed points of $\mathcal{N}=(2,2)$ non-linear sigma models with Calabi--Yau target spaces coincide with the rational Hodge structures of the middle-dimensional cohomology of the target space geometry. This gives non-trivial evidence of the general expectation in the literature that rational $\mathcal{N}=(2,2)$ supersymmetric conformal field theories associated to Calabi--Yau target spaces yield middle dimensional cohomological rational Hodge structures with complex multiplication. We exemplify our general results with the $\mathcal{N}=2$ A-type minimal model series --- which do not have a geometric origin as a non-linear sigma model --- and with two explicit $\mathcal{N}=(2,2)$ Gepner models that correspond to particular non-linear sigma models with specific Calabi--Yau threefold target spaces. 
	\end{minipage}
	\clearpage
	
	\thispagestyle{empty}
	
	{\baselineskip=8pt
		\tableofcontents} 
	
	
	\newpage
	\pagenumbering{arabic}

\section{Introduction}
The classification of two-dimensional unitary conformal field theories is a daunting problem. Such conformal field theories possess a left- and right-moving Virasoro algebra $\operatorname{Vir} \times \overline{\operatorname{Vir}}$ as its symmetry algebra,  and they can be organised according to their central charge $c$ --- the unique central element of the Virasoro algebra $\operatorname{Vir}$. Unitary two-dimensional conformal field theories with central charges in the range $0 \le c < 1$ are classified in terms of the discrete and countable series of Virasoro minimal models \cite{Friedan:1983xq}. These theories are  referred to as minimal because their Hilbert spaces are constructed from a finite number of Virasoro highest weight representations. 

Already for the central charge $c=1$ there are uncountably many unitary two-dimensional conformal field theories. The free boson with a circular target space $S^1$ with radius $R$ yields a conformal field theory with central charge $c=1$, where distinct radii $R$ (suitable normalised) in the real interval $[1,\infty)$ realise inequivalent --- and hence uncountably many --- conformal field theories with central charge $c=1$. As the conformal field theory of a boson on a circle $S^1$ has a free Lagrangian description, it can be solved exactly. Nevertheless --- opposed to Virasoro minimal models --- the Hilbert space of states arises from an infinite (countable) number of Virasoro heighest weight representations. However, it is well known that if the square of the radius $R^2$ is a rational number the Virasoro algebra $\operatorname{Vir}$ is enhanced to a larger chiral algebra $\mathcal{A}$, see for instance refs.~\cite{Moore:1989vd,Ginsparg:1988ui,DiFrancesco:1997nk}. Then in terms of this extended symmetry algebra $\mathcal{A}$ the Hilbert space of states is again constructed from a finite number of highest weight representations \cite{Anderson:1988to}. 

In general, two-dimensional conformal field theories with finitely many highest weight representations with respect to its chiral symmetry algebra $\mathcal{A}$ are called rational \cite{Anderson:1988to}. For families of conformal field theories described by a moduli space $\mathcal{M}$ --- also often referred to as the conformal manifold $\mathcal{M}$ --- a challenging question to ask is whether a particular conformal field theory in the family is rational or not and what is the distribution of rational conformal field theories in $\mathcal{M}$. Finding two-dimensional rational conformal field theories is an important task as such conformal field theories can be solved with algebraic methods \cite{Belavin:1984vu,Moore:1988uz} --- even and in particular in the absence of a Lagrangian description. 

Two-dimensional non-linear sigma models with compact Riemannian target space manifolds $X$ flow at their infrared fixed point either to the vacuum or to a non-trivial unitary two-dimensional conformal field theory \cite{Friedan:1980jf,Friedan:1980jm,Zamolodchikov:1986gt}. In the latter case, the resulting conformal field theory may be a member of a whole family of conformal field theories, in which the moduli of the Riemannian manifold~$X$ parametrise (a subspace of) the conformal manifold~$\mathcal{M}$ of this family. Starting from its geometric origin, it is in general a difficult problem to determine the infrared unitary conformal field theories and their properties. To further characterise the infrared conformal field theories an important question is now to determine those infrared fixed points which correspond to conformal field theories that are rational and to describe the distribution of such rational conformal field theories in the conformal manifold~$\mathcal{M}$. In particular, a dense distribution of rational conformal field theories in $\mathcal{M}$ implies that the correlators of any member of the conformal family can (at least in principle) be approximated arbitrarily precisely by an exactly solvable rational conformal field theory in its vicinity.\footnote{Using methods of conformal perturbation theory, the question about the density of rational conformal field theories have recently been addressed from the conformal perturbation theory perspective in ref.~\cite{Benjamin:2020flm, Keller:2023ssv}.} 

The infrared fixed points of families of non-linear sigma models with a toroidal target space $X$ with a flat metric (and orbifolds thereof) admit a Lagrangian description in terms of free bosons. Therefore, these conformal field theories can be solved by means of canonical quantization. For these classes of theories --- which serve as an important theoretical laboratory to examine conformal field theories arising from non-linear sigma models --- the raised question of rationality has been solved in refs.~\cite{Moore:1998pn,Wendland:2000ye,Hosono:2002yb,Gukov:2002nw,Chen:2005gm,Benjamin:2020flm,Kidambi:2024vwl}.

Motivated from the worldsheet perspective of type~II string theories, an important class of families of conformal field theories arise from two-dimensional $\mathcal{N}=(2,2)$ supersymmetric non-linear sigma models with Calabi--Yau manifolds $X$ of complex dimension $n$. The anomaly-free axial and vector $R$-symmetries of such non-linear sigma models imply non-trivial $\mathcal{N}=(2,2)$ superconformal field theories with central charge $c = 3 n$ at their infrared fixed points, see, e.g., refs.~\cite{Distler:1992gi,Greene:1996cy}. Moreover, such conformal field theories come in families, whose conformal manifolds~$\mathcal{M}$ are parametrised by the complex structure moduli spaces of both the Calabi--Yau target space $X$ and the mirror Calabi--Yau target space $\widehat{X}$.\footnote{See for instance the review~\cite{Greene:1996cy} and references therein} For $n=1$ the target space is a complex elliptic curve $E$, and the resulting conformal field theories are $\mathcal{N}=(2,2)$ supersymmetric extensions of the free bosonic conformal field theories with target space $T^2$. These conformal field theories are rational if and only if the elliptic curve~$E$ and the mirror elliptic curve~$\widehat{E}$ are elliptic curves with complex multiplications both associated to the same imaginary quadratic number field $\mathbb{Q}(\sqrt{-d})$ \cite{Wendland:2000ye,Gukov:2002nw,Chen:2005gm,Kidambi:2024vwl,Jockers:2024ffr}. 

For higher dimensional Calabi--Yau target spaces, the notion of complex multiplication of the elliptic curves $E$ generalises to the concept of Hodge structures of complex multiplication type for the Hodge structure of the middle dimensional cohomology $H^n(X,\mathbb{C}) = \bigoplus_q H^{3-q,q}_{\overline\partial}(X)$ of the Calabi--Yau manifolds $X$. Therefore, it is natural to propose that for $\mathcal{N}=(2,2)$ non-linear sigma models the Calabi--Yau target space $X$ must be of complex multiplication type in order for the corresponding conformal field theory to be rational \cite{Gukov:2002nw,Chen:2005gm,Kidambi:2024vwl}. This proposal passes several non-trivial tests for target spaces that are Abelian varieties or orbifolds of Abelian varieties \cite{Wendland:2000ye,Gukov:2002nw,Chen:2005gm,Kidambi:2024vwl}. There are powerful mathematical tools in arithmetic geometry to address the distribution of Hodge structures of complex multiplication type in the moduli space of varieties~\cite{MR990016,MR1472499,MR1273413,MR2520786,MR2800724,MR3245009,MR4689371}. Thus, establishing a connection between rational conformal field theories and Calabi--Yau target spaces of complex multiplication type improves our understanding of the distribution of rational conformal field theories in the conformal manifold $\mathcal{M}$. 

Similar arithmetic considerations have implications for the question whether effective quantum field theories admit a UV completion \cite{Grimm:2021vpn,Bakker:2021uqw}, which is often referred to as the swampland program, see for instance the reviews~\cite{Palti:2019pca,Agmon:2022thq} and references therein. As two-dimensional unitary conformal field theories are important ingredients in the worldsheet description of string theories, which in turn furnish UV complete effective quantum field theories at low energies, the appearance of analogous arithmetic properties suggests a possible deeper connection between the rationality of conformal field theories and the swampland program. Complex multiplication property of Calabi--Yau manifolds have also been used in ref.~\cite{Grimm:2024fip} to find exact flux vacuum solutions by utilising existence of symmetries corresponding to the complex multiplication property. More generally, flux vacua and symmetries have been discussed for instance in refs.~\cite{DeWolfe:2005gy,Kanno:2017nub,Kachru:2020sio,Kachru:2020abh,Schimmrigk:2020dfl,Candelas:2023yrg,Jockers:2023zzi} in connection to arithmetic geometry, a topic which we briefly revisit in section \ref{sect:CM_and_arithmetic}. In this context flux vacua minimise the flux-induced scalar potential of the effective theories of string compactifications on Calabi--Yau manifolds with background fluxes \cite{Gukov:1999ya,Taylor:1999ii}. Determining minima of the flux-induced scalar potential in the moduli space of such theories amounts to finding solutions to transcendental equations \cite{Mayr:2000hh,Louis:2012ux,Candelas:2023yrg,Jockers:2024ocq}, which can often be determined by means of arithmetic geometry \cite{DeWolfe:2005gy,Kanno:2017nub,Kachru:2020sio,Kachru:2020abh,Schimmrigk:2020dfl,Candelas:2023yrg,Jockers:2023zzi,Grimm:2024fip}.

The goal of this work is to formulate a precise connection between two-dimensional unitary rational $\mathcal{N}=(2,2)$ conformal field theories and Hodge structures of complex multiplication type. From the $\text{(c,c)}$-rings and the $\text{(c,a)}$-rings of such superconformal field theories \cite{Lerche:1989uy}, we respectively construct canonical Hodge structures together with a rational structure, which is obtained from choosing a suitable set of Cardy boundary states of the associated boundary conformal field theories. It is these rational Hodge structures that are possibly of complex multiplication type. We show that --- under certain assumptions --- it is indeed true that such rational $\mathcal{N}=(2,2)$ conformal field theories yield Hodge structures with complex multiplication. This establishes within the given setup that complex multiplication of Hodge structures is a necessary condition for rationality of the underlying conformal field theory.

An important ingredient in our considerations is the Verlinde formula \cite{Verlinde:1988sn}, which yields a non-trivial action of the modular group $\operatorname{SL}(2,\mathbb{Z})$ on the (finitely many) irreducible representations of the chiral algebra $\mathcal{A}$. A further consequence of the Verlinde formula is that it associates to the characteristic polynomial of the modular $S$-transformation of a rational conformal field theory a Galois group, which in turn acts non-trivially on the considered set of Cardy boundary states and allows us to prove that the constructed rational Hodge structures are of complex multiplication type.

We illustrate our findings with explicit examples of unitary rational $\mathcal{N}=(2,2)$ supersymmetric conformal field theories and demonstrate that the resulting rational Hodge structures have the described complex multiplication property. First of all, we show that the rational Hodge structures for the $A$-type $\mathcal{N}=2$ minimal models series is of complex multiplication type. As these supersymmetric minimal models are not directly related to a non-linear sigma model description with a geometric target space, this class of examples illustrates that rational Hodge structures with complex multiplication are a property of the rational superconformal field theory without the need of a geometric origin. 

Nevertheless, as the rational Hodge structures of complex multiplication type stem from geometric non-linear sigma model considerations, we examine in detail two representative $\mathcal{N}=(2,2)$ Gepner models \cite{Gepner:1987qi,Gepner:1989gr} that correspond to points in the conformal manifolds $\mathcal{M}$ of families of non-linear sigma models with Calabi--Yau threefold target spaces $X$. For these examples, we show that the rational Hodge structures arising from the conformal field theory data coincide with the geometric rational Hodge structures of the middle dimensional cohomologies of the target spaces $X$. Hence, for these examples the rationality of the $\mathcal{N}=(2,2)$ Gepner model implies the complex multiplication property of the associated geometric Hodge structures, which is in accord with the general expectation formulated in refs.~\cite{Gukov:2002nw,Chen:2005gm,Kidambi:2024vwl}. However, the made assumptions in our general discussion on $\mathcal{N}=(2,2)$ rational conformal field theories are not fulfilled for all $\mathcal{N}=(2,2)$~Gepner models, as we demonstrate in an example. At the same time, this example shows that our proposed framework can be formulated more generally to setup a criterium for rationality.

The structure of this work is as follows. In section~\ref{sec:cft} we review some basic concepts of two-dimensional conformal field theories. In particular, we focus on the Galois symmetry of two-dimensional rational conformal field theories, we spell out those properties of two-dimensional $\mathcal{N}=(2,2)$ superconformal field theories that are relevant to this work, and discuss some features of boundary conformal field theories used in later sections. Starting from a two-dimensional rational $\mathcal{N}=(2,2)$ superconformal field theory, we construct a corresponding Hodge structure in section~\ref{sect:Hodge_structures_from_RCFT}. Using boundary states of the associated boundary conformal field theories, we obtain rational Hodge structures and we discuss their polarisations. We show in detail that two-dimensional rational $\mathcal{N}=(2,2)$ superconformal field theories --- obeying certain technical assumptions --- give rise to Hodge structures of complex multiplication type. We compare these findings to Hodge structures arising from $\mathcal{N}=(2,2)$ superconformal field theories that are obtained from infrared fixed-points of two-dimensional $\mathcal{N}=(2,2)$ supersymmetric non-linear sigma models and that are generically non-rational. In section~\ref{sect:minimal_models}, we illustrate our general construction with $\mathcal{N}=2$ superconformal minimal models and present explicit examples. In section~\ref{sec:exgepnermodels}, we study two rational $\mathcal{N}=(2,2)$ superconformal field theories that arise as Gepner models. While the first example --- corresponding to the Fermat quintic Calabi--Yau threefold --- fulfills all our technical assumptions spelt out in section~\ref{sect:Hodge_structures_from_RCFT}, the second Gepner model --- associated to a degree eight hypersurface in the weighted projective space $\IP^4_{1,1,2,2,2}$ --- illustrates that the general idea of relation of Galois symmetry and complex multiplication covers a broader class of rational $\mathcal{N}=(2,2)$ superconformal field theories, as one of our technical assumptions is not fulfilled in this particular case. Finally, in section~\ref{sec:exgepnermodels}, we present a summary of our results that transitions into a discussion about arithmetic properties of two-dimensional conformal field theories from a more general vantage point.  To three appendices we have relegated some mathematical definitions and some supplementary material about Hodge structures of $\mathcal{N}=2$ superconformal minimal models.

\newpage
\subsection*{Notation}
Below we list some symbols that appear throughout the text. We also try to adhere to the following conventions for these symbols to make the formulae more readable: Matrices are typically denoted by symbols in a roman font and vectors are usually denoted by bra-ket notation when the vectors correspond to states in a conformal field theory and by symbols in a boldface font otherwise. 

\bigskip

{
	\renewcommand{\arraystretch}{1.35}
	\centering
	\begin{tabularx}{0.99\textwidth}{|>{\hsize=.16\hsize\textwidth=\hsize}X|
			>{\hsize=0.76\hsize\textwidth=\hsize}X|>{\hsize=0.07\hsize\textwidth=\hsize}X|}
		\hline
		\textbf{Symbol} & \hfil\textbf{Definition/Description}\hfil & \hfil \textbf{Ref.}\\[3pt]
		\hline	\hline
		
		$\cA$ & A subalgebra $\cA \subseteq \cA_{\text{max}}$ of the chiral algebra $\cA_{\text{max}}$  & \eqref{eq:DefcA} \\[4pt]	\hline			
		$\text{HWR}(\mathcal{A})$ & The set of highest-weight representations of the algebra $\cA$ & \eqref{eq:setsofreps} \\[4pt]	\hline	

		$\HWR$ & The set of highest-weight representations of the algebra $\cA$ associated to in the Hilbert space $\cH$ of the relevant CFT. & \eqref{eq:setsofreps} \\[4pt]	\hline	
		$\text{HWR}(\mathcal{A})_\text{R}$ & The set of highest-weight representations of the algebra $\cA$ in the Ramond sector & \eqref{eq:Intersection_matrix} \\[4pt]	\hline
		
		$\cH_i$ & The Hilbert space associated to the representation $i \in \HWR$ & \eqref{eq:Hspace} \\[4pt]	\hline
					
		$\text{(c,c)}_{q_L=q_R}$ & The subring of the chiral-chiral ring formed by fields with equal right- and left-moving $\mtU(1)$-charges, $q_L = q_R$. & p.~\pageref{def:cc} \\[4pt]	\hline	
		$\cR$ & The set of representations whose highest-weight states are RR~ground states related to $\text{(c,c)}_{q_L=q_R}$-states by the action of the spectral flow. & \eqref{eq:V_C_definition} \\[4pt]	\hline	
		$\mtS$ & The modular S-matrix. & \eqref{eq:Smat} \\[4pt]	\hline			
		$\cU$ & The spectral flow operator. & \eqref{eq:spec1} \\[4pt]	\hline	
		$\cC$ & The charge conjugation operator. &  \eqref{eq:cCdef} \\[4pt]	\hline		$\ish{i}$ & An Ishibashi state in the representation $i \in \HWR$.
		& \eqref{eq:defish} 	\\[4pt] \hline	  										
		$\ket{\alpha}_\cN$ & The RR~ground state corresponding to a representation $\alpha \in \cR$, normalised so that ${{\vphantom{\braket{\alpha|\beta}}}}_{\cN} \! \braket{ \alpha | \beta }_\cN = \delta_{\alpha \beta}$. & \eqref{eq:normalised_states_definition} \\[4pt] \hline  			
		$\ket{B_A}$ & The boundary state associated to the representation $A \in \text{HWR}(\mathcal{A})$. & \eqref{eq:boundarystate} \\[4pt] \hline 
		$\ket{B_A}_\cP$ & The boundary state associated to the representation $A \in \text{HWR}(\mathcal{A})_\text{R}$, projected to the space spanned by the RR~ground states $\ket{\alpha}$, $\alpha \in \cR$  & \eqref{eq:projected_boundary_state_definition} \\[4pt] \hline 
		$\cB$ & The set of representations corresponding to a basis of boundary states.  & \eqref{eq:V_Q_definition} \\[4pt] \hline 		
		$\Sigma$ & The intersection matrix $\Sigma_{AB} \defineas {\vphantom{\ket{}}}_\cP \! \bra{B_A}\ee^{\ii \pi J_0} \ket{B_B}_{\cP}$, $A,B \in \cB$ & \eqref{eq:Intersection_matrix} \\[4pt] \hline 				 		
		$\IQ(\text{S})$ & The field extension of $\IQ$ generated by the S-matrix elements $\{\text{S}_{ij} \; | \; i,j \in \text{HWR}(\mathcal{A})\}$. & \eqref{eq:Q(S)_definition} \\[4pt] \hline  
		$\IQ(\text{S}_\alpha)$ & The field extension of $\IQ$ generated by the S-matrix elements $\{\text{S}_{\alpha A} \; | \; A \in \cB \}$ for a fixed representation $\alpha \in \cR$. & \eqref{eq:subextension} \\[4pt] \hline 
		
		$(V_\IQ, V_\IC)$ & The rational Hodge structure constructed from conformal field theory data & \SS \ref{sect:Rational_Hodge_structs_from_RCFTs} \\[4pt] \hline 	
		$W^\alpha_\IQ$ & The Hodge substructure associated to a representation $\alpha \in \cR$ & \eqref{eq:W_Q_first_expression} \\[4pt] \hline 			    
	\end{tabularx}
}

\newpage
\section{A Review of Two-Dimensional Unitary Rational Conformal Field Theories}\label{sec:cft}
\vskip-10pt
In this section, we review some aspects of rational conformal field theories in order to set the stage for the analysis performed in the subsequent sections. We first introduce the notion of two-dimensional unitary rational conformal field theories and review their modular properties captured by what is known as the Verlinde formula. The Verlinde formula further implies a Galois action on the modular $S$-matrices of rational conformal field theories. Then we specialise to unitary $\mathcal{N}=(2,2)$ superconformal field theories, as this class of theories are relevant for this work. Finally, we introduce boundary states in the context of the associated boundary conformal field theories following the Cardy construction.

\subsection{Unitary rational conformal field theories} \label{sec:UniRCFT}
A two-dimensional conformal field theory consists of quantum fields $\phi_\alpha(z,\overline z), \alpha \in I$ --- labelled by an index set $I$ --- that in general depend on the left-moving holomorphic coordinate $z$ and the right-moving anti-holomorphic coordinate $\overline z$ of the (Euclidean) two-dimensional space-time. The conformal field theory is equipped with the operator product algebra by virtue of the operator product expansions among the quantum fields $\phi_\alpha(z,\overline z)$, see, e.g. the textbooks \cite{DiFrancesco:1997nk,Ginsparg:1988ui,Blumenhagen:2009zz}.

The subset $J\subset I$ of holomorphic quantum fields $\mathcal{O}_\alpha(z)$, $\alpha\in J$, forms a closed subalgebra $\mathcal{A}_\text{max}$ of the operator product algebra because the operator product of two holomorphic fields expands again into holomorphic fields only. The subalgebra $\mathcal{A}_\text{max}$ is called the chiral algebra of the conformal field theory. Analogously, there is the anti-chiral subalgebra $\overline{\mathcal{A}}_\text{max}$ of the operator product algebra, which concerns the operator product expansion of anti-holomorphic quantum fields. In this work, we concentrate on left-right-symmetric conformal field theories for which the chiral and anti-chiral algebras $\mathcal{A}_\text{max}$ and $\overline{\mathcal{A}}_\text{max}$ are isomorphic. Note that any conformal field theory contains with the identity field and the holomorphic energy momentum tensor $T(z)$ at least two holomorphic fields, which give rise to the Virasoro algebra $\operatorname{Vir}$ contained in the chiral algebra $\mathcal{A}_\text{max}$. That is to say, $\operatorname{Vir}$ is the smallest possible chiral algebra of any two-dimensional conformal field theory.

The Hilbert space $\cH$ of physical states\footnote{We will use the words ``state'' and ``vector'' often interchangeably as a state refers to a vector in a physical Hilbert space.} of a unitary conformal field theory decomposes into irreducible representations of the algebra $\mathcal{A}_\text{max} \times \overline{\mathcal{A}}_\text{max}$. A unitary conformal field theory is called rational \cite{Harvey:1987da,Moore:1988qv,Moore:1989vd}, if the states in its Hilbert space $\cH$ decompose into finite sets $\text{HWR}_\cH(\cA_\text{max})$ and $\text{HWR}_\cH(\overline{\cA}_\text{max})$ of irreducible representations with respect to the algebras $\cA_\text{max}$ and $\overline{\cA}_\text{max}$. 

More generally, we may consider a subalgebra $\cA$ of $\cA_\text{max}$, containing the Virasoro algebra $\operatorname{Vir}$, i.e.
\begin{equation} \label{eq:DefcA}
   \operatorname{Vir} \; \subseteq \; \cA \; \subseteq \; \cA_\text{max} \ ,
\end{equation}
such that the states of the rational unitary conformal field theory still assemble themselves into finitely many representations of $\HWR$ and $\HWRbar$.\footnote{Neither of the inclusions of algebras in eq.~\eqref{eq:DefcA} are required to be proper. For instance, the unitary minimal model conformal field theories possess finitely many Virasoro highest weight representations, such that $\cA = \operatorname{Vir}$ is consistent with our requirements.} In terms of such a subalgebra the Hilbert space $\cH$ of the rational unitary conformal field theory becomes
\begin{equation} \label{eq:Hspace}
  \cH \= \bigoplus_{\substack{i \in \HWR \\ \ib \in \HWRbar}} m_{i, \ib} \, \cH_i \otimes \cH_{\ib} \ ,
\end{equation}
where $m_{i,\ib}$ are non-negative integral multiplicities and $\cH_i$ (and $\cH_{\ib}$) furnish the spaces of physical states associated to the representations $i$ (and $\ib$). These spaces are constructed from highest weight representations upon acting with the negative modes of the quantum fields of the algebra $\cA$ (and the algebra $\overline{\cA}$). For ease of notation, we use $i$ (and $\ib$) both for the irreducible highest weight representation and for the associated space of states $\cH_i$ (and $\cH_{\ib}$). For the rational conformal field theories studied in this work the collection of irreducible representations $\HWR$ and $\HWRbar$ are assumed to be isomorphic with respect to the isomorphic algebras $\cA$ and~$\overline{\cA}$. 

Note that the highest weight representations $\HWR$ that appear in the construction of the Hilbert space $\cH$ of a given unitary rational conformal field theory may furnish a proper subset of the set of all highest weight representation $\HWRrep$ of the algebra $\cA$, i.e.
\begin{equation} \label{eq:setsofreps}
  \HWR \; \subset \; \HWRrep \ .
\end{equation}
The set of representations $\HWR$ depends on the detailed definition of the unitary rational conformal field theory under consideration, whereas the set of representations $\HWRrep$ is determined by the symmetry algebra $\cA$ alone.

The quantum fields associated to highest weight representations of the algebras~$\cA$ and $\overline{\cA}$ are called $\cA$-primary fields $\varphi_a(z,\overline z)$ or short primary fields, where the index $a$ ranges over the highest weight representations, which is a finite set for rational conformal field theories. The primary fields have the two-point correlation functions
\begin{equation}
  \left\langle \varphi_a(z,\overline z) \varphi_b(w,\overline w) \right\rangle \=  \frac{c_{ab}}{(z-w)^{2 h_{L,a}} (\overline z - \overline w)^{2 h_{R,a}}} \ ,
\end{equation}  
where $h_{L/R,a}$ are the left and right conformal weights of the primary field $\varphi_a(z,\overline z)$ and the symmetric constants $c_{ab}$ are only non-vanishing for $h_{L/R,a} = h_{L/R,b}$. For unitary rational conformal field theories, the two-point correlation constants $c_{ab}$ furnish a non-degenerate finite-dimensional quadratic pairing, which allows us to assign to each primary field $\varphi_a(z,\overline z)$ a dual primary field $\varphi^a(z,\overline z) = \sum_b c^{ab} \varphi_b(z,\overline z)$ in terms of the inverse matrix $c^{ab}$ of $c_{ab}$. As a result, every unitary rational conformal field theory is equipped on the set of primary fields $\varphi_a(z,\overline z)$ with an involutive duality homomorphism $\mtC$
\begin{equation} \label{eq:charge_conjugation_definition}
  \mtC: \begin{array}{l} \varphi_a(z,\overline z) \; \mapsto\; \varphi^a(z,\overline z) \ , \\[0.5ex] \varphi^a(z,\overline z) \; \mapsto \; \varphi_a(z,\overline z) \ , \end{array}  \qquad \mtC^2 \= \mtI \ .
\end{equation}  
This duality involution $\mtC$ acting on primary fields is known as the charge conjugation operator, as it reverses the charges of the primary fields with respect to any spin-1 currents in the conformal field theory. Note that the unique identity primary field $\varphi_0 \defineas \mtI$ of the unitary conformal field theory --- labelled by the representation index $0$ --- is self-dual with respect to the charge conjugation operators $\mtC$, that is $\mtC(\varphi_0) = \varphi_0$.

\subsection{Modularity of unitary rational conformal field theories}
As a two-dimensional conformal field theory ought to be well-defined on any (Euclidean) two-dimensional space-time geometry, it must be consistent in particular on any two-dimensional torus~$T^2$ with a choice of complex structure realizing local holomorphic and anti-holomorphic coordinates $z$ of the quantum fields $\phi_\alpha(z,\overline z)$. For a given algebra $\cA$ with finite set $\HWRrep$ of irreducible highest-weight representation, let us assume that there is a rational conformal field theory with $\HWR = \HWRrep$. Then $\HWRrep$ furnishes a finite-dimensional unitary representation $\rho: \operatorname{SL}(2,\IZ) \to \operatorname{U}( \dimHWR )$, with $\dimHWR \defineas \left| \HWRrep \right|$. In particular, the unique irreducible representation associated to the identity primary $\varphi_0$ is labelled by the representation index $0$.

The modular group $\operatorname{SL}(2,\IZ)$ is generated by the $2\times 2$-matrices
\begin{equation}
  S \= \begin{pmatrix} 0 & -1 \\ 1 & 0 \end{pmatrix} \ , \qquad
  T \= \begin{pmatrix} 1 & 1 \\ 0 & 1 \end{pmatrix} \ ,
\end{equation}
which obey the relations $S^4 = \mtI_{2{\times}2}$ and $(ST)^3 = S^2$. The unitary representation $\rho$ is specified by the images $\rho(S) \asdefine \mtS$ and $\rho(T) \asdefine \mtT$ in $\operatorname{U}(\dimHWR )$, which are referred to as the modular $T$- and $S$-matrices of the rational conformal field theory. Using the formulation of ref.~\cite{Schellekens:1990xy}, the $S$-matrices in the image of the representation $\rho$ is symmetric and squares to the charge conjugation operator $\mtC$,\footnote{The action of the square $S^2 = -\mtI_{2\times 2}$ of the modular $2\times 2$ matrix $S$ on the one-cycles of the torus $T^2$ reverses the orientation of all homology one-cycles. As a consequence the unitary transformation $\mtS^2 = \rho(-\mtI_{2\times 2})$ multiplies all spin-1 currents with $(-1)$, which implies the relation to the charge conjugation operator $\mtS^2 = \mtC$.} i.e.
\begin{equation} \label{eq:Smat}
   \mtS^T \= \mtS \ , \quad \mtS^2 \= \mtC \ ,
\end{equation}
which also implies that $\mtS^* \mtS = \mtI_{\dimHWR \times \dimHWR}$ and $\mtS = \mtC \mtS^*$.  From the latter condition we readily infer with the self-duality of the identity representation $0$ that the entries $\mtS_{0i}$ of the symmetric matrix $\mtS$ for any representation $i \in \HWRrep$ are real and non-vanishing. For later convenience, we define the charge conjugation operator
\begin{equation} \label{eq:cCdef}
  \cC: \HWRrep \; \to \; \HWRrep~, \qquad  i \mapsto \cC(i) \ ,
\end{equation}  
which is induced from the unitary charge conjugation operator $\mtC$.

The existence of the representation $\rho$ of the modular group $\operatorname{SL}(2,\IZ)$ of the conformal field theory means that, on the level of the characters $\chi_i(\tau)$ of the representations $i \in \HWRrep$, the modular $T$- and $S$-transformations of the characters are
\begin{equation} \label{eq:CharTrans}
   \chi_i\left(\tau+1\right) \= \sum_{j \in \HWRrep} \mtT_{ij}\, \chi_j(\tau) \ , \qquad
    \chi_i\left(-\frac{1}{\tau}\right) \= \sum_{j \in \HWRrep} \mtS_{ij} \, \chi_j(\tau) \ .
\end{equation}
Here $\mtT_{ij}$ and $\mtS_{ij}$ are the entries of the finite-dimensional unitary modular matrices $\mtT$ and $\mtS$, and $\tau$ is the complex structure parameter in the upper half plane.\footnote{For extended chiral algebras $\cA$, the characters $\chi_i$ typically depend on further parameters. For simplicity and ease of notation, we do not spell out these dependences here.}

The operator-state-correspondence assigns to states in the representations $i \in \HWRrep$ quantum fields, which, in virtue of the operator product expansion, give a ring structure to the set of representations $\HWRrep$ known as the fusion ring \cite{Belavin:1984vu}. For rational conformal field theories --- i.e. for a finite set $\HWRrep$ of representations $i$ --- the fusion product is conveniently expressed in terms of the fusion ring structure constants $N^i_{jk}$, namely
\begin{equation} \label{eq:fusion}
  j \times k \= \sum_{i \in \HWRrep} N_{jk}^i \ i \ ,
\end{equation}  
with the integral structure constants $N_{jk}^i \in \IZ$. The fusion product now states that the operator product expansion of the quantum fields $\phi_j$ and $\phi_k$ associated to the representations $j$ and $k$ expands into a linear combination of quantum fields of the representations $i\in \HWRrep$ with multiplicities~$N_{jk}^i$.\footnote{The fusion product is independent of the chosen quantum fields assigned to the representations $i$ and $j$. Note further that the fusion product originates from the interaction of quantum fields and is thus a property of the quantum field theory. It should not be confused with the tensor product of highest weight representations. For a mathematical exposition to fusion rings, see for instance refs.~\cite{Segal:2002ei,MR1029428,MR1186962}.}

The modular consistency of a conformal field theory imposes strong constraints on the fusion ring. For rational conformal field theories these modular consistency conditions give rise to a remarkable and far-reaching connection between the structure constants $N^i_{jk}$ of the fusion ring and the unitary modular $S$-transformation matrix $\mtS$, as introduced by Verlinde in the seminal work \cite{Verlinde:1988sn}. This relationship is expressed in the Verlinde formula that reads \cite{Dijkgraaf:1988tf,Moore:1988uz}
\begin{equation} \label{eq:Verlinde_formula}
	N_{ak}^i \= \sum_{\ell \in \HWRrep} \mtS_{k\ell} \ \frac{\mtS_{a\ell}}{\mtS_{0\ell}} \ \mtS^*_{\ell i} \ ,
\end{equation}
Upon defining the matrices $[\mtN_a]^i_j  \defineas N_{aj}^i$ and using the symmetry and the unitarity $\mtS^\dagger = \mtS^* = \mtS^{-1}$ of the modular $S$-matrix $\mtS$, the Verlinde formula is conveniently expressed in the matrix form
\begin{equation} \label{eq:Verlinde_formula_matrix}
	\mtN_a \= \mtS \, \Lambda_a \, \mtS^{-1} \quad \text{with} \quad 
	\Lambda_a \; \defineas \;
	\diag\left(\lambda_{a,0}, \lambda_{a,i_1}, \ldots, \lambda_{a,i_{\dimHWR-1}} \right) \ , \quad
	\lambda_{a,i} \; \defineas \; \frac{\mtS_{a,i}}{\mtS_{0,i}} \ ,
\end{equation}
where $0,i_1,\ldots,i_{\dimHWR-1}$ label the $\dimHWR$ representations of $\HWRrep$. 

A representation $\cJ \in \HWRrep$ is called a simple current if the fusion product of $\cJ$ with any other representation $i \in\HWRrep$ yields a single representation $\cJ(i) \in \HWRrep$, i.e. $\cJ \times i \asdefine \cJ(i)$ \cite{Schellekens:1990xy}.\footnote{The discussion on simple currents in this subsection applies equally well for the subset of representations~$\HWR$.} It follows that the set of all simple currents form an Abelian group with respect to the fusion product $\times$, where the neutral element is the identity representation $0 \in \HWRrep$ associated to the identity primary field $\varphi_0$. Furthermore, the monodromy charge of a simple current $\cJ$ is defined as the map
\begin{equation} \label{eq:monodromy_charge_definition}
	\Phi_{\cJ}:\ \HWRrep \; \to \; \IR/\IZ~,~~i \; \mapsto \; [ h_\cJ + h_i - h_{\cJ(i)} ] \ ,
\end{equation}
where $h_\cJ$, $h_i$ and $h_{\cJ(i)}$ are the conformal weights of the representations $\cJ$, $i$, and $\cJ(i)$, respectively. Note that the monodromy charge $\Phi_{\cJ}$ of a simple current $\cJ$ of order $N$, i.e. $\cJ^N = \mtI$, applied to any representation $i \in \HWRrep$ takes values in $\IQ/\IZ$, namely\footnote{For rational conformal field theories, any simple current $\cJ$ is of finite order, due to the finiteness of the set $\HWRrep$.}
\begin{equation}
	\Phi_{\cJ}(i) \; \in \; \IQ/\IZ \qquad \text{with} \qquad  N  \Phi_{\cJ}(i) \= [0] \in \IQ/\IZ \ .
\end{equation}  
The transformation behaviour of the modular $S$-matrix with respect to the action of a simple current $\cJ$ is given by \cite{Fuchs:1996dd,Brunner:2000nk}
\begin{equation} \label{eq:S-matrix_transformation_simple_current}
	\mtS_{i\cJ(j)} \= \me^{2\pi \ii \Phi_J(i)} \mtS_{ij} \qquad \text{for any} \qquad i,j\in \HWRrep \ .
\end{equation}
\subsection{The Galois group of a rational conformal field theory} \label{sect:RCFTs_and_Galois_Groups}
Following the arguments of refs.~\cite{DeBoer:1990em,Coste:1993af}, the Verlinde formula~\eqref{eq:Verlinde_formula_matrix} implies that we can associate a Galois group to any rational conformal field theory. It is this Galois group that plays a central role in the forth-coming considerations of this work.

The characteristic polynomials of the $\dimHWR \times \dimHWR$ fusion matrices $\mtN_a$ define a set of $\dimHWR$ degree-$\dimHWR$ polynomials $P_a \in \IZ[X] \subset \IQ[X]$ with rational (in fact integral) coefficients as
\begin{align} \label{eq:N_matrix_characteristic_polynomial}
	P_a(X) \; \defineas \; \det(\mtN_a - X \, \mtI_{\dimHWR \times \dimHWR} ) \ .	
\end{align}
As the Verlinde formula \eqref{eq:Verlinde_formula_matrix} shows that the matrices $\mtN_a$ are similar to the diagonal matrices $\Lambda_a$, the roots of the characteristic polynomials $P_a(X)$ --- that is to say the eigenvalues of $\mtN_a$ --- are the algebraic numbers $\lambda_{a,i}$, which give rise to the finite field extension
\begin{equation}
	\IQ(\lambda) \; \defineas \; \IQ\left(\{\lambda_{a,i} \}_{a,i\in{\HWRrep}} \right) \ .
\end{equation}
By construction, $\IQ(\lambda)$ is the splitting field of the product polynomial $\prod_{a\in \HWRrep} P_a(X) \in \IQ[X]$, which is hence a normal extension of $\IQ$. As any finite field extension of $\IQ$ is separable, $\IQ(\lambda)$ is a Galois extension of $\IQ$. 

Let $\sigma \in \operatorname{Gal}(\IQ(\lambda)/\IQ)$ be an element of the Galois group of the field extension $\IQ(\lambda)$. Due to the relation $P_a( \sigma(\lambda_{a,i})) = \sigma( P_a(\lambda_{a,i}) ) = 0$, the root $\lambda_{a,i}$ is mapped to another root $\sigma(\lambda_{a,i})$ of the polynomial~$P_a$, we get
\begin{equation} \label{eq:Gal_first}
	\sigma(\lambda_{a,i}) \= \lambda_{a,\varsigma_a(i)} \ ,
\end{equation}
where for a given $a\in\HWRrep$ the permutation $\varsigma_a$ of the symmetric group $\operatorname{Sym}(\HWRrep)$ permutes the representations $\HWRrep$ that label the highest weight representations of the chiral algebra $\cA$. The non-trivial claim is now that the permutation $\varsigma_a \in \operatorname{Sym}(\HWRrep)$ is in fact independent of the representation $a \in \HWRrep$ \cite{DeBoer:1990em},\footnote{The independence of $a$ can be argued for by observing that the Verlinde formula~\eqref{eq:Verlinde_formula} implies for the roots $\lambda_{a,i}$ the relation $\lambda_{a,i}\lambda_{b,i} = \sum_{c\in\HWRrep} N_{ab}^c \lambda_{c,i}$. The action of the Galois group element~$\sigma$ yields the relation $\lambda_{a,\varsigma_a(i)}\lambda_{b,\varsigma_b(i)} = \sum_c N_{ab}^c \lambda_{c,\varsigma_c(i)}$, which is only consistent if the permutation $\varsigma_a$ is independent of the index $a$ \cite{DeBoer:1990em}.}
such that the elements $\sigma$ of the Galois group~$\operatorname{Gal}(\IQ(\lambda)/\IQ)$ act on the roots as
\begin{equation} \label{eq:Galois_action_on_HWR}
	\sigma(\lambda_{a,i}) \= \lambda_{a,\varsigma(i)} \qquad \text{for all} \qquad a \in \HWRrep  \ ,
\end{equation}
in terms of a permutation $\varsigma$ of the symmetric group $\operatorname{Sym}(\HWRrep)$ associated to the Galois group element $\sigma$. 

The complex conjugate roots $\lambda_{a,i}^*$ are in the normal field extension~$\IQ(\lambda)$ because $P_a(\lambda_{a,i}) = 0$. Using $\mtS \mtS^* =\mtI_{\dimHWR \times \dimHWR}$, $\mtS^T =\mtS$ and $\mtS_{0,i} \in \IR\setminus\{0\}$, we observe
\begin{equation} \label{eq:Ssquare_inv}
	\frac{1}{\mtS_{0,i}^2} \= \sum_{a\in\HWRrep} \frac{\mtS_{a,i}}{\mtS_{0,i}} \left(\frac{\mtS_{a,i}}{\mtS_{0,i}}\right)^*  
	\= \sum_{a\in\HWRrep} \lambda_{a,i} \lambda_{a,i}^* \ .
\end{equation}
That is to say, the squares $\mtS_{0,i}^2$ are in the field extension $\IQ(\lambda)$ as well. The definition~\eqref{eq:Verlinde_formula_matrix} of $\lambda_{a,i}$ further implies for all squares of the entries of the matrix $\mtS$ that
\begin{equation}
   \mtS_{a,i}^2 \in \IQ(\lambda) \ , \qquad \text{for all} \qquad a,i \in\HWRrep  \ .
 \end{equation}
In order to determine the Galois action on the squares $\mtS_{a,i}^2$, we first note that complex conjugation $ \cdot^*: \IQ(\lambda) \to \IQ(\lambda)$ is a $\IQ$-automorphism that realises a group element $\iota$ of the Galois group $\operatorname{Gal}(\IQ(\lambda)/\IQ)$,\footnote{If the field $\IQ(\lambda)$ is totally real, then $\iota$ is the identity element of $\operatorname{Gal}(\IQ(\lambda)/\IQ)$. Otherwise the involution $\iota$ generates a subgroup $\IZ/2\IZ \subset \operatorname{Gal}(\IQ(\lambda)/\IQ)$.} which according to eq.~\eqref{eq:Galois_action_on_HWR} is given by $\lambda_{a,i}^* = \iota( \lambda_{a,i} )= \lambda_{a,\iota(i)}$. From eq.~\eqref{eq:Ssquare_inv} we therefore deduce for any $\sigma \in \operatorname{Gal}(\IQ(\lambda)/\IQ)$ the action on $\mtS_{0,i}^{-2}$ as
\begin{equation}
    \sigma\left(\frac1{\mtS_{0,i}^2}\right)
    \= \sum_{a\in\HWRrep} \lambda_{a,\varsigma(i)} \lambda_{a,\iota \circ \varsigma \circ \iota(i)}^* 
    \= \sum_{a\in \HWRrep} \frac{ \mtS_{a,\varsigma(i)} \mtS_{a,\iota \circ \varsigma \circ \iota(i)}^*}{\mtS_{0,\varsigma(i)} \mtS_{0,\iota \circ \varsigma \circ \iota(i)}^*} 
    \= \frac{\delta_{\varsigma(i),\iota \circ \varsigma \circ \iota(i)}}{\mtS_{0,\varsigma(i)} \mtS_{0,\iota \circ \varsigma \circ \iota(i)}} \ ,
\end{equation}
where we use $\mtS_{0,i} \in \IR \setminus \{0\}$. As the inverse squares $\mtS_{0,i}^{-2}$ are non-vanishing, their images with respect to the $\IQ$-automorphism~$\sigma$ are non-zero for any index~$i$ as well, that is, $\delta_{\varsigma(i),\iota \circ \varsigma \circ \iota(i)}=1$ for all~$i\in\HWRrep$. This means that the elements $\varsigma$ and $\iota \circ \varsigma \circ \iota$ of the symmetric group $\operatorname{Sym}(\HWRrep)$ are identical so that $\sigma(\mtS_{0,i}^2) = \mtS_{0,\varsigma(i)}^2$, which together with the definition of $\lambda_{a,i}$ and the equation~\eqref{eq:Galois_action_on_HWR} yields for any $\sigma \in \operatorname{Gal}(\IQ(\lambda)/\IQ)$ the Galois action \cite{DeBoer:1990em}
\begin{equation} \label{eq:GalonS}
    \sigma( \mtS_{a,i}^2 ) \= \mtS_{a,\varsigma(i)}^{2} \qquad \text{for all} \qquad a\in\HWRrep  \ .
\end{equation}
The presented arguments show that complex conjugation via the Galois group element $\iota$ generates a normal subgroup (which is trivial if the field $\IQ(\lambda)$ is totally real) of the Galois group $\operatorname{Gal}(\IQ(\lambda)/\IQ)$. Such fields are called Kroneckerian. This means that the field $\IQ(\lambda)$ is either totally real or that there exists an intermediate totally real and normal field extension $\IL/\IQ$ such that $\IQ(\lambda)/\IL$ is a totally imaginary quadratic field extension --- in other words, $\IQ(\lambda)$ is a CM-field over $\IL$.\footnote{In latter case, by the fundamental theorem of Galois theory, $\IL$ arises as the invariant field with respect to the normal subgroup $\IZ/2\IZ$ of the Galois group $\operatorname{Gal}(\IQ(\lambda)/\IQ)$.}

The Galois group $\operatorname{Gal}(\IQ(\lambda)/\IQ)$ is actually Abelian, which can be argued for as follows \cite{DeBoer:1990em}. Firstly, as the matrix $\mtS$ is symmetric the definition~\eqref{eq:Verlinde_formula_matrix} of $\lambda_{a,i}$ implies
\begin{equation} \label{eq:LambdaSym}
    \lambda_{a,i} \= \lambda_{i,a} \frac{\lambda_{a,0}}{\lambda_{i,0}} \qquad \text{for all} \qquad a,i \in \HWRrep  \ .
\end{equation}
Secondly, taking the square root the action~\eqref{eq:GalonS} of the Galois group $\operatorname{Gal}(\IQ(\lambda)/\IQ)$ yields 
\begin{equation} \label{eq:LambdaRoot}
   \lambda_{\varsigma(a),i} \lambda_{i,0} \= \pm \lambda_{\varsigma(i),a} \lambda_{a,0} \qquad \text{for all} \qquad 
   \sigma \in \operatorname{Gal}(\IQ(\lambda)/\IQ) \ .
\end{equation}
Then, for any two elements $\sigma_1,\sigma_2\in \operatorname{Gal}(\IQ(\lambda)/\IQ)$ one finds \cite{DeBoer:1990em}
\begin{equation}
\begin{aligned}
  \sigma_1 \sigma_2(\lambda_{a,i}) &\=  \sigma_1\sigma_2\left( \lambda_{i,a} \frac{\lambda_{a,0}}{\lambda_{i,0}}\right)
  \= \sigma_1\left( \lambda_{i,\varsigma_2(a)} \frac{\lambda_{a,\varsigma_2(0)}}{\lambda_{i,\varsigma_2(0)}}\right)
  \= \sigma_1\left(\frac{\lambda_{\varsigma_2(a),i}}{\lambda_{\varsigma_2(0),i}} \frac{\lambda_{a,0}}{\lambda_{\varsigma_2(a),0}}\lambda_{\varsigma_2(0),a} \right) \\
 & \= \frac{\lambda_{\varsigma_2(a),\varsigma_1(i)}}{\lambda_{\varsigma_2(0),\varsigma_1(i)}}
     \cdot \sigma_1\left( \frac{\lambda_{\varsigma_2(0),a}}{\lambda_{\varsigma_2(a),0}}\lambda_{a,0}\right)
     \= \frac{\lambda_{\varsigma_2(a),\varsigma_1(i)}}{\lambda_{\varsigma_2(0),\varsigma_1(i)}}
       \frac{\lambda_{\varsigma_2(0),a}}{\lambda_{\varsigma_2(a),0}}\lambda_{a,0} \\
  &\= \sigma_2 \left( \frac{\lambda_{\varsigma_1(i),a}}{\lambda_{\varsigma_1(i),0}} \lambda_{a,0}\right)
  \= \sigma_2(\lambda_{a,\varsigma_1(i)}) \= \sigma_2\sigma_1(\lambda_{a,i}) \ ,
\end{aligned}
\end{equation}  
which proves that the Galois group $\operatorname{Gal}(\IQ(\lambda)/\IQ)$ is indeed Abelian. This derivation uses extensively the identity~\eqref{eq:LambdaSym}, and in the second line the identity~\eqref{eq:LambdaRoot} is used for the representation $i=0$ together with the property $\sigma(\pm 1) = \pm 1$ for any $\mathbb{Q}$-automorphism $\sigma \in \operatorname{Gal}(\IQ(\lambda)/\IQ)$.

Let us finally introduce the field extension $\IQ(\mtS)$ generated by the entries of the modular S-matrix elements $\mtS_{a,i}$ given by \cite{Coste:1993af}
\begin{align} \label{eq:Q(S)_definition}
	\IQ(\mtS) \; \defineas \; \IQ\left(\{\mtS_{a,i}\}_{a,i \in\HWRrep}\right) \; \supset \; \IQ(\lambda) \; \supset \; \IQ~.
\end{align}
This is a Galois extension of the field $\IQ(\lambda)$, since $\IQ(\mtS)$ is the splitting field of the polynomials
\begin{equation}
  P_{a,i} \= x^2 - (\mtS_{a,i})^2 \ \in \ \IQ(\lambda)[x] \ ,
\end{equation}  
which are separable for non-vanishing entries $\mtS_{a,i}$. Furthermore, the field extension $\IQ(\mtS)$ is Galois over $\IQ$ as well, because any $\IQ$-automorphism $\sigma$ of any field containing $\IQ(\mtS)$ is closed on the subfield $\IQ(\lambda)$,\footnote{This follows from the fact that the roots of $\lambda_{a,i}$ arise from polynomials over $\IQ$.} i.e., $\sigma(\IQ(\lambda)) \subset \IQ(\lambda)$, which implies for $\mtS_{0,i}^2 \in \IQ(\lambda)$, for any $i \in \HWRrep$, that $\sigma(\mtS_{0,i}) = \epsilon_\sigma(i) \mtS_{0,\varsigma(i)}$, where $\epsilon_\sigma(i) \in \{ -1,+1\}$. Due to the relation $\mtS_{a,i} = \mtS_{0,i} \lambda_{a,i}$, we further find
\begin{equation} \label{eq:GaloisOnS}
  \sigma(\mtS_{a,i}) \= \epsilon_\sigma(i) \, \mtS_{a,\varsigma(i)} \qquad \text{with} \qquad  \epsilon_\sigma(i) \in \{ -1,+1\}
  \qquad \text{for all} \qquad i\in \HWRrep  \ .
\end{equation}
Here, the permutation $\varsigma \in \operatorname{Sym}(\HWRrep)$ of eq.~\eqref{eq:Galois_action_on_HWR} is obtained from the restriction of the $\IQ$-automorphism $\sigma$ to the subfield $\IQ(\lambda)$. Note that in this equation the sign $\epsilon_\sigma(i)$ is independent of the representation~$a\in\HWRrep$. 

Finally, we conclude that $\sigma(\IQ(\mtS)) \subset \IQ(\mtS)$ for any $\IQ$-automorphism $\sigma$ as defined above, and therefore the extension $\IQ(\mtS)$ of $\IQ$ is normal and separable and hence Galois \cite{Coste:1993af}. Furthermore, due to the symmetry $\mtS_{a,i} = \mtS_{i,a}$ it is straightforward to verify with eq.~\eqref{eq:GaloisOnS} that the Galois group $\operatorname{Gal}(\IQ(\mtS)/\IQ)$ is also Abelian \cite{Coste:1993af}.

\subsection{\texorpdfstring{$\mathcal{N}=(2,2)$}{N=(2,2)} supersymmetric rational conformal field theories}  \label{sect:N=(2,2)_RSCFT}
In this section, we briefly review some salient points of $\cN=(2,2)$ rational superconformal field theories, mostly to set the notation for the discussion in later sections. This standard material is covered in many review articles and textbooks on two-dimensional superconformal field theories, see, for instance, refs.~\cite{Gepner:1987qi,Warner:1989dj,Gepner:1989gr,Greene:1996cy,Blumenhagen:2009zz}.

In addition to the Virasoro generators $L_m$, $m\in \mathbb{Z}$, of the Virasoro algebra $\operatorname{Vir}$, which are the Laurent modes of the spin-two energy momentum tensor $T(z)=\sum_m L_m z^{-m-2}$, the $\cN=2$ superconformal algebra extends the Virasoro algebra $\operatorname{Vir}$ by the modes $J_m$, $m\in \mathbb{Z}$, which form the Laurent modes of a spin-one $\mtU(1)$ current $J(z)=\sum_m J_m z^{-m-1}$, and by the Grassmannian --- that is to say fermionic --- modes $G^\pm_r$, which form the Laurent modes of a pair of spin-$\frac32$ currents $G^\pm(z)=\sum_r G_r^\pm \,z^{-r-\frac32}$. The spin-$\frac32$ currents $G^\pm(z)$ have charge $\pm 1$ with respect to the $\mtU(1)$~current $J(z)$. Note that the energy momentum tensor current $T(z)$, the spin-$\frac32$ current $G^\pm(z)$ and the $R$-symmetry current $J(z)$ combine into a two-dimensional $\mathcal{N}=2$ super energy momentum tensor. 

Depending on a choice of spin structure on the punctured plane $\IC^*$, the currents $G^\pm(z)$ have anti-periodic boundary conditions --- called the Ramond (R) sector --- or periodic boundary conditions --- called the Neveu--Schwarz (NS) sector, which respectively yields integral modes $G_r^\pm, r\in \IZ$ or half-integral modes $G_r^\pm, r \in \IZ +\frac12$, i.e.,
\begin{equation}
\begin{aligned}
  &\text{Ramond (R) sector:} &&G^\pm(\ee^{2\pi \ii } z) = - G^\pm(z) &&\Rightarrow &&\left\{G^\pm_r \right\}_{r \in \IZ} \ ,\\
  &\text{Neveu-Schwarz (NS) sector:} &&G^\pm(\ee^{2\pi \ii } z) =  + G^\pm(z)  &&\Rightarrow &&\left\{G^\pm_r \right\}_{r \in \IZ+\frac12} \ .
\end{aligned}
\end{equation}  
All these modes generate the $\cN=2$ superconformal algebra --- also known as the $\cN=2$ super-Virasoro algebra $\operatorname{SVir}$ --- which is given in terms of the commutation and anti-commutation relations
\begin{equation} \label{eq:SCA}
	\begin{aligned}
		&\left[L_m,L_n\right] = (m-n)L_{m+n} + \frac{c}{12}m(m^2-1)\delta_{m+n,0} \ , 
		&&\left[J_m,J_n\right] = \frac{c}{3}m \delta_{m+n,0} \ , \\
		&\left\{G_r^+,G_s^-\right\} = 2 L_{r+s} + (r-s)J_{r+s} + \frac{c}{3}\left(r^2-\frac{1}{4} \right) \delta_{r+s,0} \ , 
		&&\left\{G_r^\pm,G_s^\pm\right\} = 0 \ ,\\
		&\left[L_m,J_n\right] = -n J_{m+n} \ , \qquad \left[L_m,G_r^{\pm}\right] = \left(\frac{m}{2}-r\right) G^\pm_{m+r} \ ,
		&&\left[J_m,G^\pm_{r}\right] = \pm G^\pm_{m+r} \ .
	\end{aligned}
\end{equation}

An $\mathcal{N}=(2,2)$ supersymmetric rational conformal field theory possesses a left- and right-moving $\cN=2$~super-Virasoro algebra $\operatorname{SVir} \times \overline{\operatorname{SVir}}$ as its symmetry algebra. Note, however, that the entire super-Virasoro algebra $\operatorname{SVir}$ is not part of the chiral algebra $\cA$, as the supersymmetry currents $G^\pm(z)$ give rise to half-integral weight operators, which make $\operatorname{SVir}$ into a super-Lie algebra as opposed to an ordinary Lie algebra. Thus, the $\cN=2$~super-Virasoro algebra $\operatorname{SVir}$ is not a chiral algebra \cite{Moore:1988qv}, but instead the chiral subalgebra $\cA$ of an $\mathcal{N}=(2,2)$ supersymmetric rational conformal field theory is generated from Grassmann even operators.

The respective $\cN=2$ superconformal algebras of the Ramond sector and the Neveu--Schwarz sector $\operatorname{SVir}_\text{R}$ and $\operatorname{SVir}_\text{NS}$ are actually isomorphic \cite{Schwimmer:1986mf}. Let $\chi$ and $\chi'$ be elements of these algebras, then the isomorphism is explicitly realised as \cite{Schwimmer:1986mf,Warner:1989dj,Gepner:1989gr,Greene:1996cy}
\begin{equation} \label{eq:spec1}
  \operatorname{SVir}_R {\stackrel\simeq\longrightarrow} \operatorname{SVir}_\text{NS} \ , \
  \chi \mapsto \chi'=\cU^{-1} \, \chi \, \cU  \quad \text{with} \quad
  \left\{ 
  \begin{aligned}
    \ L'_m &= L_m + \frac12 J_m + \frac{c}{24} \delta_{m,0} \\
    \ J'_m &= J_m + \frac{c}6 \delta_{m,0}\\
    \ {G'}^\pm_r &= G^\pm_{r \pm \frac12}
  \end{aligned}
  \right. \ .
\end{equation}  
In particular, this algebra isomorphism induces an isomorphism $\cU$ on the level of representations, which is called the spectral flow operator $\cU$. In the above definition it acts by conjugation on the elements of the super-Virasoro algebra by conjugations, while it maps states from representations of the Ramond sector to states of the corresponding representation in the Neveu--Schwarz sector,~i.e.,
\begin{equation} \label{eq:spectral_flow}
  \cU: \HWR_{\text{NS}} \; \longrightarrow \; \HWR_\text{R} \ .
\end{equation}

Let us also recall the definition of the chiral ring of a $\cN=2$ super conformal algebra \cite{Lerche:1989uy}. A super-conformal primary field $\phi(z,\overline z)$ is called a chiral primary field or an anti-chiral primary field, if the operator product expansion with the spin-$\frac32$ current $G^+(z)$ or the spin-$\frac32$ current $G^-(z)$ is regular, respectively. Equivalently, due to the operator-state-correspondence between the field $\phi(z,\overline z)$ and the state $\ket{\phi}$, a super-conformal primary state $\ket{\phi}$ is called a chiral primary state or an anti-chiral primary state, if $G^+_{-1/2} \ket{\phi} = 0$ or $G^-_{-1/2} \ket{\phi} = 0$, respectively. Analogous characterizations apply to the anti-holomorphic sector with respect to the anti-holomorphic currents $\overline G^\pm(\overline z)$, such that super-conformal primary fields with regular operator product expansions with respect to both the holomorphic and the anti-holomorphic sectors are characterised as
\begin{equation} \label{eq:chiralcond}
\begin{aligned}
   \text{(c,c) state: } &G^+_{-1/2} \ket{\phi} =  \overline G^+_{-1/2} \ket{\phi}=0 \ ,
   \quad&  \text{(a,a) state: }&G^-_{-1/2} \ket{\phi} =  \overline G^-_{-1/2} \ket{\phi}=0\ , \\
    \text{(c,a) state: } &G^+_{-1/2} \ket{\phi} =  \overline G^-_{-1/2} \ket{\phi}=0 \ , 
    \quad&  \text{(a,c) state: } &G^-_{-1/2} \ket{\phi} =  \overline G^+_{-1/2} \ket{\phi}=0 \ .
\end{aligned}
\end{equation}
Here the entries of the labels $(\text{c/a}, \text{c/a})$ refer to chiral (c) and anti-chiral (a) constraints in the holomorphic and anti-holomorphic sector. A far-reaching implication now is that the operator product expansion among $(\text{c/a},\text{c/a})$ primary fields  are regular and preserve the chirality property. As result the limit of two such primary fields approaching each other is well-defined and defines a ring structure among these primary fields \cite{Lerche:1989uy}. Therefore, the $\cN=(2,2)$ superconformal field theories are equipped with four rings, namely the (c,c)-ring, the (a,a)-ring, the (c,a)-ring, and the (a,c)-ring of primary fields.

Let us also record that for a chiral (anti-chiral) primary state $\ket{\phi}$ the left- and right-moving conformal weights and $\operatorname{U}(1)$ charges, which are given by
\begin{equation}
  L_0 \ket{\phi} = h_L(\ket{\phi}) \ket{\phi} \ , \
  \overline L_0 \ket{\phi} = h_R(\ket{\phi})\ket{\phi} \ , \quad
 J_0 \ket{\phi} = q_L(\ket{\phi}) \ket{\phi} \ , \
  \overline J_0 \ket{\phi} = q_R(\ket{\phi}) \ket{\phi} \ , 
\end{equation}
obey --- as a consequence of the anti-commutators $\{ G^\pm_{-1/2}, G^\mp_{+1/2} \} = 2 L_0 \mp J_0$ ---
\begin{equation}
   h_{L/R}(\ket{\phi}) = \pm \tfrac{1}{2} \, q_{L/R}(\ket{\phi}) \ ,
\end{equation}  
where the positive (negative) sign applies for chiral (anti-chiral) states, respectively. Furthermore, for unitary $\mathcal{N}=(2,2)$ supersymmetric conformal field theories the anti-commutators $\{ G^\pm_{-3/2}, G^\mp_{3/2} \} = 2 L_0 \pm J_0 + \frac23 c$ together with the positive definite inner product on the space of physical states implies 
\begin{align} \label{eq:q_bound}
	0 \leq h_{L/R}(\ket{\phi}) \leq \tfrac{c}{6} \qquad \Longleftrightarrow \qquad 0 \le | q_{L/R}(\ket{\phi}) | \leq \tfrac{c}{3} \ ,
\end{align}
where the charges $q_{L/R}(\ket{\phi})$ are positive (negative) in the chiral (anti-chiral) states.

Finally, acting with the spectral flow operator $\cU$ on a chiral primary $\ket{\phi}$, we observe that the chirality condition implies that the corresponding state $\cU \ket{\phi}$ in the Ramond sector is annihilated by $G^+_0$ according to eq.~\eqref{eq:spec1}. Hence a chiral primary state is mapped to a Ramond ground state via the spectral flow operator. Note also that a generic state $\ket{\alpha}$ in the Neveu--Schwarz sector with $\mtU(1)$ charge $q_{L/R}(\ket{\alpha})$ flows to a state $\cU \ket{\alpha}$ in the Ramond sector with charge
\begin{equation} \label{eq:spectral_flow_charge_transformation}
	q_{L/R}(\cU \ket{\alpha}) = q_{L/R}(\ket{\alpha}) - \tfrac{c}{6} \ ,
\end{equation}
where the subscript is chosen according to whether the spectral flow is taken in the left- or right-moving algebra. For later reference, the charge conjugation operator $\cC$, which is induced by the charge conjugation operator \eqref{eq:charge_conjugation_definition} of the fields, acts on the $J_{0},\overline J_0$ eigenstates as
\begin{equation} \label{eq:charge_conjugation_charge_transformation}
	q_{L/R}(\cC \ket{\alpha}) \= -q_{L/R}(\ket{\alpha})~.
\end{equation}

\subsection{Boundary conformal field theory} \label{sect:boundary_conformal_field_theory}
Consider on the unit conformal disk of a left-right symmetric unitary rational conformal field theory with symmetry algebra $\cA \times \overline{\cA}$ as introduced in section~\ref{sec:UniRCFT}, together with an algebra automorphism $\Omega: \cA \to \cA$ of the algebra $\cA$ that restricts to the identity on the Virasoro subalgebra of $\cA$. In order to preserve the subalgebra $\cA \hookrightarrow \cA \times \overline{\cA}, \, W \mapsto (W,\Omega(\overline{W}))$ at the boundary of the conformal unit disk, the boundary state $\ket{B}$ --- which in radial quantization is a closed string state at the unit circle --- must obey the conditions \cite{Ishibashi:1988kg,Ishibashi:1988tf,Cardy:1989ir} 
\begin{equation} \label{eq:Wconstraints}
  \left( W_m - (-1)^{h(W)} \, \Omega(\overline W_{-m})  \right) \ket{B} \= 0 \qquad \text{for all} \qquad m \in \IZ \ .
\end{equation}  
Here $W_m$ are the Laurent modes of the symmetry generator $W(z) = \sum_{m} W_m z^{-m-h(W)}$ of conformal weight $h(W)$ of the symmetry algebra $\cA$. In particular, for the Virasoro generators $L_m$ there is the constraint $(L_m - \overline{L}_{-m})\ket{B} = 0$ for all integers $m$. 

Given an irreducible representation $i \in \HWR$ of the algebra $\cA$ with Hilbert space $\cH_i$, a solution to the constraints~\eqref{eq:Wconstraints} is realised by the Ishibashi boundary state \cite{Ishibashi:1988kg}
\begin{equation} \label{eq:defish}
   \ish{i} \= \sum_{N=0}^{+\infty} \sum_{k=1}^{d_i(N)} \ket{ i; N, k} \otimes U \overline{\ket{i; N, k}} \ ,
\end{equation}   
where $\ket{i; N,k}$ and $\overline{\ket{i; N,k}}$ are, respectively, $d_i(N)$-dimensional orthonormal bases for the states of the isomorphic Hilbert spaces $\cH_i$ and $\overline{\cH_i}$ at level $N$. In particular, at level zero, $\ket{i; 0,1} \defineas \ket{i}$ is the highest weight state of the representation $i\in \HWR$. Moreover, $U: \overline{{\cH}_{i}} \to \overline{\cH_{i}}$ is the anti-unitary operator that acts by complex conjugation on the state $\overline{\ket{i}}$ and conjugates the modes of the chiral symmetry operators of $\overline{\cA}$ according to
\begin{equation}
   U \overline{W}_m U^{-1} \= (-1)^{h(W)} \Omega(\overline{W}_m) \qquad \text{for all} \qquad m \in \IZ \ .
\end{equation}   
The vanishing inner products $\bra{i; N,\ell} \otimes \bra{U(\bar\imath; N, \ell)} \left(W_m - (-1)^{h(W)} \, \Omega(\overline W_{-m})\right) \ish{i}$ for all labels $\ell=1,\ldots,d_i(N)$ at any given level $N$ imply that the Ishibashi boundary state~\eqref{eq:defish} is indeed a solution to the constraint~\eqref{eq:Wconstraints} \cite{Cardy:1989ir}. 

While the Ishibashi boundary states~$\ish{i}$ for all $i \in \HWR$ fulfill the constraints~\eqref{eq:Wconstraints}, a boundary state $\ket{B}$ must be compatible with modular $S$-transformations that transforms a propagating closed string boundary states $\ket{B}$ into the partition function of the open string with the imposed boundary conditions \cite{Cardy:1989ir}. Specifically, one equates the matrix element $M_{AB}(q)$ for the overlap between the boundary state $\ket{B_A}$ and the boundary states that arises from propagating $\ket{B_B}$ for an imaginary time $t$ with the open-string partition function $Z_{AB}^\text{open}(\wt q)$ of the open-string states with boundary conditions labelled by $A$ and $B$, i.e.
\begin{equation} \label{eq:Cardy1}
  M_{AB}(q) \= Z_{AB}^\text{open}(\wt q) \ .
\end{equation}
Here the matrix element $M_{AB}(q)$ and the $S$-transformed open-string partition function $Z_{AB}^\text{open}(\wt q)$ are respectively given by \cite{Cardy:1989ir}
\begin{equation} \label{eq:DefM}
  M_{AB}(q) \= \bra{B_A} q^{L_0 + \overline{L}_0 - \frac{c}{12}} \ket{B_B} \qquad \text{with} \qquad q \; \defineas \; \ee^{-2 \pi t} \ ,
\end{equation}
and
\begin{equation} \label{eq:OpenPart}
  Z_{AB}^\text{open}(\wt q) \= \operatorname{Tr}_{\mathcal{H}_{AB}} \wt q^{L_0 - \frac{c}{24}} \qquad \text{with} \qquad \wt q \; \defineas \; \ee^{-\frac{2\pi}t} \ ,
\end{equation}
with $\cH_{AB}$ being the Hilbert space of open-string states with boundary conditions $A$ and $B$.

In order to fulfil the boundary conditions~\eqref{eq:Wconstraints}, the boundary states $\ket{B_A}$ are realised as linear combinations of Ishibashi boundary states $\ish{i}$, i.e 
\begin{align}
\label{eq:boundarystate}
	\ket{B_A} = \sum_{i \in \HWR} b_{A,i} \ish{i}~,
\end{align}
which yield the matrix~elements
\begin{equation}
   M_{AB}(q) \= \sum_{j \in \HWR} b^*_{A,j} b_{B,j} \, \chi_j(2 \ii t ) \ ,
\end{equation}
in terms of the characters~$\chi_j(\tau)$ of the representations $j \in \HWR$. On the other hand, the open-string partition function~\eqref{eq:OpenPart} expands also into representations of the algebra $\cA$, namely
\begin{equation}
  Z_{AB}^\text{open}(\wt q) \= \sum_{j \in \HWR} n_{A,B}^j \,  \chi_j\left( \tfrac{\ii}{t} \right) \ ,
\end{equation}  
in terms of integral multiplicities $n_{A,B}^j$. Imposing eq.~\eqref{eq:Cardy1} leads for any boundary conditions $A$ and $B$, together with the transformation modular behaviour~\eqref{eq:CharTrans}, to the Cardy conditions~\cite{Cardy:1989ir}
\begin{equation} \label{eq:Cardy2}
    b^*_{A,j} b_{B,j} \= \sum_{i \in \HWR}  n_{A,B}^i \, S_{ji}  \qquad \text{for all} \qquad j \in \HWR \ .
\end{equation}

For conformal field theories with $\HWRrep=\HWR$ the constraints~\eqref{eq:Cardy2} are fulfilled upon identifying the multiplicities $n_{A,B}^j$ with the fusion structure constants $N_{A,B}^j$ defined in eq.~\eqref{eq:fusion}. Then the Cardy boundary states $\ket{B_A}$ for a boundary label $A$ enjoy the expansion \cite{Cardy:1989ir}
\begin{equation} \label{eq:Cardy_solution}
   \ket{B_A} \= \nu \sum_{i \in \HWRrep} \frac{\mtS_{Ai}}{\sqrt{\mtS_{0i}}} \ish{i} \ ,
\end{equation}
with an overall normalisation $\nu$, which must be chosen such that the condition~\eqref{eq:Cardy2} is fulfilled. If, however, $\HWR$ is a proper subset of $\HWRrep$, the states given by the sum over $\HWR$ of the form
\begin{equation} \label{eq:Cardy_solution_generalised}
	\ket{B_A} \= \nu \sum_{i \in \HWR} \frac{\mtS_{Ai}}{\sqrt{\mtS_{0i}}} \ish{i} \ ,
\end{equation}
may not fulfill the Cardy condition~\eqref{eq:Cardy2}. However, those states $\ket{B_A}$ with $\frac{\mtS_{Ai}}{\sqrt{\mtS_{0i}}} = 0$ for all representations~$i \in \HWRrep \setminus \HWR$ --- or more generally those rational linear combinations of states $\sum_A c_A \ket{B_A}$ with $\sum_A c_A \frac{\mtS_{Ai}}{\sqrt{\mtS_{0i}}} = 0$ for all $i \in \HWRrep \setminus \HWR$ --- realise boundary states as they obey the Cardy conditions~\eqref{eq:Cardy2}. In the following, we refer to legitimate boundary states of the form~\eqref{eq:Cardy_solution_generalised} as \emph{generalised Cardy boundary states}.

In the presence of fermions, the definition of the matrix element~\eqref{eq:DefM} and the open string partition function~\eqref{eq:OpenPart} depends on a choice of spin structure on $S^1$ \cite{Govindarajan:2000my,Hori:2000ck}. Namely, in the computation of the open-string partition function~\eqref{eq:OpenPart} the trace is obtained from fermions with anti-periodic boundary conditions, which in the closed string channel yields fermions in the Neveu-Schwarz sector with anti-periodic boundary conditions. The open-string trace with the insertion of the operator $(-1)^F$ (in terms of the fermion number operator~$F$) imposes periodic fermionic boundary conditions, which corresponds to periodic closed-string fermions in the Ramond sector. As a result the open- and closed-string channels get identified as \cite{Douglas:1999hq,Brunner:1999jq,Govindarajan:2000my,Hori:2000ck}
\begin{equation} \label{eq:CardyFermions}
  M_{AB}^{(\text{R})}(q) \= \operatorname{Tr}_{\mathcal{H}_{AB}} (-1)^F \wt q^{L_0 - \frac{c}{24}} \ , \qquad 
  M_{AB}^{(\text{NS})}(q) \= \operatorname{Tr}_{\mathcal{H}_{AB}} \wt q^{L_0 - \frac{c}{24}}  \ ,
\end{equation}
where the superscripts $(\text{R})$ and $(\text{NS})$ refer accordingly to the Ramond and Neveu--Schwarz sectors of the closed string matrix elements. 

Let us consider Cardy boundary states for left-right symmetric $\cN=(2,2)$ supersymmetric rational conformal field theories, which possess a left- and right-moving $\cN = 2$ super-Virasoro algebra $\operatorname{SVir} \times \overline{\operatorname{SVir}}$. BPS boundary states preserve an $\cN =2$ subalgebra $\operatorname{SVir} \hookrightarrow \operatorname{SVir} \times \overline{\operatorname{SVir}}$, which is determined by a choice of algebra automorphism $\Omega: \operatorname{SVir} \to \operatorname{SVir}$. 

For BPS boundary states --- which preserve half of the supersymmetry at the boundary and hence a subalgebra $\operatorname{SVir}$ of $\operatorname{SVir} \times \overline{\operatorname{SVir}}$ --- there are A-type boundary conditions, which arise from an algebra automorphism mapping $J$ to $-J$ and consequentially $G^\pm$ to $\ee^{\mp \ii\alpha} G^{\mp}$ for the phase $\alpha$, and there are B-type boundary conditions, which arise from an algebra automorphism mapping $J$ to $J$ and therefore $G^\pm$ to $\ee^{\pm \ii\beta} G^\pm$ for the phase $\beta$.\footnote{Note that in general the conformal field theory has boundary states for different values of the phase $\alpha$.} In addition to the Virasoro constraints $(L_m - \overline{L}_{-m})\ket{B}=0$, an A-type or B-type boundary state $\ket{B}$ is therefore further constrained. Namely, it must fulfil --- due to the conformal weights $h(J)=1$ and $h(G^\pm)=\frac32$ together with eq.~\eqref{eq:Wconstraints} --- the conditions \cite{Brunner:1999jq,Govindarajan:2000my,Hori:2000ck}
\begin{equation} \label{eq:ABtypeBdry}
\begin{aligned}
  &\text{A-type: } &&(J_m - \overline{J}_{-m})\ket{B} \= 0 \ ,\quad &&(G_r^\pm + \ii \me^{\mp \ii\alpha} \overline{G}^{\mp}_{-r})\ket{B} \= 0 \ , \\
  &\text{B-type:}  &&(J_m + \overline{J}_{-m})\ket{B} \= 0 \ ,\quad  &&(G_r^\pm + \ii \me^{\pm \ii\beta}\overline{G}^{\pm}_{-r})\ket{B} \= 0 \ ,
\end{aligned}    
\end{equation}
for all integral or half-integral modes $m$ and $r$.  

For such BPS boundary conditions the positive energy eigenstates in the open-string Hilbert space~$\cH_{AB}$ --- as measured by the open-string Hamiltonian $H_{\text{open}}=L_0 -\frac{c}{24}$ --- come in massive supersymmetric multiplets with an equal number of bosonic and fermionic states at the same positive energy eigenvalue. Hence, in the supersymmetric setting the positive energy eigenstates in the trace of the open-string partition function $\operatorname{Tr}_{\mathcal{H}_{AB}} (-1)^F \wt q^{L_0 - \frac{c}{24}}$ cancel, the partition function becomes independent of the parameter $\wt q$, and it turns into the index $\operatorname{Tr}_{\mathcal{H}_{AB}} (-1)^F$ of the open-string ground states. According to eq.~\eqref{eq:CardyFermions}, the matrix element $M_{AB}^{(\text{R})}(q)$ is independent on $q=\ee^{-2\pi t}$ as well. 

The bra states in the matrix element $M_{AB}^{(\text{R})}(q)$  associated to the boundary label~$A$ must be chosen with care in order to agree with the boundary conditions of the open-string fermionic modes of the index~$\operatorname{Tr}_{\mathcal{H}_{AB}} (-1)^F$. Let $\ket{B_A}$ be an A-type boundary state that fulfils the boundary constraint~\eqref{eq:ABtypeBdry} with the phase $\alpha$. Then the Hermitian conjugate of this constraint yields the A-type constraint~$\bra{B_A} (G_r^\pm + \ii \me^{\mp \ii(\alpha - \pi)} \overline{G}^{\mp}_{-r}) = 0$ with the new phase $\alpha - \pi$, and hence the bra state $\bra{B_A}$ preserves another linear combination of left-right moving supercharges than the ket state $\ket{B_A}$. However, by acting from the right on this conjugate constraint with the operator $\ee^{\ii \pi J_0}$ where $J_0$ denotes the left-moving zero mode of the $R$-symmetry current $J(z)$, one obtains the A-type constraint $\bra{B_A}\ee^{\ii \pi J_0} (G_r^\pm + \ii \me^{\mp i\alpha} \overline{G}^{\mp}_{-r}) = 0$ with the original phase $\alpha$ \cite{Brunner:1999jq,Hori:2000ck}.\footnote{Assigning the ket boundary state $\ket{B_A}$ to a D-brane $p$, the phase shift of $\alpha$ by $\pi$ indicates that the bra boundary state $\bra{B_A}$ corresponds to boundary state of the anti-D-brane $\overline{p}$ \cite{Witten:1998cd}. The multiplication of the bra boundary state $\bra{B_A}$ with $(-1)^{F_L}$ amounts to changing the anti-D-brane $\overline{p}$ back to the D-brane $p$.} A similar argument applies for the B-type boundary states. Hence, to compare with the open-string index calculation, the closed-string matrix element $M_{AB}^{(\text{R})}(q)$ is given by\footnote{As the product of operators $(-1)^{F_L} \ee^{\ii \pi J_0}$ in terms of the left-moving fermion number operator $F_L$ is central~\cite{Witten:1993jg}, one can alternatively define an index by ${\vphantom{\ket{}}}_\cP \! \bra{B_A} (-1)^{F_L} \ket{B_B}_\cP$~\cite{Brunner:1999jq,Govindarajan:2000my}. As compared to the insertion of $\ee^{\ii \pi J_0}$ the contributions to the index may differ by a phase. In this work we use the insertion $e^{i \pi J_0}$.}
\begin{equation}
    M_{AB}^{(\text{R})}(q) \= \bra{B_A} \ee^{\ii\pi J_0}  q^{L_0 + \overline{L}_0 - \frac{c}{12} } \ket{B_B} \ .
\end{equation}
Since this matrix element is independent of the parameter $q$, it can be computed in the limit $t \to +\infty$. In this limit, none of the positive closed string energy modes with respect to the closed string Hamiltonian $H_\text{closed}=L_0 + \overline{L}_0 -\frac{c}{12}$ contributes. As a result, we find, for the open-string index, the relationship 
\begin{equation} \label{eq:IndexFL}
   \operatorname{Tr}_{\mathcal{H}_{AB}} (-1)^F \=   {\vphantom{\ket{}}}_\cP \! \bra{B_A} \ee^{\ii\pi J_0} \ket{B_B}_\cP \ .
\end{equation}
where the states $\ket{B_A}_\cP$ are the BPS boundary states $\ket{B_A}$ projected to their Ramond--Ramond~(RR) ground states. 

Let us further introduce the projected boundary states $\BP{A}$ that are labelled by the highest-weight representations $A \in \HWRrep_{\text{R}}$, where $\HWRrep_{\text{R}}$ is the set of the highest-weight representations in the Ramond sector of the super-Virasoro algebra $\operatorname{SVir}$, c.f. eq.~\eqref{eq:spectral_flow}. With respect to these representations, we define the intersection matrix
\begin{align} \label{eq:Intersection_matrix}
	\wt \Sigma \= [\wt \Sigma_{A,B}]_{A,B \in \HWRrep_{\text{R}}}~, \qquad 
	\wt \Sigma_{A,B} \; \defineas \;  {\vphantom{\ket{}}}_\cP \! \bra{B_A} \ee^{\ii \pi J_0} \ket{B_B}_{\cP} \= 
	\operatorname{Tr}_{\mathcal{H}_{AB}} (-1)^F~,
\end{align}
Note that this set of representations furnish a set of mutually compatible boundary states, which implies that all entries of the intersection matrix enjoy the interpretation of an open-string index such that the intersection matrix is valued in $\IQ$ (even $\IZ$).

The distinction between A-type and B-type boundary states is intimately link to the mirror automorphism of the $\cN=2$ super-Virasoro $\operatorname{SVir} \times \overline{\operatorname{SVir}}$ \cite{Greene:1990ud}, which acts on the right-moving super-Virasoro algebra~$\overline{\operatorname{SVir}}$ by mapping the right-moving $\mtU(1)$-current $\overline{J}$ to $-\overline{J}$ and the spin-$\frac32$ currents $\overline{G}^\pm$ to $\overline{G}^\mp$. Moreover, the mirror automorphism maps the constraint for A-type boundary states to the constraint for B-type boundary states according to eq.~\eqref{eq:ABtypeBdry} and vice versa \cite{Hori:2000ck}. Therefore, mirror symmetry exchanges A-type boundary states with B-type boundary states. In this work, we mainly focus on A-type boundary states $\ket{B}$, which implies that such boundary states arise from a linear combination of states with equal left- and right-moving $\operatorname{U}(1)$ charges, among which there can in particular be $\text{(c,c)}$-primary states with equal left- and right-moving $\operatorname{U}(1)$ charges. Applying the mirror automorphism to such A-type boundary states with a (c,c)-primary states maps them to B-type boundary states with a $\text{(c,a)}$-primary state.

By the operator-state-correspondence the (c,c)-primary states form the (c,c)-ring. In the following the $\text{(c,c)}_{q_L=q_R}$-ring refers to the subring of (c,c)-ring elements with equal left- and right-moving $\operatorname{U}(1)$~charges. \label{def:cc}

For the studied supersymmetric conformal field theory and its associated boundary conformal field theory with A-type boundary conditions, we often assume that the algebra $\cA$ with its irreducible representations~$\HWR$ can be chosen as subalgebra of $\cA_\text{max}$ with the following properties:\footnote{The conditions on the algebra~$\cA$ spelt out with eq.~\eqref{eq:DefcA} automatically guarantee fulfilment of the first property.}
\begin{enumerate} \label{p:assumptions}
\item The irreducible representations~$\text{HWR}(\cA_\text{max})$ of the algebra $\cA_\text{max}$ decomposes into a finite set~$\HWR$ of irreducible representations with respect to the subalgebra $\cA$.
\item  Each irreducible representation of $\HWR$ contains at most a single chiral super-Virasoro primary state.
\item Each $\text{(c,c)}_{q_L=q_R}$ primary state belongs to a diagonal representation of $\HWR\times\HWRbar$. 
\item The intersection matrix $\wt\Sigma$ defined in eq.~\eqref{eq:Intersection_matrix} has maximal rank, which --- as a consequence --- coincides with the number of $\text{(c,c)}_{q_L=q_R}$ primary states.
\end{enumerate}

\newpage
\section{Hodge Structures of CM Type from Rational Conformal Field Theories}\label{sect:Hodge_structures_from_RCFT}
\vskip-10pt
We begin by showing how to construct a rational Hodge structure given a rational conformal field theory. This construction is motivated by the fact that in the case of a non-linear sigma models on a complex Kähler manifold $X$, the construction presented in this section reduces to the well-known identification between the chiral ring of the conformal field theory and certain sheaf-valued cohomology theories (see e.g. ref.~\cite{Greene:1996cy} and references therein). However, in addition to this classical construction, we emphasise the role of the boundary conformal field theory and the rational structure given by the boundary states, which makes it possible to discuss \textit{rational} Hodge structures, and not only real Hodge structures. We then proceed to define a Hodge structure of complex multiplication type, and lay out the general idea for obtaining such Hodge structures purely from conformal field theory considerations. Some explicit settings where these Hodge structures can be constructed are discussed as examples in following sections.

\subsection{Real and rational Hodge structures} \label{sect:Rational_Hodge_structures}
A \textit{rational Hodge structure of weight $m$} is defined as a tuple $(V_\IQ, V_\IC)$ of a $\IQ$-vector space $V_\IQ$ together with a decomposition of its complexification:\footnote{Such Hodge structures are often called \textit{pure} to distinguish them from mixed Hodge structures. All Hodge structures we consider in this paper are pure.}
\begin{align}
	V_\IC \defineas V_\IQ \otimes_\IQ \IC \= \bigoplus_{p+q=m} V_\IC^{p,q} 
\end{align}
such that under complex conjugation the subspaces behave as
\begin{align} \label{eq:Hodge_structure_complex_conjugation}
	\overline{V_\IC^{p,q}} \= V_\IC^{q,p}~.
\end{align}
The dimensions of the graded pieces are called the \textit{Hodge numbers} $h^{p,q} \defineas \dim_\IC V_\IC^{p,q}$. We call the array $[h^{m,0},h^{m-1,1},\dots,h^{0,m}]$ the \textit{Hodge type} of the Hodge structure. Following the common convention, sometimes we also refer the list of pairs $(p,q)$ with $h^{p,q}\neq 0$ as the Hodge type.

A \textit{Hodge substructure} is a $\IQ$-vector subspace $W_\IQ \subset V_\IQ$ that is compatible with the Hodge decomposition of $V_\IQ$, that is
\begin{align} \label{eq:Hodge_substructure_condition}
	W_\IC \defineas W_\IQ \otimes_\IQ \IC \= \bigoplus_{p+q=m} W_\IC \cap V_\IC^{p,q}
\end{align}
A Hodge structure is called \textit{simple} if it has no non-trivial Hodge substructures.

A \textit{real Hodge structure of weight $m$} and its Hodge substructures are defined in completely analogous way, by replacing the rational vector space $V_\IQ$ by a real vector space $V_\IR$ everywhere in the definition.

A rational Hodge structure of weight $m$ is said to be \textit{polarised} if there exists a \textit{polarisation}, a non-degenerate bilinear form $Q : V_\IQ \times V_\IQ \to \IQ$, which, extended to $V_\IC$ by linearity, satisfies the following conditions: 
\begin{align} \label{eq:polarisation_1}
	Q(v,w) &\= (-1)^m Q(v,w)~,\\[5pt] \label{eq:polarisation_2}
	Q(v,w) &\= 0~, \qquad  \text{if }~v \in V_\IC^{p,q}~,~~ w \in V_\IC^{p',q'}~,~\text{ and } p + p' \neq m~,\\[5pt] \label{eq:polarisation_3}
	\ii^{p-q} \; Q(v,\overline{v}) & \; > \;0~, \qquad \text{if }~v \in V_\IC^{p,q}~,~\text{ and } v \neq 0~.
\end{align}
Note that it is clear that a Hodge substructure of a polarised Hodge structure is also polarised, with the polarisation given by the restriction of $Q$ to the substructure.

\subsection{Real Hodge structures from conformal field theories} \label{sect:real_Hodge_structures}
In this section we show that given any conformal field theory with $\cN=(2,2)$ supersymmetry and finite-dimensional $\text{(c,c)}$-ring, we can associate with it two Hodge structures which we call the A- and B-type Hodge structures. When the theory admits a geometric interpretation, they correspond to the horizontal and vertical Hodge structures, respectively. These are constructed using the states of the chiral-chiral ring $\text{(c,c)}$, and the chiral-anti-chiral ring $\text{(c,a)}$, respectively. 

\subsubsection*{A-type Hodge structures}
\vskip-5pt
The definition of the A-type Hodge structure is based on the $\text{(c,c)}$-ring considered as the complexified vector space $V_\IC$. For simplicity, we further restrict to the subring $\text{(c,c)}_{q_L=q_R}$ formed by the states whose left-moving $\mtU(1)$ charge $q_L$ is equal to the right-moving charge $q_R$,\footnote{We could also consider the full $\text{(c,c)}$-ring, which would give rise to a mixed Hodge structure (see e.g. ref.~\cite{Greene:1996cy}).} and make the assumption that this ring is finite-dimensional.\footnote{For superconformal field theories arising from non-linear sigma models, this is always the case when the target space is a smooth and compact Calabi--Yau manifold \cite{Greene:1996cy}.} Since we have restricted to the case where the right- and left-moving states appear on equal footing, we mostly only discuss the left-moving sector. It is left implicit that any action on the left-moving states is accompanied by an action on the right-moving states such that action is well-defined one the relevant subring. 

As reviewed in section~\ref{sec:cft}, the spectral flow operator $\cU$ acting on a highest-weight state in $\text{(c,c)}_{q_L=q_R}$ gives a RR~ground state $\ket{\alpha}$. Define $\cR \ni \alpha$ as the finite set of representations corresponding to these states. Then we define the vector space $V_\IC$ as the space spanned by these states:
\begin{align} \label{eq:V_C_definition}
	V_\IC \; \defineas \; \langle \; \ket{\alpha}\; \rangle_\IC^{\alpha \in \cR}~.
\end{align}
We can always choose the basis states to be orthogonal, so it makes sense to define normalised RR~ground states $\ket{\alpha}_\cN$, $\alpha \in \cR$, which we take to satisfy 
\begin{align} \label{eq:normalised_states_definition}
	{{\vphantom{\braket{\alpha|\beta}}}}_{\cN} \! \braket{ \alpha | \beta }_\cN \= \delta_{\alpha \beta}~.
\end{align}
We define the complex conjugation by introducing a complex antilinear endomorphism $\iota: V_\IC \to V_\IC$, which acts on the elements of $V_\IC$ as
\begin{align} \label{eq:Hodge_structure_complex_conjugation_defition}
	\iota \left( \sum_{\alpha \in \cR} a_\alpha \ket{\alpha}_\cN \right) \defineas \sum_{\alpha \in \cR} \text{sgn}(\mtS_{0\cC(\alpha)}) a_\alpha^* \ket{\cC(\alpha)}_\cN~.
\end{align}

Here $\cC$ is the the charge conjugation operator~\eqref{eq:cCdef} and $a_\alpha$ are any complex numbers. The sign is included so that this definition agrees with the one obtained from geometric considerations when a geometric interpretation exists (see eq.~\eqref{eq:complex_conjugation_rational_structure_compatibility}). From the expressions \eqref{eq:spectral_flow_charge_transformation} and \eqref{eq:charge_conjugation_charge_transformation}, it follows that the $\mtU(1)$-charges of the state $\iota(\ket{\alpha}_\cN)$ are related to the charge of the original basis state by\footnote{As the charges $q_{L/R}$ of the RR~ground states $\ket{a}$ and the normalised ground states $\ket{\alpha}_\cN$ are equal, for ease of notation the subscript $\cN$ is dropped  when referred to the charge of the normalised states $\ket{\alpha}_\cN$.}
\begin{align} \label{eq:complex_conjugation_R-states_U(1)_charge}
	q_L(\iota(\ket{\alpha})) \= - q_L(\ket{\alpha})~, \qquad q_R(\iota(\ket{\alpha})) \= - q_R(\ket{\alpha})~.
\end{align}
By tracing how the conformal weights transform under $\cU$ and $\cC$, it is easy to show that $\cU^{-1}\iota(\ket{\alpha})$ is a chiral state, implying that $\alpha \in \cR$ and thus $\iota(\ket{\alpha}) \in V_\IC$. Therefore, it is possible to consistently define the real vector space $V_\IR$ as being spanned over $\IR$ by those elements of $V_\IC$ that are fixed by the action of the complex conjugation map $\iota: V_\IC \to V_\IC$:
\begin{align} \label{eq:V_R_definition}
	V_\IR \defineas \left\langle v \in V_\IC \; \middle | \; \iota(v) = v \right\rangle_\IR~.
\end{align}
From the definition it immediately follows that the complex vector space $V_\IC$ is obtained as the complexification $V_\IR \otimes_\IR \IC$ of the real vector space $V_\IR$.

We wish to define the graded vector spaces $V^{p,q}_\IC$ by assigning each basis state $\ket{\alpha}_\cN$ to one of the $V_\IC^{p,q}$. To make these assignment, we utilise the $\mtU(1)$ charge $q_L(\ket{\alpha})$ of the state $\ket{\alpha}_\cN$, and define the pair $(p,q)$ giving the Hodge type of the vector $\ket{\alpha}_\cN$ as a function of $q_L(\ket{\alpha})$.

To obtain a Hodge structure, the assignment of the pair $(p,q)$ to each state must satisfy the compatibility condition \eqref{eq:Hodge_structure_complex_conjugation}. By the virtue of the relation \eqref{eq:complex_conjugation_R-states_U(1)_charge}, $q_L(\iota(\ket{\alpha})) = -q_L(\ket{\alpha})$ this is equivalent to requiring
\begin{align}
	p(-q_L(\ket{\alpha})) \= q(q_L(\ket{\alpha})) \qquad \text{for all } \alpha \in \cR~.
\end{align}
We also need that $p$ and $q$ are non-negative integers, in addition to which we require that these integers are ``minimal" in the sense that
\begin{equation} \label{eq:weight_minimality_condition}
	\text{gcd}(\{p(q_L(\ket{\alpha})),~q(q_L(\ket{\alpha}))\; | \; \alpha \in \cR \})=1\ , 
\end{equation}	
and
\begin{equation}\label{eq:twist_minimality_condition}
	p(q_L(\ket{\alpha})) = q(q_L(\ket{\beta})) = 0 \quad \text{for some } \alpha, \beta \in \cR \ .
\end{equation}

The significance of the first condition \eqref{eq:weight_minimality_condition} is exemplified by considering a situation where we have two states $\ket{\alpha}, \ket{\beta} \in \cR$ with $q_L(\ket{\alpha})=-c/6$ and $q_L(\ket{\beta}) = c/6$. In this case, we could define a weight-1 Hodge structure $(V_\IQ,V_\IC)$ of type $[1,1]$ with $\ket{\alpha} \in V_\IC^{(1,0)}$ and $\ket{\beta} \in V_\IC^{(0,1)}$. However, without the above condition, we could equally well define a weight-2 Hodge structure $(U_\IQ,U_\IC)$ of type $[1,0,1]$ with $\ket{\alpha} \in U_\IC^{(2,0)}$ and $\ket{\beta} \in U_\IC^{(0,2)}$. The condition \eqref{eq:weight_minimality_condition} requires that we pick the Hodge structure $(V_\IQ,V_\IC)$ of lower weight. 

The purpose of the second condition \eqref{eq:twist_minimality_condition} is similar. To illustrate it, consider a case where we have one state $\ket{\alpha}$, $\alpha \in \cR$ with $q_L(\ket{\alpha})=0$. Without imposing this condition, we could either define a weight-0 Hodge structure $(V_\IQ,V_\IC)$ of the Hodge type $[1]$ with $\ket{\alpha} \in V_\IC^{(0,0)}$ or a weight-2 Hodge structure $(U_\IQ,U_\IC)$ of the type $[0,1,0]$ with $\ket{\alpha} \in U_\IC^{(1,1)}$. These two are trivially related by tensoring (\textit{twisting}) with the dual $\IQ(-1)$ of the \textit{Hodge-Tate structure}, which is a one-dimensional Hodge structure of type $(1,1)$ (see e.g. ref.~\cite{Green2012a}). The condition \eqref{eq:twist_minimality_condition} ensures that we consider only the Hodge structure $(V_\IQ,V_\IC)$ of lower weight. 

Finally, we require that the functions $p(q_L)$ and $q(q_L)$ be linear in $q_L(\ket{\alpha})$.\footnote{Restricting to linear functions can be justified a posteriori as it turns out that for Gepner models, the Hodge structures obtained in this way are equal to those obtained from the corresponding Calabi--Yau geometry (see section~\ref{sect:relation_to_geometry}).} A pair $(p,q)$ of linear functions satisfying these properties are given by\footnote{In terms of the $\mtU(1)$ charges of the states in the $\text{(c,c)}_{q_L = q_R}$-ring, these formulae would read simply $p(q_L) = \ell c/3-l q_L$ and $q(q_L) = \ell q_L$.}
\begin{align} \label{eq:Hodge_structure_pq_definition}
	p(q_L) \defineas \ell\left(\frac{c}{6} - q_L\right)~, \qquad q(q_L) \defineas   \ell \left(\frac{c}{6} + q_L\right)~,
\end{align}
where we have defined the auxiliary quantity $\ell$, which ensures the integrality of $p$ and $q$, by
\begin{align}
	\ell \defineas \text{lcm}\left( \left\{ \text{denom}\big( c/6 \pm q_L(\ket{\alpha}) \big) \; \middle| \; \alpha \in \cR  \right\} \right).
\end{align}
Furthermore, these are the unique such pair of linear function, up to exchanging $p$ with $q$ which corresponds to complex conjugation, and does therefore not affect the resulting Hodge structure in any essential way. The choice we have made here is such that for Gepner models it corresponds to the Hodge structure of the corresponding Calabi--Yau geometry, as we show in section \ref{sect:relation_to_geometry}.

The positivity of $p$ and $q$ follows from the bound \eqref{eq:q_bound} together with \eqref{eq:spectral_flow_charge_transformation}, which imply that $|q_L(\ket{\alpha})| \leq c/6$ for all $\alpha \in \cR$. The fact that $p$ and $q$ attain the value $0$ for some states can be seen by recalling that the state $\cU \ket{0}$, where $\ket{0}$ denotes the ground state, saturates the bound $q_L(\cU \ket{0}) = -c/6$.  Its conjugate $\iota(\cU \ket{0})$ then satisfies the bound $q_L(\iota(\cU \ket{0})) = c/6$.

The graded vector spaces $V_\IC^{p,q}$ are then defined as
\begin{align} \label{eq:Hodge_structure_graded_vector_spaces}
	V^{p,q}_\IC \defineas  \left\{ \ket{\alpha} \; \middle| \; \alpha \in \cR~, \quad \frac{c}{6} - q_L(\ket{\alpha}) = \frac{p}{\ell}~, \quad \frac{c}{6} + q_L(\ket{\alpha}) = \frac{q}{\ell} \right\}~.
\end{align}
With this definition, the pair $(V_\IR,V_\IC)$ defines a Hodge structure of weight $m = p+q = \ell c/3$.

\subsubsection*{B-type Hodge structures}
\vskip-5pt
We define the B-type Hodge structures to be related to the A-type Hodge structure by the mirror automorphism \cite{Greene:1990ud}, which reverses the sign of the right-moving $\mtU(1)$ current. This implies that the construction of the B-type Hodge structures is based on the $\text{(c,a)}$-ring, though for simplicity we again restrict to the subring $\text{(c,a)}_{q_L=-q_R}$, where the left-moving $\mtU(1)$-charge is negative of the right-moving charge. Since the dependence of A-type Hodge structures in the right-moving states is given by the left-moving states, it comes as no surprise that the construction of B-type Hodge structures is almost identical.

Let $\cS$ be the set of representations whose highest-weight state are RR~ground states that are obtained the spectral flow from  $\text{(c,a)}_{q_L=-q_R}$.\footnote{When discussing B-type Hodge structures, we take the spectral flow operator to act consistently with mirror symmetry so that applying $\cU$ to a state in the $\text{(c,a)}$-ring increases, rather than decreases, the right-moving $\mtU(1)$-charge.} Then the associated B-type Hodge structure is given by the complex vector space
\begin{equation}
   V_\IC \defineas \langle \; \ket{\beta} \; \rangle_\IC^{\beta \in \cS} \ ,
\end{equation}
together with the graded pieces
\begin{equation}
    V_\IC^{p,q} \defineas  
    \left\{ \ket{\beta} \; \middle| \; \beta \in \cS~, \quad \frac{c}{6} - q_L(\ket{\beta}) = \frac{p}{\ell}~, 
    \quad \frac{c}{6} + q_L(\ket{\beta}) = \frac{q}{\ell} \right\} \ .
\end{equation}
With these definitions, the complex conjugation map \eqref{eq:Hodge_structure_complex_conjugation_defition} maps $V^{p,q}_\IC = \overline{V_\IC^{q,p}}$, so that we indeed obtain a Hodge structure of weight $m = \frac{\ell c}3$.

\subsection{Rational Hodge structures from rational conformal field theories} \label{sect:Rational_Hodge_structs_from_RCFTs}
For rational conformal field theories, as briefly reviewed in section \ref{sec:UniRCFT}, it is possible to explicitly construct boundary states in practice, which provide extra information that we can use to consistently define a rational structure on the real vector space $V_\IR$ constructed in the previous section so that we obtain a rational Hodge structure based on a rational vector space $V_\IQ$. 

In this section, we assume, for simplicity, that sufficiently many boundary states are given by the Cardy construction and are thus of the form \eqref{eq:Cardy_solution}. This is known to not to be the most general form of a boundary state, and indeed finding the most general boundary states in a generic rational conformal field theory is a difficult problem. The construction we present in this section can in many cases be easily generalised, as we explicitly illustrate via an example in section \ref{sect:Gepner_model_octic}, where we present how our construction works for a Gepner model whose boundary states are not all given by the Cardy construction. In addition, in this section, we only consider A-type Hodge structures, as B-type structures can be obtained from these via the mirror automorphism in a manner described~above. 

\subsubsection*{Rational structure from boundary states}
\vskip-5pt
To construct the rational vector space $V_\IQ$, consider the projection of the boundary states \eqref{eq:Cardy_solution} to the space $V_\IC$ spanned by the RR~ground states, consider the projection operator 
\begin{align}
	\cP_\cR \= \sum_{\alpha \in \cR } \ket{\alpha}_{\cN} {\vphantom{\ket{}}}_{\cN} \! \bra{\alpha}~,
\end{align}
which projects the boundary states to their R-ground states. The \textit{projected boundary states} $\BP{A}$, $A \in \HWRrep_{\text{R}}$, introduced in eqs.~\eqref{eq:IndexFL} and \eqref{eq:Intersection_matrix} can be written in terms of this operator as
\begin{align} \label{eq:projected_boundary_state_definition}
	\ket{B_A}_\cP \; \defineas \; \cP_\cR \ket{B_A} \= \sum_{\alpha \in \cR} \ket{\alpha}_{\cN} {\vphantom{\ket{}}}_{\cN} \! \braket{\alpha | B_A}~.
\end{align}  

We now define the space $V_\IQ$ to be the space spanned by these boundary states over $\IQ$:
\begin{align} \label{eq:V_Q_definition}
	V_\IQ \; \defineas \; \big\langle \; \ket{B_A}_\cP \; \big\rangle_\IQ^{A \in \HWRrep_\text{R}}~.
\end{align}
To explicitly see how this space is related to the complex vector space $V_\IC$ spanned by the RR~ground states $\ket{\alpha}_\cN$, choose any subset $\cB \subset \HWRrep_\text{R}$ with $|\cB| = |\cR|$ such that the restriction $\Sigma \defineas \wt \Sigma|_{\cB}$ of $\Sigma$ to $\cB$ is a matrix of full rank. Then the corresponding set $\{\ket{B_A}_\cP | A \in \cB \}$ of projected boundary states forms a basis of $V_\IC$ and $V_\IQ$.\footnote{\label{foot:independence_of_boundary_states}That this set forms a basis of $V_\IC$ follows from the matrix $\wt \Sigma$ is of full rank. To see that these form a basis of $V_\IQ$, by using the identity of $V_\IC$ in the form \eqref{eq:boundary_projection}, we see that $\ket{B_M}_{\cP} = \sum_{B \in \cB} \ket{B_A}_{\cP} \Sigma^{AB} \wt \Sigma_{BM}$ for any $M \in \HWR_{\text{R}}$.}

In terms of this basis, we can write the identity operator of $V_\IC$ as the projection operator to the complexification of $V_\IQ$, that is: 
\begin{align} \label{eq:boundary_projection}
	\mtI \= \cP_\cB \; \defineas \; \ket{B_A}_\cP \Sigma^{AB} {\vphantom{\ket{}}}_\cP \! \bra{B_B} \me^{\pi \ii J_{0}} \; \defineas \; \ket{B_A}_\cP {\vphantom{\ket{}}}_\cP \! \bra{B^A}\me^{\pi \ii J_{0}}~,
\end{align} 
where $\Sigma^{AB}$ are the entries of the inverse matrix of $\Sigma$, and in the last equation we used $\Sigma^{AB}$ to raise the indices of the bra-vectors: ${\vphantom{\ket{}}}_\cP \! \bra{B^A} \defineas \Sigma^{AB} {\vphantom{\ket{}}}_\cP \! \bra{B_B} \in V_\IQ$.
 
By using this, we can express the normalised basis states $\ket{\alpha}_\cN$ of $V_\IC$ in terms of the basis of projected boundary states: 
\begin{align} \label{eq:R-state_boundary_expansion}
	\ket{\alpha}_\cN \= \cP_\cB \ket{\alpha}_\cN \= \me^{\pi \ii q_L(\ket{\alpha})} \overline{{\vphantom{\ket{}}}_\cN \! \braket{\alpha|B_A}_\cP} \ket{B^A}_\cP
\end{align}
This shows explicitly that $V_\IQ$ defined by eq.~\eqref{eq:V_Q_definition} is a rational vector space of $\dim_\IQ V_\IQ = \dim_\IC V_\IC$ and that $V_\IC$ is the complexification of $V_\IQ$.

\subsubsection*{Compatibility with complex conjugation}
\vskip-5pt
We claim that the pair $(V_\IQ,V_\IC)$ gives a Hodge structure with complex conjugation given again by eq.~\eqref{eq:Hodge_structure_complex_conjugation_defition} so that the rational structure given by the boundary states is compatible with the real structure $V_\IR$ constructed in the previous section \ref{sect:real_Hodge_structures}. In particular, this immediately implies that we can again introduce the graded vector spaces $V_\IC^{p,q}$ by using the definition \eqref{eq:Hodge_structure_graded_vector_spaces} from the case of the real Hodge structure. 

To do this, we just need to check that any state $v \in V_\IQ$ is fixed under the complex conjugation map \eqref{eq:Hodge_structure_complex_conjugation_defition}. It is enough to show that every projected boundary state $\ket{B_A}_\cP$, $A \in \cB$, is kept fixed. Substituting in the form of the generalised Cardy solution \eqref{eq:Cardy_solution_generalised} in  eq.~\eqref{eq:projected_boundary_state_definition}, the projected boundary state can be expressed as
\begin{align}
 	\ket{B_A}_\cP \= \cP_\cR \ket{B_i} \= \nu \sum_{\alpha \in \cR} \frac{\mtS_{A\alpha}}{\sqrt{\mtS_{0\alpha}}} \ket{\alpha}_{\cN}~,
\end{align}
where $\nu \in \IR$ is an overall normalisation constant independent of the representations $A$ or $\alpha$, chosen so that the intersection matrix $\Sigma$ is rational.\footnote{The coefficient $\nu$ can always be chosen to be in $\IR$ since a complex phase cancels out of the definition of the intersection matrix $\Sigma$.} The complex conjugation map $\iota$ acts on this as 
\begin{align}
	\iota(\ket{B_A}_\cP) &\= \nu \sum_{\alpha \in \cR} \text{sgn}(\mtS_{0\cC(\alpha)}) \frac{\mtS_{A\alpha}^*}{\sqrt{\mtS_{0\alpha}}^*} \ket{\cC \alpha}_\cN \= \nu \sum_{\beta \in \cR} \text{sgn}(\mtS_{0\beta}) \frac{\mtS_{A \cC(\beta)}^*}{\sqrt{\mtS_{0 \cC (\beta)}}^*} \ket{\beta}_\cN~,
\end{align}
where we have noted that $\cC$ restricts to an invertible map $\cR \to \cR$, so we can equally well sum over the images of the map. Using the transformation properties of the modular S-matrix under the action of the charge conjugation operator $\cC$ (see eq.~\eqref{eq:Smat})
\begin{align}
	\mtS_{i \cC(j)} \= \sum_{k \in \HWRrep}\mtS_{ik} \mtC_{kj} \= \mtS^*_{ij}~,
\end{align}
Recalling that $\mtS_{0\beta} \in \IR$, we have that
\begin{align}
	\sqrt{\mtS_{0\beta}}^* \= \text{sgn}(\mtS_{0\beta}) \sqrt{\mtS_{0\beta}}~.
\end{align}
Therefore the projected boundary states are indeed invariant under the complex conjugation,
\begin{align} \label{eq:complex_conjugation_rational_structure_compatibility}
	\iota(\ket{B_A}_\cP) \= \nu \sum_{\beta \in \cR} \frac{\mtS_{A \beta}}{\sqrt{\mtS_{0\beta }}} \ket{\beta}_\cN \= \ket{B_A}_\cP~,
\end{align}
and we have shown that we can define a rational Hodge structure by setting
\begin{align} 
	V_\IQ \defineas \big\langle  \ket{B_A}_\cP  \big\rangle_\IQ^{A \in \HWR_\text{R}}~, \quad 
		V^{p,q}_\IC \defineas  \left\{ \; \ket{\alpha} \; \middle|  \alpha \in \cR\,,  \ \frac{c}{6} + q_L(\ket{\alpha}) = \frac{p}{\ell}\,, \ \frac{c}{6} - q_L(\ket{\alpha}) = \frac{q}{\ell} \right\}~.
\end{align}

\subsubsection*{Polarisation from the Witten index}
\vskip-5pt
As demonstrated in the following, for theories where all states in the $\text{(c,c)}_{q_L = q_R}$-ring have integral $\mtU(1)$-charges and that are subject to assumptions made in the previous section, the intersection matrix~\eqref{eq:Intersection_matrix} --- related to the open-string Witten index~\eqref{eq:IndexFL} --- gives rise to a polarisation of the constructed rational Hodge structure. However, for theories with fractional $\mtU(1)$-charges the intersection matrix~\eqref{eq:Intersection_matrix} does not yield a polarisation. We define in the following a twisted version of the open-string Witten index~\eqref{eq:IndexFL}. The  intersection matrix corresponding to it yields, in some cases, a polarisation for the constructed rational Hodge structure. Even if this construction is not applicable, the corresponding Hodge structure can still be polarisable, as we show in section \ref{sect:minimal_model_examples}.

The relationship to the open-string index~\eqref{eq:IndexFL} enables us to construct a $\IQ$-bilinear function $V_\IQ \times V_\IQ \to \IQ$ on the rational vector space $V_\IQ$ of boundary states. This index can be twisted by a unitary transformation $\cV$  central to the $\mathcal{N}=2$ super-Virasoro algebra \cite{Witten:1993jg}, which in particular implies $[ \ee^{\ii \pi J_0}, \cV ] = 0$. We further require that the operator $\cV$ restricted to the rational vector space~$V_\IQ$ yields a linear map $\left.\cV\right|_{\IQ}: V_\IQ \to V_\IQ$. This means that the restricted operator $\left. \cV \right|_\IQ$ realises an orthogonal transformation on $V_{\IQ}$. Such an operator $\cV$ allows us to define a twisted Witten index that gives rise to the $\IQ$-bilinear function $\Sigma: V_\IQ \times V_\IQ \to \IQ$ with
\begin{equation} \label{eq:SigmaRat}
   \Sigma( \; \ket{B_A}_\cP, \ket{B_B}_\cP) \defineas  {\vphantom{\ket{}}}_\cP \! \bra{B_A} \ee^{\ii \pi J_{0}} \cV \ket{B_B}_{\cP} \ .
\end{equation}

In order for $\Sigma$ to yield a polarisation for the Hodge structure $(V_\IQ,V_\IC^{p,q})$ of weight $m=\frac{\ell c}3$, the following conditions have to be fulfilled: First of all, the pairing $\Sigma$ must obey condition~\eqref{eq:polarisation_1}, which is to say $\Sigma( \; \ket{B_B}_\cP, \ket{B_A}_\cP) = (-1)^{-\frac{\ell c}3} \Sigma( \; \ket{B_A}_\cP, \ket{B_B}_\cP)$. This, together with the fact that $\ell$ has been chosen such that $\frac{\ell c}{6} + \ell q_L(\ket{\alpha}) \in \IZ$ for all $\alpha \in \cR$ (according to eq.~\eqref{eq:Hodge_structure_graded_vector_spaces}), yields the condition 
\begin{equation} \label{eq:Wittentwist}
     \ee^{2 \ii \pi (J_0+\frac{\ell c}6)} \cV^{-1} = \ee^{2 \ii \pi (\ell-1) J_0} \cV^{-1} =  \cV \  \quad
     \Longrightarrow \quad  \cV^2 = \ee^{2 \ii \pi (\ell-1) J_0} \ .
\end{equation}

In order for the Witten index~$\Sigma$ twisted by $\cV$ to yield a polarisation of $(V_\IQ,V_\IC^{p,q})$, in addition to eq.~\eqref{eq:Wittentwist}, we still need to check the remaining two conditions~\eqref{eq:polarisation_2} and \eqref{eq:polarisation_3}. Let us first extend the rational paring $\Sigma$ to the complex vector space $V_\IC$. Since the Hermitian inner product on the Hilbert space of states is complex anti-linear in the first argument, the definition~\eqref{eq:SigmaRat} naturally extends to a paring $\Sigma$ on $V_\IC$ that is complex anti-linear in the first argument. Therefore, to define a polarisation, we consider the closely related function
\begin{equation} \label{eq:polarisation_definition}
  Q(\ket{\alpha},\ket{\beta}) \; \defineas \; \Sigma(\iota(\ket{\alpha}),\ket{\beta}) 
    = \iota(\bra{\alpha}) \ee^{\ii \pi J_{0}} \cV  \ket{\beta}~, \qquad \ket{\alpha}, \ket{\beta} \in V_\IC~,
\end{equation}
where the appearance complex anti-linear complex conjugation map~$\iota: V_\IC \to V_\IC$ renders the first argument of $Q$ complex linear. Since $\iota$ keeps $V_\IQ$ fixed, $Q$ is a complex bilinear extension of $\Sigma: V_\IQ \to V_\IQ$ to $V_\IC$.

Next, we consider the bilinear pairing $Q$ evaluated for the normalised states $\ket{\alpha}_\cN$, which form a basis of $V_\IC$. From eq.~\eqref{eq:Hodge_structure_complex_conjugation_defition}, it follows that 
\begin{equation} \label{eq:polarisation_basis}
  Q( \; \ket{\alpha}_\cN, \ket{\beta}_\cN) 
  =  \text{sgn}(\mtS_{0 \cC(\alpha)}) \, \ee^{\ii \pi q_L(\ket{\beta})} \, {\vphantom{\ket{}}}_\cN \! \bra{ \cC(\alpha)} \cV \ket{\beta}_\cN \ .
\end{equation}
Due to eq.~\eqref{eq:normalised_states_definition} and the commutation relation $[J_0,\cV]=0$, it further follows that $Q( \; \ket{\alpha}_\cN, \ket{\beta}_\cN)$ is only non-vanishing if the states $\ket{\cC(\alpha)}_\cN$ and $\ket{\beta}_\cN$ have the same charges. This implies, with eq.~\eqref{eq:charge_conjugation_charge_transformation}, that $q_L(\ket{\alpha}) = -q_L(\ket{\beta})$ and hence $\ket{\alpha}_\cN \in V_\IC^{p,q}$ and $\ket{\beta}_\cN \in V_\IC^{q,p}$.  Therefore, the condition \eqref{eq:polarisation_2} is indeed fulfilled.

Finally, we calculate for any graded state $\ket{\alpha} \in V_\IC^{p,q}$
\begin{equation}
  \ii^{p-q} Q( \ket{\alpha}, \iota(\ket{\alpha}) ) 
  = \ii^{2\ell q_L(\ket{\alpha})} \overline{Q(\iota(\ket{\alpha}), \ket{\alpha})}
  =  \overline{\bra{\alpha} \ee^{-\ii \pi (\ell-1) J_0} \cV \ket{\alpha}} \ ,
\end{equation}  
Thus, in order for the Hodge structure $(V_\IQ, V_\IC)$ to be polarised by $Q$, the operator $\ee^{-\ii \pi (\ell-1) J_0} \cV$ must be positive definite, i.e., 
\begin{equation} \label{eq:poscond}
   \bra{\alpha} \ee^{-\ii \pi (\ell-1) J_0} \cV \ket{\alpha} > 0 \quad \text{for any non-zero $\ket\alpha \in V^{p,q}_\IC$} \ .
\end{equation}

A solution to eq.~\eqref{eq:Wittentwist} and the positivity condition~\eqref{eq:poscond} is given by the unitary operator
\begin{equation} \label{eq:solTwistedIndex}
  \cV =  \ee^{\ii \pi (\ell-1) J_0} \ .
\end{equation}
However, $\cV$ only commutes with the spin-$\frac32$ currents $G^\pm$ for odd integers $\ell$, and hence is only central to the $\mathcal{N}=2$ super-Virasoro algebra in this case. In section \ref{sect:minimal_models},  we explicitly confirm that for A-type $\mathcal{N}=2$ minimal models of odd levels $k$, which give rise to odd integers $\ell$, the Witten index twisted with the operator $\ee^{\ii \pi (\ell-1) J_0}$ restricts to a well-defined operator on $V_\IQ$, which indeed fulfils all requirements of a polarisation. In particular, if all states in the $\text{(c,c)}_{q_L = q_R}$-ring have integral $\mtU(1)$-charges so that $\ell=1$, the operator $\cV$ is the identity operator. Then the untwisted open-string Witten index realises a well-defined polarisation.

\subsection{Hodge structures of complex multiplication type} \label{sect:Hodge_structures_with_CM}
We now turn to the definition of rational Hodge structures of complex multiplication (CM) type. There exist several equivalent notions in the literature, and for the purposes of this paper, we use one that is often convenient for explicit computations. Some other definitions are discussed in appendix~\ref{app:complex_multiplication}. The somewhat abstract discussion in this section is exemplified in appendix~\ref{app:CM_for_elliptic_curves}.

We first define the concept of a \textit{Hodge endomorphism}. Let $(V_\IQ,V_\IC)$ be a Hodge structure of weight~$m$. Then Hodge endomorphisms are vector space endomorphisms $\varphi: V_\IQ \to V_\IQ$ that preserve the Hodge decomposition in the sense that their complex linear extensions $\varphi_\IC$ satisfy
\begin{align} \label{eq:Hodge_endomorphism_condition}
	\varphi_\IC(V_\IC^{p,q}) \subseteq V_\IC^{p,q}~.
\end{align}
The Hodge endomorphisms form a $\IQ$-algebra with the multiplication being the function composition. We denote this algebra as
\begin{align}
	\text{End}_\text{Hdg}(V_\IQ) \defineas \left\{ \varphi \in \text{Hom}(V_\IQ, V_\IQ) \; \right| \; \left. \varphi_\IC(V_\IC^{p,q}) \subseteq V_\IC^{p,q} \right\}~.
\end{align}
We say that an irreducible Hodge structure $(V_\IQ,V_\IC)$ has an \textit{$E$-multiplication}\footnote{We learned this terminology from Matt Kerr.} if there exists an embedding of a number field $E \hookrightarrow \text{End}_\text{Hdg}(V_\IQ)$, which is a $\IQ$-algebra morphism. If $[E:\IQ] = \dim_\IQ V_\IQ$, and $(V_\IQ,V_\IC)$ is polarisable, we say that the Hodge structure $(V_\IQ,V_\IC)$ is of \textit{complex multiplication (CM) type}. In this case one can in fact show that $\text{End}_{\text{Hdg}}(W_\IQ^\alpha)$ is commutative, and in fact isomorphic to a \textit{CM field} \cite{Green2012a}.\footnote{Recall that a CM field $\IE$ is defined as a quadratic totally imaginary field extension $\IE/\IL$ of some totally real field $\IL$ (see e.g. ref.~\cite{Lang1983a}). $\IL$ is a totally real field if every complex embedding $\psi: \IL \to \IC$ has $\Im(\IL) \subset \IR$. A totally imaginary field is one which cannot be embedded in $\IR$. A useful result is that every number field Galois over $\IQ$ is either totally real or totally imaginary.}

If a Hodge structure $(V_\IQ,V_\IC)$ is not irreducible, it may be decomposed into finitely many irreducible substructures $W_\IQ^\alpha$
\begin{align}
	V_\IQ \= \bigoplus_{\alpha \in \cI} W^{\alpha}_\IQ~, \qquad \text{where} \qquad W_\IC \= \bigoplus_{p+q=m} W_\IC^{\alpha,(p,q)}~.
\end{align} 
To be able to distinguish between polarised and non-polarised cases, we say here that the Hodge structure $(V_\IQ,V_\IC)$ has \textit{sufficiently many complex multiplications} if every irreducible Hodge substructure $W_\IQ^\alpha$ has an $E^\alpha$ multiplication with $E^\alpha$ a number field of degree $[E^\alpha:\IQ] = \dim_\IQ W_\IQ^\alpha$. If, in addition, every irreducible substructure $W_\IQ^\alpha$ is of CM type in the sense defined above, we say that $(V_\IQ,V_\IC)$ is of CM type.

\subsection{Hodge structures of CM type from rational conformal field theories} \label{sect:CM_Hodge_Structures_from_RCFTs}
In this section, we will show that the Hodge structure $(V_\IQ,V_\IC)$ constructed in section \ref{sect:Rational_Hodge_structs_from_RCFTs} have always sufficiently many complex multiplications if the corresponding RCFT satisfies a compatibility condition of the Galois action. This condition, which we will spell out in detail in the following subsections, essentially states that for any given representation $\alpha \in \cR \subset \HWR$, the suitably-defined action of the Galois group we call $\text{Gal}(\IQ(\mtS_{\alpha})/\IQ)$ on the representations $\HWR$ actually closes on the set of representation $\cR$. In particular, the arguments presented in this section are enough to show that, if the Hodge structure is polarised and satisfies this condition, it is of CM~type.\footnote{Recall from section \ref{sect:Rational_Hodge_structs_from_RCFTs} that a polarisation exists for theories with integral $\mtU(1)$ charges in the $\text{(c,c)}_{q_L=q_R}$-ring, but also for many other examples.}

The structure of the argument is as follows: We first show that for any $\alpha \in \cR$ there exists a tower of Galois extensions $\IQ(S) \supseteq \IQ(\mtS_{\alpha}) \supseteq \IQ(\mtS_\alpha)^{\text{Stab}(\alpha)} \supseteq \IQ$, which we will define in detail below. We then show that, the Galois action corresponding to $\IQ(\mtS_\alpha)^{\text{Stab}(\alpha)}$ maps suitably normalised the basis states $\ket{\alpha}_\cS$, $\alpha \in \cR$ of $V_\IC$ to another basis state. As these basis states have definite $\mtU(1)$ charges, they belong to one of the spaces $V_\IC^{p,q}$ defined in eq.~\eqref{eq:Hodge_structure_graded_vector_spaces}. It is then shown that the space $W^{\alpha}_\IC$ spanned by the Galois images of $\ket{\alpha}_\cS$ gives the complex space corresponding to a Hodge substructure, and in this way we are able to find how the Hodge structure splits. Finally, we argue that it is possible construct enough Hodge endomorphisms and the embedding of a number field $E \hookrightarrow \text{End}_\text{Hdg}(V_\IQ)$ by considering multiplication of $\ket{\alpha}_\cS$ by the primitive generator~$\zeta$ of $ \IQ(\mtS_\alpha)^{\text{Stab}(\alpha)} = \IQ(\zeta)$.

\subsubsection*{The Galois groups and their actions}
\vskip-5pt
Let $(V_\IQ,V_\IC)$ be a Hodge structure constructed as in section \ref{sect:Rational_Hodge_structs_from_RCFTs}, and let $\ket{B_A}$, $A \in \cB$ be a basis of $V_\IQ$, and let $\ket{\alpha}_\cN$, $\alpha \in \cN$ be the normalised RR~ground states corresponding to the states in the $\text{(c,c)}_{q_L=q_R}$-ring. 

We first show that we can find states $\intscal{\alpha}$, which are proportional to the normalised states $\ket{\alpha}_\cN \in V_\IC$ such that $\intscal{\alpha}$ in fact live in the vector space $V_\IQ \otimes_{\IQ} \IQ(S)$, where $\IQ(S)$ is the field extension generated by the elements $\mtS_{ij}$ of the modular S-matrix, as defined in eq.~\eqref{eq:Q(S)_definition}. To see this, substitute in the formula~\eqref{eq:R-state_boundary_expansion} the expression~\eqref{eq:Cardy_solution} for a boundary state given by the Cardy construction: 
\begin{align}
	\ket{\alpha}_\cN \= \me^{\pi \ii q_L(\ket{\alpha})} \nu \sum_{A \in \cB} \frac{\mtS_{\alpha A}^*}{\sqrt{\mtS_{0 \alpha}}^{\; *}} \ket{B^A}_\cP~,
\end{align}
where we used the fact that the normalised boundary states satisfy the relation ${\vphantom{\ket{}}}_{\cN} \! \braket{\alpha|i}\! \rangle \= \delta_{\alpha i}$ with the Ishibashi states $\ish{i}$, $i \in \HWR$.\footnote{This follows from the normalisation for the Ishibashi states chosen in \eqref{eq:defish}, together with the assumption that we have picked the algebra $\cA$ so that each irreducible representation contains at most a single super-Virasoro primary state in the $\text{(c,c)}_ {q_L=q_R}$-ring, see page~\pageref{p:assumptions}.} Thus, for any $\alpha \in \cR$, we can define the states
\begin{align} \label{eq:intermediate_scaled_states}
	\intscal{\alpha} \defineas \me^{-\pi \ii q_L(\ket{\alpha})} \nu^{-1} \sqrt{\mtS_{0 \alpha}}^{\; *} \ket{\alpha}_\cN \= \sum_{A \in \cB} \mtS_{\alpha A}^* \ket{B^A}_\cP \in V_\IQ \otimes_{\IQ} \IQ(S)~.
\end{align}
In fact, it turn out to be useful to define the subextensions $\IQ(\mtS_\alpha) \subseteq \IQ(\mtS)$, where
\begin{align}
\label{eq:subextension}
	\IQ(\mtS_\alpha) \; \defineas \; \IQ\left( \left\{ \mtS_{\alpha A} \; | \; A \in \cB \right\}\right) \= \IQ\left( \left\{ \mtS_{\alpha A} \; | \; A \in \HWR_\text{R} \right\}\right)~,
\end{align}
so that in fact
\begin{align}
	\intscal{\alpha} \in V_\IQ \otimes_\IQ \IQ(\mtS_\alpha)~.
\end{align}
The extension $\IQ(\mtS_\alpha)$ is a Galois extension since it is a subextension of the Galois extension $\IQ(\mtS)$ with an Abelian Galois group. From elementary Galois theory (see e.g. Theorem VI.1.10 of ref.~\cite{Lang2002a}), it follows that there is a surjective restriction morphism
\begin{align} \label{eq:restriction_morphism}
\cdot \; |_{\IQ(\mtS_\alpha)} :  \text{Gal}(\IQ(\mtS)/\IQ) \to \text{Gal}(\IQ(\mtS_\alpha)/\IQ)~,
\end{align}
which means that we can express any element $\rho \in \text{Gal}(\IQ(\mtS_\alpha)/\IQ)$ as a restriction $\sigma|_{\IQ(\mtS_\alpha)}$ of an element $\sigma \in \text{Gal}(\IQ(\mtS)/\IQ)$.

The action of the Galois group $\text{Gal}(\IQ(\mtS_\alpha)/\IQ)$ can be extended to act on $V_\IQ \otimes_\IQ \IQ(\mtS_\alpha)$ in a natural way, by defining that every $\rho \in \text{Gal}(\IQ(\mtS_\alpha)/\IQ)$ acts trivially on the elements of $V_\IQ$. Since $\intscal{\alpha}$ can be interpreted as an element of $V_\IQ \otimes_\IQ \IQ(\mtS_\alpha)$, we have a natural action of $\rho$ on it: Let $\sigma \in  \text{Gal}(\IQ(\mtS)/\IQ)$ be any element of the Galois group of $\IQ(\mtS)$ such that it restricts to $\rho$, so that $\sigma|_{\IQ(\mtS_\alpha)} = \rho$. Then, recalling the formulae~\eqref{eq:GaloisOnS}, the action of $\rho$ on the S-matrix elements $\mtS_{\alpha A}$ is given by
\begin{align} \label{eq:rho_action_S}
	\rho(\mtS_{\alpha A}) \= \epsilon_{\sigma}(\alpha) \mtS_{\varsigma(\alpha) A} \asdefine \epsilon_\rho(\alpha) \mtS_{\varrho(\alpha)A}~, \qquad \text{where} \qquad \epsilon_\rho(\alpha) \in \{-1,+1\}~,
\end{align}
and we defined new notation $\epsilon_\rho(\alpha) \defineas \epsilon_\sigma(\alpha)$ and $\varrho(\alpha) \defineas \varsigma(\alpha)$ to emphasise that we only consider the Galois action of elements $\rho$ of $\text{Gal}(\IQ(\mtS_\alpha)/\IQ)$ and not of the full group $\Gal(\IQ(\mtS)/\IQ)$ studied in section \ref{sect:RCFTs_and_Galois_Groups}. 
Using the above formula, the action on the states $\intscal{\alpha}$ is
\begin{align} \label{eq:R_Galois_action}
	\rho \intscal{\alpha} \= \sum_{A \in \cB} \rho(\mtS_{\alpha A}^*) \ket{B^A}_\cP \= \sum_{A \in \cB} \epsilon_{\rho}(\alpha) \mtS_{\varrho(\alpha) A}^* \ket{B^A}_\cP \= \epsilon_{\rho}(\alpha) \intscal{\varrho(\alpha)}~,
\end{align}
where we have also implicitly used the fact that since the Galois group $\Gal(\IQ(S)/\IQ)$ is Abelian, the Galois action commutes with the complex conjugation. Note that the last expression makes sense only if the representation $\varrho(\alpha)$ also belongs to the set $\cR$, and thus there is a corresponding suitably-normalised highest-weight state $\intscal{\varrho(\alpha)}$. This is true in all cases known to us, but we do not know a general proof for this, so take this as an assumption. 

Let us make this assumption more precise. For a given $\alpha \in \cR$, we have described above the Galois group~$\operatorname{Gal}(\IQ(\mtS_\alpha)/\IQ)$ and the map $\varrho$. Using this, we can define the set of representations
\begin{equation}
  \cR_\alpha \; \defineas \; \left\{ \ \varrho(\alpha) \  \middle| \   \rho \in \operatorname{Gal}(\IQ(\mtS_\alpha)/\IQ) \ \right\} \= \text{Im}(\varrho) 
  \ \subset \ \HWR \ .
\end{equation}  
Then, by definition, $\varrho$ is a well-defined map
\begin{equation} \label{eq:compatibility_R_consequence}
	\varrho: \cR_\alpha \to \cR_\alpha \ .
\end{equation}
The assumption is that the set of representations $\cR_\alpha$ is actually a subset of $\cR$, that is,
\begin{equation} \label{eq:compatibility_R}
   \cR_\alpha \subset \cR \quad \text{for any} \quad \alpha \in \cR \ .
\end{equation}   
We refer to this condition as compatibility of R-ground states with the Galois action.
Note also that, by construction, $\IQ(S_\alpha) = \IQ(S_\beta)$ for any $\beta \in \cR_\alpha$, which implies $\text{Gal}(\IQ(\mtS_\alpha)/\IQ) = \text{Gal}(\IQ(\mtS_{\beta})/\IQ)$.

\subsubsection*{Splitting of the Hodge structure}
\vskip-5pt
Given a state $\intscal{\alpha}$, $\alpha \in \cR$, let us define $W_\IQ^{\alpha}$ as the smallest subset $W_\IQ^\alpha \subset V_\IQ$ such that the state~$\intscal{\alpha}$ belongs to its complexification $W_\IC^\alpha \defineas W_\IQ^\alpha \otimes_\IQ \IC$. We will now provide an explicit construction of these spaces and argue that these provide Hodge substructures of the Hodge structure $(V_\IQ,V_\IC)$ constructed in section \ref{sect:Rational_Hodge_structs_from_RCFTs}.

By the primitive element theorem the field extension $\IQ(\mtS_\alpha)$ can be generated by a single element $\zeta \in \IQ(\mtS_\alpha)$ with a minimal polynomial of degree $n \defineas \left[\IQ(\mtS_\alpha):\IQ\right] = |\text{Gal}(\IQ(\mtS_\alpha)/\IQ)|$. Hence every modular S-matrix element $\mtS_{\alpha A}^*$ can be expanded as a rational linear combination of powers of $\zeta$, and the state $\intscal{\alpha}$  can be written as 
\begin{align} \label{eq:cc_element_generator_expansion}
	\intscal{\alpha} \= \sum_{k=0}^{n-1} \zeta^k \ket{Q_k^{\alpha}} \quad \text{with} \quad \ket{Q_k^{\alpha}} \in V_\IQ~,
\end{align}
where the states $\ket{Q_k^{\alpha}}$ are rational linear combinations of the boundary states $\ket{B_A}$, $A \in \cB$, uniquely defined by the expression above. Clearly then the space $W^\alpha_\IQ$ can be written as the $\IQ$-span
\begin{align} \label{eq:W_Q_first_expression}
	W_\IQ^{\alpha} \= \langle \; \ket{Q_0^{\alpha}}, \dots, \ket{Q_{n-1}^{\alpha}} \; \rangle_\IQ~,
\end{align}
and its complexification $W_\IC^\alpha$ contains any of the Galois conjugates $\rho \intscal{\alpha}$ of $\intscal{\alpha}$, and in particular their complex conjugates.

In fact, the Galois conjugates of $\intscal{\alpha}$ span the complexification $W^\alpha_\IC$. To prove this, we will show that the number of linearly independent Galois conjugates of $\intscal{\alpha}$ equals the dimension of the vector space $W^\alpha_\IQ$.\footnote{This is equal to the complex dimension of the complexification since the boundary states $\ket{B_A}$, $A \in \cB$, and hence the states $\ket{Q^\alpha_k}$ are linearly independent over $\IC$ (see footnote \ref{foot:independence_of_boundary_states}).} Note that from eq.~\eqref{eq:W_Q_first_expression}, it follows that $\dim_\IQ W_\IQ^{\alpha} \leq n = |\Gal(\IQ(\mtS_\alpha)/\IQ)|$. Therefore, if the Galois conjugates $\rho \intscal{\alpha}$ of $\intscal{\alpha}$ are linearly independent for all $\rho \in \Gal(\IQ(\mtS_\alpha)/\IQ)$, we are done, since in this way we get $n = |\Gal(\IQ(\mtS_\alpha)/\IQ)|$ independent vectors in $W_\IC^\alpha$.

Assume then that there exists a set non-trivial $\IQ$-automorphisms $\{\rho_0,\dots,\rho_{n'}\} \subset \Gal(\IQ(\mtS_\alpha)/\IQ)$, such that $\intscal{\alpha}$ is linearly dependent over $\IC$. That is, for some complex constants $c_s$,\footnote{Note that this is equivalent to the seemingly more general condition $\rho \intscal{\alpha} = \sum_{s=0}^{n'} \rho_s \intscal{\alpha}$ with $\rho \neq \rho_s$.}
\begin{align}
	\intscal{\alpha} \= \sum_{s=0}^{n'} c_s \rho_s \intscal{\alpha} \= \sum_{s=0}^{n'} c_s \epsilon_{\rho_s}(\alpha) \intscal{\varrho_s(\alpha)}~,
\end{align}
By definition $\intscal{\alpha}$ and $\intscal{\beta}$ are only linearly dependent over $\IC$ if $\alpha = \beta$. Therefore, for the above equality to hold, we must have that for at least one $\rho_*$, it holds that $\varrho_*(\alpha) = \alpha$. Let us study the subgroup of $\Gal(\IQ(\mtS_a)/\IQ)$ whose elements have this property:
\begin{align} \label{eq:Stabiliser_group_definition}
	\text{Stab}(\alpha) \defineas \{ \rho_* \in \Gal(\IQ(\mtS_\alpha)/\IQ) \; | \; \varrho_*(\alpha) = \alpha \} 
\end{align}
From eq.~\eqref{eq:rho_action_S}, it follows that for any $\rho_* \in \text{Stab}(\alpha)$
\begin{align} \label{eq:trivial_Galois_actions}
	\rho_*(\mtS_{\alpha A}) \= \mtS_{\alpha A} \qquad \text{or} \qquad \rho_*(\mtS_{\alpha A}) \= -\mtS_{\alpha A}~ \qquad \quad \text{for all } A \in \cB~.
\end{align}
Since $\mtS_{\alpha A}$, $A \in \cB$ generate $\IQ(\mtS_\alpha)$ over $\IQ$, the former being true implies that $\rho_*$ is the identity element, whereas the latter condition implies that $\rho_*$ is a non-trivial element of order two. In fact $\text{Stab}$ is of order two, as if we have two morphisms $\rho_*$ and $\rho_*'$ satisfying the latter condition, it follows that $\rho_* \circ \rho_*' = I$, implying that $\rho_*' = \rho_*^{-1} = \rho_*$. Therefore $\text{Stab}(\alpha)$ is generated by $\rho_*$ that is either of order 2 or trivial:
\begin{align}
	\text{Stab}(\alpha) \= \langle \rho_* \rangle \qquad \text{with} \qquad \text{ord}(\rho_*) \leq 2~.
\end{align}
Since $\Gal(\IQ(\mtS_\alpha)/\IQ)$ is Abelian, $\text{Stab}(\alpha)$ is normal, and we can define the quotient group 
\begin{align}
	\frac{\Gal(\IQ(\mtS_\alpha)/\IQ)}{\text{Stab}(\alpha)} \= \frac{\Gal(\IQ(\mtS_\alpha)/\IQ)}{\langle \rho_* \rangle}~.
\end{align}
If the subgroup $\text{Stab}(\alpha)$ is trivial, then by definition, all Galois conjugates of $\intscal{\alpha}$ are linearly independent, which is a contradiction. Let us therefore concentrate on the case where $\text{ord}(\rho_*) = 2$ in which case $\rho_*$ satisfies $\rho_*(\mtS_{\alpha A}) = - \mtS_{\alpha A}$ for all $A \in \cB$, and thus $\rho_* \intscal{\alpha} = - \intscal{\alpha}$.

Using the fundamental theorem of Galois theory we then obtain a tower of field extensions
\begin{align}
	\IQ(\mtS_\alpha) \supset \IQ(\mtS_\alpha)^{\langle \rho_* \rangle} \supset \IQ~, \quad \text{where} \quad \left[\IQ(\mtS_\alpha):\IQ(\mtS_\alpha)^{\langle \rho_* \rangle}\right] \= 2~,~~\left[\IQ(\mtS_\alpha)^{\langle \rho_* \rangle}:\IQ\right] \= \frac{n}{2}~.
\end{align}
Applying the primitive element theorem twice tell us that
\begin{align}
	\IQ(\mtS_\alpha) \= \IQ(\mtS_\alpha)^{\langle \rho_* \rangle}(\xi) \= \IQ(\eta,\xi)~,
\end{align}
where we can choose $\xi$ to satisfy $\xi^2 + \kappa = 0$ for some $\kappa \in \IQ(\eta)$,\footnote{Since the extension $\IQ(\xi,\eta)/\IQ(\eta)$ is of degree 2, we know that $\xi$ must satisfy a quadratic polynomial with coefficients in $\IQ(\eta)$. By redefining $\xi$ by a shift by a constant required to complete the square, we can always take $\xi$ to satisfy the equation $\xi^2 + \kappa = 0$, where $\kappa \in \IQ(\eta)$.} and the minimal polynomial of $\eta$ a polynomial of order $n/2$. By definition, $\rho_*$ leaves the elements of $\IQ(\mtS_\alpha)^{\langle \rho_* \rangle} = \IQ(\eta)$ invariant, so since it is not trivial, it must act non-trivially on $\xi$, that is to say $\rho_*(\eta) = \eta$ and $\rho_*(\xi) = - \xi$. Therefore, due to $\rho_* \intscal{\alpha} = - \intscal{\alpha}$ we can write the state $\intscal{\alpha}$ as 
\begin{equation} \label{eq:alpha_S_expansion}
	\intscal{\alpha} = \xi \sum_{a=0}^{n/2-1} \eta^a \ket{Q_a^\alpha}~, \quad \text{where} \quad \ket{Q_{a}^\alpha} \in V_\IQ \ .
\end{equation}
Thus, we see that in fact $W_\IQ^\alpha$ is given by
\begin{align} \label{eq:W_Q_span_non-trivial_Stab}
	W_\IQ^\alpha \= \langle \; \ket{Q^\alpha_{a}} \; \rangle_\IQ~,
\end{align}
and therefore $\dim_\IQ W_\IQ^\alpha \leq n/2 = |\Gal(\IQ(\mtS_\alpha)/\IQ)/\langle \rho_* \rangle|$. 

Now we note that $\xi^{-1}\intscal{\alpha} \in V_\IQ \otimes_\IQ \IQ(\mtS_\alpha)^{\text{Stab}(\alpha)}$. We can, as earlier, define a natural action of $\Gal(\IQ(\mtS_\alpha)^{\text{Stab}(\alpha)}/\IQ)$ on the vectors in $V_\IQ \otimes_\IQ \IQ(\mtS_\alpha)^{\text{Stab}(\alpha)}$, by demanding that the action keeps $V_\IQ$ fixed. By the the restriction theorem \eqref{eq:restriction_morphism}, this action coincides with that obtained by restricting the action of $\Gal(\IQ(\mtS_\alpha)/\IQ)$ to $V_\IQ \otimes_\IQ \IQ(\mtS_\alpha)^{\text{Stab}(\alpha)}$.

Now consider the Galois conjugates of 
\begin{align} \label{eq:Galois_conjugates_theta}
  \theta \left(  \xi^{-1} \intscal{\alpha} \right)~, \qquad \text{with} \qquad \theta \in \Gal(\IQ(\mtS_\alpha)^{\text{Stab}(\alpha)}/\IQ) \simeq \frac{\Gal(\IQ(\mtS_\alpha)/\IQ)}{\langle \rho \rangle}~.
\end{align}
All these Galois conjugates are linearly independent, as can be seen by the following simple argument: Let $\rho \in \Gal(\IQ(\mtS_\alpha)/\IQ)$ be an automorphism such that it restricts to $\theta$, that is $\rho|_{\IQ(\mtS_\alpha)^{\text{Stab}(\alpha)}} = \theta$. Then we have that
\begin{equation}  \label{eq:Galois_conjugation_theta_rho}
  \theta(\xi^{-1} \intscal{\alpha} )= \rho( \xi^{-1} \intscal{\alpha} ) 
  = \rho(\xi^{-1}) \rho ( \intscal{\alpha} ) 
  =  \rho(\xi^{-1}) \epsilon_{\rho}(\alpha) \intscal{\varrho(\alpha)} \ .
\end{equation}
Using this result, we can, as before, argue that to have that the states given by the Galois conjugates are linearly dependent, there must exist a non-trivial $\theta \in \Gal(\IQ(\mtS_\alpha)^{\text{Stab}(\alpha)}/\IQ)$ such that the corresponding $\rho$ keeps the representation $\alpha$ fixed, that is $\varrho(\alpha) = \alpha$. But if such $\theta$ existed, then by definition $\rho \in \text{Stab}(\alpha)$, and hence $\rho$ acts trivially on $\IQ(\mtS_\alpha)^{\text{Stab}(\alpha)}$. From this it follows that $\theta \= \rho|_{\IQ(\mtS_\alpha)^{\text{Stab}(\alpha)}} = I$, which is a contradiction. Thus we have shown that all Galois conjugates~\eqref{eq:Galois_conjugates_theta} are linearly independent over $\IC$. 

This shows that the Galois conjugates give $|\Gal(\IQ(\mtS_\alpha)^{\text{Stab}(\alpha)}/\IQ)| = |\Gal(\IQ(\mtS_\alpha)/\IQ)/\text{Stab}(\alpha)| = n/2$ independent vectors in $W_\IC^\alpha$. It follows from eq.~\eqref{eq:W_Q_span_non-trivial_Stab} that $\dim_\IQ W_\IQ^\alpha \leq n/2$ so the Galois conjugates necessarily span the complexification $W^\alpha_\IC$, and in fact $\dim W_\IQ^\alpha = n/2$.

To unify the notation, we introduce the conveniently scaled states
\begin{align} \label{eq:R_state_generator_expansion}
	\ket{\alpha}_\cS \defineas \left\{ \begin{aligned}
		\phantom{\xi^{-1}}\intscal{\alpha} \quad \text{if} \quad |\text{Stab}(\alpha)|=1\\
		\xi^{-1} \intscal{\alpha} \quad \text{if} \quad |\text{Stab}(\alpha)|=2
	\end{aligned} \right\} \= \sum_{k=0}^{n-1} \zeta^k \ket{Q_k^{\alpha}}~, 
\end{align}
where $\ket{Q_k^{\alpha}} \in V_\IQ$, $\IQ(\mtS_\alpha)^{\text{Stab}(\alpha)} = \IQ(\zeta)$, and $n = |\Gal(\IQ(\mtS_\alpha)^{\text{Stab}(\alpha)}/\IQ)|$. The main utility of these states is that, irrespective of the structure of $\text{Stab}(\alpha)$, these belong to $V_\IQ \otimes_\IQ \IQ(\mtS_\alpha)^{\text{Stab}(\alpha)}$. Thus we have shown that in every case we can write $W_\IC^\alpha$ as
\begin{align} \label{eq:W_IC_generators}
	W_\IC^{\alpha} \= \langle \; \theta \ket{\alpha}_\cS \; \rangle_\IC^{\theta \in \Gal(\IQ(\mtS_\alpha)^{\text{Stab}(\alpha)}/\IQ)}~.
\end{align}
From this it directly follows that the space $W^{\alpha}_\IQ$ satisfies the condition \eqref{eq:Hodge_substructure_condition} required of a Hodge substructure, as by eqs.~\eqref{eq:R_Galois_action} and \eqref{eq:Galois_conjugation_theta_rho} for every $\theta \in \Gal(\IQ(\mtS_\alpha)^{\text{Stab}(\alpha)}/\IQ)$
\begin{align}
	\theta \ket{\alpha}_\cS \= \left\{ \begin{aligned}
		\epsilon_{\rho}(\alpha) \intscal{\varrho(\alpha)} \quad \text{if} \quad |\text{Stab}(\alpha)|=1\\
		\rho(\xi^{-1}) \epsilon_{\rho}(\alpha) \intscal{\varrho(\alpha)} \quad \text{if} \quad |\text{Stab}(\alpha)|=2
	\end{aligned} \right\} \propto \ket{\varrho(\alpha)}_\cN,
\end{align}
where again $\rho \in \Gal(\IQ(\mtS_\alpha)/\IQ)$ is an automorphism that restricts to $\theta$, and as before, we have defined  $\epsilon_\theta(\alpha) \defineas \epsilon_\rho(\alpha)$ and $\vartheta(\alpha) \defineas \varrho(\alpha)$ to emphasise that we are considering automorphisms $\theta \in \Gal(\IQ(\mtS_\alpha)^{\text{Stab}(\alpha)}/\IQ)$. By definition $\ket{\varrho(\alpha)}_\cN$ belongs to one of the subspaces $V^{p,q}$ determined by its charges, which shows that the condition \eqref{eq:Hodge_substructure_condition} is satisfied.

\subsubsection*{Existence of Hodge endomorphisms}
\vskip-5pt
We finish the proof by showing that the Hodge structure $(V_\IQ,V_\IC)$ has enough complex multiplications, and is thus of CM type if it is polarised. This requires that we prove that every Hodge substructure $W^\alpha_\IQ$ admits an $E^\alpha$-multiplication by a number field $E ^\alpha \simeq \IQ(\mtS_\alpha)^{\text{Stab}(\alpha)}$ of degree $[E^\alpha:\IQ] = \dim_\IQ W^\alpha_\IQ$. We do this by constructing Hodge endomorphisms acting on the irreducible pieces $W^\alpha_\IQ$ as maps induced on the rational vector space $W^\alpha_\IQ$ by multiplying the state $\ket{\alpha}_\cS$ by any $\xi \in \IQ(\mtS_\alpha)^{\text{Stab}(\alpha)} = \IQ(\zeta)$, where we have introduced $\zeta$ to denote the generator of the field $\IQ(\mtS_\alpha)^{\text{Stab}(\alpha)}$ discussed in the previous subsection. These endomorphisms will then be in obvious correspondence with the elements of the number field $\IQ(\zeta)$, allowing us to explicitly construct the embedding $E^\alpha \simeq \IQ(\zeta) \hookrightarrow \text{End}_{\text{Hdg}}(W_\IQ^\alpha)$.

To see how we can define the endomorphism $\varphi^\alpha_\xi: W_\IQ^\alpha \to W_\IQ^\alpha$ corresponding to the multiplication of $\ket{\alpha}_\cS$ by $\xi$, recall (see e.g. ref.~\cite{Lang2002a}) that a field extension $\IQ(\zeta)$ can be thought of as a $\IQ$-vector space whose basis we can take to be given by the powers of $\zeta$:
\begin{align}
	\IQ(\zeta) \= \langle 1, \zeta, \dots, \zeta^{n-1} \rangle_\IQ~.
\end{align}
Multiplication by $\xi$ can be then thought of as a $\IQ$-linear map $\xi \cdot * : \IQ(\zeta) \to \IQ(\zeta)$, and thus it can be represented by a $\IQ$-valued matrix $\mtT(\xi) \in \text{Mat}(n {\times} n, \IQ)$ so that
\begin{align} \label{eq:multiplication_matrix}
	\zeta^k \; \mapsto \; \xi \zeta^{k} \= \sum_{s=0}^{n-1} \mtT(\xi)^k_{\phantom{k}s} \zeta^s~.
\end{align}
Using the form~\eqref{eq:R_state_generator_expansion} of $\ket{\alpha}_\cS$, we can write the state $\xi \ket{\alpha}_\cS$ as
\begin{align} \label{eq:zeta_multiplication_R_state}
	\xi \ket{\alpha}_\cS \= \sum_{k,s=0}^{n-1} \zeta^s \mtT(\xi)^k_{\phantom{k}s} \ket{Q_k^\alpha}~.
\end{align}
We can therefore define $\varphi_\xi^\alpha: W_\IQ^\alpha \to W_\IQ^\alpha$ as the $\IQ$-linear map corresponding to the $\IQ$-valued matrix $\mtT(\xi)$, that is
\begin{align} \label{eq:Hodge_endomorphism_zeta}
	\varphi_\xi^\alpha(\ket{Q_k^\alpha}) \; \defineas \; \mtT(\xi)_{\phantom{s}k}^s \ket{Q^\alpha_s}~.
\end{align}
Then from the above equation \eqref{eq:zeta_multiplication_R_state}, it follows directly that the $\IC$-linear extension of $\varphi_\xi^\alpha$ maps 
\begin{align}
	\varphi_{\xi,\IC}^\alpha \left(\ket{\alpha}_\cS \right) \= \xi \ket{\alpha}_\cS~,
\end{align}
so it indeed corresponds to multiplying the state $\ket{\alpha}_\cS$ by $\xi$.

We can easily find the action of the $\IC$-linear map $\varphi_{\xi,\IC}^{\alpha}$ on the full space $W_\IC^\alpha$ by using the fact that by eq.~\eqref{eq:W_IC_generators} this space has a basis given by the Galois conjugates $\theta \ket{\alpha}_\cS$, $\theta \in \Gal(\IQ(\mtS_\alpha)^{\text{Stab}(\alpha)}/\IQ)$, of the state $\ket{\alpha}_\cS$. Since we have defined the Galois conjugation so that it leaves the states in $W_\IQ^\alpha$ invariant, we can now simply use the properties of the field automorphism $\theta$ to find the action of $\varphi_{\xi,\IC}^\alpha$ on $\rho \ket{\alpha}_\cS$:
\begin{align}
	\begin{split}
		\varphi_{\xi,\IC}^{\alpha}\left(\theta \ket{\alpha}_\IC \right) & = \sum_{k=0}^{n-1} \theta(\zeta^k) \varphi_\xi^{\alpha}(\ket{Q^\alpha_k}) = \sum_{k,s=0}^{n-1} \theta(\zeta^k) \mtT(\xi)^s_{\phantom{s}k}\ket{Q^\alpha_s} = \sum_{k,s=0}^{n-1} \theta\left(\mtT(\xi)^s_{\phantom{s}k}\zeta^k\right) \ket{Q^\alpha_s} \\
		& = \sum_{s=0}^{n-1} \theta (\xi \zeta^{s}) \ket{Q^\alpha_s} = \theta(\xi) \sum_{s=0}^{n-1} \theta(\zeta^{s}) \ket{Q^\alpha_s} = \theta(\xi) \theta  \ket{\alpha}_\cS~.
	\end{split}
\end{align}
Thus we see that in the basis of $W_\IC^\alpha$ spanned by the Galois conjugates $\theta \ket{\alpha}_\cS$, $\theta \in \Gal(\IQ(\mtS_\alpha)^{\text{Stab}(\alpha)}/\IQ)$, of $\ket{\alpha}_\cS$, the map $\varphi_{\xi,\IC}^\alpha$ is diagonalised, acting on the basis state $\theta \ket{\alpha}_\cS$ by multiplication by $\theta(\zeta)$. Since every $\theta \ket{\alpha}_\cS$ belongs to one of the subspaces $W_\IC^{\alpha,(p,q)}$ and together the conjugates in fact span these spaces, this means that $\varphi_{\xi,\IC}^\alpha$ preserves the Hodge structure, 
\begin{align}
	\varphi_{\xi,\IC}^\alpha(W_\IC^{\alpha,(p,q)}) \subseteq W_\IC^{\alpha,(p,q)}~.
\end{align}
Let us denote the algebra generated by $\varphi_\xi^\alpha$, $\xi \in \IQ(\zeta)$ by $E^\alpha$. Any two Hodge endomorphisms $\varphi_{\xi}^\alpha, \varphi_{\eta}^\alpha \in E^\alpha$ clearly satisfy the relations
\begin{align}
	\varphi_{\xi}^\alpha \circ \varphi_\eta^{\alpha} \= \varphi_{\xi \eta}^\alpha~, \qquad \varphi_{\xi}^\alpha + \varphi_\eta^\alpha \= \varphi_{\xi + \eta}^\alpha~.
\end{align}
From these it follows that $E^\alpha$ is isomorphic to the field $\IQ(\zeta)$, with the isomorphism given by $\varphi^\alpha_{\xi} \mapsto \xi$. Thus we see that each $W^\alpha_\IQ$ has $E^\alpha$-multiplication with $E^\alpha \simeq \IQ(\zeta)$. This finishes the proof that for any rational conformal field theory satisfying the compatibility condition \eqref{eq:compatibility_R}, the Hodge structure defined in section \ref{sect:Rational_Hodge_structs_from_RCFTs} has sufficiently many complex multiplications. In particular, the Hodge structure is of CM type if it is polarised.

\subsection{Hodge structures of non-linear sigma model Calabi--Yau target spaces} \label{sect:relation_to_geometry}
To motivate the construction of Hodge structures from conformal field theories, we connect our discussion on conformal field theories to two-dimensional non-linear sigma models with geometric target spaces. It is well established that a two-dimensional $\mathcal{N}=(2,2)$ non-linear sigma model with a complex $n$-dimensional Calabi--Yau target space $X$ realises a $\mathcal{N}=(2,2)$ superconformal field theory with central charge $c=3n$ at its infrared fixed point \cite{Martinec:1988zu,Greene:1988ut,Distler:1992gi,Witten:1993yc}. In the following we relate the constructed Hodge structures of the $\mathcal{N}=(2,2)$ superconformal field theories to the Hodge structures of the middle cohomology $H^n_{\text{sing}}(X,\IQ)$ of the Calabi--Yau manifold~$X$. 

\subsubsection*{Chiral ring elements and cohomology of Calabi--Yau manifolds}
\vskip-5pt
Let $g(\,\cdot\,,\,\cdot\,)$ be the K\"ahler metric of the Calabi--Yau manifold $X$ and $R(\,\cdot\,,\,\cdot\,)\, \cdot \,$ the Riemann tensor of the Levi--Civita connection of the Calabi--Yau metric  $g(\,\cdot\,,\,\cdot\,)$. The map $\varphi: \Sigma \hookrightarrow X$ realises the non-linear sigma model scalar fields embedding the two-dimensional oriented worldsheet $\Sigma$ into the target space $X$. The Grassmann-odd fields $\Psi$ and $\wt\Psi$ are the fermionic superpartners of the scalar fields $\varphi$, which --- for a choice of spin structure on $\Sigma$ --- are respectively sections of the chiral and anti-chiral spin bundles with values in the complexified tangent bundles of $TX_\IC|_{\varphi(\Sigma)}=\left( T^{(1,0)}X\oplus T^{(0,1)}X\right)|_{\varphi(\Sigma)}$ restricted to the embedded worldsheet $\varphi(\Sigma)$. Then the $\mathcal{N}=(2,2)$ non-linear sigma model action~$S_\text{NL$\sigma$M}$ reads \cite{Witten:1991zz,Distler:1992gi,Greene:1996cy}
\begin{equation} \label{eq:NLSM}
  S_\text{NL$\sigma$M} \= \frac{1}{4\pi \alpha'} \int_\Sigma \dd^2 z \left[ \frac12 \varphi^*g(\partial_z,\partial_{\bar z})
  + g( \Psi , D_{\bar z} \Psi) + g( \wt\Psi, D_{z} \wt\Psi )
  + \frac14 g( R(\Psi,\Psi)\wt \Psi, \wt\Psi) \right] \ .
\end{equation}
Here $\alpha'$ is the coupling constant of dimension $[\text{length}]^2$, $(z,\bar z)$ are local holomorphic worldsheet coordinates together with their worldsheet tangent vectors $(\partial_z,\partial_{\bar z})$ of the complexified tangent vector space $T_{(z,\bar z)}\Sigma \otimes \mathbb{C}$, and the Dirac operators $D_z$ and $D_{\bar z}$ act on the spin bundles of the fermionic fields $\Psi$ and $\wt\Psi$.

While the non-linear sigma model action \eqref{eq:NLSM} with Calabi--Yau target manifold $X$ is not conformal, it has an anomaly-free $\mtU(1)_L \times \mtU(1)_R$ $R$-symmetry \cite{Witten:1991zz,Distler:1992gi,Greene:1996cy}. With respect to this $R$-symmetry, the scalar fields~$\varphi$ are neutral, and the chiral and anti-chiral fermionic fields $\Psi$ and $\wt\Psi$ respectively split into section $\Psi^{(1,0)}$ and $\Psi^{(0,1)}$ of charges $(\mp1,0)$ and into sections $\wt\Psi^{(1,0)}$ and $\wt\Psi^{(0,1)}$ of charges $(0,\mp1)$. Under the renormalisation group flow, the non-linear sigma model~\eqref{eq:NLSM} flows in the infrared to a conformal fixed point, at which the supersymmetry generators enhance to the left- and right-moving spin-$\frac32$ currents $G^\pm$ and the generators of the $\mtU(1)_{L/R}$ $R$-symmetry augment to the left- and right-moving $\operatorname{U}(1)$ currents $J$. Thus the infrared fixed point becomes a non-trivial $\mathcal{N}=(2,2)$ superconformal field theory with central charge $c= 3 \dim_\IC X = 3n$ \cite{Greene:1996cy}.

In order to quantise the fermionic fields, we expand the left-moving fermionic fields into a basis of holomorphic tangent and co-tangent section and the right-moving fields into a basis of anti-holomorphic tangent and co-tangent sections of the Calabi--Yau target manifold $X$ as\footnote{At any given point of the K\"ahler manifold $X$ the Hermitian metric $g( \, \cdot \,,\, \cdot \,)$ maps anti-holomorphic tangent vectors to holomorphic co-tangent vectors and holomorphic tangent vectors to anti-holomorphic co-tanget vectors.}
\begin{equation}
\begin{aligned}
   &\sum_{i} \Psi^i(z) \partial_i  \; \defineas \; \Psi^{(1,0)} \ ,   
   \quad&& \sum_j \Psi_j(z) \, \dd \varphi^j \; \defineas \;  g( \Psi^{(0,1)}, \,\cdot\, ) \ , \\[1ex]
   &\sum_{\bar\imath} \wt\Psi^{\bar\imath}(\bar z) \partial_{\bar\imath} \; \defineas \;  \wt\Psi^{(0,1)} \ , 
   \quad&& \sum_{\bar\jmath} \wt\Psi_{\bar\jmath}(\bar z) \, \dd \bar{\varphi}^{\bar\jmath} \; \defineas \;  g( \wt\Psi^{(1,0)},\,\cdot\, ) \  .
\end{aligned}
\end{equation}
In the Neveu--Schwarz sector with periodic boundary conditions on the complex plane, upon canonically quantising the fermionic fields, the constant modes $\psi^i$, $\psi_j$ of the fermionic left-movers $\Psi^i(z)$, $\Psi_j(z)$ and the constant modes $\wt\psi^{\bar\imath}$, $\wt\psi_{\bar\jmath}$ of the right mover $\wt\Psi^{\bar\imath}(\bar z)$, $\wt\Psi_{\bar\jmath}(\bar z)$ obey the anti-commutation relations
\begin{equation}
   \{ \psi^i,\psi_j \} \= \delta^i_j \ , \qquad
   \{ \wt\psi^{\bar\imath},\wt\psi_{\bar\jmath} \} \= \delta^{\bar\imath}_{\bar\jmath}  \  ,  
\end{equation}
where all other anti-commutators among these modes vanish. These anti-commutators realise a left- and right-moving Clifford algebras that yield Grassmann-odd ladder operators on the Hilbert space of states of the $\mathcal{N}=(2,2)$ supersymmetric non-linear sigma model. By defining the zero modes $\psi^i$ and $\wt\psi_{\bar\jmath}$ to be lowering operators and $\psi_j$ and $\wt\psi^{\bar\imath}$ to be raising operators, we construct a Clifford vacuum state $\ket0$ by the conditions
\begin{equation}
   \psi^i \ket0 \= \wt\psi_{\bar\jmath} \ket0 = 0 \quad  \text{for all $i$, $\bar\jmath$} \ .
\end{equation}
Furthermore, excited states $\ket{\phi^{(p,q)}}$ in this Hilbert space are given by
\begin{equation} \label{eq:pqstates}
  \ket{\phi^{(p,q)}} \= \sum_{\substack{i_1,\ldots,i_p\\ \bar\jmath_1,\ldots,\bar\jmath_q} }  \phi(\varphi,\bar\varphi)^{i_1,\ldots,i_p}_{\bar\jmath_1,\dots,\bar\jmath_q}\ 
  \psi_{i_1} \cdots \psi_{i_p} \,
  \wt\psi^{\bar\jmath_1} \cdots \wt\psi^{\bar\jmath_q} \ket0 \ ,
\end{equation}
where the superscript $(p,q)$ labels the $\mtU(1)_L \times \mtU(1)_R$ charges of these states. Since the Grassmann-odd zero modes $\psi_i$ and $\wt\psi^{\bar\jmath}$ transform, respectively, as holomorphic coordinate vector fields~$\partial_i$ and as anti-holomorphic one-form differentials~$\dd \bar\varphi^{\bar\jmath}$ on the Calabi--Yau manifold $X$, these states are in one-to-one correspondence to smooth global sections of the bundle $\Omega^{(0,q)}(X) \otimes \Lambda^p T^{(1,0)}X$ over $X$, 
that is to say
\begin{equation} \label{eq:qsecs}
 \sum_{\substack{i_1,\ldots,i_p \\ \bar\jmath_1,\ldots,\bar\jmath_q}} 
  \phi(\varphi,\bar\varphi)^{i_1,\ldots,i_p}_{\bar\jmath_1,\dots,\bar\jmath_q}\ 
  (\partial_{i_1} \wedge \cdots \wedge \partial_{i_p}) \otimes
  (\dd\bar\varphi^{\bar\jmath_1} \wedge \cdots \wedge \dd\bar\varphi^{\bar\jmath_q} ) \ 
  \in \ C^{\infty}(\Omega^{(0,q)}(X) \otimes \Lambda^p T^{(1,0)}X) \ .
\end{equation}

The right-moving zero mode $\wt Q^+$ of the supersymmetry generator reads \cite{Witten:1991zz,Greene:1996cy}
\begin{equation}
  \wt Q^+ \=  \sum_{\bar\imath} \wt\psi^{\bar\jmath} \partial_{\bar\imath} \ .
\end{equation}
On the one hand, the operator $\wt Q^+$ acts on the states~\eqref{eq:pqstates} as
\begin{equation}
   {\wt Q^+} \ket{\phi^{(p,q)}} \= (-1)^p \sum_{\substack{i_1,\ldots,i_p\\ \bar\jmath_1,\ldots,\bar\jmath_q}}  
   \partial_{\bar\jmath_0} \phi(\varphi,\bar\varphi)^{i_1,\ldots,i_p}_{\bar\jmath_1,\dots,\bar\jmath_q}\ 
  \psi_{i_1} \cdots \psi_{i_p} \,
  \wt\psi^{\bar\jmath_0}\wt\psi^{\bar\jmath_1} \cdots \wt\psi^{\bar\jmath_q} \ket0 \ .
\end{equation}   
Hence, on the level of the sections~\eqref{eq:qsecs}, the operator $\wt Q^+$  becomes the Dolbeault operator $\bar\partial: \Omega^{(0,q)}(X) \otimes \Lambda^p T^{(1,0)}X \to \Omega^{(0,q+1)}(X) \otimes \Lambda^p T^{(1,0)}X$. On the other hand, the operator $\wt Q^+$ flows in the infrared to the mode $\overline{G}^+_{-1/2}$ of the spin-$\frac32$ current $\overline{G}^+(\bar z)$ of the resulting two-dimensional $\mathcal{N}=(2,2)$ superconformal field theory. Therefore, according to eq.~\eqref{eq:chiralcond} the $\bar\partial$-closed sections~\eqref{eq:qsecs} correspond in the infrared to anti-chiral primary states, which --- as a consequence of $\{ \overline{G}^+_{-1/2},\overline{G}^+_{-1/2} \} = 0$ --- are only non-zero if the $\bar\partial$-closed sections~\eqref{eq:qsecs} are not $\bar\partial$-exact. Hence, the anti-chiral primary states map to sheaf cohomology elements \cite{Witten:1991zz,Greene:1996cy}
\begin{equation}
  H^{(0,q)}_{\bar\partial}(X,\Lambda^p T^{(1,0)}X) \;  \simeq \; H^{(n-p,q)}_{\bar\partial} (X) \ , 
\end{equation}  
where the isomorphism of these cohomology groups is realised by contracting the sheaf of wedge products of tangent vectors with the unique holomorphic $(n,0)$-form of the complex $n$-dimensional Calabi--Yau manifold $X$. 

Repeating the above arguments with the left-moving zero mode $Q^+$ \cite{Greene:1996cy}, one arrives at the condition that anti-chiral states are chiral if and only if the associated $\bar\partial$-closed $(p,q)$-forms are co-closed as well. That is to say elements of the (c,c)-ring with charge $(p,q)$ are in one-to-one correspondence to harmonic $(p,q)$-forms in $\mathcal{H}^{(n-p,q)}(X)$, which are again isomorphic to the Dolbeault cohomology groups $H^{(n-p,q)}_{\bar\partial} (X)$. Thus the renormalization group flow to the infrared fixed-point of the $\mathcal{N}=(2,2)$ non-linear sigma model induces the isomorphism \cite{Witten:1991zz,Greene:1996cy}
\begin{equation} \label{eq:RGDolbeault}
   \operatorname{RG}: \ H^{(n-p,q)}_{\bar\partial} (X) \xrightarrow{~\simeq~} 
   \cH^{(p,q)}_\text{(c,c)} \ .
 \end{equation}   
Here $\cH^{(p,q)}_\text{(c,c)}$ denotes the subspace of (c,c)-states with $U(1)$ charges $(p,q)$ of the Hilbert space of states $\cH$ of the infrared $\mathcal{N}=(2,2)$ superconformal field theory.

\subsubsection*{Hodge Structures of Two-Dimensional $\mathcal{N}=(2,2)$ non-Linear Sigma Models}
\vskip-5pt
Given the correspondence between the (c,c)-ring of $\mathcal{N}=(2,2)$ non-linear sigma models and the Dolbeault cohomology of the $n$-dimensional Calabi--Yau target space according to eq.~\eqref{eq:RGDolbeault}, we can now spell out the Hodge structure of weight $n$ discussed in sect.~\ref{sect:real_Hodge_structures} for a $\mathcal{N}=(2,2)$ superconformal field theory. Applying to the subring $\text{(c,c)}_{q_L=q_R}$ the renormalization group flow isomorphism $\operatorname{RG}$ of eq.~\eqref{eq:RGDolbeault} combined with the spectral flow operator $\cU$, we arrive at the isomorphism
\begin{equation}
  \cU \circ \operatorname{RG}: \ H^n(X,\IC) \; \simeq \; \bigoplus_{r=0}^n H^{(n-r,r)}_{\bar\partial} (X) \xrightarrow{~\simeq~} \cR \ ,
\end{equation}
in terms of the previously defined set $\cR$ of RR~ground states. The isomorphism $\cU \circ \operatorname{RG}$ is a graded map of vector spaces, as a $(n-r,r)$-form Dolbeault cohomology representative $\alpha$ is mapped to a RR ground state $\ket{\alpha}$ with charges  $q_L(\ket{\alpha})= r - \frac{n}2$ and $q_R(\ket{\alpha}) = r - \frac{n}2$, such that the RR~ground state $\ket{\alpha}$ is an element of the graded vector space~$V^{(n-r,r)}_\IC$ defined in eq.~\eqref{eq:Hodge_structure_graded_vector_spaces}, and we obtain the graded isomorphisms 
\begin{equation} \label{eq:nonIso}
  \cU \circ \operatorname{RG}: \ H^{(n-r,r)}_{\bar\partial} (X) \xrightarrow{~\simeq~} V^{(n-r,r)}_\IC \ .
\end{equation}
with the real structure
\begin{equation}
  \cU \circ \operatorname{RG}: \ H^n(X,\IR) \xrightarrow{~\simeq~} V_\IR \ .
\end{equation}
Hence, for two-dimensional $\mathcal{N}=(2,2)$ supersymmetric non-linear sigma models $\cU \circ \operatorname{RG}$ is an isomorphism between the Hodge structure of the complex and real middle dimensional cohomology groups $(H^n(X,\IC), H^n(X,\IR))$ and the Hodge structure of the RR~ground states $(V_\IC,V_\IR)$ defined in eqs.~\eqref{eq:V_C_definition} and \eqref{eq:V_R_definition}. 

The rational singular cohomology group $H^n_{\text{sing}}(X,\IQ)$ of the Calabi--Yau manifold $X$ gives the complex middle dimensional cohomology group a canonical rational structure $(H^n(X,\IC), H^n_\text{sing}(X,\IQ))$, where $H^n_{\text{sing}}(X,\IQ)$ is Poincar\'e dual to the middle-dimensional homology group $H_n(X,\IQ)$. Their $n$-cycle homology elements represent (rational linear combinations of) Lagrangian submanifolds known as A-type branes \cite{Ooguri:1996ck}, which serve as A-type boundary conditions of the two-dimensional non-linear sigma model space-time, i.e. $\varphi|_{\partial\Sigma}: \partial\Sigma \to \cB_A \subset X$ \cite{Hori:2000ck,Hori:2003ic}. A set of a A-branes, which yields a basis $\gamma_A$, $A=1,\ldots,\dim_\IQ H_n(X,\IQ)$, of homology $n$-cycles classes in $H_n(X,\IQ)$, is Poincar\'e dual to a basis $[\gamma_A]$ of $H^n_{\text{sing}}(X,\IQ)$. Such a basis enjoys on the level of cohomology the expansion 
\begin{equation}
  [\gamma_A]   \= \sum_{\alpha \in \cR} c^\alpha(A) \,\omega_{\alpha} \ ,
\end{equation}  
where $\omega_\alpha \defineas (\cU \circ \operatorname{RG})^{-1}( \ket\alpha ) \in H^n(X,\IC)$ is obtained from the RR~ground state elements $\ket\alpha$ associated to the subring $\text{(c,c)}_{q_L=q_R}$ of the two-dimensional $\mathcal{N}=(2,2)$ supersymmetric non-linear sigma model. 

As argued in detail in ref.~\cite{Ooguri:1996ck}, the isomorphism $\cU \circ \operatorname{RG}$ maps the cohomology class $[\gamma_a]$ to the projected boundary state $\ket{B_A}_\cP$, such that the above expansion becomes 
\begin{equation}
  \ket{B_A}_\cP \= \sum_{\alpha \in \cR} c^\alpha(A) \,\ket\alpha \ .
\end{equation}  
Hence, the canonical rational structure on the middle dimensional cohomology $H^n(X,\IC)$ of the non-linear sigma model target space~$X$ becomes the previously introduced rational structure~$V_\IQ$ given in eq.~\eqref{eq:V_Q_definition} in terms of projected boundary states of the conformal field theory at the infrared fixed-point. 

Note that the expansion coefficients $c^\alpha(A)$ depend continuously on the (complex structure) moduli of the two-dimensional $\mathcal{N}=(2,2)$ supersymmetric non-linear sigma model \cite{Ooguri:1996ck}. While the conformal field theories obtained from such non-linear sigma models give rise to the rational Hodge structure $(V_\IC,V_\IQ)$, for generic values of the complex structure moduli these Hodge structures are not of CM~type.

\newpage
\section{Hodge Structures of CM Type from \texorpdfstring{$\cN{=}2$}{N=2} Minimal Models} \label{sect:minimal_models}
\vskip-10pt
As a first example, we show that the Hodge structure constructed in section \ref{sect:Hodge_structures_from_RCFT} has sufficiently many complex multiplications for all $\cN=2$ minimal models with left-right symmetric spectrum, which implies together with a polarisation that we have a Hodge structure of CM~type. Given the proof in the previous section \ref{sect:CM_Hodge_Structures_from_RCFTs}, to establish this, it would suffice just to check that we have enough boundary states and that the compatibility condition \eqref{eq:compatibility_R} of the Galois action with R-ground states is satisfied. However, to illustrate the discussion in section \ref{sect:CM_Hodge_Structures_from_RCFTs} in more concrete terms, we keep the discussion here mostly self-contained, rederiving many results of section \ref{sect:CM_Hodge_Structures_from_RCFTs} explicitly. Apart from providing an important class of examples, this section also serves to illustrate that our construction is independent of the $\mathcal{N}=2$ superconformal field theory having a geometric origin. 

The finitely many super Virasoro modules of the $\cN=2$ minimal model at level $k$ and central charge $c=\frac{3k}{k+2}$ are most conveniently realised through the coset construction introduced in refs.~\cite{Goddard:1986ee, DiVecchia:1986fwg}. In the following, all conformal field theory quantities, like the S-matrix, are written in terms of irreducible highest weight representations of the chiral algebra 
\begin{equation} \label{eq:cosetalgebra}
\mathcal{A} \= \frac{\widehat{\mathfrak{su}}(2)_k \otimes \widehat{\mathfrak{u}}(1)_2}{\widehat{\mathfrak{u}}(1)_{k+2}}~,
\end{equation}
which yield the representations
\begin{align} \label{eq:minmal_model_HWRs}
	\HWR = \big\{ (l,m,s) \in \IZ \times \IZ/2(k{+}2)\IZ \times \IZ/4\IZ \; \big| \; 0 \leq l \leq k~,~~l{+}m{+}s \equiv 0 \!\! \mod 2 \big\}/\sim~,
\end{align}
where the equivalence relation is obtained from the coset construction.
\begin{equation} \label{eq:field_identication_equivalence}
	(l,m,s) \; \sim \; (k-l,m+k+2,s+2)~. 
\end{equation}
In the following, we often choose explicit representative for $m$ and $s$ such that $ -k{-}1 {\leq} m {\leq} k{+}2$ and $ s \in \{0, 2, \pm 1 \}$. When $s\in \{0, 2\}$, the corresponding state is in the NS sector, while $s\in \{ \pm1 \}$ corresponds to the R sector. 

The spectral flow operator $\cU$ corresponds to the highest-weight state of the representation $(0,-1,-1)$, and the supercurrents $G^\pm$ belong to the representation $(0,0,2)$. We also utilise the following fusion products (see e.g. ref.~\cite{Blumenhagen:2009zz})
\begin{align}
	(0,-1,-1) \times (l,m,s) \= (l,m-1,s-1)~, \qquad (0,0,2) \times (l,m,s) \= (l,m,s+2)~.
\end{align}
The states in the $\text{(c,c)}_{q_L=q_R}$-ring defined in section \ref{sect:real_Hodge_structures} are exactly the chiral states, which are labelled by $\ket{l,-l,0}$. From the above fusion rules we see that under the action of the spectral flow operator $\cU$, these map into the states $\ket{l,-(l+1),-1}$, so the set $\cR$ of representations defined above eq.~\eqref{eq:V_C_definition} is given by
\begin{align}
	\cR \= \left\{ \; (l,-(l{+}1),-1) \; \middle| \; 0 \leq l \leq k \; \right\}~.
\end{align}
For ease of notation we abbreviate in the following the representation labels in the set $\cR$ as
\begin{equation} \label{eq:Rl}
\Rl \defineas (l,-(l{+}1),-1) \in \cR \ .
\end{equation}
Later, in section \ref{sec:exgepnermodels}, we will make significant use of the fractional parts of the conformal weights and $\mtU(1)$ charges of the highest-weight states, so we will make use of the following identities:
\begin{align} \label{eq:h_q_minimal_model}
	h(\ket{l,m,s}) \; \equiv \; \frac{l(l+2)-m^2}{4(k{+}2)} + \frac{s^2}{8} \mod 1~, \qquad 	q_L(\ket{l,m,s}) \; \equiv \; -\frac{m}{k{+}2} + \frac{s}{2} \mod 1~.
\end{align}
It turns out that the above expression for the $\mtU(1)$ charge is exact for the states in $\cR$, so that 
\begin{equation} \label{eq:minimal_model_R-states_U(1)_charge}
	q_L(\ket{\Rl}) \=   -\frac{k-2 l}{4+2k}\,.
\end{equation}
For future reference, we note also the monodromy charges which follow from the definition \eqref{eq:monodromy_charge_definition} of the monodromy charge together with the formulae \eqref{eq:h_q_minimal_model} for the conformal weights
\begin{equation}
	\Phi_{(0,0,2)}(( l,  m,  s)) \; \equiv \; \frac{s}{2} \mod 1~.
\end{equation}

Note that the combination $(l,m,s) \oplus (l,m,s+2)$ realises a full irreducible $\cN=2$ super Virasoro module~\cite{Gepner:1987qi}. The individual $(l,m,s)$ representations correspond to primary fields with respect to the bosonic subalgebra of the $\cN=2$ Virasoro algebra. Note that it is this bosonic subalgebra that is the chiral algebra $\cA=\cA_{\text{max}}$ according to the definition given earlier, c.f. sections~\ref{sec:UniRCFT} and~\ref{sect:N=(2,2)_RSCFT}. 

The modular S-matrix elements are given by
\begin{align} \label{eq:S-matrix_minimal_model}
	\begin{split}
		\mtS_{(l,m,s),(L,M,S)} &\= \+\frac{1}{(k+2)} \sin\left(\frac{\pi}{k+2}(l+1)(L+1)\right) \exp\left(\ii \frac{\pi}{k+2} m M \right) \exp\left(-\ii \frac{\pi}{2} s S\right)\\[5pt]
		&\= -\frac{\ii}{2(k+2)} \left( \zeta_{2(k+2)}^{(l+1)(L+1)} -  \zeta_{2(k+2)}^{-(l+1)(L+1)}\right) \zeta_{2(k+2)}^{m M} \me^{-\ii \pi s S/2}~,
	\end{split}
\end{align}
where
\begin{align}
\zeta_n \; \defineas \; \me^{2\pi\ii/n}~.	
\end{align}
For purposes of illustration, we now go through the steps of the general construction presented in the previous section for the diagonal $\mathcal{N}=2$ minimal models, where diagonal refers to a decomposition into left- and right-moving irreducible representations of the $\cN=2$ super-Virasoro algebra.
 
\subsection{The complex vector space \texorpdfstring{$V_\mathbb{C}$}{VC} and rational vector space \texorpdfstring{$V_\mathbb{Q}$}{VQ}}
For the $\mathcal{N}=2$ minimal model of level $k$ the complex vector space $V_\mathbb{C}$ is given in terms of the representation labels~\eqref{eq:Rl} by 
\begin{equation}  \label{eq:mmvpq}
	V_\IC   \defineas \; \left\langle \; \ket{\Rl}_\mathcal{N} \ \middle| \ 0 \leq l \leq k \; \right\rangle_\IC \ , \quad
	V_\IC^{p,q} \defineas \;  \left \langle \; \ket{\Rl}_\cN \ \middle| \ k-l = p~,\   l = q \; \right\rangle_\IC~.
\end{equation}

To obtain this from eq.~\eqref{eq:Hodge_structure_graded_vector_spaces}, we have used the expression \eqref{eq:minimal_model_R-states_U(1)_charge} for the $\mtU(1)$-charge of the states, and we have taken $\ell =k+2$ ensuring that $p$ and $q$ are integral and minimal in the sense explained in section \ref{sect:real_Hodge_structures}.

In section \ref{sect:real_Hodge_structures}, we have shown that this defines a real Hodge structure of weight $k$ if we define the complex conjugation on $V_\IC$ by the complex antilinear map $\iota: V_\IC \to V_\IC$ which acts on the basis vectors $\ket{\Rl}_\cN$ by (cf. eq.~\eqref{eq:Hodge_structure_complex_conjugation_defition})
\begin{align}
	\iota\left( \ket{\Rl}_\cN \right) \= \text{sgn}(\mtS_{(0,0,0),(l,l+1,1)}) \ket{\Rla{k-l}}_\mathcal{N} \= \ket{\Rla{k-l}}_\mathcal{N} ~,
\end{align}
where we have used that
\begin{equation}  
	\cC(\Rl) \= \Rla{k-l}~, \qquad \text{and} \qquad \text{sgn}(\mtS_{(0,0,0),(l,l+1,1)}) \= \text{sgn}\left( \sin \left(\frac{1+l}{k+2} \pi \right) \right) \= 1~.
\end{equation} 
Then the real vector space $V_\IR$ is defined as the space spanned by vectors $v \in V_\IC$ invariant under the action of $\iota$. To check that the pair $(V_\IR,V_\IC)$ defines a real Hodge structure, note that for any $p \geq 0$, we can write any vector $v \in V_\IC^{k-p,p}$ as $v = a \ket{\Rla{p}}_\cN$ for some $a \in \IC$. Under $\iota$, this gets mapped to the state 
\begin{align}
	\iota(v) \= a^* \ket{\Rla{k-p}}_\cN \in V_\IC^{p,k-p}~.
\end{align}
Therefore $\iota$ indeed maps $V^{p,q}_\IC \to V_\IC^{q,p}$ for any $p$, as required by eq~\eqref{eq:Hodge_structure_complex_conjugation}, verifying that $(V_\IR,V_\IC)$ is a real Hodge structure.

As in section \ref{sect:Rational_Hodge_structs_from_RCFTs}, since $\cN=2$ minimal models are rational, we can construct a rational Hodge structure by defining the rational vector space $V_\IQ$ as the space spanned by the projected boundary states~\eqref{eq:projected_boundary_state_definition} labelled by highest-weight representations $(L,M,S) \in \HWR_{\text{R}}$, that is the labels obey the same conditions~\eqref{eq:minmal_model_HWRs} with $S \in \{-1,1\}$.

In terms of the modular S-matrix values, the correctly normalised projected boundary states are given by~\cite{Brunner:1999jq,Hori:2000ck}
\begin{equation}
	\ket{B_{(L,M,S)}}_{\cP} \=  \sqrt{2} \, \sum\limits_{l =0}^k \frac{\mtS_{(L,M,S),\Rl}}{\sqrt{\mtS_{(0,0,0),\Rl}}} \ket{\Rl} _{\mathcal{N}}~.
\end{equation}
For $\cN=2$ minimal models, it is always possible to choose a subset $\cB \subset \HWR_{\text{R}}$ of representations such that $|\cB|=k+1 = \dim_\IC V_\IC$ and the matrix  $\Sigma$, defined by
\begin{align}
	\Sigma_{(L,M,S),(L',M',S')} \; \defineas \; \BBP{{(L,M,S)}} \me^{\ii \pi J_0} \BP{{(L',M',S')}}~,
\end{align}
is of full rank when $(L,M,S),(L',M',S') \in \cB$ \cite{Brunner:1999jq,Hori:2000ck}. As discussed in footnote \ref{foot:independence_of_boundary_states}, this is enough to show that the set
\begin{equation} \label{eq:mmvqbasis}
	\{ \; \ket{B_{(L,M,S)}}_\mathcal{P} \; \mid \; (L,M,S) \in \mathcal{B} \; \}~,
\end{equation} 
forms a basis of $V_\IQ$.

Recalling the discussion around~\eqref{eq:polarisation_definition}, for $\cN=2$ minimal models, we can construct a polarisation $Q: V_\IQ \to V_\IQ$ from a twisted Witten index by setting, as in eq.~\eqref{eq:polarisation_definition},
\begin{equation} \label{eq:polarisation_minimal_model}
	Q(\ket{\alpha},\ket{\beta}) \; \defineas \; \iota(\bra{\alpha}) \ee^{\ii \pi (k+2) J_{0}}  \ket{\beta}~, \qquad \ket{\alpha}, \ket{\beta} \in V_\IC~,
\end{equation}
if $\cV = \me^{\pi \ii (k+1) J_0}$ is central to the $\cN=2$ super-Virasoro algebra and defines a map $V_\IQ \to V_\IQ$. From section \ref{sect:Rational_Hodge_structs_from_RCFTs}, we know that $\cV$ can only be central if the level $k$ is odd. In these cases $\cV$ indeed gives a well-defined map $V_\IQ \to V_\IQ$, which follows from the formula \cite{Govindarajan:2000my,Hori:2003ic,Hori:2000ck} 
\begin{align}
\me^{\ii (k+1) \pi J_0} \BP{{(L,M,S)}} \= \BP{{(L,M+k+1,S+k+1)}}~,
\end{align}
where we have to take into account the equivalence relation~\eqref{eq:field_identication_equivalence}.

For minimal models with even level $k$, we cannot obtain a polarisation from the Witten index. Nevertheless, for the first few even levels $k$, we have explicitly checked that a polarisation can be constructed. At the end of this section, we will illustrate the construction of a polarisation in the case $k=4$.

\subsection{Hodge substructures and Hodge endomorphisms} \label{sect:minimal_model_Hodge_substructures}
We proceed to explicitly construct the Hodge substructures of the Hodge structure associated with the $\mathcal{N}=2$ minimal models at level $k$. To do so, we first find the field extensions $\IQ(\mtS_{\Rl})$, the stabiliser group $\text{Stab}(\Rl)$ and the fixed field $\IQ(\mtS_\Rl)^{\text{Stab}(\Rl)}$ for every representation $\Rl$ defined in eq.~\eqref{eq:Rl}.\footnote{Note that since $\mathbb{Q}(\mtS)$ is Galois and $\text{Gal}(\mathbb{Q}(\mtS)/\mathbb{Q})$ is Abelian, all subfields are Galois over $\mathbb{Q}$.} The stabiliser group turns out to be trivial in every case, so to illustrate the case with non-trivial stabilisers, in appendix \ref{app:Minimal_Model_NS-states} we consider the case where the boundary states are taken to be in the NS-sector. This analysis ultimately leads to the same conclusion, but involves non-trivial stabiliser groups. 

In order to construct the number field $\IQ( \mtS_\Rl)$, we observe that the modular S-matrix~\eqref{eq:S-matrix_minimal_model} for the representations $\Rl \in \cR$, c.f. eq.~\eqref{eq:Rl}, can be written as
\begin{equation} \label{eq:S-matrix_generators_k}
	\begin{aligned}
		\mtS_{\Rl,(L,M,S)} &\= \frac{(-1)^{\frac{S+1}{2}+1}}{2(k+2)} \left(\zeta_{k+2}^{l+1} \right)^{\frac{L-M+1}{2}} \left( 1 -  \zeta_{k+2}^{-(l+1)(L+1)}\right)~.
	\end{aligned}	
\end{equation}
Since $(L,M,S) \in \HWR_{\text{R}}$, which implies that $S+1$ and $L-M+1$ are even, it is clear from this formula that $\IQ(S_\Rl) \subseteq \IQ(\zeta_{k+2}^{l+1})$. To see that in fact $\IQ(S_\Rl) = \IQ(\zeta_{k+2}^{l+1})$, it is then sufficient to note that
\begin{equation}
	\frac{\mtS_{\Rl,(0,1,1)}}{\mtS_{\Rl,(0,3,1)}}\= \zeta_{k+2}^{l+1}~.
\end{equation}
Thus we have that
\begin{align} \label{eq:kappa_definition}
	\IQ( \mtS_\Rl) \= \IQ(\zeta_\kappa) \qquad \text{with} \qquad \kappa \; \defineas \;  \frac{k+2}{\gcd\left( l+1 , k+2\right)}~.
\end{align}
For the construction of the number field $\IQ( \mtS_{\Rl})^{\text{Stab}(\Rl)}$, we first determine the action of the Galois group  $\Gal( \IQ(\mtS_\Rl)/\IQ)$ on the representations $\Rl \in \cR$. The utilised key results are that the Galois groups of Abelian field extensions generated by the primitive $n$'th root of unity $\zeta_n$ is isomorphic to the multiplicative group of units of the ring $\IZ/n \IZ$, that is to say
\begin{equation} \label{eq:GalCorresp}
	\Gal(\IQ(\zeta_n)/\IQ) \; \simeq \; (\IZ/n \IZ)^\times \ .
\end{equation}
As a consequence of B\'ezout's lemma the elements $[a]$ of $(\IZ/n \IZ)^\times$ are represented by integers $a$ coprime to $n$. Under the isomorphism~\eqref{eq:GalCorresp}, the element $[a] \in (\IZ/n \IZ)^\times$ corresponds to the Galois automorphism $\rho_a \in \Gal(\IQ(\zeta_n)/\IQ)$, which acts on the primitive root of unity $\zeta_n$ as
\begin{equation} \label{eq:minimal_model_rho_definition}
	\rho_a \; : \; \zeta_n \longmapsto \zeta_n^a \ .	
\end{equation}
Let $\rho_a \in \Gal(\IQ(\zeta_{\kappa})/\IQ)$ be the Galois automorphism corresponding to $[a] \in (\IZ/\kappa \IZ)^\times$. It transforms the primitive root $\zeta_{k+2}^{l+1}$ as
\begin{equation}
	\rho_a(\zeta_{k+2}^{l+1}) \= \zeta_{k+2}^{a(l+1)} \= \zeta_{k+2}^{l'+1}~,
\end{equation}
where $l'$ is the unique integer such that
\begin{align} \label{eq:l_prime_definition}
	0 \leq l' \leq k \qquad \text{and} \qquad l' \equiv a(l+1) -1 \!\! \mod k+2~.
\end{align}  
It is a priori clear that such an integer exists in the range $0 \leq l' \leq k+1$. To see that $l' \neq k+1$, note that by definition $a$ is invertible modulo $k{+}2$, so $l' = k{+}1$ would imply $l{+}1 \equiv 0 \mod k{+}2$, which is not possible for $0 \leq l \leq k$. As a consequence, the Galois automorphism $\rho_a$ acts on the modular S-matrix~\eqref{eq:S-matrix_generators_k} as
\begin{equation} \label{eq:minimal_model_S-matrix_Galois_action}
	\rho_a( \mtS_{\Rl,(L,M,S)} ) \=  \mtS_{\Rla{l'},(L,M,S)} \ ,
\end{equation}
which on the set of representations $\cR$ induces the action
\begin{equation} \label{eq:FuncVarSigma}
	\varrho_a:\ \cR \to \cR~, \qquad \Rl \mapsto \Rla{l'} 
\end{equation}
From this we can also easily show that the stabiliser subgroup is $\text{Stab}(\Rl)$ trivial. For a given representation $\Rl \in \cR$ we need to calculate the conjugacy classes  $[a] \in (\IZ/ \kappa \IZ)^\times$ with $\varrho_a(\Rl) = \Rl$, which, according to eq.~\eqref{eq:l_prime_definition}, amounts to solving the equation
\begin{equation} \label{eq:stabiliser_congruence}
	(a-1)(l+1) \= \lambda (k+2) \ \Longleftrightarrow \
	(a-1)\frac{(l+1)}{\gcd(l+1,k+2)} \= \lambda \frac{(k+2)}{\gcd(l+1,k+2)} 
	\quad \text{for $\lambda\in \IZ$} \ .
\end{equation}
Any solution to this equation is for the form
\begin{equation}
	a-1 \= m \frac{k+2}{\gcd(l+1,k+2)}
	\ \Longleftrightarrow \
	a \= 1+ m\,\kappa  
	\quad \text{with $m \in \IZ$} \ .
\end{equation}
As the classes $[a]$ are represented by integers $a$ modulo $\kappa$, $[a] = [1+\kappa] = [1]$ and the stabiliser group $\text{Stab}(\Rl)$ is is trivial. Consequently, the invariant number field $\IQ( \mtS_{\Rl})^{\text{Stab}(\Rl)}$ is given, in all cases, by
\begin{equation}
	\IQ( \mtS_{\Rl})^{\text{Stab}(\Rl)} \= \IQ( \mtS_{\Rl}) \= \IQ(\zeta_{\kappa}) \ ,
\end{equation}
and the conveniently scaled vectors $\ket{\Rl}_\cS$ defined in eq.~\eqref{eq:R_state_generator_expansion} in terms of the states $\ket{\Rl}_\cI$ (see eq.~\eqref{eq:intermediate_scaled_states}) are given simply by
\begin{equation} \label{eq:scaledminimalmodel}
	\ket{\Rl}_\cS \; \defineas \; \ket{\Rl}_\mathcal{I} \in V_\mathbb{Q} \otimes_\mathbb{Q} \mathbb{Q}(\zeta_{\kappa})
\end{equation}
On these states, the field automorphisms $\theta_a \in \Gal(\IQ(\zeta_{\kappa})/\IQ)$ corresponding to the elements $[a] \in (\IZ/\kappa\IZ)^\times$ act as
\begin{align} \label{eq:minimal_model_states_Galois_transformation}
	\theta_a \ket{\Rl} \= \ket{\vartheta_a(\Rl)} = \ket{\Rla{l'}}~, \quad \text{with} \quad 0 \leq l' \leq k~\text{ and }~l' \equiv a(l+1) -1 \mod k+2~.
\end{align}

\subsubsection*{Hodge substructures and their endomorphisms}
\vskip-5pt
The rest of the proof follows now exactly as in section \ref{sect:CM_Hodge_Structures_from_RCFTs}. To keep this section self-contained, we retrace the main steps of the proof here. More details on the argument made here can be found in section~\ref{sect:CM_Hodge_Structures_from_RCFTs}. We first find the Hodge substructure $W^{\Rl}$ corresponding to the representation $\Rl$. As in eq.~\eqref{eq:R_state_generator_expansion}, we can now expand the rescaled vector eq.~\eqref{eq:scaledminimalmodel} in powers of $\zeta_\kappa$:
\begin{equation} \label{eq:minimal_model_state_generator_expansion}
	\ket{\Rla{l}}_\mathcal{S} \= \sum_{m=0}^{d-1} \zeta_\kappa^{m} \, \ket{Q_m^{\Rl}} \, ,
\end{equation}
where $\ket{Q_m^{\Rl}} \in V_\mathbb{Q}$ and $d \defineas [\IQ(\zeta_\kappa):\IQ]$. Following eq.~\eqref{eq:W_Q_first_expression}, we then define the subspaces
\begin{equation} \label{eq:WQbasis}
	W_\mathbb{Q}^{\Rl} \; \defineas \; \left\langle \; \ket{Q_0^{\Rl}}, \dots, \ket{Q_{d-1}^{\Rl}} \;  \right \rangle_\mathbb{Q} \subset V_\IQ~.
\end{equation}
From the expansion \eqref{eq:minimal_model_state_generator_expansion}, it is clear that the complexification $W_\IC^{\Rl}$ of this space contains all Galois conjugates $\theta \ket{\Rl}_\mathcal{S}$ with $\theta \in \text{Gal}(\IQ(\zeta_\kappa)/\IQ)$. We have seen above in eq.~\eqref{eq:minimal_model_states_Galois_transformation}, that each Galois conjugate corresponds to a different representation, and hence $\theta \ket{\Rl}_\mathcal{S}$ are linearly independent over $\IC$. There are exactly $d = [\IQ(\zeta_\kappa):\IQ] = |\Gal(\IQ(\zeta_\kappa)/\IQ)|$ such states, so they form a basis of the complexification:
\begin{equation} \label{eq:W_C_basis}
	W_{\mathbb{C}}^{\Rl} \= \left \langle \; \theta \ket{\Rl}_\mathcal{S} \; \right \rangle^{\theta \in \text{Gal}(\mathbb{Q}(\zeta_\kappa)/\mathbb{Q})}_\mathbb{C}~.
\end{equation}  
From eq.~\eqref{eq:minimal_model_states_Galois_transformation}, comparing to the definition \eqref{eq:mmvpq} of the subspaces $V_\IC^{p,q}$, it also follows that each Galois conjugate $\theta \ket{\Rl}_\mathcal{S}$ lies in one of these spaces. Thus, $W_\mathbb{Q}^{\Rl}$ gives a Hodge substructure.

Finally, we establish the existence of sufficiently many Hodge endomorphisms on these Hodge substructures. We do this by considering the multiplication by $\zeta_\kappa$ on $\IQ(\zeta_\kappa)$. Recall that the field extension $\mathbb{Q}(\zeta_\kappa)$ can be regarded as a $\mathbb{Q}$-vector space with the basis 
\begin{equation}
	\mathbb{Q}(\zeta_\kappa) \= \braket{1, \zeta_\kappa, \dots, \zeta_\kappa^{d-1}}_\mathbb{Q}~.
\end{equation} 
Multiplication by $\zeta_\kappa$ is a linear map on this $\IQ$-vector space that is described by 
\begin{equation}
	\zeta_\kappa^m \mapsto \zeta_\kappa^{m+1} \; \asdefine \; \sum_{n=0}^{d-1} \mtT(\zeta_\kappa)_{\phantom{m}n}^m \, \zeta_\kappa^n~,
\end{equation}
where the matrix $\mtT(\zeta_\kappa) \in \text{Mat}(d {\times} d, \IQ)$ defined by the equality is called \textit{the companion matrix} of the minimal polynomial of $\zeta_\kappa$. 

We can now define an endomorphism $\varphi^{\Rl}_{\zeta_\kappa}$ on the rational vector space $W_\mathbb{Q}^{\Rl}$ by defining its action on the basis in eq.~\eqref{eq:WQbasis} as
\begin{equation} \label{eq:Hodge_endomorphism_varphi_zeta_definition}
	\varphi^{\Rl}_{\zeta_\kappa}  \left(\ket{Q_m^{\Rl}} \right) \; \defineas \; \mtT(\zeta_\kappa)^{s}_{\phantom{s}m}  \ket{Q_s^{\Rl}}~.
\end{equation}
A small amount of algebra reveals the action of the complex linear extension $\varphi^{\Rl}_{\zeta_\kappa,\IC}$ on the basis of the complexification $W_\IC^{\Rl}$ given in eq.~\eqref{eq:W_C_basis}. For any $\theta \in \Gal(\IQ(\zeta_\kappa)/\IQ)$,
\begin{equation} \label{eq:Hodge_endomorphism_action_minimal_model}
	\begin{aligned}
		\varphi^{\Rl}_{\zeta_\kappa, \mathbb{C}}  \left( \theta \ket{\Rl}_\mathcal{S}\right) =  \sum_{m,n=0}^{d-1} \theta \left(\mtT(\zeta_\kappa)_{\phantom{m}n}^m \, \zeta_\kappa^{n}\right)   \ket{Q_m^{\Rl}} =  \theta(\zeta_\kappa) \sum_{m=0}^{d-1} \theta \left(  \zeta_\kappa^{m}\right)   \ket{Q_m^{\Rl}} = \theta (\zeta_\kappa) \theta \ket{\Rl}_\mathcal{S}~.
	\end{aligned}
\end{equation}
We conclude that, in the basis of eq.~\eqref{eq:W_C_basis}, the complex linear extension $\varphi^{\Rl}_{\zeta_\kappa, \IC}$ is diagonalised. In particular, it preserves the Hodge decomposition,
\begin{equation}
	\varphi^{\Rl}_{\zeta_\kappa, \IC} \left(W_{\IC}^{\Rl, (p,q)} \right) \subseteq W_{\IC}^{\Rl, (p,q)}~,
\end{equation} 
and is therefore a Hodge endomorphism. It generates a commutative algebra $E^{\Rl}$ of Hodge endomorphisms isomorphic to $\IQ(\zeta_\kappa)$, as can be seen by noting that
\begin{align}
	\alpha \varphi^{\Rl}_{\zeta_\kappa} &\= \varphi^{\Rl}_{\alpha \zeta_\kappa}~, \qquad
	\left(\varphi^{\Rl}_{\zeta_\kappa} \right)^n \= \varphi^{\Rl}_{\zeta_\kappa^n}~, \qquad \varphi^{\Rl}_{\alpha \zeta_\kappa^n} + \varphi^{\Rl}_{\beta \zeta_\kappa^m} \= \varphi^{\Rl}_{\alpha \zeta_\kappa^n + \beta \zeta_\kappa^m}~
\end{align}
for all $\alpha \in \IQ$, $m,n \in \IZ$. From this, we see that the natural map
\begin{align}
	\varphi^{\Rl}_{\zeta_\kappa} \; \longmapsto \; \zeta_\kappa
\end{align}
gives in fact rise to an isomorphism between $E^{\Rl}$ and $\IQ(\zeta_\kappa)$. We therefore conclude that the Hodge structures constructed in this section for the diagonal $\mathcal{N}=(2,2)$ minimal models admit sufficiently many complex multiplications.

For odd values of $k$, we have seen above (see eq.~\eqref{eq:polarisation_minimal_model}) that a polarisation can always be constructed, implying that the corresponding Hodge structure is of CM type.

\subsection{Examples} \label{sect:minimal_model_examples}
To make the discussion in this section even more concrete, we work out few examples in detail for a fixed level $k$. We start with the $k=3$ minimal model for which we construct the Hodge structure in detail, and show how the Galois action works. This example will also be of central importance to the discussion of Gepner models in the following section \ref{sec:exgepnermodels}. For $k=4$ and $k=6$ we concentrate on studying polarisability of the associated Hodge structure, constructing a polarisation explicitly for these minimal model. Finally, we list a few more examples for low levels $k$, for which we give the central data associated to the Hodge structures.
\subsubsection*{Example: $k=3$}
\vskip-10pt
Let us investigate the Hodge structure for the $k=3$ minimal model explicitly. The complex vector space $V_{\mathbb{C}}$ defined in eq.~\eqref{eq:V_C_definition} is given by
\begin{align} \label{eq:V_C_k=3_minimal_model}
	\begin{split}
	V_{\mathbb{C}} \= \langle \; \ket{\Rla{0}}_{\mathcal{S}}, \ket{\Rla{1}}_{\mathcal{S}}, \ket{\Rla{2}}_{\mathcal{S}}, \ket{\Rla{3}}_{\mathcal{S}} \; \rangle_\IC~,
	\end{split}
\end{align}
with the graded pieces given by
\begin{align}
\ket{\Rla{0}}_\cS \in V_\IC^{3,0}~, \quad \ket{\Rla{1}}_\cS \in V_\IC^{2,1}~, \quad \ket{\Rla{2}}_\cS \in V_\IC^{1,2}~, \quad \ket{\Rla{3}}_\cS \in V_\IC^{0,3}~.
\end{align}
We can pick $\cB = \cR$ so that the rational vector space $V_\mathbb{Q}$ of eq.~\eqref{eq:V_Q_definition} can be written as
\begin{equation} \label{eq:qbasisk=3}
V_\mathbb{Q} = \langle \; \ket{B_{\Rla{0}}}_{\cP}, \ket{B_{\Rla{1}}}_{\cP},\ket{B_{\Rla{2}}}_{\cP},\ket{B_{\Rla{3}}}_{\cP} \; \rangle_\IQ~,
\end{equation}
With this choice, the intersection matrix is given by
\begin{equation}
	\Sigma \= \begin{pmatrix}
		1 & 1 & 1 & 1  \\
		0 & 1 & 1 & 1  \\
		0 & 0 & 1 & 1  \\
		0 & 0 & 0 & 1  \\
	\end{pmatrix}.
\end{equation}
In the rational basis of eq.~\eqref{eq:qbasisk=3} the polarisation is given by
\begin{equation}
Q(\; \ket{B_A}_\mathcal{P}, \ket{B_B}_\mathcal{P}) \; \asdefine \; \mtQ_{A,B} \quad \text{with} \qquad \mtQ \= \begin{pmatrix}0&0&-1&0 \\ 0&0&-1&-1\\1&1&0&0\\0&1&0&0  \end{pmatrix}.
\end{equation}
Since $k+2 = 5$ is prime for $k=3$, for every $l$ the field $\IQ( \mtS_\Rl)$ is simply the cyclotomic field generated by the fifth root of unity:
\begin{align}
	\IQ( \mtS_\Rl) \= \IQ(\zeta_\kappa) \= \IQ(\zeta_5)~.
\end{align}
The minimal polynomial of the generator $\zeta_5$ is
\begin{align}
	1 + x + x^2 + x^3 + x^4~,
\end{align}
so the field extension $\IQ(\zeta_5)/\IQ$ has degree $[\IQ(\zeta_5):\IQ] = 4$, and its Galois group is the cyclic group $(\IZ/5 \IZ)^\times \simeq \IZ_4 \simeq C_4$, which can be taken to be generated by the field automorphism~${\theta_2: \zeta_5 \mapsto \zeta_5^2}$.

As in eq.~\eqref{eq:minimal_model_state_generator_expansion}, the R-ground states $\ket{\Rl}$ can be expanded in terms of rational vectors $\ket{Q_k^{\Rl}}$, which are defined so that
\begin{align}
	\ket{\Rl}_{\mathcal{S}}  \= \sum_{k=0}^3 \zeta_5^k \ket{Q^{\Rl}_k} \in V_\IQ \otimes_\IQ \IQ(\zeta_5) ~.
\end{align}
For instance, the states $\ket{Q_k^{\Rla{0}}}$ can be expressed in terms of the boundary states of eq.~\eqref{eq:qbasisk=3} as 
\begin{equation}
\begin{aligned}
\label{eq:k=3rrdiagbasis}
\ket{Q^{\Rla{0}}_0} &\= 
   \tfrac{1}{10} \ket{B_{\Rla{0}}}_{\cP} + \tfrac{1}{10} \ket{B_{\Rla{3}}}_{\cP}\ , \\  
\ket{Q^{\Rla{0}}_1}&\= \tfrac{1}{10} \ket{B_{\Rla{0}}}_{\cP} + \tfrac{1}{10} \ket{B_{\Rla{2}}}_{\cP}-\tfrac{1}{10} \ket{B_{\Rla{3}}}_{\cP} \ ,  \\  
\ket{Q^{\Rla{0}}_2}&\= \tfrac{1}{10} \ket{B_{\Rla{0}}}_{\cP} + \tfrac{1}{10} \ket{B_{\Rla{1}}}_{\cP}- \tfrac{1}{10}  \ket{B_{\Rla{2}}}_{\cP}\ , \\
\ket{Q^{\Rla{0}}_3}&\= \tfrac{1}{5} \ket{B_{\Rla{0}}}_{\cP} - \tfrac{1}{10} \ket{B_{\Rla{1}}}_{\cP}  \ .
\end{aligned}
\end{equation}
Picking now $\ket{Q^{\Rla{0}}_k}$ as the basis vectors, the scaled R-ground states $\ket{\Rl}_\cS$ can be written as
\begin{align}
 	\ket{\Rla{0}}_{\mathcal{S}} \= 
 		\left( \begin{matrix}
		1 \\ \zeta_5 \\ \zeta_5^2 \\ \zeta_5^3
		\end{matrix} \right), \qquad 
	\ket{\Rla{1}}_{\mathcal{S}} \= 
		\left( \begin{matrix}
		1 \\ \zeta_5^2 \\ \zeta_5^4 \\ \zeta_5
		\end{matrix} \right), \qquad 
	\ket{\Rla{2}}_{\mathcal{S}} \= 
		\left( \begin{matrix}
		1 \\ \zeta_5^3 \\ \zeta_5 \\ \zeta_5^4
		\end{matrix}  \right)~, \qquad
 	\ket{\Rla{3}}_{\mathcal{S}} \= 
 		\left( \begin{matrix}
		1 \\ \zeta_5^4 \\ \zeta_5^3 \\ \zeta_5^2
		\end{matrix} \right),	
\end{align}
From this we can immediately read off the action of $\theta_2$ on the vectors $\ket{\Rl}_\cS$:
\begin{align} \nonumber
	\theta_2 \ket{\Rla{0}}_\cS = \ket{\Rla{1}}_\cS~,~~ 	\theta_2 \ket{\Rla{1}}_\cS = \ket{\Rla{3}}_\cS~,~~ 	\theta_2 \ket{\Rla{2}}_\cS  = \ket{\Rla{0}}_\cS~,~~ 	\theta_2 \ket{\Rla{3}}_\cS = \ket{\Rla{2}}_\cS~.
\end{align}
which agrees with the action \eqref{eq:minimal_model_states_Galois_transformation} derived in the previous section. In this way we also see that the Hodge structure is irreducible, since (cf. eq.~\eqref{eq:W_C_basis})
\begin{align}
	V_ \IC \= \langle \; \theta (\ket{\Rla{0}}_{\mathcal{S}}) \; \rangle_\IC^{\theta \in \text{Gal}(\IQ(\zeta_5)/\IQ)} \= W^{\Rla{0}}_\IC~,
\end{align}
and have previously shown that the Hodge substructures corresponding to $W_\IC^{\Rla{0}}$ are irreducible.

The matrix $\mtT(\zeta_5)$ representing the multiplication by $\zeta_5$ in $\IQ(\zeta_5)$ is given, in the basis $\{1,\zeta_5,\zeta_5^2,\zeta_5^3\}$,~as
\begin{align}
	\mtT(\zeta_5) \= \left(
	\begin{array}{cccc}
		0 & 1 & 0 & 0 \\
		0 & 0 & 1 & 0 \\
		0 & 0 & 0 & 1 \\
		-1 & -1 & -1 & -1 \\
	\end{array}
	\right).
\end{align}
By the previous discussion, $\mtT(\zeta_5)$ represents a Hodge endomorphism $\varphi^{\Rla{0}}_{\zeta_5}: V_\IQ \to V_\IQ$ (cf. eq.~\eqref{eq:Hodge_endomorphism_varphi_zeta_definition}). A direct computation reveals that it acts on the basis of eq.~\eqref{eq:V_C_k=3_minimal_model} as
\begin{align}
	\begin{split}
\mtT(\zeta_5)  \ket{\Rla{0}}_{\mathcal{S}} &\= \zeta_5^{\phantom{4}} \ket{\Rla{0}}_{\mathcal{S}}~, \qquad 
\mtT(\zeta_5)  \ket{\Rla{1}}_{\mathcal{S}} \= \zeta_5^2 \ket{\Rla{1}}_{\mathcal{S}}~,\\[5pt]
\mtT(\zeta_5)  \ket{\Rla{2}}_{\mathcal{S}} &\= \zeta_5^3 \ket{\Rla{2}}_{\mathcal{S}}~, \qquad
\mtT(\zeta_5)  \ket{\Rla{3}}_{\mathcal{S}} \= \zeta_5^4 \ket{\Rla{3}}_{\mathcal{S}}~,  
	\end{split}
\end{align}
which is exactly the result \eqref{eq:Hodge_endomorphism_action_minimal_model}, and shows that the Hodge type is indeed preserved by the complexification of this endomorphism. A simple computation also shows that $\mtI, \mtT(\zeta_5), \mtT(\zeta_5)^2$, and $\mtT(\zeta_5)^3$ are independent over $\IQ$, and that the minimal polynomial of $\mtT(\zeta_5)$ is
\begin{align}
	\mtI + \mtT(\zeta_5) +\mtT(\zeta_5)^2 + \mtT(\zeta_5)^3 + \mtT(\zeta_5)^4 \= 0~.	
\end{align}
These powers of $\mtT(\zeta_5)$ obviously commute with each other, so $\mtT(\zeta_5)$ generates over $\IQ$ a commutative subalgebra $E^{\Rla{0}} \subseteq \text{End}_\text{Hdg}(V_\IQ)$ of the algebra of Hodge endomorphisms. The Hodge structure $(V_\IQ,V_\IC)$ is of CM type as $E^{\Rla{0}}$ is in fact isomorphic to $\IQ(\zeta_5)$ with the isomorphism given by
\begin{align}
	\mtT(\zeta_5) \; \longmapsto \; \zeta_5~.
\end{align}

\subsubsection*{Existence of polarisation for $k=4$ and $k=6$}
\vskip-5pt
In this section we show that the Hodge structure corresponding to the $k=4$ minimal model is polarisable even though we have seen above that we cannot construct a polarisation in this case from the twisted Witten index \eqref{eq:polarisation_minimal_model}. However, we can explicitly construct a family of polarisations. 

In the case $k=4$, the complex vector space $V_\IC$ and its graded pieces are given by
\begin{align} \label{eq:k=4_V_C_basis}
	V_{\mathbb{C}} \= \langle \; \ket{\Rl}_\cN \; \rangle_\IC^{l  \in \{0,1, \dots, 4 \}}~, \qquad \text{with} \qquad \ket{\Rla{l}}_\cN \in V_\IC^{k-l,l}~.
\end{align}
As before, we can choose $\cB = \cR$ so that the rational vector space $V_\IQ$ is given by
\begin{equation}
V_\mathbb{Q} \= \langle \; \ket{B_\Rl}_\cP \; \rangle_\mathbb{Q}^{l \in \{0,\dots,4\}}~,
\end{equation}
and the intersection matrix is again an upper triangular matrix
\begin{equation}
\Sigma \= \begin{pmatrix}
	1 & 1 & 1 & 1 & 1 \\
	0 & 1 & 1 & 1 & 1 \\
	0 & 0 & 1 & 1 & 1 \\
	0 & 0 & 0 & 1 & 1 \\
	0 & 0 & 0 & 0 & 1
\end{pmatrix}.
\end{equation}
To construct a polarisation, we find the most general bilinear form $Q_\IC$ on $V_\IC$ that satisfies the conditions~\eqref{eq:polarisation_1}-\eqref{eq:polarisation_3}. In the basis $\{\ket{\Rla{0}}_\cN,\dots,\ket{\Rla{4}}_\cN\}$, any such form is associated to a~matrix of the form
\begin{equation} \label{eq:V_C_bilinear_form_matrix}
	\mtQ \= \begin{pmatrix}
		0&0&0&0&a \\ 
		0&0&0&-b&0\\
		0&0&c&0&0\\
		0&-b&0&0&0\\
		a&0&0&0&0\\  
	\end{pmatrix}, \qquad \text{with} \qquad a,b,c \; > \; 0~.
\end{equation}
This gives a polarisation if and only if it arises as a complex bilinear extension of a bilinear form on $V_\IQ$. Therefore we check if we can find values for the coefficients $a,b,c$ such that this condition is satisfied. Consider the change-of-basis matrix from the boundary basis $\{\ket{B_\Rla{0}},\dots,\ket{B_\Rla{4}}\}$ to the normalised basis of R-states $\{\ket{\Rla{0}}_\cN,\dots,\ket{\Rla{4}}_\cN\}$, which takes the form
\begin{align}
	\mtB \= \frac{1}{\sqrt{6}}
	\begin{pmatrix}
		\+\;\,-\ii \zeta_{12} &
		-\ii \sqrt{3}\,\zeta_{12}^2 &
		2 &
		-\ii \sqrt{3}\,\zeta_{12}^4 &
		\+\;\,-\ii \zeta_{12}^5 \\[6pt]
		-\ii \sqrt[4]{3}\,\zeta_{12}^2 &
		-\ii \sqrt[4]{3}\,\zeta_{12}^4 &
		0 &
		\+\ii \sqrt[4]{3}\,\zeta_{12}^8 &
		\+\ii \sqrt[4]{3}\,\zeta_{12}^{10} \\[6pt]
		\sqrt{2} &
		0 &
		\sqrt{2} &
		0 &
		\sqrt{2} \\[6pt]
		-\ii \sqrt[4]{3}\,\zeta_{12}^4 &
		\+\ii \sqrt[4]{3}\,\zeta_{12}^8 &
		0 &
		-\ii \sqrt[4]{3}\,\zeta_{12}^4 &
		\+\ii \sqrt[4]{3}\,\zeta_{12}^8 \\[6pt]
		\+\;\,-\ii \zeta_{12}^5 &
		\+\ii \sqrt{3}\,\zeta_{12}^{10} &
		2 &
		\+\ii \sqrt{3}\,\zeta_{12}^8 &
		\+\;\,-\ii \zeta_{12}
	\end{pmatrix} \ .
\end{align}
The matrix $\mtQ$ in the basis $\{\ket{B_\Rla{0}},\dots,\ket{B_\Rla{4}}\}$ is thus given by
\begin{align} \label{eq:Q_matrix_rational_basis}
	\mtB^T \mtQ \mtB \= \frac{1}{6}
	\begin{pmatrix}
		2 a - 2 \sqrt{3} b + 2 c & 3 a - \sqrt{3} b & 2 a + 2 c & -2 \sqrt{3} b & -a - \sqrt{3} b + 2 c \\[5pt]
		3 a - \sqrt{3} b & 6 a - 2 \sqrt{3} b & 6 a & 3 a - \sqrt{3} b & -2 \sqrt{3} b \\[5pt]
		2 a + 2 c & 6 a & 4 a + 2 c & 6 a & 2 a + 2 c \\[5pt]
		-2 \sqrt{3} b & 3 a - \sqrt{3} b & 6 a & 6 a - 2 \sqrt{3} b & 3 a - \sqrt{3} b \\[5pt]
		-a - \sqrt{3} b + 2 c & -2 \sqrt{3} b & 2 a + 2 c & 3 a - \sqrt{3} b & 2 a - 2 \sqrt{3} b + 2 c
	\end{pmatrix} \ .
\end{align}
The corresponding bilinear form $Q_\IC$ restricts to a bilinear form $Q$ on $V_\IQ$ if all entries of the above matrix are in $\IQ$, which is the case if and only if $a,c \in \mathbb{Q}$ and $b \in \sqrt{3} \IQ$. Since we are always free to pick such constants, we have shown that the Hodge structure for the $k=4$ model is polarisable.

We could have also argued for the existence of the polarisation by noting that the Hodge structure of the $k=4$ minimal model splits into three Hodge substructures 
\begin{align}
	W_\IC^\Rla{0} = \langle \; \ket{\Rla{0}}_\cN~, \ket{\Rla{4}}_\cN \, \rangle_\IC~, \hskip7pt W_\IC^\Rla{1} = \langle \; \ket{\Rla{1}}_\cN~, \ket{\Rla{3}}_\cN \,  \rangle_\IC~, \hskip7pt
	W_\IC^\Rla{2} = \langle \; \ket{\Rla{2}}_\cN \,  \rangle_\IC~, \quad
\end{align}
and recalling that these one- and two-dimensional Hodge substructures are always polarisable. However, the argument presented in this section can be generalised mutatis mutandis for more complicated Hodge structures, such as those arising from the case $k=6$ with the only essential difference being that the analogous of the matrix \eqref{eq:Q_matrix_rational_basis} is valued in a higher-degree field extension of $\IQ$, and to ensure that $Q_\IC$ restricts to a bilinear form on $V_\IQ$, the entries of $\mtQ$ have to be chosen to be suitable elements of this extension of $\IQ$.

For instance, in the case $k=6$, we find that the most general form of $\mtQ$ is given by
\begin{align} \notag
\mtQ \=
{\footnotesize
\left(
\begin{array}{ccccccc}
	0 & 0 & 0 & 0 & 0 & 0 & 2 a_1 \sin\!\frac{\pi}{8} {+} 2 a_2 \cos\!\frac{\pi}{8} \\
	0 & 0 & 0 & 0 & 0 & \sqrt{2} b & 0 \\
	0 & 0 & 0 & 0 & 2 a_1 \cos\!\frac{\pi}{8} {-} 2 a_2 \sin\!\frac{\pi}{8} & 0 & 0 \\
	0 & 0 & 0 & d & 0 & 0 & 0 \\
	0 & 0 & 2 a_1 \cos\!\frac{\pi}{8} {-} 2 a_2 \sin\!\frac{\pi}{8} & 0 & 0 & 0 & 0 \\
	0 & \sqrt{2} b & 0 & 0 & 0 & 0 & 0 \\
	2 a_1 \sin\!\frac{\pi}{8} {+} 2 a_2 \cos\!\frac{\pi}{8} & 0 & 0 & 0 & 0 & 0 & 0
\end{array}
\right)
}\ ,
\end{align}
with conditions that $a_1,a_2,b,d \in \IQ$ and that $a_1$ and $a_2$ are chosen so that
\begin{equation}
	a_2 < 0 \ , \qquad a_2 \cot\frac{\pi}{8} < a_1 < -a_2 \tan\frac{\pi}{8} \ .
\end{equation}

\subsubsection*{Further examples}
Further examples can easily be worked out in a manner completely analogous to the two examples presented above. We take in all cases $\cB = \cR = \{\Rla{0},\dots,\Rla{k}\}$, which gives the intersection~matrix
\begin{align}
	\Sigma \= 
\begin{pmatrix}
	1 & 1 & 1 & \cdots & 1 \\
	0 & 1 & 1 & \cdots & 1 \\
	0 & 0 & 1 & \cdots & 1 \\
	\vdots & \vdots & \vdots & \ddots & \vdots \\
	0 & 0 & 0 & \cdots & 1
\end{pmatrix}.
\end{align}
The Hodge structure $(V_\IQ,V_\IC)$ splits into substructures $W_\IQ^{\Rl}$ whose complexifications $W_\IC^\Rl$ are of the form
\begin{align}
	W_\IC^{\Rl} \= \langle \; \ket{\Rla{i_1}}_\cS, \dots \ket{\Rla{i_{\phi(\kappa)}}} \; \rangle_\IC~,
\end{align}
where $\phi(\kappa)$ is the Euler phi-function giving the degree of the field extension ${\IQ(\zeta_\kappa) = \IQ(\mtS_\Rl)^{\text{Stab}(\Rl)}}$. Recall that the integer $\kappa$ depends on $k$ and $l$ via eq.~\eqref{eq:kappa_definition}. The factorisations are indicated in table \ref{tab:minimal_model_examples}, where the index sets $\{i_1,\dots,i_{\phi(\kappa)}\}$ corresponding to each irreducible Hodge substructure are displayed.

\begin{table}[H]
	\begin{center}
		\renewcommand{\arraystretch}{1.35}
		\begin{tabular}{|c|l|l|l|}
			\hline
			\multicolumn{1}{|c|}{$k$} & \multicolumn{1}{c|}{$l$} & \multicolumn{1}{c|}{$\{i_1,\dots,i_{\phi(\kappa)}\}$} & \multicolumn{1}{c|}{$\kappa$} \\ \hline \hline
			\multirow{3}{*}{13} 
			& 0 & $\{0, 1, 3, 6, 7, 10, 12, 13\}$ & 15 \\ \cline{2-4} 
			& 2 & $\{2, 5, 8, 11\}$ &  5 \\ \cline{2-4} 
			& 4 & $\{4, 9\}$ &  3 \\ 
			\hline \hline
			\multirow{1}{*}{11} 
			& 0 & $\{0, 1, 2, 3, 4, 5, 6, 7, 8, 9, 10, 11\}$ & 13 \\ \cline{2-4} 
			\hline \hline 			
			\multirow{1}{*}{9} 
			& 0 & $\{0, 1, 2, 3, 4, 5, 6, 7, 8, 9\}$ & 11 \\ \cline{2-4} 
			\hline \hline
			\multirow{3}{*}{8} 
			& 0 & $\{0, 2, 6, 8\}$ 	&  10\\ \cline{2-4} 
			& 1 & $\{1, 3, 5, 7\}$ 	&  5 \\ \cline{2-4}			
			& 4 & $\{4\}$	 		&  1 \\ \cline{2-4} 
			\hline \hline 				
			\multirow{2}{*}{7} 
			& 0 & $\{0, 1, 3, 4, 6, 7\}$ & 9 \\ \cline{2-4} 
			& 2 & $\{2, 5\}$ &  3 \\ \cline{2-4} 
			\hline								
		\end{tabular}
	\hskip5pt
		\begin{tabular}{|c|l|l|l|}
			\hline
			\multicolumn{1}{|c|}{$k$} & \multicolumn{1}{c|}{$l$} & \multicolumn{1}{c|}{$\{i_1,\dots,i_{\phi(\kappa)}\}$} & \multicolumn{1}{c|}{$\kappa$} \\ \hline \hline
			\multirow{3}{*}{6} 
			& 0 & $\{0, 2, 4, 6\}$ &  8\\ \cline{2-4} 
			& 1 & $\{1, 5\}$ &  4 \\ \cline{2-4}			
			& 3 & $\{3\}$	 &  2 \\ \cline{2-4} 
			\hline \hline 					
			\multirow{1}{*}{5} 
			& 0 & $\{0, 1, 2, 3, 4, 5\}$ & 7 \\ \cline{2-4} 	
			\hline \hline		
			\multirow{3}{*}{4} 
			& 0 & $\{0, 4\}$ &  6\\ \cline{2-4} 
			& 1 & $\{1, 3\}$ &  3 \\ \cline{2-4}			
			& 2 & $\{2\}$	 &  2 \\ \cline{2-4} 
			\hline \hline 
			\multirow{1}{*}{3} 
			& 0 & $\{0, 1, 2, 3\}$ &  5\\ \cline{2-4} 
			\hline \hline 
			\multirow{2}{*}{2} 
			& 0 & $\{0, 2\}$ &  4\\ \cline{2-4} 
			& 1 & $\{1\}$ &  2\\ \cline{2-4} 						
			\hline 										
		\end{tabular}	
	\vskip10pt
	\capt{5.4in}{tab:minimal_model_examples}{The data associated to the Hodge structures constructed for minimal models with different levels $k$. The second column displays labels $l$ associated to each irreducible Hodge substructures $W_\IQ^\Rla{l'}$, which we take to be the smallest value of $l$ for which $\ket{\Rl}_\cS \in W_\IC^{\Rla{l'}}$. The third column displays all values of $l$ such that $\ket{\Rl}_\cS \in W_\IC^{\Rla{l'}}$, while the last column gives the integer $\kappa$, which determines the number field $\IQ(\mtS_\Rl) = \IQ(\zeta_\kappa)$.}
	\end{center}
\end{table}

\newpage
\section{Example: Gepner Models} \label{sec:exgepnermodels} 
\vskip-10pt
Having illustrated the construction of Hodge structures of CM type for $\mathcal{N}=2$ minimal models, we now study Gepner models \cite{Gepner:1987qi,Gepner:1989gr}, which are closely related to the minimal models, but admit geometric realisations as fixed points of non-linear sigma models on Calabi--Yau threefolds. The first example is the Gepner model which corresponds to the Fermat quintic. This model is built out of a tensor product of minimal models of odd levels, so it can be verified that it has an algebra $\cA$ which satisfies the properties (i)-(iv) set out on page~\pageref{p:assumptions}. Therefore we can readily apply to proof of section \ref{sect:Hodge_structures_from_RCFT} needing only to check the compatibility condition \eqref{eq:compatibility_R}. 

We include a second Gepner model to illustrate that the construction presented in this paper can in fact be easily extended further. The second example does not have an algebra satisfying assumption (iv), due to existence of multiple copies of isomorphic representations $i \in \HWR$ with $\text{(c,c)}_{q_L=q_R}$-states in the associated Hilbert spaces $\cH_i$ for every algebra $\cA$ satisfying the other assumptions (i)-(iii). In such a case, to construct enough boundary states, the isomorphic representations need to be ``resolved'', and one need to study boundary states constructed using a generalised $\mtS$-matrix \cite{Fuchs:1996dd,Fuchs:1999zi,Fuchs:1999xn,Fuchs:2000gv}. However, our construction of section \ref{sect:Hodge_structures_from_RCFT} goes through for these generalised quantities with minimal modifications, as will be demonstrated in section \ref{sect:Gepner_model_octic}

\subsection{The Gepner model \texorpdfstring{$(3,3,3,3,3)$}{(3,3,3,3,3)} corresponding to the Fermat quintic threefold}
\label{sect:Fermat_quintic}
The first example we consider is the Gepner model corresponding to the conformal fixed point of the supersymmetric non-linear sigma model on the Fermat quintic threefold. This Gepner model of total central charge $c=9$ is a \textit{simple current extension} of a tensor product $(k=3) \otimes \dots \otimes (k=3)$ of five copies of A-type $\mathcal{N}=2$ minimal models at level $k=3$. 
 
The chiral algebra of the tensor product theory contains the product algebra $\cA_1 \times \dots \times \cA_5$, where $\cA_i$ refers to the chiral algebra of the $i$'th tensor product factor, each isomorphic to eq.~\eqref{eq:cosetalgebra} with $k=3$. In particular, the chiral algebra contains the fields
\begin{equation} \label{eq:QuinticTJ}
  T(z)  \; \defineas \; \sum_{i=1}^5 \mtI \otimes \cdots \otimes  T_i(z) \otimes  \cdots \otimes \mtI \ , \qquad
  J(z)  \; \defineas \; \sum_{i=1}^5 \mtI \otimes \cdots \otimes  J_i(z) \otimes  \cdots \otimes \mtI
\end{equation} 
corresponding to the energy-momentum tensor and the $\mtU(1)$-currents in terms of the energy-momentum tensors $T_i(z)$, the $\mtU(1)$-currents $J_i(z)$, and the identity operators $\mtI$ of the individual minimal model factors $i=1,\ldots,5$. Then the conformal weight and the left- and right-moving $\mtU(1)$-charges are given by
\begin{align}
	h \= h_1 + \dots + h_5~, \qquad  q_{L/R} \= q_{L/R,1} + \dots + q_{L/R,5}~.
\end{align}
Here $h_i$  and $q_{L/R,i}$ are the conformal weights and the $\mtU(1)$-charges with respect to $T_i(z)$ and $J_i(z)$ for $i=1,\ldots,5$, respectively. Furthermore, the tensor product theory contains the diagonal spin-$\frac32$ currents
\begin{equation} \label{eq:QuinticG}
  G^\pm(z)  \; \defineas \; \sum_{i=1}^5 \mtI \otimes \cdots \otimes  G_i^\pm(z) \otimes  \cdots \otimes \mtI \ , 
\end{equation} 
where $G_i^\pm(z)$, $i=1,\ldots,5$, are the spin-$\frac32$ currents of the tensor product factors. 

Note, however, that the product algebra $\cA_1 \times \dots \times \cA_5$ is not the chiral algebra of the tensor product theory. For instance, we have the simple currents 
\begin{align}
	\cJ_2,\dots,\cJ_5,\cJ \in \text{HWR}(\cA_1 \times \dots \times \cA_5)
\end{align}	
explicitly given by
\begin{align} \label{eq:simple_currents_J_i}
	\cJ_i &\=
	(0,0,2) \otimes (0,0,0)  \otimes (0,0,0) \otimes \overbrace{(0,0,2)}^{\text{$i$th pos.}} \otimes (0,0,0)~,~~ i \in \{2, \dots, 5\}~, \\[6pt] \label{eq:simple_current_J}
	\cJ_{\phantom{i}} &\=
	(0,-2,0) \otimes (0,-2,2) \otimes (0,-2,2) \otimes (0,-2,2) \otimes (0,-2,2)~,
\end{align}
which belong to the chiral algebra of the tensor product theory but not to the product algebra  $\cA_1 \times \dots \times \cA_5$. Given these currents, we can consider the simple current extension theory \cite{Schellekens:1989am,Schellekens:1989dq,Schellekens:1990xy}, which is a conformal field theory with the chiral algebra
\begin{align} \label{eq:chiral_algebra_simple_current_extension}
	\cA_{\text{max}}  \= \left\langle \cI, \cJ_2, \dots, \cJ_5, \cJ \; \middle| \; \cI \in \cA_1 \times \dots \times \cA_5 \right\rangle~.
\end{align}
As explained in ref.~\cite{Schellekens:1990xy}, simple current extension amounts to essentially orbifolding the tensor product theory. The monodromy charges of these simple currents are given by
\begin{align}  \label{eq:monodromy_charges_J_i}
	&\Phi_{\cJ_i}((l_1,m_1,s_1) \otimes \dots \otimes (l_5,m_5,s_5)) \= \left[ \frac{s_1}2 + \frac{s_i}2 \right] \ \in \ \IR/\IZ \ , \\[6pt] \label{eq:monodromy_charge_J}
	&\Phi_{\cJ_{\phantom{i}}}((l_1,m_1,s_1) \otimes \dots \otimes (l_5,m_5,s_5)) \= \left[ q_L((l_1,m_1,s_1) \otimes \dots \otimes (l_5,m_5,s_5)) + \frac{s_1}2 \right] \ \in \ \IR/\IZ ~,
\end{align}
the highest-weight representations of the chiral algebra \eqref{eq:chiral_algebra_simple_current_extension} are obtained by restricting to states in $\text{HWR}(\cA_1 \times \dots \times \cA_5)$ that have vanishing monodromy charges \eqref{eq:monodromy_charges_J_i} and \eqref{eq:monodromy_charge_J}, and identifying states which are related by the action of the simple currents \eqref{eq:simple_currents_J_i} and \eqref{eq:simple_current_J}. That is to say
\begin{align} \label{eq:Quintic_Gepner_HWRs}
	\text{HWR}_\cH(\cA_\text{max}) \= \{ i \in \text{HWR}(\cA_1 \times \dots \times \cA_5) \; | \; \Phi_{\cJ_2}(i) = \dots = \Phi_{\cJ_5}(i) = \Phi_{\cJ}(i) = [0] \}/\sim~,
\end{align}
where
\begin{equation} \label{eq:simple_current_equivalence}
  i \; \sim \; \cJ_2(i) \; \sim \; \ldots \; \sim \; \cJ_5(i) \; \sim \; \cJ(i)~.
\end{equation} 
This simple current construction ensures that the diagonal currents defined in eqs.~\eqref{eq:QuinticTJ} and \eqref{eq:QuinticG} realise the super-Virasoro algebra $\operatorname{SVir}$ of the extended theory of central charge $c=9$, which is well-defined and hence modular invariant \cite{Gepner:1987qi,Gepner:1989gr,Fuchs:2000gv}.

The projection onto fields with integral monodromy charge with respect to the currents $\cJ_i$ amounts to keeping only fields where all tensor factors are in the R sector, or fields where all tensor factors are in the NS sector. This projection is often referred to as \textit{fermion alignment}. The projection with respect to $\cJ$ ensures that we only have fields with integral $\mtU(1)$ charge in the NS sector, and onto fields with half-integral $\mtU(1)$ charge in the R sector. This projection is often referred to as \textit{GSO projection}.

The full spectrum of the Gepner model corresponding to the Fermat quintic threefold is now described by the Hilbert space $\cH$ that is obtained from the diagonal theory of the maximal extended chiral super-algebra, which is the maximal extended chiral algebra $\cA_{\text{max}}$ combined with the spin-$\frac32$ currents $G^\pm(z)$ defined in eq.~\eqref{eq:QuinticG}.
	
For constructing enough boundary states to satisfy the condition (iv) on p. \ref{p:assumptions}, we study the subalgebra $\mathcal{A} \subset \mathcal{A}_\text{max}$, which is given by
\begin{align} \label{eq:A_algebra_simple_current_extension}
	\cA  \= \left\langle \cI, \cJ_2, \dots, \cJ_5 \; \middle| \; \cI \in \cA_1 \times \dots \times \cA_5 \right\rangle~.
\end{align}
We consider the highest weight representations of $\cA$ that are given by
\begin{align} \label{eq:Quintic_Gepner_HWRssub}
	\HWRrep \= \{ i \in \text{HWR}(\cA_1 \times \dots \times \cA_5) \; | \; \Phi_{\cJ_2}(i) = \dots = \Phi_{\cJ_5}(i)  = [0] \}/\simeq~,
\end{align}
where the equivalence relation $\simeq$ is defined by
\begin{equation} \label{eq:simple_current_equivalence2}
	i \; \simeq \; \cJ_2(i) \; \simeq \; \ldots \; \simeq \; \cJ_5(i) ~.
\end{equation} 
However, not all of these representations appear in the Hilbert space $\cH$ of the Gepner model, since by eq.~\eqref{eq:Quintic_Gepner_HWRs} such representations have to have vanishing monodromy charge with respect to $\cJ$. Thus we have that the set $\HWR$ is given by
\begin{align}
	\HWR \= \{ i \in \HWRrep \; | \; \Phi_\cJ(i) = [0] \}~.
\end{align}
To keep the notation compact, we often denote the representations in $\HWR$ and the corresponding heighest weight states by\footnote{\label{foot:highest_weight_state_labeling} Note that the label $(\mathbf{l},\mathbf{m},\mathbf{s})$ of a highest weight state $\ket{\mathbf{l},\mathbf{m},\mathbf{s}}$ is not unique, as highest weight states with representation labels that are related by the action of simple current yield, by definition, the same highest weight state.}
\begin{equation}
(\mathbf{l},\mathbf{m},\mathbf{s}) \; \defineas \; (l_1,m_1,s_1) \otimes \dots \otimes (l_5,m_5,s_5)~, \qquad \ket{\mathbf{l},\mathbf{m},\mathbf{s}} \; \defineas \; \ket{l_1,m_1,s_1} \otimes \dots \otimes \ket{l_5,m_5,s_5}~.
\end{equation}
To make sure that such a label indeed corresponds to a representation in $\HWR$, we have to take into account the restrictions on the labels $l_i,m_i,s_i$ arising from the condition that the monodromy charges vanish. It is also useful to keep in mind that the identifications \eqref{eq:simple_current_equivalence2} imply that several different values of labels may correspond to the same representation.

The modular S-matrix of the Gepner model with respect to the algebra $\mathcal{A}$ is given by~\cite{Gato-Rivera:1991bqv,Brunner:1999jq,Recknagel:1997sb,Fuchs:2000gv}
\begin{align}
	\text{S}_{(\mathbf{l},\mathbf{m},\mathbf{s}),(\mathbf{L},\mathbf{M},\mathbf{S})} \=16 \prod_{i=1}^5 \text{S}_{(l_i,m_i,s_i),(L_i,M_i,S_i)}~,
\end{align}
where the $\mtS$-matrix elements appearing on the right-hand side are those of a $k=3$ minimal model, given explicitly in eq.~\eqref{eq:S-matrix_minimal_model}. Note that this formula is well-defined: if $(\mathbf{l}',\mathbf{m}',\mathbf{s}') = \cJ'((\mathbf{l},\mathbf{m},\mathbf{s}))$ for some $\cJ' \in \langle \cJ_2, \dots, \cJ_5 \rangle$, then by eq.~\eqref{eq:S-matrix_transformation_simple_current} 
\begin{align}
	\text{S}_{(\mathbf{l}',\mathbf{m}',\mathbf{s}'),(\mathbf{L},\mathbf{M},\mathbf{S})} \= \me^{2\pi \ii \Phi_{\cJ'}((\mathbf{l},\mathbf{m},\mathbf{s}))} \mtS_{(\mathbf{l},\mathbf{m},\mathbf{s}),(\mathbf{L},\mathbf{M},\mathbf{S})} \= \mtS_{(\mathbf{l},\mathbf{m},\mathbf{s}),(\mathbf{L},\mathbf{M},\mathbf{S})}~.
\end{align}
To obtain the last equality we have recalled that in constructing the simple current extension theory, we have restricted to representations $(\mathbf{l},\mathbf{m},\mathbf{s})$ with vanishing monodromy charge for all $\cJ' \in \langle \cJ_2, \dots, \cJ_5, \cJ \rangle$.

To construct the Hodge structure for the quintic Gepner model, we first construct the complex vector space $V_\IC$. To do this, we find the $\text{(c,c)}_{q_L = q_R}$-ring which is a purely combinatorial problem. It amounts to finding states $\ket{\mathbf{l},\mathbf{m},\mathbf{s}}$ such that
\begin{equation}\label{eq:ccringGepner}
	h_L(\ket{\mathbf{l},\mathbf{m},\mathbf{s}}) \= \frac{1}{2} q_L(\ket{\mathbf{l},\mathbf{m},\mathbf{s}})~,
\end{equation} 
First, we note that this BPS condition can only be fulfilled if every tensor factor is a $(\text{c,c})$-state of a $k=3$ minimal model. Then the sought $\text{(c,c)}_{q_L = q_R}$-ring states $\ket{\mathbf{l},\mathbf{m},\mathbf{s}}$ with $(\mathbf{l},\mathbf{m},\mathbf{s}) \in \HWR$ are obtained from tensor products of minimal model $(\text{c,c})$-states for which the monodromy charges \eqref{eq:monodromy_charges_J_i} and \eqref{eq:monodromy_charge_J} vanish. The conformal weight and the $\mtU(1)$-charge for these particular states are
\begin{equation}	\label{eq:gepnerconfweightcharge}
  h_L(\ket{\mathbf{l},\mathbf{m},\mathbf{s}}) \= \sum_{i=1}^5 \frac{l_i}{10} \ , \qquad	
  q_L(\ket{\mathbf{l},\mathbf{m},\mathbf{s}}) \=  \sum_{i=1}^5 \frac{l_i}{5} \ ,
\end{equation}
where the labels $l_i$ are unambiguous in the equivalence class $(\mathbf{l},\mathbf{m},\mathbf{s})$. The representations in the set $\cR$ (see eq.~\eqref{eq:V_C_definition}) can be obtained by application of the spectral flow operator $\cU$ and are given~by
\begin{align} \label{eq:cR_Quintic_Gepner_model}
	\cR \= \left\{ \Rla{l_1} \otimes \dots \otimes \Rla{l_5} \in \HWR \right\}~,
\end{align}
with the corresponding states listed explicitly in table \ref{tab:chiral}. From this table we see that the complex vector space $V_\IC$ is 204-dimensional. 

\begin{table}[H] 
	\begin{center}
		\renewcommand{\arraystretch}{1.3}
		\begin{tabular}{|l|c|c|}
			\hline
			\vrule height12pt width0pt depth6pt \hfil State & $(p,q)$ & Number \\\hline \hline
			
			\vrule height14pt width0pt depth6pt $\ket{\Rla{3}}^{\phantom{\otimes2}} \otimes \ket{\Rla{2}}^{\phantom{\otimes2}} \otimes \ket{\Rla{0}}^{\otimes 3}$ & $(1,2)$ & 20 \\ \hline
			\vrule height14pt width0pt depth6pt $\ket{\Rla{3}}^{\phantom{\otimes2}} \otimes \ket{\Rla{1}}^{\otimes 2} \otimes \ket{\Rla{0}}^{\otimes 2}$ & $(1,2)$ & 30 \\ \hline
			\vrule height14pt width0pt depth6pt $\ket{\Rla{2}}^{\otimes 2} \otimes \ket{\Rla{1}}^{\phantom{\otimes2}} \otimes \ket{\Rla{0}}^{\otimes 2}$ & $(1,2)$ & 30 \\ \hline
			\vrule height14pt width0pt depth6pt $\ket{\Rla{2}}^{\phantom{\otimes2}} \otimes \ket{\Rla{1}}^{\otimes 3} \otimes \ket{\Rla{0}}$ & $(1,2)$ & 20 \\ \hline
			\vrule height14pt width0pt depth6pt $\ket{\Rla{1}}^{\otimes 5}$ & $(1,2)$ & 1 \\ \hline					
			\vrule height14pt width0pt depth6pt $\ket{\Rla{3}}^{\otimes 3} \otimes \ket{\Rla{1}}^{\phantom{\otimes2}} \otimes \ket{\Rla{0}}^{\phantom{\otimes2}}$ & $(2,1)$ & 20 \\ \hline	
			\vrule height14pt width0pt depth6pt $\ket{\Rla{3}}^{\otimes 2} \otimes \ket{\Rla{2}}^{\otimes 2} \otimes \ket{\Rla{0}}$ & $(2,1)$ & 30 \\ \hline
			\vrule height14pt width0pt depth6pt $\ket{\Rla{3}}^{\otimes 2} \otimes \ket{\Rla{2}}^{\phantom{\otimes2}} \otimes \ket{\Rla{1}}^{\otimes 2}$ & $(2,1)$ & 30 \\ \hline
			\vrule height14pt width0pt depth6pt $\ket{\Rla{3}}^{\phantom{\otimes2}} \otimes \ket{\Rla{2}}^{\otimes 3} \otimes \ket{\Rla{1}}^{\phantom{\otimes2}}$ & $(2,1)$ & 20 \\ \hline
			\vrule height14pt width0pt depth6pt $\ket{\Rla{2}}^{\otimes 5}$ & $(2,1)$ & 1 \\ \hline							
			\vrule height14pt width0pt depth6pt $\ket{\Rla{3}}^{\otimes 5}$ & $(0,3)$ & 1 \\ \hline	
			\vrule height14pt width0pt depth6pt $\ket{\Rla{0}}^{\otimes 5}$ & $(3,0)$ & 1 \\ \hline
		\end{tabular}		
		\vskip10pt
		\capt{5.2in}{tab:chiral}{The basis states of the complex vector space $V_\IC$ of the Hodge structure associated to the Gepner model $(3,3,3,3,3)$. The first column gives the state, up to a permutation of the factors of the tensor product. The second column lists the Hodge type of the corresponding state, with the third column indicates how many distinct chiral states can be obtained by the permutations of the tensor factors of the displayed state.}
	\end{center}
	\vskip-30pt	
\end{table}
The boundary states take the form
\begin{align}
	\ket{B_{(\mathbf{L},\mathbf{M},\mathbf{S})}} \= \nu \!\! \sum_{(\mathbf{l},\mathbf{m},\mathbf{s}) \in \HWR} \frac{\mtS_{(\mathbf{L},\mathbf{M}, \mathbf{S}),(\mathbf{l},\mathbf{m},\mathbf{s})}}{\sqrt{\mtS_{(0,0,0),(\mathbf{l},\mathbf{m},\mathbf{s})}}} \ish{\mathbf{l},\mathbf{m},\mathbf{s}} \quad \text{with } \quad (\mathbf{L},\mathbf{M},\mathbf{S}) \in \text{HWR}(\mathcal{A})_{\text{R}} \ ,
\end{align}
with an overall normalisation constant~$\nu$ chosen to fulfill the Cardy condition \cite{Recknagel:1997sb,Fuchs:2000gv}. Theses states are  generalised Cardy boundary states in the sense of eq.~\eqref{eq:Cardy_solution_generalised}. The projected boundary states, which span the rational vector space $V_\IQ$, are consequently given by
\begin{align}
	\ket{B_{(\mathbf{L},\mathbf{M},\mathbf{S})}}_{\mathcal{P}} \= \nu \sum_{(\mathbf{l},\mathbf{m},\mathbf{s}) \in \mathcal{R}} \frac{\mtS_{(\mathbf{L},\mathbf{M}, \mathbf{S}),(\mathbf{l},\mathbf{m},\mathbf{s})}}{\sqrt{\mtS_{(0,0,0),(\mathbf{l},\mathbf{m},\mathbf{s})}}} \ket{\mathbf{l},\mathbf{m},\mathbf{s}} _{\mathcal{N}}~.
\end{align}
It is now possible to choose the set of representations $\cB \subset \HWRrep_\text{R}$ (see below eq.~\eqref{eq:V_Q_definition} for the relevant definitions) such that $(L_i,M_i,S_i) \in \text{HWR}(\mathcal{A}_i)_{\text{R}}$ \cite{Brunner:1999jq}. Together with the decomposition indicated in table \ref{tab:chiral}, the pair $(V_\IC,V_\IQ)$ gives a Hodge structure. 

Utilising the proof given in section \ref{sect:CM_Hodge_Structures_from_RCFTs}, this Hodge structure is of CM type if we can show that the compatibility condition~\eqref{eq:compatibility_R} is satisfied. To recap, we know that all $\rho \in \Gal(\IQ(\mtS_\alpha)/\IQ)$ act on the modular $\mtS$-matrix as
\begin{equation}
	\rho(\text{S}_{(\mathbf{l},\mathbf{m},\mathbf{s}) (\mathbf{L},\mathbf{M},\mathbf{S})}) \= \epsilon_{\rho}((\mathbf{l},\mathbf{m},\mathbf{s})) \text{S}_{ \varrho((\mathbf{l},\mathbf{m},\mathbf{s})) (\mathbf{L},\mathbf{M},\mathbf{S})}~,
\end{equation}
where $\varrho \in \text{Sym}(\HWR)$ is a permutation of representations. To establish the compatibility condition, we need to show that $\varrho((\mathbf{l},\mathbf{m},\mathbf{s})) \in \cR$.

Recall that $\mtS$-matrix entries of the Gepner model are, up to an integral prefactor, given by a product of $k=3$ minimal model $\mtS$-matrix entries and the elements $(\mathbf{l},\mathbf{m},\mathbf{s}) \in \cR$ are tensor products of the representations $\Rla{l_i}$ of minimal models as shown in eq.~\eqref{eq:cR_Quintic_Gepner_model}. Therefore, from \eqref{eq:kappa_definition} it follows that $\IQ(S_{(\mathbf{l},\mathbf{m},\mathbf{s})}) \subseteq \IQ(\mtS_{\Rla{l_1}}) \cdots \IQ(\mtS_{\Rla{l_5}}) = \IQ(\zeta_5)$. To see that equality in fact holds, note that
\begin{align}
 \zeta_5 \=  \frac{\text{S}_{\Rla{0}^{\otimes 3} \otimes \Rla{2}\otimes \Rla{3} , \,\Rla{0}^{\otimes 2} \otimes \Rla{2}^{\otimes 2}\otimes \Rla{1}  }}{\text{S}_{\Rla{0}^{\otimes 3} \otimes \Rla{2}\otimes \Rla{3} , \,\Rla{0}^{\otimes 2} \otimes \Rla{1} \otimes \Rla{2}^{\otimes 2}  }} \ .
 \end{align}
Let $\rho_a \in \Gal(\IQ(\mtS_{(\mathbf{l},\mathbf{m},\mathbf{s})})/\IQ) = \Gal(\IQ(\zeta_5)/\IQ)$ be an element corresponding to $[a] \in (\IZ/5\IZ)^\times$ as defined in \eqref{eq:minimal_model_rho_definition}. We can use the formula \eqref{eq:minimal_model_S-matrix_Galois_action} for the Galois action on the minimal model $\mtS$-matrices to find the action of $\rho_a$ on the $\mtS$-matrix elements generating $\IQ(\mtS_{(\mathbf{l},\mathbf{m},\mathbf{s})})$:
\begin{align}
	\rho_a \left(\mtS_{\Rla{l_1} \otimes \cdots \otimes \Rla{l_5}, (\mathbf{L},\mathbf{M},\mathbf{S})} \right) \= 16 \prod_{i=1}^5 \rho_a(\mtS_{\Rla{l_i},(L_i,M_i,S_i)}) \= \mtS_{\Rla{l_1'} \otimes \cdots \otimes \Rla{l_5'}, (\mathbf{L},\mathbf{M},\mathbf{S})}~,
\end{align} 
where $l_i'$ are the unique integers such that
\begin{align} \label{eq:l_prime_definition_Gepner}
	0 \leq l_i' \leq 3 \qquad \text{and} \qquad l_i' \equiv a(l_i+1) -1 \!\! \mod 5~.
\end{align}  
and we have used that we chose the set of representations $\cB$ such that $(L_i,M_i,S_i) \in \text{HWR}(\cA_i)_{\text{R}}$ for $(\mathbf{L},\mathbf{M},\mathbf{S}) \in \cB$.\footnote{Naturally the conclusion is independent of the choice of basis. Had we chosen a basis where some of the tensor factors are in NS-sector, the only significant difference would have been appearance of overall factors of $\ii$ in the expressions for the individual minimal model $\mtS$-matrix elements. We have treated the Galois action in such case explicitly in appendix \ref{app:Minimal_Model_NS-states}. Using the results of this appendix, it is not difficult to see how the argument could be made to work with a basis that involves tensor product factors which are in the NS-sector.}

While we know that $\Rla{l_1'} \otimes \cdots \otimes \Rla{l_5'} \in \HWRrep$, to show that in fact $\Rla{l_1'} \otimes \cdots \otimes \Rla{l_5'} \in \cR \subset \HWR$, we have to show that this representation has a vanishing monodromy charge with respect to $\cJ$. By eqs.~\eqref{eq:monodromy_charge_J} and \eqref{eq:gepnerconfweightcharge} for the monodromy charge and the $\mtU(1)$-charge, this follows simply by noting that for all $a \in (\IZ/5\IZ)^\times$
\begin{align}
	\Phi_\cJ(\Rla{l_1'} \otimes \cdots \otimes \Rla{l_5'}) \; \equiv \; \sum_{i=1}^5 \frac{l_i'+1}{5} \; \equiv \; a \sum_{i=1}^5 \frac{ l_i+1}{5} \; \equiv \; 0 \mod 1~,
\end{align}
where we have used the formula \eqref{eq:l_prime_definition_Gepner} to arrive at the second equality, and the fact that by definition $\Rla{l_1} \otimes \dots \otimes \Rla{l_5} \in \cR$ has vanishing monodromy charge to obtain the third. This shows that $\Rla{l_1'} \otimes \dots \otimes \Rla{l_5'} \in \HWR$. Since we have argued previously that any such state belong also to $\cR$, this is also sufficient to show that $\rho_a$ maps any $\Rla{l_1} \otimes \dots \otimes \Rla{l_5} \in \cR$ into an element of $\cR$, and the compatibility condition is satisfied.

As discussed in section \ref{sect:Rational_Hodge_structs_from_RCFTs}, the Hodge structure is polarised, with the polarisation provided by the Witten index \eqref{eq:polarisation_definition}. Twisting is not required since the $\mtU(1)$ charges of the states in the {$\text{(c,c)}_{q_L=q_R}$-ring} of this theory are integral. Therefore the Hodge structure we have constructed is of CM type with the CM-field $\IQ(\zeta_5)$. To be specific, the Hodge structure splits into 51 irreducible Hodge substructures of dimension four, and the Hodge endomorphism algebra of each irreducible piece is isomorphic to $\IQ(\zeta_5)$.

This Gepner model corresponds to the conformal fixed point of the $\mathcal{N}=(2,2)$ non-linear sigma model on the Fermat quintic threefold~\cite{Gepner:1987vz,Gepner:1987qi,Greene:1988ut,Martinec:1988zu,Witten:1993yc}
\begin{align}
	\left\{ \ x_1^5 + x_2^5 + x_3^5 + x_4^5 + x_5^5 \= 0\  \right\} \; \subset \; \IP^5~.
\end{align}
Thus we have obtained a physics-based derivation of the known result that the Fermat quintic admits complex multiplication, with the CM-field $\IQ(\zeta_5)$ (see, for example, ref.~\cite{Borcea}).

\subsection{The Gepner model \texorpdfstring{$(6,6,2,2,2)$}{(6,6,2,2,2)} corresponding to \texorpdfstring{a threefold $\mathbb{P}^4_{1,1,2,2,2}[8]$}{an Octic threefold}} \label{sect:Gepner_model_octic}
As the second example, we consider the Gepner model $(6,6,2,2,2)$ given by a simple current extension of the tensor product $(k_1=6) \otimes (k_2=6) \otimes (k_3=2) \otimes (k_4=2) \otimes (k_5=2)$ of A-type $\cN=2$ minimal models with the indicated levels $k_i$. As discussed in the beginning of the section, this case does not strictly fit into the general framework we have set out in sections \ref{sec:cft} and \ref{sect:Hodge_structures_from_RCFT}. Rather than providing a detailed example of our construction, the purpose of this section is to substantiate our claim that further generalisations of the constructions presented in this paper are possible. Consequently, we will be rather brief about the conformal field theory constructions underlying this case, contenting ourselves with presenting the bare necessities required to understand the construction of Hodge structures and Galois actions. We refer the interested reader to refs.  \cite{Schellekens:1990xy,Fuchs:1996dd,Fuchs:1999zi,Fuchs:1999xn,Fuchs:2000gv} for further details on the conformal field theory aspects of the Gepner model presented in this section.

The simple current extension is again taken with respect to the currents
\begin{align}
	\cJ_i &\=
	(0,0,2) \otimes (0,0,0)  \otimes (0,0,0) \otimes \overbrace{(0,0,2)}^{\text{$i$th pos.}} \otimes (0,0,0)~,~~ i \in \{2, \dots, 5\}~, \\[6pt]
	\cJ_{\phantom{i}} &\=
	(0,-2,0) \otimes (0,-2,2) \otimes (0,-2,2) \otimes (0,-2,2) \otimes (0,-2,2)~,
\end{align}
which generate the algebra
\begin{align}
	\wt \cA \= \langle \cA_1 \times \dots \times \cA_5, \cJ_2, \dots, \cJ_5, \cJ \rangle \; \subseteq \; \cA_{\text{max}} \ .
\end{align}
The distinct highest-weight representations are
\begin{align}
	\text{HWR}_\cH(\wt \cA) \= \{ i \in \text{HWR}(\cA_1 \times \dots \times \cA_5) \; | \; \Phi_{\cJ_2}(i) = \dots = \Phi_{\cJ_5}(i) = \Phi_{\cJ}(i) = [0] \}/\sim~,
\end{align}
where
\begin{equation} \label{eq:Octic_equivalence_relation1}
	i \; \sim \; \cJ_2(i) \; \sim \; \ldots \; \sim \; \cJ_5(i) \; \sim \; \cJ(i)~.
\end{equation} 
However, as in the previous section, we find it useful to organise the fields of the simple current extension theory in terms of representations of a subalgebra 
\begin{align}
	 \cA \= \langle \cA_1 \times \dots \times \cA_5, \cJ_2, \dots, \cJ_5 \rangle \; \subseteq \; \wt \cA~.
\end{align}
The distinct highest-weight representations of $\cA$ which appear in the simple current extended theory are given by 
\begin{align}
	\HWR \= \{ i \in \text{HWR}(\cA_1 \times \dots \times \cA_5) \; | \; \Phi_{\cJ_2}(i) = \dots = \Phi_{\cJ_5}(i) = \Phi_{\cJ}(i) = [0] \}/\simeq~,
\end{align}
where the equivalence relation $\simeq$ is now
\begin{equation} \label{eq:Octic_equivalence_relation2}
	i \; \simeq \; \cJ_2(i) \; \simeq \; \ldots \; \simeq \; \cJ_5(i)~.
\end{equation}
We denote the highest-weight representations in $\HWR$ by\footnote{As explained in footnote \ref{foot:highest_weight_state_labeling}, the labels $\mathbf{l}$, $\mathbf{m}$, $\mathbf{s}$ are not unique for a given state.} 
\begin{align}
	(\mathbf{l},\mathbf{m},\mathbf{s}) \; \defineas \; (l_1,m_1,s_1) \otimes \dots \otimes (l_5,m_5,s_5)~.
\end{align}
In contrast to the previous example, the simple current
\begin{equation} \label{eq:fixed_point_simple_current}
	\cL \= \cJ^{-4} \cJ_2 \= (0,8,2) \otimes (0,8,2) \otimes (0,0,0) \otimes(0,0,0) \otimes (0,0,0) \in \langle \cJ_2, \dots, \cJ_5, \cJ \rangle
\end{equation}
acts trivially on exactly those representations $(l_1,m_1,s_1) \otimes \dots \otimes (l_5,m_5,s_5) \in \text{HWR}_\cH(\cA_1 \times \dots \times \cA_5)$ for which $l_1=l_2=3$, that is to say
\begin{multline}
\cL((3,m_1,s_1) \otimes (3,m_2,s_2)\otimes(l_3,m_3,s_3) \otimes \dots \otimes (l_5,m_5,s_5)) \\
  \= (3,m_1,s_1) \otimes (3,m_2,s_2)\otimes(l_3,m_3,s_3) \otimes\dots \otimes (l_5,m_5,s_5)~.
\end{multline}
It is not difficult to see that in the present case, $\cL$ is the only simple current which acts trivially on some representations.

As discussed in ref.~\cite{Schellekens:1990xy}, the existence of the current $\cL$ with such fixed representations implies that to construct a modular invariant partition function, we must include the corresponding representations of $\cA$ (that is, the equivalence classes of such representations under $\simeq$ of eq.~\eqref{eq:Octic_equivalence_relation2}) with multiplicity in the Hilbert space $\cH$ of the theory. The multiplicity of the Hibert space $\cH_{(\mathbf{l},\mathbf{m},\mathbf{s})}$, $(\mathbf{l},\mathbf{m},\mathbf{s}) \in \HWR$ is given by the order of the stabiliser group 
\begin{align} \label{eq:Octic_stabiliser_definition}
	 \text{Stab}_\cA((\mathbf{l},\mathbf{m},\mathbf{s})) \; \defineas \; \left\{ \cI \in  \langle \cJ_2, \dots, \cJ_5, \cJ \rangle  \; \middle | \; \cI(i) = i \right\}~,
\end{align}
where $i \in \text{HWR}(\cA_1 \times \dots \times \cA_5)$ is any representative of the equivalence class $(\mathbf{l},\mathbf{m},\mathbf{s}) \in \HWR$. For the model $(6,6,2,2,2)$ the only stabiliser groups that appear are cyclic, being the trivial group $1$ or $\IZ_2$, generated by $\cL$.

Following refs.~\cite{Fuchs:1999zi,Fuchs:1999xn}, we find it useful to distinguish different copies of isomorphic representations by labeling each with a character $\psi \in  \text{Stab}_\cA((\mathbf{l},\mathbf{m},\mathbf{s}))^*$, that is a group homomorphism ${\psi: \text{Stab}_\cA((\mathbf{l},\mathbf{m},\mathbf{s})) \to \IC^*}$. Since the only non-trivial stabiliser groups that appear are isomorphic to $\IZ/2\IZ$, so the only charachters we need to consider are the trivial character we denote by $\mathbf{1}^+$, and the character $\mathbf{1}^-$, defined by the relation $\mathbf{1}^-(\cL) = -1$. 

We denote the representations and the corresponding highest-weight states by
\begin{align}
	(\mathbf{l},\mathbf{m},\mathbf{s})_\psi \; \defineas \; ((l_1,m_1,s_1) \otimes \cdots \otimes (l_5,m_5,s_5),\psi)~, \qquad 
	\ket{(\mathbf{l},\mathbf{m},\mathbf{s})_\psi} \ .
\end{align}

The set $\cR$ is again defined as the set of representations decorated by characters related by spectral flow to elements of the $(\text{c,c})_{q_L = q_R}$-ring. As in the quintic case, it turns out that these states are given by tensor products of the minimal model representations $\Rla{l_i} \in \text{HWR}(\cA_i)$ defined in eq.~\eqref{eq:Rl}, decorated by characters. 

We normalise the corresponding R-ground states so that they are orthonormal, i.e.
\begin{align}
	{{\vphantom{\braket{(\textbf{l},\textbf{m},\textbf{s})_\psi|(\textbf{l}',\textbf{m}',\textbf{s}')_\phi}}}}_{\cN} \!\braket{(\textbf{l},\textbf{m},\textbf{s})_\psi|(\textbf{l}',\textbf{m}',\textbf{s}')_\phi}_\cN  \= \delta_{\psi \phi} \,\delta_{(\textbf{l},\textbf{m},\textbf{s}), (\textbf{l}',\textbf{m}',\textbf{s}')} \ .
\end{align}

\subsubsection*{Boundary states and the generalised $\mtS$-matrix}
\vskip-5pt
In the present case, we have representations $(\mathbf{l},\mathbf{m},\mathbf{s})_{\bm{1}^+}$, $(\mathbf{l},\mathbf{m},\mathbf{s})_{\bm{1}^-} \in \cR$ which differ only by the character $\psi \in \text{Stab}_{\cA}((\mathbf{l},\mathbf{m},\mathbf{s}))^*$. For instance the representations
\begin{align}
(\Rla{3} \otimes \Rla{3} \otimes \Rla{1} \otimes \Rla{0} \otimes \Rla{0})_{\bm{1}^+}~,~~(\Rla{3} \otimes \Rla{3} \otimes \Rla{1} \otimes \Rla{0} \otimes \Rla{0})_{\bm{1}^-} \in \cR
\end{align}
have this property, and are thus isomorphic as representations of $\cA$. Since the modular $\mtS$-matrix has only one entry for each isomorphic representation of $\cA$, we cannot construct enough boundary states to obtain an intersection matrix $\Sigma$ of rank equal to the number of $(\text{c,c})_{q_L=q_R}$ primary states. Therefore this case does not satisfy the assumption (iv) on page~\pageref{p:assumptions}.

However, it is possible to proceed with only minor modifications, as we can obtain an intersection matrix of high enough rank, which is enables us to construct the rational vector space $V_\IQ$ as in section \ref{sect:Rational_Hodge_structs_from_RCFTs}, if we consider a generalised $\mtS$-matrix $\mtS_{\{\textbf{L},\textbf{M},\textbf{S}\}_\Psi, (\textbf{l},\textbf{m},\textbf{s})_\psi}$  \cite{Fuchs:1999zi,Fuchs:1999xn}. The first index corresponds to boundary labels $\{\textbf{L},\textbf{M},\textbf{S}\}_\Psi$, where
\begin{align} \label{eq:boundary_label_definition}
 \{\textbf{L},\textbf{M},\textbf{S}\} \in \text{HWR}_\cH(\cA_1 \times \dots \times \cA_5)/\sim~ \quad \text{and} \quad \Psi \in  \text{Stab}_\cA(\{\textbf{L},\textbf{M},\textbf{S}\})^*~,
\end{align}
where $\sim$ denotes the equivalence relation \eqref{eq:Octic_equivalence_relation1}, and the stabiliser of $\{\textbf{L},\textbf{M},\textbf{S}\}$ is defined completely analogously to eq.~\eqref{eq:Octic_stabiliser_definition}. Note that in contrast to the set $\HWR$, there is no restriction to representations with vanishing monodromy charge.

In the present case, the generalised $\mtS$-matrix reads
\begin{align} \label{eq:Octic_Generalised_S-matrix}
\begin{split}
	\mtS_{\{\textbf{L},\textbf{M},\textbf{S}\}_\Psi, (\textbf{l},\textbf{m},\textbf{s})_\psi} \= 128& \prod_{i=1}^5 \mtS^{(i)}_{(L_i,M_i,S_i),(l_i,m_i,s_i)} - 8 \, \delta^S_{\{\textbf{L},\textbf{M},\textbf{S}\}, (\textbf{l},\textbf{m},\textbf{s})} \prod_{i=3}^5 \mtS^{(i)}_{(L_i,M_i,S_i),(l_i,m_i,s_i)} \times \\
	& \times \left( 12 \prod_{i=1}^{2} \mtS^{(i)}_{(L_i,M_i,S_i),(l_i,m_i,s_i)} + \Psi(\cL) \psi(\cL) \prod_{i=1}^2 e^{\pi \ii (m_i M_i /8 -s_i S_i /2)} \right)~,
\end{split}
\end{align}
where $\mtS^{(i)}_{(L_i,M_i,S_i),(l_i,m_i,s_i)}$ denotes the S-matrix of the $i$'th minimal model factor, and the $\delta^S_{\{\textbf{L},\textbf{M},\textbf{S}\}, (\textbf{l},\textbf{m},\textbf{s})}$ signifies that the second term only contributes when both representations $\{\textbf{L},\textbf{M},\textbf{S}\}$ and $(\textbf{l},\textbf{m},\textbf{s})$ have non-trivial stabilisers: recalling the discussion around eq.~\eqref{eq:fixed_point_simple_current}, we can write
\begin{align} \notag
	\delta^S_{\{\textbf{L},\textbf{M},\textbf{S}\}, (\textbf{l},\textbf{m},\textbf{s})} \; \defineas \; 
		\prod_{i=1}^{2} \delta_{l_i,3} \delta_{L_i,3} \=
			\begin{cases}
				1 \qquad \text{if } \text{Stab}_\cA(\{\textbf{L},\textbf{M},\textbf{S}\}),~ \text{Stab}_\cA((\textbf{l},\textbf{m},\textbf{s}) \text{ non-trivial} \\[5pt]
				0 \qquad \text{otherwise}~.
			\end{cases}
\end{align}
With this matrix, it turns out that we can apply the construction of section \ref{sect:Hodge_structures_from_RCFT} mutatis mutandis: The boundary states are labeled by the boundary labels $\{\textbf{L},\textbf{M},\textbf{S}\}$ defined in eq.~\eqref{eq:boundary_label_definition}, and are given by
\begin{align}
\ket{B_{\{\textbf{L},\textbf{M},\textbf{S}\}_\Psi}} \= \nu \sum_{\substack{(\textbf{l},\textbf{m},\textbf{s}) \\ \psi \in \text{Stab}_{\cA}((\mathbf{l},\mathbf{m},\mathbf{s}))^*}}\frac{\mtS_{\{\textbf{L},\textbf{M},\textbf{S}\}_\Psi, (\textbf{l},\textbf{m},\textbf{s})_\psi}}{\sqrt{\mtS_{\{\mathbf{0},\mathbf{0},\mathbf{0}\}_{\mathbf{1}^+},(\textbf{l},\textbf{m},\textbf{s})_\psi}}} \ish{(\textbf{l},\textbf{m},\textbf{s})_\psi}~.
\end{align}
Here $\ish{(\mathbf{l},\mathbf{m},\mathbf{s})_\psi}$ are the generalised Ishibashi states \cite{Fuchs:1999zi,Fuchs:1999xn} which are labeled by the representations $(\mathbf{l},\mathbf{m},\mathbf{s})_\psi \in \HWR$, and which are normalised such that
\begin{align} \label{eq:Octic_Ishibashi_normalisation}	{\vphantom{\braket{(\textbf{l},\textbf{m},\textbf{s})_\psi|[\textbf{l}',\textbf{m}',\textbf{s}']_\phi}}}_{\cN} \!\braket{(\textbf{l},\textbf{m},\textbf{s})_\psi|(\textbf{l}',\textbf{m}',\textbf{s}')_\phi}\! \rangle \= |\text{Stab}_\cA((\textbf{l},\textbf{m},\textbf{s}))| \; \delta_{\psi \phi}\,\delta_{(\textbf{l},\textbf{m},\textbf{s}), (\textbf{l}',\textbf{m}',\textbf{s}')} \ .
\end{align}
The projected boundary states take the form which is similar to eq.~\eqref{eq:projected_boundary_state_definition}, albeit not completely analogous, due to the different normalisation \eqref{eq:Octic_Ishibashi_normalisation} of Ishibashi states here:
\begin{align} \label{eq:Octic_projectec_boundary_states}
	\begin{split}
	\ket{B_{\{\textbf{L},\textbf{M},\textbf{S}\}_\Psi}}_\cP &\; \defineas \;  \sum_{\substack{(\textbf{l},\textbf{m},\textbf{s}) \in \HWR \\ \psi \in \text{Stab}_\cA((\textbf{l},\textbf{m},\textbf{s}))^*}}  \ket{(\textbf{l},\textbf{m},\textbf{s})_\psi}_\cN {\vphantom{\ket{(\textbf{l},\textbf{m},\textbf{s})_\psi}}}_{\cN} \! \braket{(\textbf{l},\textbf{m},\textbf{s})_\psi | B_{\{\textbf{L},\textbf{M},\textbf{S}\}_\Psi}} \\
	&\=	\nu \sum_{\substack{(\textbf{l},\textbf{m},\textbf{s}) \in \HWR \\ \psi \in \text{Stab}_{\cA}((\mathbf{l},\mathbf{m},\mathbf{s}))^*}} |\text{Stab}_\cA{((\textbf{l},\textbf{m},\textbf{s}))}|
	\frac{\mtS_{\{\textbf{L},\textbf{M},\textbf{S}\}_\Psi, (\textbf{l},\textbf{m},\textbf{s})_\psi}}{\sqrt{\mtS_{\{\mathbf{0},\mathbf{0},\mathbf{0}\}_{\mathbf{1}^+},(\textbf{l},\textbf{m},\textbf{s})_\psi}}} \ket{(\textbf{l},\textbf{m},\textbf{s})_\psi}_\cN~.
	\end{split}
\end{align}

We can choose a set $\cB$ of boundary labels $\{\textbf{L},\textbf{M},\textbf{S}\}_{\Psi}$ such that $(L_i,M_i,S_i) \in \text{HWR}(\cA_i)_\text{R}$. For this choice of $\cB$, the intersection matrix 
\begin{align} \label{eq:intersection_matrix_Octic}
	\Sigma_{AB} \defineas {\vphantom{\ket{}}}_\cP \! \bra{B_A} \me^{2\pi \ii J_0} \ket{B_B}_{\cP}~, \qquad A,B \in \cB~.
\end{align}	
is of maximal rank, valued in $\IQ$, and antisymmetric. This also allows us to take the rational vector space $V_\IQ$ to be defined by the span of the projected boundary states \eqref{eq:Octic_projectec_boundary_states}. The conveniently normalised states of eq.~\eqref{eq:intermediate_scaled_states} now take the form
\begin{align} \label{eq:intermediate_scaled_states_Octic}
	\begin{split}
	\intscal{(\textbf{l},\textbf{m},\textbf{s})_\psi} \; &\defineas \; \me^{-\pi \ii q_L(\ket{(\textbf{l},\textbf{m},\textbf{s})_\psi})} \frac{1}{|\text{Stab}_\cA{((\textbf{l},\textbf{m},\textbf{s}))}| \nu} \sqrt{\mtS_{\{\mathbf{0},\mathbf{0},\mathbf{0}\}_{\mathbf{1}^+},(\textbf{l},\textbf{m},\textbf{s})_\psi}}^{\; *} \ket{(\textbf{l},\textbf{m},\textbf{s})_\psi}_\cN \\[5pt]
	&\= \sum_{\{\textbf{L},\textbf{M},\textbf{S}\}_{\Psi} \in \cB} \mtS_{\{\textbf{L},\textbf{M},\textbf{S}\}_\Psi, (\textbf{l},\textbf{m},\textbf{s})_\psi}^* \ket{B^{\{\textbf{L},\textbf{M},\textbf{S}\}_{\Psi}}}_\cP \in V_\IQ \otimes_{\IQ} \IQ(S)~,
	\end{split}
\end{align}
where, compared to the definition \eqref{eq:intermediate_scaled_states}, we have taken into account the different normalisation of the Ishibashi states. However, the last expression is completely analogous to eq.~\eqref{eq:intermediate_scaled_states}. It is this latter relation that is the key to proving the CM property in section \ref{sect:CM_Hodge_Structures_from_RCFTs}.

These relations imply that, for the construction of CM-type Hodge structures presented in sections~\ref{sect:Rational_Hodge_structs_from_RCFTs} and \ref{sect:CM_Hodge_Structures_from_RCFTs}, we can use the projected generalised boundary states to define the rational vector space $V_\IQ$, with the graded spaces $V_\IC^{(p,q)}$ constructed from the states $\ket{(\mathbf{l},\mathbf{m},\mathbf{s})_\psi}$ with $(\mathbf{l},\mathbf{m},\mathbf{s})_\psi \in \cR$ as before. 

\subsubsection*{Galois action on the generalised $\mtS$-matrix}
\vskip-5pt
Since we are using the generalised $\mtS$-matrix, the proof of refs.~\cite{DeBoer:1990em,Coste:1993af} of existence of Galois symmetry and the properties of the Galois action on the $\mtS$-matrix, reviewed in section \ref{sect:RCFTs_and_Galois_Groups}, does not apply directly. However, we can check the requisite properties directly by using the explicit formula~\eqref{eq:Octic_Generalised_S-matrix} for the generalised $\mtS$-matrix. From this formula it is manifest that the $\mtS$-matrix elements are algebraic numbers, lying in a field generated by roots of unity, so they generate a cyclotomic field extension $\IQ(\mtS) = \IQ(\zeta_{32})$, and hence there is an action of an Abelian Galois group $\Gal(\IQ(\mtS)/\IQ) \simeq (\IZ/32\IZ)^\times$ on the generalised $\mtS$-matrix.

It is not difficult, although somewhat tedious, to show that this Galois action satisfies
\begin{equation} \label{eq:Galois_transformation_generalised_S-matrix}
	\sigma(\mtS_{\{\textbf{L},\textbf{M},\textbf{S}\}_\Psi,(\mathbf{l},\mathbf{m},\mathbf{s})_\psi}) \= \epsilon_\sigma((\mathbf{l},\mathbf{m},\mathbf{s})_\psi) \, \mtS_{\{\textbf{L},\textbf{M},\textbf{S}\}_\Psi,\varsigma((\mathbf{l},\mathbf{m},\mathbf{s})_\psi)} \ \text{with} \  \epsilon_\sigma((\mathbf{l},\mathbf{m},\mathbf{s})_\psi) \in \{ -1,+1\} \ ,
\end{equation}
for all $\sigma \in \Gal(\IQ(\mtS)/\IQ)$ and all representations $(\mathbf{l},\mathbf{m},\mathbf{s})_\psi \in \HWR$. Here $\varsigma$ is a permutation of all such representations, which is independent of the boundary label $\{\textbf{L},\textbf{M},\textbf{S}\}_\Psi$. 

To see this, we note at first that when the representation $(\mathbf{l},\mathbf{m},\mathbf{s})_\psi \in \HWR$ has a trivial stabiliser $\text{Stab}_\cA((\mathbf{l},\mathbf{m},\mathbf{s}))$, the generalised $\mtS$-matrix of eq.~\eqref{eq:Octic_Generalised_S-matrix} reduces to a product of $\mtS$-matrices of minimal models:
\begin{align}
	\mtS_{\{\textbf{L},\textbf{M},\textbf{S}\}_\Psi,(\mathbf{l},\mathbf{m},\mathbf{s})_\psi} \= 128 \prod_{i=1}^5 \mtS^{(i)}_{(L_i,M_i,S_i),(l_i,m_i,s_i)}~.
\end{align}
Therefore, under the action of $\sigma_a \in \Gal(\IQ(\mtS)/\IQ)$ corresponding to $[a] \in (\IZ/32\IZ)^\times$, the generalised $\mtS$-matrix transforms as
\begin{align} \label{eq:generalised_S-matrix_Galois_transformation}
	\sigma_a \left(\mtS_{\{\textbf{L},\textbf{M},\textbf{S}\}_\Psi,(\mathbf{l},\mathbf{m},\mathbf{s})_\psi}\right) \= \left(\prod_{i=1}^5 \epsilon_{\sigma_a^{(i)}}((l_i,m_i,s_i))\right) 128 \prod_{i=1}^5 \mtS^{(i)}_{(L_i,M_i,S_i),(l_i',m_i',s_i')}~,
\end{align}
where $\sigma_a^{(i)} \defineas \sigma_a|_{\IQ(\mtS^{(i)})}$ is the restriction of $\sigma_a$ to the number field $\IQ(\mtS^{(i)})$ generated by the modular $\mtS$-matrix elements of the $i$'th minimal model factor. Here $l_i',m_i',s_i'$ are the unique integers such that
\begin{align} \nonumber
	0 \leq &\;l_i' \leq k_i~, &\qquad l_i &\;\equiv\; a(l_i+1)-1 \mod k_i{+}2~,\\ \label{eq:minimal_model_representations_Galois_transformation}
	{-}k_i{-}1 \leq &\;m_i' \leq k_i{+}2~,  &\qquad  m_i &\;\equiv\; a m_i \mod 2(k_i{+}2)~,\\ \nonumber
	-1 \leq &\;s_i' \leq 2~, &\qquad s_i' &\;\equiv\; a s_i \mod 4~.
\end{align}
To derive these, one can use arguments similar to those used in section \ref{sect:minimal_model_Hodge_substructures}. See also the discussion around eq.~\eqref{eq:NS_minimal_model_Galois_action} for the exact arguments that can be used to derive these formulae.

Now we only have to check that the right-hand side of eq.~\eqref{eq:generalised_S-matrix_Galois_transformation} is, up to sign, an element of the generalised $\mtS$-matrix of the form $\mtS_{\{\textbf{L},\textbf{M},\textbf{S}\}_\Psi,(\mathbf{l}',\mathbf{m}',\mathbf{s}')_\psi}$ for a representation $(\mathbf{l}',\mathbf{m}',\mathbf{s}')_\psi \in \HWR$. For this, it is sufficient to show that 
\begin{align}
	\varsigma_a((\mathbf{l},\mathbf{m},\mathbf{s})) \; \defineas \; (l_1',m_1',s_1') \otimes \dots \otimes (l_5',m_5',s_5') \in \text{HWR}_\cH(\cA_1 \times \dots \times \cA_5)
\end{align}
is a representation with vanishing monodromy chagres under $\cJ_i$ and $\cJ$, and that this representation does not have a non-trivial stabiliser. 

To check the vanishing of the monodromy charge under $\cJ_i$, we note that from the above formulae~\eqref{eq:minimal_model_representations_Galois_transformation}, it follows that $s_j' = a s'_j + 4n_j$ for some $n_j \in \IZ$. Thus we have that
\begin{align}
	\Phi_{\cJ_i}(\varsigma_a((\mathbf{l},\mathbf{m},\mathbf{s}))) \= \left[ \frac{s_1'}{2} + \frac{s_i'}{2} \right]  \= 
	\left[ a \left( \frac{s_1}{2} + \frac{s_i}{2} \right) + 2(n_1+n_i) \right] \= [0] \ \in \ \IR/\IZ \ ,
\end{align}
where we have used $\Phi_{\cJ_i}((\mathbf{l},\mathbf{m},\mathbf{s}))= \left[\frac{s_1}{2} + \frac{s_i}{2}\right] = [0]$.
The proof that $\Phi_{\cJ}(\varsigma_a((\mathbf{l},\mathbf{m},\mathbf{s}))) = [ 0 ]$ is completely analogous.

To see that the representation $\varsigma_a((\mathbf{l},\mathbf{m},\mathbf{s}))$ does not have a non-trivial stabiliser, that is, $l_1' \neq 3$ or $l_2' \neq 3$, it suffices to note that
\begin{align}
	a(3+1) -1 \; \equiv \; 3 \mod 8
\end{align}
for all $a \in (\IZ/32\IZ)^\times$. Therefore the representations with non-trivial stabilisers get mapped to other such representations, which also implies that representations with trivial stabilisers are mapped to other representations with trivial stabilisers under the Galois action.

Thus we have that
\begin{align} \label{eq:Octic_S-matrix_transformation_trivial_stabiliser}
	\sigma_a \left(\mtS_{\{\textbf{L},\textbf{M},\textbf{S}\}_\Psi,(\mathbf{l},\mathbf{m},\mathbf{s})_\psi}\right) \= \epsilon_{\sigma_a}((\mathbf{l},\mathbf{m},\mathbf{s})_\psi)\mtS_{\{\textbf{L},\textbf{M},\textbf{S}\}_\Psi,[\mathbf{l}',\mathbf{m}',\mathbf{s}']_\psi}~,
\end{align}
with the overall sign given by
\begin{align} \label{eq:Octic_overall_sign_trivial_stabiliser}
	 \epsilon_{\sigma_a}((\mathbf{l},\mathbf{m},\mathbf{s})_\psi) \= \prod_{i=1}^5 \epsilon_{\sigma_a^{(i)}}((l_i,m_i,s_i))~.
\end{align}
In the case where the stabiliser of $(\mathbf{l},\mathbf{m},\mathbf{s})$ is non-trivial, that is $l_1 = l_2 = 3$, the expression for the generalised $\mtS$-matrix consists of two terms, the first of which again a product of the modular $\mtS$-matrices of minimal models, and the second one is given by
\begin{align}
	-8  \Psi(\cL) \psi(\cL)\prod_{i=1}^2 e^{\pi \ii (m_i M_i /8 -s_i S_i /2)} \prod_{i=3}^5 \mtS^{(i)}_{(L_i,M_i,S_i),(l_i,m_i,s_i)}~.
\end{align}
The transformation behaviour of the product of $\mtS$-matrices of minimal models can be found using the same arguments as above, which show that
\begin{align} \label{eq:S_matrix_non-trivial_stabiliser_extra_term}
	\varsigma_a((\mathbf{l},\mathbf{m},\mathbf{s})) \= (l_1',m_1',s_1') \otimes \dots \otimes (l_5',m_5',s_5')  \in \text{HWR}_\cH(\cA_1 \times \dots \times \cA_5)~,
\end{align} 
where $l_i'$, $m_i'$, and $s_i'$ are still given by eq.~\eqref{eq:minimal_model_representations_Galois_transformation}, and the overall sign $\epsilon_{\sigma_a}((\mathbf{l},\mathbf{m},\mathbf{s})_\psi)$ is also given by eq.~\eqref{eq:Octic_overall_sign_trivial_stabiliser}. We have also shown above that this transformation maps representations with non-trivial stabilisers to other such representations. Under the action of $\sigma_a \in \Gal(\IQ(\mtS)/\IQ)$, the second term \eqref{eq:S_matrix_non-trivial_stabiliser_extra_term} transforms to
\begin{align} \label{eq:Octic_extra_term_transformation}
	\prod_{i=3}^5 \epsilon_{\sigma_a^{(i)}}((l_i,m_i,s_i)) \left(-8\Psi(\cL) \psi(\cL)\prod_{i=1}^2 e^{\pi \ii (m_i' M_i /8 -s_i' S_i /2)} \prod_{i=3}^5 \mtS^{(i)}_{(L_i,M_i,S_i),(l_i',m_i',s_i')}\right)~,
\end{align}
so all we have to do to establish the formula \eqref{eq:Octic_S-matrix_transformation_trivial_stabiliser} for the transformation of the generalised $\mtS$-matrix in the case of representations with non-trivial stabilisers, we have to show that
\begin{align} \label{eq:epsilon_condition_l=3}
	\prod_{i=1}^2 \epsilon_{\sigma_a^{(i)}}((3,m_i,s_i)) \= 1~.
\end{align}
This follows straightforwardly by using the form \eqref{eq:S-matrix_minimal_model} of the $\mtS$-matrix of minimal model, from which we see that for $k=6$, that is $i=1,2$, and $l=3$, the matrix elements vanish for $L$ odd, and the condition becomes empty. For $L$ even the expression reduces to
\begin{align}
		\mtS^{(i)}_{(3,m,s),(L,M,S)} &\= \frac{(-1)^{L/2}}{8} \zeta_{2(k+2)}^{m M} \me^{-\ii \pi s S/2}~, \qquad \text{for} \qquad i =1,2~,
\end{align}
from which we can read $\epsilon_{\sigma_a^{(i)}}((3,m,s)) = 1$, which implies that the condition \eqref{eq:epsilon_condition_l=3} is satisfied. This finishes the proof of the Galois transformation property \eqref{eq:Galois_transformation_generalised_S-matrix} of the generalised $\mtS$-matrix.

\subsubsection*{Compatibility condition}
\vskip-5pt
Given the above observations, the proof given in section \ref{sect:CM_Hodge_Structures_from_RCFTs} still works mutatis mutandis, implying that the Hodge structure $(V_\IQ,V_\IC)$, the construction of which was explained in the previous subsection, is of CM type provided the obvious analogue of the compatibility condition \eqref{eq:compatibility_R} is satisfied. That is to say, if for all $(\mathbf{l},\mathbf{m},\mathbf{s})_\psi \in \cR$, we define
\begin{align} \label{eq:Octic_field_extension}
	\IQ(\mtS_{(\mathbf{l},\mathbf{m},\mathbf{s})_\psi}) \; \defineas \; \IQ\left(\left\{ \mtS_{\{\mathbf{L},\mathbf{M},\mathbf{S}\}_\Psi, (\mathbf{l},\mathbf{m},\mathbf{s})_\psi} \; \middle| \; \{\mathbf{L},\mathbf{M},\mathbf{S}\}_\Psi \in \cB \right\}  \right)~,
\end{align}
the compatibility condition is the statement that every $\rho \in \Gal(\IQ(\mtS_{(\mathbf{l},\mathbf{m},\mathbf{s})_\psi})/\IQ)$ acts on the $\mtS$-matrix so that
\begin{align} \label{eq:compatibility_condition_(6,6,2,2,2)}
	\rho(\mtS_{\{\textbf{L},\textbf{M},\textbf{S}\}_\Psi,(\mathbf{l},\mathbf{m},\mathbf{s})_\psi}) = \epsilon_{\rho}((\mathbf{l},\mathbf{m},\mathbf{s})_\psi) \mtS_{ \{\textbf{L},\textbf{M},\textbf{S}\}_\Psi,\varrho((\mathbf{l},\mathbf{m},\mathbf{s})_\psi)}~, \qquad \text{for all} \qquad \{\textbf{L},\textbf{M},\textbf{S}\}_\Psi \in \cB~, 
\end{align}
where $\epsilon_\sigma((\mathbf{l},\mathbf{m},\mathbf{s})_\psi) \in \{\pm 1\}$ is a sign dependent only on the choice of $\rho$ and on the representation $(\mathbf{l},\mathbf{m},\mathbf{s})_\psi$, and $\varrho \in \text{Sym}(\cR)$ so that $\varrho((\mathbf{l},\mathbf{m},\mathbf{s})_\psi) \in \cR$.

We will finish this section by showing that this condition is satisfied. The proof proceeds largely along the same lines as for the quintic, with the only essential difference being that the expression \eqref{eq:Octic_Generalised_S-matrix} does not quite reduce to a product of minimal model $\mtS$-matrix elements, so we need to analyse the action of the Galois group on the additional character-dependent term separately.

Depending on the representation $(\mathbf{l},\mathbf{m},\mathbf{s})_\psi \in \cR$, the field extension $\IQ(S_{(\mathbf{l},\mathbf{m},\mathbf{s})_\psi})$ defined in eq.~\eqref{eq:Octic_field_extension} is either $\IQ(\zeta_8)$ or $\IQ(\ii)$:
\begin{align}
	\IQ(\mtS_{(\mathbf{l},\mathbf{m},\mathbf{s})_\psi}) \= 
	\begin{cases}
		\IQ(\ii) \qquad \hskip6pt \text{if} \,\,\, l_1 \in \{1,5 \} \, \,\text{or}\,\,\, l_2 \in \{1,5  \} \, \,\text{or}\,\,\, l_1 = l_2 =3 ~,\\[5pt]
	\IQ(\zeta_8) \qquad \text{otherwise}~.
	\end{cases}
\end{align}
The corresponding Galois groups $\Gal(\IQ(\mtS_{(\mathbf{l},\mathbf{m},\mathbf{s})_\psi})/\IQ)$ are isomorphic to $\IZ_2$ and $\IZ_2 \times \IZ_2$, respectively. It is most convenient to study the action of the Galois group on the representations $(\mathbf{l},\mathbf{m},\mathbf{s})_\psi \in \cR$ separately in the two cases. We study the second case first.

\subsubsection*{Case 2: $\mathbb{Q}(\mtS_{(\mathbf{l},\mathbf{m},\mathbf{s})_\psi})=\mathbb{Q}(\zeta_8)$}
\vskip-5pt
In this case the Galois group $\Gal(\IQ(\zeta_8)/\IQ) \simeq \IZ_2 \times \IZ_2$ is generated by two elements $\rho_3$ and $\rho_5$ that act as
\begin{align}
	\rho_3(\zeta_8) \= \zeta_8^3~, \qquad \rho_5(\zeta_8) \= \zeta_8^5~.
\end{align}
On $\IQ(\ii) \subset \IQ(\zeta_8)$, where the $\mtS$-matrix elements of the last three minimal model factors are valued, these restrict to $\theta_3 \in \Gal(\IQ(\ii)/\IQ)$ and $\theta_1 \in \Gal(\IQ(\ii)/\IQ)$, respectively, which map
\begin{align}
	\theta_3(\ii) \= - \ii~, \qquad \theta_1(\ii) \= \ii~.
\end{align}
Since this case does not include the representations $(\textbf{l},\textbf{m},\textbf{s})$ with $l_1 = l_2 = 3$, which is the condition to have a non-trivial stabiliser, the relevant generalised $\mtS$-matrix elements are simply products of minimal model $\mtS$-matrix elements. Therefore we can immediately deduce the action of the above two generators on the representations $(\Rla{l_1} \otimes \cdots \otimes \Rla{l_5})_\psi \in \cR$. For instance, for $\rho_3$, we have that
\begin{align}
	\begin{split}
	\rho_3 \left( \mtS_{\{\textbf{L},\textbf{M},\textbf{S}\}_\Psi, (\Rla{l_1} \otimes \cdots \otimes \Rla{l_5})_\psi} \right) &\=  \prod_{i=1}^2 \rho_3\left(\mtS^{(i)}_{(L_i,M_i,S_i),\Rla{l_i}}\right) \prod_{i=3}^5 \theta_3 \left(\mtS^{(i)}_{(L_i,M_i,S_i),\Rla{l_i}}\right)\\
	&\=  \mtS_{\{\textbf{L},\textbf{M},\textbf{S}\}_\Psi, (\Rla{l_1'} \otimes \cdots \otimes \Rla{l_5'})_\psi}~,
	\end{split}
\end{align}
where $l_i'$ are the unique integers such that
\begin{align}
	\begin{split}
	&0 \leq l_i' \leq 6 \qquad \text{and} \qquad l_i' \equiv 3(l_i+1) -1 \!\! \mod 8~, \qquad \text{for} \qquad i=1,2~,\\
	&0 \leq l_i' \leq 2 \qquad \text{and} \qquad l_i' \equiv 3(l_i+1) -1 \!\! \mod 4~, \qquad \text{for} \qquad i=3,4,5~.
	\end{split}
\end{align} 
The same argument, applied for $\rho_5$ shows that it maps the representations $(\Rla{l_1} \otimes \cdots \otimes \Rla{l_5})_\psi \in \cR$~as
\begin{align}
	\varrho_5((\Rla{l_1} \otimes \cdots \otimes \Rla{l_5})_\psi) \= 	(\Rla{l_1'} \otimes \Rla{l_2'} \otimes \Rla{l_3} \otimes \Rla{l_3} \otimes \Rla{l_5})_\psi~,
\end{align}
where $l_i'$ are again unique integers determined by
\begin{align}
		0 \leq l_i' \leq 6 \qquad \text{and} \qquad l_i' \equiv 5(l_i+1) -1 \!\! \mod 8~.
\end{align} 

\subsubsection*{Case 1: $\mathbb{Q}(\mtS_{(\mathbf{l},\mathbf{m},\mathbf{s})_\psi})=\mathbb{Q}(i)$}
\vskip-5pt
In the first case, the Galois group $\Gal(\IQ(\ii)/\IQ) \simeq \IZ_2$ is generated by a single element $\rho_3$, which maps 
\begin{align}
	\rho_3(\ii) \= -\ii~.
\end{align}
If either of the stabiliser groups $\text{Stab}_\cA(\Rla{l_1} \otimes \dots \otimes \Rla{l_5})$, $\Rla{l_1} \otimes \dots \otimes \Rla{l_5} \in \cR$, or $\text{Stab}(\{\mathbf{L},\mathbf{M},\mathbf{S}\})$, $\{\mathbf{L},\mathbf{M},\mathbf{S} \} \in \cB$, is trivial, then the generalised $\mtS$-matrix \eqref{eq:Octic_Generalised_S-matrix} reduces to a product of minimal model $\mtS$-matrices, and using an argument identical to that employed in the first case, we find that $\rho_3 \in \Gal(\IQ(\ii)/\IQ)$ acts on the representations $(\Rla{l_1} \otimes \cdots \otimes \Rla{l_5})_\psi \in \cR$~as
\begin{align}
	\varrho_3((\Rla{l_1} \otimes \cdots \otimes \Rla{l_5})_\psi) \= 	(\Rla{l_1'} \otimes \Rla{l_2'} \otimes \Rla{l_3'} \otimes \Rla{l_4'} \otimes \Rla{l_5'})_\psi~,
\end{align}
where $l_i'$ are the unique integers determined by
\begin{align}
	\begin{split}
		&0 \leq l_i' \leq 6 \qquad \text{and} \qquad l_i' \equiv 3(l_i+1) -1 \!\! \mod 8~, \qquad \text{for} \qquad i=1,2~,\\
		&0 \leq l_i' \leq 2 \qquad \text{and} \qquad l_i' \equiv 3(l_i+1) -1 \!\! \mod 4~, \qquad \text{for} \qquad i=3,4,5~.
	\end{split}
\end{align} 
If both of the aforementioned stabiliser groups are non-trivial, then there is an additional term 
\begin{align} \label{eq:Octic_R_extra_term}
- 8 \Psi(\cL) \psi(\cL) \prod_{i=1}^2 e^{\pi \ii (m_i M_i /8 + S_i /2)} \prod_{i=3}^5 \mtS^{(i)}_{(L_i,M_i,S_i),(l_i,m_i,s_i)}~,
\end{align}
which is not a product of minimal model $\mtS$-matrices. Note that the only representations in $\cR$ with non-trivial stabilisers are of the form $(\Rla{3} \otimes \Rla{3} \otimes \Rla{l_3} \otimes \Rla{l_4}  \otimes \Rla{l_5})_\psi$, and therefore have $m_1=m_2=-4$. The fact that the boundary label has a non-trivial stabiliser implies that $L_1 = L_2 = 3$.  Since we have chosen the set of boundary labels $\cB$ such that $(L_i,M_i,S_i) \in \text{HWR}(\cA_i)_\text{R}$, we have that $S_1$ and $S_2$ are odd, which in turn implies that $M_1$ and $M_2$ are even. Hence, the additional term takes the form
\begin{align}
	- 8 \Psi(\cL) \psi(\cL) \prod_{i=1}^2 (-1)^{(M_i - 1) /2} \prod_{i=3}^5 \mtS^{(i)}_{(L_i,M_i,S_i),(l_i,m_i,s_i)}~,
\end{align}
Thus under the action of $\rho_3$, this term transforms to
\begin{align}
	\prod_{i=1}^5 \epsilon_{\rho_3}(\Rla{l_i}) \left( - 8 \Psi(\cL) \psi(\cL) \prod_{i=1}^2 (-1)^{(M_i - 1) /2} \prod_{i=3}^5 \mtS^{(i)}_{(L_i,M_i,S_i),\varrho_3(l_i,m_i,s_i)} \right)
\end{align}
where we have used that $\epsilon_{\rho_3}(\Rla{3})^2 = 1$. Therefore the additional term \eqref{eq:Octic_R_extra_term} transforms in the same way as the product of minimal model $\mtS$-matrices, so that the corresponding transformation on the representations is
\begin{align}
	\varrho_3((\Rla{3} \otimes \Rla{3} \otimes \Rla{l_3} \otimes \Rla{l_4} \otimes \Rla{l_5})_\psi) \= (\Rla{3} \otimes \Rla{3} \otimes \Rla{l_3'} \otimes \Rla{l_4'} \otimes \Rla{l_5'})_\psi~,
\end{align}
and the overall sign is given by the product $	\prod_{i=1}^5 \epsilon_{\rho_3}(\Rla{l_i})$. Thus we have shown that the compatibility condition is satisfied in all cases.

This is enough to show that the proof given in section \ref{sect:CM_Hodge_Structures_from_RCFTs} can be used to show that the Hodge structure $(V_\IQ,V_\IC)$ admits sufficiently many complex multiplications. Since the intersection matrix $\Sigma$ given in eq.~\eqref{eq:intersection_matrix_Octic} is antisymmetric, it can be shown that this Hodge structure is indeed polarisable, and hence the Hodge structure is in fact of CM type. To be slightly more detailed, the proof of section \ref{sect:CM_Hodge_Structures_from_RCFTs} implies that the Hodge structure splits into 27 Hodge substructures $W^{(\mathbf{l},\mathbf{m},\mathbf{s})_\psi}_\IQ$ corresponding to representations $(\mathbf{l},\mathbf{m},\mathbf{s})_\psi \in \cR$ with $\IQ(\mtS_{(\mathbf{l},\mathbf{m},\mathbf{s})_\psi}) = \IQ(\zeta_8)$ with Hodge endomorphism algebra is isomorphic to $\IQ(\zeta_8)$, and 33 Hodge substructures corresponding to representations with $\IQ(\mtS_{(\mathbf{l},\mathbf{m},\mathbf{s})_\psi}) = \IQ(\ii)$, with Hodge endomorphism algebra isomorphic to $\IQ(\ii)$.

This Gepner model corresponds to the degree eight hypersurface \cite{Gepner:1987vz,Gepner:1987qi,Greene:1988ut,Martinec:1988zu,Witten:1993yc}
\begin{align}
	\left\{ \ x_1^8 + x_2^8 + x_3^4+x_4^4+x_5^4=0 \right\} \; \subset \;  \mathbb{P}^4_{1,1,2,2,2}~.
\end{align}
Therefore, the results of this section, can be viewed as a physics proof that this geometry admits complex multiplication. 

\newpage
\section{Summary and Discussion}
\vskip-10pt
In these concluding remarks we summarise our results and offer some outlook about broadening the scope of our analysis in section~\ref{sect:sum_outlook}. As our findings and the existing literature on the topic indicate a more general relationship between two-dimensional conformal field theories and arithmetic geometry, we contemplate about a motivic perspective on conformal field theories in general in section~\ref{sect:CM_and_arithmetic}, which we anticipate to further investigate in the future.

\subsection{Summary and outlook} \label{sect:sum_outlook}
\vskip-5pt
In this paper, we have provided a construction that associates to a unitary $\cN=(2,2)$ supersymmetric conformal field theory a corresponding Hodge structure. We have also shown that the notion of Galois symmetry \cite{Coste:1993af,DeBoer:1990em} of rational conformal field theories is intimately linked to these Hodge structures being of CM type.

One of our main motivations for introducing this construction is that it allows us to study the conjectured equivalence between rational conformal field theories and complex multiplication \cite{Gukov:2002nw,Chen:2005gm,Kidambi:2024vwl,Kidambi:2022wvh,Okada:2022jnq} as a purely conformal field theoretic question. To this end, in section~\ref{sect:Hodge_structures_with_CM}, we used the Galois symmetry to show that the rational Hodge structures we define have always sufficiently many complex multiplications, subject to a technical assumption that the action of the relevant Galois group closes on the $\text{(c,c)}_{q_L=q_R}$-ring, in the sense explained in eq.~\eqref{eq:compatibility_R}. Thus these Hodge structures are also of CM type if they are polarised. We show that in many cases a polarisation for this Hodge structure can be obtained from a twisted Witten index defined in eq.~\eqref{eq:SigmaRat}. Such cases include in particular theories with integral $\mtU(1)$-charges, and thus the theories with a known geometric interpretation. It turns out that the polarisation also exists even in many cases where it is not provided by the twisted Witten index. 

In section~\ref{sect:real_Hodge_structures}, we show that any unitary $\cN=(2,2)$ supersymmetric conformal field theory with a finite $\text{(c,c)}$-ring has an associated real Hodge structure. The key idea of the construction is that the $\mtU(1)$-charges of the states in the $\text{(c,c)}_{q_L=q_R}$-ring can be identified with the grading of the Hodge structure. The real structure is given by the complex conjugation map whose action on the conformal field theory states is given by application of the charge conjugation operator. The real Hodge structure can be refined to give a rational Hodge structure if there are enough boundary states in the boundary conformal field theory. Then, as shown in section~\ref{sect:Rational_Hodge_structs_from_RCFTs}, one can define a rational vector space by taking the $\IQ$-linear combinations of the boundary states. It is shown that this is compatible with the real structure, meaning that the complex conjugation map introduced in section~\ref{sect:real_Hodge_structures} keeps the boundary states fixed. 

While one of the main utilities of this construction of Hodge structures is that their definition does not require a description of the conformal field theory as arising from a non-linear sigma model on a target Calabi--Yau geometry, the construction of the Hodge structure is such that in this setting it reproduces the geometric Hodge structure, as shown in section~\ref{sect:relation_to_geometry}.

We exemplify the general construction by presenting three different (families of) theories for which we can construct CM-type Hodge structures: $\cN=2$ minimal models, the Gepner model $(5,5,5,5,5)$ corresponding to the quintic threefold, and the Gepner model $(6,6,2,2,2)$, which corresponds to the Calabi--Yau manifold given as the hypersurface in the projective space $\mathbb{P}^4_{1,1,2,2,2}[8]$.\footnote{Strictly speaking, the general construction of section \ref{sect:Hodge_structures_from_RCFT} does not apply for the last example as to obtain enough boundary states, we have to use the `generalised S-matrix' of eq.~\eqref{eq:Octic_Generalised_S-matrix}. However, as we show in section \ref{sect:Gepner_model_octic}, the arguments presented in section~\ref{sect:Hodge_structures_from_RCFT} remain valid mutatis mutandis.} For these models, we use known expressions for the boundary states to construct the rational Hodge structure. Then, by using the explicit form of the S-matrices of these theories, we are able to find the relevant Galois groups, verify that their actions on the S-matrices satisfy the compatibility condition~\eqref{eq:compatibility_R}, which is enough to show that the Hodge structures are of CM type.

Since the construction presented in this paper can also be applied to models which do not have a geometric sigma model interpretation, such as minimal models, it is an interesting question whether the original conjecture of ref.~\cite{Gukov:2002nw} can be extended to cover these non-geometric rational conformal field theories. For instance, we have shown that all $\cN=2$ minimal models have a corresponding Hodge structure with sufficiently many complex multiplications. We have seen the twisted Witten index \eqref{eq:SigmaRat} gives rise to polarisation for all minimal models of odd level. Additionally, we have examined several examples of minimal models of even level, and found a polarisation for each of these examples. This raises the question whether all $\cN=(2,2)$ rational conformal field theories have associated A- and B-type Hodge structures which are of CM type and compatible in an appropriate sense.\footnote{It has been pointed out for instance in refs.~\cite{Chen:2005gm, Kidambi:2024vwl,Kidambi:2022wvh,Okada:2022jnq} that there is a subtlety in the conjectured equivalence between rationality and complex multiplication \cite{Gukov:2002nw} in the sense that one must be careful in stating which Hodge structures have to be of CM type to obtain an equivalence.}

Due to the fact that the construction presented is purely conformal field theoretic, we can only probe one direction of the conjectured equivalence between the conformal field theories and geometries with complex multiplication. If we were able to show that the condition of eq.~\eqref{eq:compatibility_R} was satisfied for all $\cN=(2,2)$ rational conformal field theories, we would have shown that every such theory has an associated Hodge structure with sufficiently many complex multiplications. However, the claim that every geometry with complex multiplication has a corresponding rational conformal field theory would still remain elusive. 

While not sufficient for a proof, the connection of rationality to Galois symmetry might give some rough heuristics on the question of existence of the conjectured equivalence. Namely, the decomposition of the spectrum into finitely many irreducible representations under $\cA$ was the crucial ingredient in showing that the S-matrix elements are algebraic numbers and thus define a Galois extension. If the theory has infinitely many representation, the characteristic polynomials \eqref{eq:N_matrix_characteristic_polynomial} of the fusion matrices would be of infinite degree, and therefore allowing for existence of transcendental $S$-matrix elements.

To substantiate the relationship among rationality of conformal field theories, Galois symmetries of the Verma modules of the chiral algebras of conformal field theories, and CM type Hodge structures associated to conformal field theories, further studies of rational two-dimensional rational conformal field theories are important. The theories discussed here are based on two-dimensional $\mathcal{N}=(2,2)$ supersymmetric rational conformal field theories that obey the assumptions spelt out on page~\pageref{p:assumptions}.

Firstly, these assumptions are possibly not all independent in the sense that starting from a more axiomatic definition of two-dimensional conformal field theories, some of the made assumptions may follow from a suitable axiomatic approach. Secondly, it would be relevant to see, which of these assumptions can be relaxed, while Hodge structures of CM type still indicate rationality of the underlying conformal field theory. By presenting a Gepner model that does not obey the last assumption in section~\ref{sect:Gepner_model_octic}, we have exemplified that a more general framework for Hodge structures of CM type can be formulated.

In order to identity the essence of arithmetic properties of rational conformal field theories, it is important to systematically examine broader classes of rational two-dimensional conformal field theories. In this context it would be worthwhile to study rational conformal field theories based on Kazama--Suzuki models \cite{Kazama:1988uz}, asymmetric orbifolds of rational conformal field theories \cite{Narain:1986qm}, and rational realizations of conformal field theories based on the Shatashvili--Vafa construction \cite{Shatashvili:1994zw,Roiban:2001cp}. Moreover, in this work supersymmetry plays an important to define Hodge structures for two-dimensional conformal field theories. Therefore, it would be interesting to examine what arithmetic data can still be defined for rational two-dimensional conformal field theories in the absence of supersymmetry.

\subsection{Conformal field theory, complex multiplication, and arithmetic} \label{sect:CM_and_arithmetic}
\vskip-5pt
Complex multiplication is also intrinsically related to arithmetic geometry whose relation to physics has seen recent increase in interest. For instance, arithmetic geometry has been studied in connection to BPS black hole solutions and D-branes (e.g. refs.~\cite{Bonisch:2022mgw,Candelas:2019llw}), flux vacua (e.g. refs.~\cite{Kachru:2020abh,Kachru:2020sio,Candelas:2023yrg,Schimmrigk:2020dfl}), and most importantly for this work, conformal field theories and complex multiplication \cite{Schimmrigk:2008mp,Schimmrigk:2006aa,Kidambi:2022wvh,Kidambi:2024vwl,Kondo:2018mha,Kondo:2019jpi}. Indeed, some connections between conformal field theories and arithmetic geometry have been drawn in these works. In the remaining of this section, we aim to very briefly explain some central results and ideas of these works, and speculate on possible connections to the construction of Hodge structures presented in this paper. The topics covered in this section are very wide and intricate, so we will necessarily be very sketchy, aiming only to provide the reader with a rough overview of some important ideas. For more discussion of the relevant concepts of arithmetic geometry, aimed at physicists, see for example refs.~\cite{Kachru:2020sio,Bonisch:2022mgw,Jockers:2023zzi}.

Studying spaces as sets of solutions of polynomial equations over finite fields, such as  $\IF_p \simeq \IZ/p\IZ$ integers modulo a prime $p$, is a central aspect of arithmetic geometry. For instance, a basic question in arithmetic geometry could be, how many points does an elliptic curve defined over $\IQ$, i.e. defined by polynomials with rational coefficients, have over $\IF_p$. That is, how many tuples $(X:Y:Z)$, up to projective equivalence, satisfy the congruence
\begin{align}
 ZY^2 - (X^3 + A Z^2X + B Z^3) \; \equiv \; 0 \mod p~.
\end{align}
For any non-singular projective variety $X$, point counts $\# X(\IF_{p^n})$ over finite fields $\IF_{p^n}$ with $p^n$ elements can be collected into a generating function, called the \textit{Hasse--Weil zeta function}, defined by
\begin{align} \label{eq:zeta_definition}
	\zeta_p(X;T) \; \defineas \; \exp \left( \sum_{n=1}^\infty \frac{\# X(\IF_{p^n}) T^n}{n} \right).
\end{align}
The utility of this function is that it satisfies some useful and intriguing properties which were set out as so-called \textit{Weil conjectures} \cite{Weil1949a}, by now proven \cite{Dwork1960a,Grothendieck1995a,Deligne1974a,Deligne1980a}. For instance, as a function of $T$, $\zeta_p(X;T)$ is a rational function of $T$, taking the form
\begin{align} \label{eq:zeta_function}
	\zeta_p(X;T) \= \frac{P^{(1)}_p(T)\cdots P^{(2d-1)}_p(T)}{P^{(0)}_p(T)\cdots P^{(2d)}_p(T)}~,
\end{align}
where the degrees of the polynomials are given by the Betti numbers of $X$, $\deg_T P^{(i)}_p(T) = b_i(X)$, and the roots $\lambda_{i,j}$ of $P^{(i)}_p(T)$ are complex numbers with absolute value $|\lambda_{i,j}| = p^{i/2}$. 

It is a remarkable fact that the polynomials $P_i(T)$ can be seen as arising from a representation of the absolute Galois group $\Gal(\overline{\IQ}/\IQ)$ acting on certain cohomology groups $H^i(X)$ of the manifold $X$. According to modularity theorems and conjectures, such representations correspond to automorphic representations, which, in simplest terms, implies that we can use the polynomials $P^{(i)}_p(T)$ to define a corresponding L-function, which
\begin{align}
	L^{(i)}(X;s) \; \defineas \; \prod_{p \text{ prime}} \left(P^{(i)}_p(p^{-s}) \right)^{-1}~,
\end{align}
where the polynomials $P^{(i)}_p(p^{-s})$ are exactly those appearing in the zeta function \eqref{eq:zeta_function}, except possibly for a finite number of `bad' primes. One can also define L-functions for various modular forms. Then various modularity conjectures and theorems relate the L-functions of the variety $X$ to L-functions of modular forms. 

Consider again the special case of elliptic curves $E$ defined over $\IQ$. For these, the zeta function takes the form
\begin{align}
	\zeta_p(E;T) \= \frac{1-a_pT + pT^2}{(1-T)(1-p^2T)}~.
\end{align}
and the L-function of $E$ is given by (up to factors corresponding to `bad' primes)
\begin{align}
	L(E,s) \= \prod_{p \text{ prime}} (1-a_p p^{-s} + p^{1-2s})~.
\end{align}
For an ordinary modular form
\begin{align}
	f(q) \= \sum_{i=0}^\infty c_i \; q^i~,
\end{align}
the L-function can be defined, up to factors corresponding to bad primes, as
\begin{align}
	L(f,s) \= \prod_{p \text{ prime}} (1 - c_p p^{-s} + p^{1-2s})^{-1}~.
\end{align}
and the modularity theorem for elliptic curves defined by polynomials with coefficients in $\IQ$ states that for every such elliptic curve $E$, there exists a weight-2 modular form $f$ so that
\begin{align}
	L(E,s) \= L(f,s)~.
\end{align}
In the case of elliptic curves, complex multiplication has an interpretation as a `symmetry' of the curve, so it stands to reason that there is a corresponding `symmetry' property of the modular form. In this case it is the existence of a \textit{self-twist} of the corresponding modular form $f$, which means that for all but possibly `bad' primes $c_p = \chi(p) c_p$, where $\chi(p)$ is the Kronecker character of imaginary quadratic field, taking values $\pm 1$. In particular, this implies that $a_p = 0$ for all $p$ for which $\chi(p)=-1$, so half of the coefficients $a_p$~vanish. 

The above is a first indication that complex multiplication is intrinsically linked to arithmetic geometry. Indeed, Deuring's theorem (see e.g. ref.~\cite{Silverman}) shows that the L-functions of elliptic curves with Hodge structure of CM-type can be expressed in terms of L-functions of (algebraic) \textit{Hecke characters}. Given a number field $K$, its ring of integers $\cO_K$ and an ideal $\fm \subset \cO_K$ the Hecke character $\chi$ of modulus $\fm$ can be thought of as functions from the set of ideals of $\cO_K$ to $\IC$ which satisfy properties analogous to those of the familiar Dirichlet characters. Namely, we require that $\chi(\cO_K) = 1$, and that for any ideals $\fa,\fb \subseteq \cO_K$, $\chi(\fa \fb) = \chi(\fa) \chi(\fb)$ and $\chi(\fa) \neq 0$ only if $(\fa,\fm) = 1$. In addition, $\chi$ is subject to a few technical conditions which we will not attempt to describe here.

The L-function of a Hecke character $\chi$ also be defined in terms of Euler factors:
\begin{align}
	L_\chi(s) \; \defineas \; \prod_{\substack{\fp \subseteq \cO_K\\\text{prime}}} \left(1 - \chi(\fp) N(\fp)^{-s}\right)^{-1}~,
\end{align}
where $N(\fp) \defineas |\cO_K/\fp|$ denotes the ideal norm. A special case of Deuring's theorem states that for an elliptic curve $E/\IQ$ with complex multiplication, there exists a Hecke character $\chi$ such~that
\begin{align} \label{eq:Deuring}
	L_\chi(s) \= L(E,s)~.
\end{align}
For instance, let us consider the elliptic curve with the LMFDB \cite{LMFDB} label \textbf{576.i1}:
\begin{align}
	y^2 \= x^3 - 3x~,
\end{align}
which has a Hodge structure of complex multiplication type, with the corresponding CM-field being~$\IQ(i)$, which has the ring of integers $\IZ[i]$.

Given a prime ideal $\fp \subseteq \IZ[i]$, let $\pi$ denote its unique \textit{primary generator}, that is, an element $\pi$ such that $\fp = (\pi)$ and $\pi \equiv 1 \!\! \mod 2+2\ii$. For each such primary generator, we define the 4th power residue symbol $\left( \smallfrac{\alpha}{\pi} \right)_4$ as the unique 4th root of unity satisfying
\begin{align}
	\left( \frac{\alpha}{\pi} \right)_4 \; \equiv \; \alpha^{(N(\fp)-1)/4} \mod \pi
\end{align}
if $\pi \nmid \alpha$. Otherwise we define $\left( \smallfrac{\alpha}{\pi} \right)_4 = 0$. One can show that with these definitions
\begin{align}
	\chi(\fp) \; \defineas \; \overline{\left( \frac{3}{\pi} \right)_4} \pi
\end{align}
defines a Hecke character. Furthermore it can be shown that this character satisfies eq.~\eqref{eq:Deuring}.\footnote{To see this, one can, for instance, derive the following result using Jacobi sums: \begin{align}
		a_p = \begin{cases}
			0 & \text{ if } p \equiv 3 \mod 4~,\\
			\overline{\left( \frac{3}{\pi} \right)_4} \pi + \left( \frac{3}{\pi} \right)_4 \overline{\pi} & \text{ otherwise}~,
		\end{cases}
\end{align}
and then proceed as in eqs.~\eqref{eq:p=5_Euler_factor} and \eqref{eq:p=7_Euler_factor}.
} To get an idea how the relation given by Deuring's theorem works, consider, for instance $p=5$, which splits in $\IZ[i]$ into $5 = (1+2\ii)(1-2\ii)$, the numerator of the zeta function is given by
\begin{align} \label{eq:p=5_Euler_factor}
\begin{split}
	P_5^{(1)}(p^{-s}) &\= 1 - 4 p^{-s} + p^{1-2s} \= (1 -(2{-}\ii)p^{-s})(1 - (2{+} \ii)p^{-s})\\ &\= \left( 1+ \chi((1+2\ii))N((1+2\ii))^{-s} \right)\left( 1+ \chi((1-2\ii))N((1-2\ii))^{-s} \right) ~,
\end{split}
\end{align}
For $p=7$, which is inert in $\IZ[i]$, the Euler factor is given by
\begin{align} \label{eq:p=7_Euler_factor}
	P^{(1)}_7(p^{-s}) \= 1 + p^{1-2s} \= 1+\chi((p))N((p))^{-s}~,
\end{align}
in accordance with the fact that half of the coefficients $a_p$ vanish. Every Euler factor behaves as one of the examples above, depending on whether the corresponding prime is split or inert in $\IZ[i]$. 

These observations are believed to readily generalise to higher-dimensional Calabi--Yau varieties with complex multiplication. This can be tested, for instance, for the one-parameter family of mirror quintic threefolds, which are given by $\IZ_5^3$ quotients of
\begin{align}
	x_1^5 + x_2^5 + x_3^5 + x_4^5 + x_5^5 + \psi x_1 x_2 x_3 x_4 x_5 \= 0 \; \subset \; \IP^5~,
\end{align}
where $\psi$ is the complex structure parameter. In particular, at $\psi = 0$ this gives the mirror quintic threefold. We have seen in section \ref{sect:Fermat_quintic} that the Fermat quintic has a Hodge structure of CM type, and by studying how taking the quotient affects the Hodge structure (or by using B-type Hodge structures), it is not difficult to show that the mirror Fermat quintic has also a 4-dimensional Hodge structure which is of CM type with the CM-field~$\IQ(\zeta_5)$ with its ring of integers given by $\IZ[\zeta_5]$.

From the Weil conjectures, it follows that the zeta function of a one-parameter Calabi--Yau threefold $X$ is of the form
\begin{align}
	\zeta_p(X;T) \= \frac{R_p(T)}{(1-T)(1-pT)^{h_{11}}(1-p^2T)^{h_{11}}(1-p^3T)}~,
\end{align}
where $R_p(T)$ is a degree-4 polynomial. For the first few primes, the zeta functions of the family of mirror quintic threefolds were computed in refs.~\cite{Candelas:2000fq,Candelas:2004sk}. It was found that for the mirror Fermat quintic, analogously to the elliptic curve case, the form of the polynomial $R_p(T)$ depends on how the rational prime $p$ factorises in $\IZ[\zeta_5]$: Let $\rho$ denote the residue degree least integer such that $p^\rho \equiv 1 \mod 5$, so that $\rho$ is the residue degree of any prime divisor in $\IZ[\zeta_5]$ of $p$. Then, for $\rho$~even,
\begin{align}
	R_p(T) \= (1-p^{3\rho/2} T^\rho)^{4/\rho}~,
\end{align}
whereas for $\rho = 1$, the polynomial $R_p(T)$ takes the form
\begin{align}
	R_p(T) \= 1 + a_p T + b_p p T^2 + a_p p^3 T^3 + p^6 T^4 \= (1 + \gamma_+ p T + p^3 T^2)(1 + \gamma_- p T + p^3 T^2)~,
\end{align}
where $a_p, b_p \in \IZ$, and $\gamma_\pm \in \IZ[\sqrt{5}]$. Further, as expected, over the ring of integers $\IZ[\zeta_5]$, the polynomial factorises into linear factors. In fact, the polynomials $R_p(T)$ computed in ref.~\cite{Candelas:2004sk} correspond exactly to the Euler factors of a Hecke character, see also for related work refs.~\cite{Watkins:2011,Watkins:2018}.

Given the relevance of arithmetic geometry to complex multiplication, and the fact that the relevant arithmetic data can be found by considering a Galois action on certain cohomology groups, it is natural to ask whether conformal field theory data can be used to find this Galois action. In this work, we have essentially constructed the singular cohomology using the boundary states, and the de Rham cohomology by using the states in the $(\text{c},\text{c})_{q_L=q_R}$-ring. These are two of the three `classical' realisations of a \textit{motive}, roughly speaking an algebraically defined piece of cohomology. The third realisation, which encodes the arithmetic data is the étale cohomology (see for example ref.~\cite{Kachru:2020sio} for a brief physicist-oriented review). It is an interesting question whether étale cohomology, or another arithmetically interesting cohomology theory,\footnote{In practice, it may be easier to construct a cohomology theory that admits a more explicit description, such as the Monsky-Washnitzer cohomology (see e.g. ref.~\cite{vanderPut1986a}).} could also be constructed from the conformal field theory data. 

While existence of a natural construction of this type may seem unlikely, there are concrete indications that the conformal field theory might encode arithmetic data. For instance, in ref.~\cite{Schimmrigk:2006aa}, the coefficients $a_p$ of elliptic curves with corresponding Gepner models was computed using conformal field theory data, so-called string functions. In refs.~\cite{Kondo:2018mha,Kondo:2019jpi} this data is obtained for elliptic curves of complex multiplication type by considering traces of conformal field theory operators over subspaces of the Hilbert space of states of the theory. It would be interesting to see if there constructions can be generalised and systematised to a larger class of conformal field theories.

\vfill
\section*{Acknowledgements}
\vskip-10pt
We are grateful to Ilka Brunner, Suresh Govindarajan, Volker Schomerus, and Johannes Walcher for sharing their knowledge and insights on aspects of conformal field theories. We thank Philip Candelas and Xenia de la Ossa for discussions on arithmetic geometry of Fermat quintic. It is a pleasure to acknowledge fruitful conversations with and useful comments by Mohamed Elmi, and Albrecht Klemm. We are grateful to Duco van Straten and Matt Kerr for insightful comments on Hodge structures and complex multiplication. We wish to thank Thomas Grimm, Damian van de Heisteeg, and David Prieto for comments on the work and discussion on applications of Hodge loci to string theory problems. We thank Abhiram Kidambi for extensive discussions on the topic of complex multiplication and on fruitful comments related to this work. We thank Taizan Watari for instruction on the literature on complex multiplication and conformal field theories.

We would like to thank the Mainz Institute for Theoretical Physics (MITP) of the Cluster of Excellence PRISMA+ (Project ID 390831469) for support and express our gratitude for its hospitality during the workshop ``The Arithmetic of Calabi--Yau Manifolds''. We would like to acknowledge the ICTP meetings ``Workshop on Number Theory and Physics'' and ``Workshop on Explicit Arithmetic Geometry'', during which much progress on the present work was made. PK is supported by the Emil Aaltonen Foundation. MS is supported by the Australian Research Council Discovery Project DP250101828.

\newpage

\appendix

\section{Definitions of CM-type Hodge structures} \label{app:complex_multiplication}
\vskip-10pt
Since there exist several equivalent definitions for Hodge structures of CM type in the literature, we briefly review some other common useful definitions in this appendix, and discuss their relation to the definition given in section \ref{sect:Hodge_structures_with_CM}, which we use in the main body of the paper. We keep the discussion very brief, and refer the reader interested in further aspects of the theory of complex multiplication to, for example, refs.~\cite{Lang1983a,Green2012a} and references therein.

For simplicity, since we follow the convention where CM-type Hodge structures are polarisable, in this section we take all Hodge structures to be polarisable. However, several results generalise to non-polarisable Hodge structures in an obvious fashion. 

\subsection{Commutative subalgebras}
For instance ref.~\cite{Okada:2023udq} defines that an irreducible polarisable Hodge structure $(V_\IQ,V_\IC)$ is of complex multiplication (CM) type if the algebra of Hodge endomorphisms $\text{End}_\text{Hdg}(V_\IQ)$ contains a commutative subalgebra $\cE$ whose dimension as a vector space over $\IQ$ is equal to the dimension of $V_\IQ$, $\text{dim}_\IQ(\cE) = \text{dim}_{\mathbb{Q}} V_{\mathbb{Q}}$. This condition is easily seen to be equivalent to the one used in the main text. To see that our condition implies this one, we just note that if $\eta: E \hookrightarrow \text{End}_\text{Hdg}(V_\IQ)$ is an embedding of a number field of degree $[E:\IQ] = \dim_\IQ V_\IQ$, the image $\eta(E) \subseteq \text{End}_\text{Hdg}(V_\IQ)$ provides a commutative subalgebra of dimension $\text{dim}_\IQ(\cE) = [E:\IQ] = \dim_\IQ V_\IQ$.

To see that the opposite implication holds, we use that $\text{End}_\text{Hdg}(V_\IQ)$ is a division algebra. Recalling that any finite-dimensional unital subalgebra of a finite-dimensional division algebra is a division algebra,\footnote{We know that the dimension of a commutative subalgebra $\cE$ is bounded by $\dim_\IQ V_\IQ$, and therefore if $\dim_\IQ \cE = V_\IQ$, it must contain the identity matrix.} we have that the subalgebra $\cE$ is a commutative division algebra, and thus a field. The field $\cE$ is isomorphic to a number field $E$, as it includes the field $\IQ \mtI$ of $\IQ$-multiples of the identity matrix as a subfield, and this field is isomorphic to $\IQ$. Thus the inclusion $E \simeq \cE \hookrightarrow \text{End}_\text{Hdg}(V_\IQ)$ provides the required embedding of a number field $E$ of degree $[E:\IQ] = \dim_\IQ V_\IQ$ to $\text{End}_\text{Hdg}(V_\IQ)$.

\subsection{Mumford--Tate group and Hodge tensors} \label{app:Mumford-Tate_group}
There is an equivalent definition of a Hodge structure of CM type via the Mumford--Tate group. While the above definition is more suited for explicit computations and suffices in particular for understanding the construction presented in this paper, we find it useful to briefly introduce the equivalent definition as well. The definitions we discuss here have in particular been used in physics context in ref.~\cite{Grimm:2024fip}, which also contains an excellent introduction to the Mumford--Tate group and Hodge tensors. A detailed mathematical discussion can be found for example in ref.~\cite{Green2012a}, which we largely follow here.

\subsubsection*{Definition of Hodge structures via Deligne tori}
\vskip-5pt
To discuss the Mumford--Tate group of a Hodge structure, we need first to give an alternative definition of a Hodge structure. Let us first define the \textit{Deligne torus} $\IS(\IR)$ and its maximal compact subgroup $\IU(\IR)$ as the $\IR$-algebraic groups given by
\begin{align}
	\IS(\IR) \; \defineas \; \left\{  \left( \begin{matrix}
		a & b \\
		-b & a
	\end{matrix} \right) \in \GL(2,\IR) \right\}~, \qquad \IU(\IR) \; \defineas \; \left\{  \left( \begin{matrix}
	a & b \\
	-b & a
\end{matrix} \right) \in \SL(2,\IR) \right\}~.
\end{align} 
Note the isomorphisms $\IS(\IR) \simeq \IC^*$ and $\IU(\IR) \simeq \mtU(1)$ which are given by
\begin{align}
	\left( \begin{matrix}
		a & b \\
		-b & a
	\end{matrix} \right) \; \longmapsto \; a + \ii b \; =: \; z~.
\end{align} 
Let $V_\IQ$ be a $\IQ$-vector space. We say that it has a \textit{Hodge structure of weight m} if there exists a non-constant homomorphism of $\IR$-algebraic groups $\wt h: \IS(\IR) \to \GL(V_\IQ \otimes \IR)$ such that for every $r \in \IR^* \subset \IS(\IR) \simeq \IC^*$ 
\begin{align} \label{eq:Hodge_structure_homomorphism_weight}
	\wt h(r) \= r^m I~,
\end{align}
where $I$ is the identity endomorphism of $V_\IR  \defineas V_\IQ \otimes \IR$. We denote the Hodge structure as $(V_\IQ,h)$. This is equivalent to the definition given in section \ref{sect:Rational_Hodge_structures} as, given a Hodge structure $(V_\IQ,V_\IC)$, we can define such a morphism by defining $\wt h: \IS(\IR) \to \GL(V_\IQ \otimes \IR)$ that for all $z \in \IS(\IR)$
\begin{align}
	\wt h(z) \; v \= z^p \overline{z}^q \; v~, \qquad \text{for} \qquad v \in V_\IC^{p,q}~.
\end{align}
It is easy to show that the induced map on $V_\IR$ is indeed a homomorphism. Conversely, given a homomorphism $\wt h: \IS(\IR) \to \GL(V_\IR)$ with the property \eqref{eq:Hodge_structure_homomorphism_weight}, we can define the spaces $V^{p,q}_\IC$ by
\begin{align} \label{eq:Hodge_decomposition_eigenspaces}
	V^{p,q}_\IC \defineas \left\{ v \in V_\IC \; \middle| \; \wt h(z) \; v \=  z^p \overline{z}^q \; v \text{ for all } z \in \IS(\IR) \right\}~.
\end{align}
It is simple to verify that over the complexification $V_\IC \defineas V_\IQ \otimes \IC$ the simultaneous eigenspaces of the automorphisms $\wt h(z) \in \GL(V_\IC)$ are all of this form with $p+q = m$. It is also often convenient to consider the restricted morphism $h: \IU(\IR) \to \SL(V_\IR)$, defined by
\begin{align}
	h \; \defineas \; \wt h \; \big|_{\IU(\IR)}~.
\end{align}

\subsubsection*{The Mumford--Tate group and complex multiplication}
\vskip-5pt
Let $(V_\IQ,\wt h)$ be a Hodge structure. Following \cite{Green2012a}, we define the \textit{Mumford--Tate group (associated to $h$)}, $\text{MT}(h)$, to be the $\IQ$-algebraic closure of the image $h(\IU(\IR)) \subseteq \GL(V_\IR)$.\footnote{As noted in \cite{Green2012a}, the group $\text{MT}(h)$ defined in this way, is often also called the \textit{special Mumford--Tate group} or the \textit{Hodge group}, and the group $\text{MT}(\wt h)$ defined as the $\IQ$-algebraic closure is called the Mumford--Tate group.} That is, $\text{MT}(h)$ is the smallest $\IQ$-algebraic subgroup of $G$ containing $h(\IU(\IR))$:
\begin{align} \label{eq:Mumford-Tate_definition}
	\text{MT}(h) \; \defineas \; h(\IU(\IR))^{\text{alg}}
\end{align}
When the Hodge structure $(V_\IQ,\wt h)$ is polarised with polarisation $Q$, the map $h$ preserves the bilinear form $Q$, as can be verified by a simple computation: if $v \in V^{p,q}$ and $w \in V^{p',q'}$, then $Q(v,w)$ vanishes and is thus preserved trivially unless $p+p'=q+q'=m$ in which case for every $z \in \IU(z)$
\begin{align}
	Q(h(z)v, h(z)w) \= Q(z^{p}\overline{z}^q v, z^{p'}\overline{z}^{q'} w) \= z^{p+p'} \overline{z}^{q+q'} Q(v,w) \= |z|^{2m} Q(v,w) \= Q(v,w)~.
\end{align}
Therefore the for polarised Hodge structures, the Mumford--Tate group is in fact a subgroup of the group $G \defineas \text{Aut}(V_\IR,Q)$ of automorphisms of $V_\IR$ which preserve the polarisation $Q$.\footnote{The groups $G$ are explicitly given by $\Sp(\dim_\IQ V_\IQ)$ for $m$ odd and $\SO(\sum_{p \text{ even}} h^{p,q}, \sum_{p \text{ odd}} h^{p,q})$ for $m$ even.}

A polarised Hodge structure $(V_\IQ,\wt h)$ is said to be of CM-type if the Mumford--Tate group $\text{MT}(h)$ is Abelian, that is, an algebraic torus.

\subsubsection*{Hodge tensors}
\vskip-5pt
To see how this definition is connected to the one in terms of Hodge endomorphisms, we introduce \textit{Hodge tensors} as elements of the \textit{tensor spaces} $T^{k,l}$ and the \textit{tensor algebra} $T^{*,*}$ of $V_\IQ$, which are defined by
\begin{align}
	T^{k,l}_\IQ \; \defineas \; V_\IQ^{\otimes k} \otimes (V_\IQ^{\otimes l})^\vee~, \qquad \text{and} \qquad T^{*,*}_\IQ \; \defineas \; \bigoplus_{k,l \geq 0} T_\IQ^{k,l}~.
\end{align}
These spaces inherit a Hodge structure of weight $m(k-l)$ from that of $V_\IQ$, as can be immediately by using the definition in terms of the representation $\wt h: \IS(\IR) \to \GL(V_\IQ \otimes \IR)$. To obtain a more explicit description of the Hodge decomposition, we note that the complexification $T_\IC^{k,l} \defineas T_\IQ^{k,l} \otimes_\IQ \IC$ is isomorphic to the space of tensors of type $(k,l)$ of the complexified spaces, $T_\IC^{k,l} \simeq V_\IC^{\otimes k} \otimes (V_\IC^{\otimes l})^\vee$. In addition, 
\begin{align} \label{eq:Hodge_representation_tensor_product}
	\begin{split}
	\wt h(z) \; v^\vee &\= z^{-p} \overline{z}^{-q} \; v^\vee \qquad \qquad \quad \hskip13pt \text{if} \qquad v^\vee \in (V_\IC^{p,q})^\vee~,\\[5pt]
	\wt h(z) \; v_1 \otimes v_2 &\= z^{p_1+p_2} \overline{z}^{q_1+q_2} \; v_1 \otimes v_2 \qquad \text{if} \qquad v_i \in V_\IC^{p_i,q_i}~.
	\end{split}
\end{align}
We can write any $t \in T^{k,l}_\IC \defineas T_\IQ^{k,l} \otimes \IC$ as a sum of vectors of the form
\begin{align}
	v \= v_1 \otimes \dots \otimes v_k \otimes v_1^\vee \otimes \dots \otimes v_l^\vee~, \qquad \text{with} \qquad v_i \in V_\IC^{p_i,q_i}~,~~v_i^\vee \in (V_\IC^{p_i', q_i'})^\vee~,
\end{align}
and by the above formulae \eqref{eq:Hodge_representation_tensor_product} each such vector satisfies
\begin{align}
	\wt h(z) \; v \= z^{r} \; \overline{z}^{s} \; v~, \qquad \text{where} \qquad  \sum_{i=1}^k p_i - \sum_{i=1}^l p'_i = r~, \qquad \sum_{i=1}^k q_i - \sum_{i=1}^l q'_i = s
\end{align}
and thus by eq.~\eqref{eq:Hodge_decomposition_eigenspaces} belongs to $(T_\IC^{k,l})^{r,s}$. Therefore we see that the complexification $T^{k,l}_\IC$ has a decomposition of the form
\begin{align} \label{eq:Hodge_tensors_Hodge_decomposition}
	T_\IC^{k,l} \= \bigoplus_{p_i,q_i,p'_i,q'_i} \bigotimes_{i=1}^k V_\IC^{p_i,q_i} \otimes \bigotimes_{i=1}^l (V_\IC^{p'_i,q'_i})^\vee~.
\end{align}
We define the algebra $\Hg^{*,*}$ of \textit{Hodge tensors} as the algebra of the Hodge classes of the tensor spaces $T^{k,l}$
\begin{align}
	\Hg^{k,l} \; &\defineas \; T_\IQ^{k,l} \cap (T_\IC^{k,l})^{p,p}~, \qquad \text{where} \qquad 2p \= m(k-l)~,\\[5pt]
	\Hg^{*,*} \; &\defineas \; \bigoplus_{k,l\geq 0} \Hg^{k,l}~.
\end{align}
\subsubsection*{Equivalence of definitions}
\vskip-5pt
To show that the definition of the CM-type Hodge structures in terms of the Mumford--Tate group agrees with that given in section \ref{sect:Hodge_structures_with_CM}, we use the result (I.B.1) of \cite{Green2012a} that the Mumford--Tate group is exactly the subgroup of $G = \text{Aut}(V,Q)$ that fixes the elements of $\Hg^{*,*}$ pointwise.

Note that the algebra of Hodge endomorphisms $\text{End}_\text{Hdg}(V_\IQ)$ is isomorphic to the space $\Hg^{1,1}$. This follows as the elements of the space $T_\IQ^{1,1}$, which has a Hodge structure of type $(0,0)$, are tensors $t \in V_\IQ \otimes V_\IQ^\vee$. The condition that $t_\IC \in T_\IC^{1,1}$ be a Hodge class is that
\begin{align}
	t_\IC \in (T^{1,1}_\IC)^{0,0} \= \bigoplus_{p_i,q_i} V_\IC^{p_i,q_i} \otimes (V_\IC^{p_i,q_i})^\vee~.  
\end{align}
Identifying the elements $t \in V \otimes V^\vee$ with maps $t: V \to V$, we see that these conditions imply that the Hodge tensors of $T_\IQ^{1,1}$ correspond exactly to maps $t: V_\IQ \to V_\IQ$ whose complexifications $t_\IC$ preserve the Hodge type in the sense of eq.~\eqref{eq:Hodge_endomorphism_condition}.

The group $\GL(V_\IR)$, and thus the Mumford--Tate group $\text{MT}(h)$, acts on $T^{1,1}_\IQ$ by adjoint action, and therefore the condition that $g \in \GL(V_\IR)$ fix the elements of $\text{Hg}^{*,*}$ implies in particular that $g$ commutes with every Hodge endomorphism, using the correspondence established above. Assume that there exists an embedding $\eta: E \hookrightarrow \text{End}_\text{Hdg}(V_\IQ)$ with $[E:\IQ] = \dim_\IQ V_\IQ$. Then $\eta(E)$ is a maximal commutative subalgebra of $\text{End}_\text{Hdg}(V_\IQ)$ and hence contains $g$, since $g$ commutes with all elements of $\eta(E)$. Thus $\text{MT}(h) \subset \eta(E)$, so $\text{MT}(h)$ must be Abelian.

The converse is also true: if the Mumford--Tate group $\text{MT}(h)$ is Abelian and non-trivial, then there exists an embedding of a number field $E$ of degree $[E:\IQ] = \dim_\IQ V_\IQ$ to the algebra of Hodge endomorphisms $\text{End}_\text{Hdg}(V_\IQ)$. In fact, in this case $\text{End}_\text{Hdg}(V_\IQ)$ is isomorphic to a CM-field. We omit the proof due to it being more lengthy, but the proof can be found in ref.~\cite{Green2012a} (Proposition~(V.2)).

\newpage

\section{Complex Multiplication for Elliptic Curves} \label{app:CM_for_elliptic_curves}
\vskip-10pt
We wish to exemplify somewhat abstract definitions in section \ref{sect:Hodge_structures_with_CM} and appendix~\ref{app:complex_multiplication} with a concrete example of elliptic curves. The example in this section has also obvious parallels to the general construction provided in section \ref{sect:CM_Hodge_Structures_from_RCFTs} and the examples of sections \ref{sect:minimal_models} and \ref{sec:exgepnermodels}. We re-derive the famous result that an elliptic curve has complex multiplication (i.e. its Hodge structure is of CM type) if and only if the corresponding lattice parameter $\tau$ belongs to a quadratic imaginary field. We will necessarily be very brief and direct a reader interested in further details of the theory of elliptic curves and their complex multiplication to refs.~\cite{Silverman, Koblitz}. 

Let $\cE_\tau$ denote the elliptic curve isomorphic to the lattice quotient
\begin{align}
	\cE_\tau \; \simeq \; \frac{\IC}{\IZ + \tau \IZ}~,
\end{align}
where we can take $\tau \in \IH$, and we denote the coordinate of $\IC$ by $x$. We choose a basis $\{\alpha_0,\alpha_1\}$ of $H^1(\cE_\tau,\IZ)$ with the corresponding dual basis of $H_1(\cE_\tau,\IZ)$ being denoted by $\{A^0,A^1\}$. We define the periods of $\cE_\tau$ as
\begin{align} \label{eq:holomorphic_differential}
	\varpi^i \defineas \int_{A^i} \dd x~, \qquad \text{so that} \qquad \dd x \= \varpi^0 \alpha_0 + \varpi^1 \alpha_1 \= \varpi^0(\alpha_0  + \tau \alpha_1)~,
\end{align}
where we have used that the lattice parameter is given by $\tau = \varpi^1/\varpi^0$.

\subsection{Complex multiplication using Hodge endomorphisms}
We find $\text{End}_\text{Hdg}(H^1(\cE_\tau,\IQ))$ by representing an arbitrary endomorphism $\varphi: H^1(\cE_\tau,\IQ) \to H^1(\cE_\tau,\IQ)$ as a matrix $\text{M} \in \text{Mat}(2{\times}2, \IQ)$ in the basis $\alpha_0,\alpha_1$. Since $H^{1,0}(\cE_\tau,\IC) = \langle \frac{1}{\varpi^0} \dd z \rangle_\IC$, the condition for the endomorphism represented by $\text{M}$ to preserve the Hodge structure can be written as
\begin{align} \label{eq:Elliptic_curve_Hodge_endomorphism_matrix_condition}
	\left( \begin{matrix}
		a & b \\
		c & d
	\end{matrix} \right) \left( \begin{matrix}
		1 \\ \tau
	\end{matrix} \right) \=  C \left( \begin{matrix}
		1 \\ \tau
	\end{matrix} \right) \qquad \text{ for some } C \in \IC~.
\end{align}
Writing out this condition gives two equations:
\begin{align}
	a - C + b \tau \= 0~, \qquad c - C \tau + d \tau \= 0~.
\end{align}
This implies that the lattice parameter $\tau$ must satisfy a quadratic equation. In particular, since $\tau \in \IH$, it has to belong an imaginary quadratic field $\IQ(\sqrt{-D})$ for some $D \in \IZ_+$. By making an $\SL(2,\IQ)$ transformation of basis if necessary, we can always take $\tau$ to be purely imaginary.\footnote{Here $\tau$ is defined by \eqref{eq:holomorphic_differential} and in general longer is equal to the lattice parameter, as in general the required basis transformation is not in $\SL(2,\IZ)$.} For $\tau = \ii \sqrt{D}$, from the above equations it follows that
\begin{align} \label{eq:Elliptic_curve_Hodge_endomorphisms}
	\text{M} \in \left\{ \left( \begin{matrix}
		a & b \\
		-Db & a
	\end{matrix} \right) \; \middle | \; a,b \in \IQ \right\} \; \simeq \;	\text{End}_\text{Hdg}(H^1(E_\tau,\IQ))~.
\end{align}
Note that we have the algebra isomorphism
\begin{align}
	\text{End}_\text{Hdg}(H^1(E_\tau,\IQ)) \simeq \IQ(\sqrt{-D})~, \qquad \text{given by } \qquad  \left( \begin{matrix}
		a & b \\
		-D b & a
	\end{matrix} \right) \mapsto a + b \sqrt{-D}~,
\end{align}
so we have the inclusion of a number field $E = \IQ(\sqrt{-D})$ of degree $[\IQ(\sqrt{-D}):\IQ]$ to the endomorphism algebra given by this isomorphism. Therefore the Hodge structure has $E$-multiplication according to the definition given in section \ref{sect:Hodge_structures_with_CM}. The polarisation for the Hodge structure is given~by
\begin{align}
	Q(v,w) \; \defineas \; - \ii \int_{E_\tau} v \wedge w~,
\end{align}
so the Hodge structure $H^1(\cE_\tau,\IQ)$ is of CM-type. We have thus shown, using Hodge endomorphism algebra, that $H^1(\cE_\tau,\IQ)$ is of CM-type if and only if $\tau \in \IQ(\sqrt{-D})$ for some $D \in \IZ_+$.

\subsection{Complex multiplication using the Mumford--Tate group}
We now re-derive this result by using the definition of CM in terms of the Mumford--Tate group (see appendix \ref{app:Mumford-Tate_group}). We first find the map $\wt h: \IS(\IR) \to \GL(H^1(\cE,\IR))$ defined in eq.~\eqref{eq:Hodge_structure_homomorphism_weight}, which gives the alternative definition of the Hodge structure. We can easily find this map explicitly since we know that it must act as
\begin{align}
	\wt h(z) \; \frac{\dd x}{\varpi^0} \= z \; \frac{\dd x}{\varpi^0}~, \qquad \wt h(z)\; \frac{\dd \overline{x}}{\overline{\varpi}^0} \= \overline{z} \; \frac{\dd \overline{x}}{\overline{\varpi}^0}~.
\end{align}
If we represent the map $\wt h(z)$ by a matrix in the basis given by $\alpha_0$ and $\alpha_1$, the above condition can be written as (cf. eq.~\eqref{eq:Elliptic_curve_Hodge_endomorphism_matrix_condition})
\begin{align}
	\left( \begin{matrix}
		a(z) & b(z) \\
		c(z) & d(z)
	\end{matrix} \right) \left( \begin{matrix}
		1 \\ \tau
	\end{matrix} \right) \= z \left( \begin{matrix}
	1 \\ \tau
\end{matrix} \right)~, \qquad \left( \begin{matrix}
a(z) & b(z) \\
c(z) & d(z)
\end{matrix} \right) \left( \begin{matrix}
1 \\ \overline{\tau}
\end{matrix} \right) \= \overline{z} \left( \begin{matrix}
1 \\ \overline{\tau}
\end{matrix} \right)~, 
\end{align}
where $a(z),b(z),c(z),d(z) \in \IR$. Solving these conditions gives
\begin{align} \label{eq:Elliptic_curve_h(z)_general}
	\wt h(z) \= \frac{1}{\tau_2} \left( \begin{matrix}
		z_1 \tau_2 - z_2 \tau_1 & z_2 \\
		-z_2(\tau_1^2 + \tau_2^2) & z_2 \tau_1 + z_1 \tau_2
	\end{matrix} \right) ~,
\end{align}
where we have denoted $\tau = \tau_1 + \ii \tau_2$, $z = z_1 + \ii z_2$. The related map $h: \IU(\IR) \to \SL(H^1(\cE_\tau,\IR))$ is given by restricting this to values of $z$ such that $|z|^2=1$. As in eq.~\eqref{eq:Mumford-Tate_definition}, the Mumford--Tate group is defined as the $\IQ$-algebraic closure $h(\IU(\IR))^{\text{alg}}$ of $h(\IU(\IR))\subseteq \SL(H^1(\cE_\tau,\IR)) \simeq \SL(2,\IR)$.

Let us first explicitly show that when $\tau \in \IQ(\sqrt{-D})$, the Mumford--Tate group is indeed Abelian, and thus the Hodge structure is of CM-type as defined in appendix \ref{app:Mumford-Tate_group}. We can again, without loss of generality, take $\tau = \ii \sqrt{D}$ so that we have
\begin{align}
	h(z) \=  \left( \begin{matrix}
		z_1 & z_2/\sqrt{D} \\
		-z_2 \sqrt{D} & z_1
	\end{matrix} \right)~.
\end{align}
Any such matrix satisfies $\IQ$-algebraic equations, so the image is in fact $\IQ$-algebraically closed and we can write it as
\begin{align}\label{eq:Elliptic_curve_MT_group}
	h(\IU(\IR))^{\text{alg}} \= \left\{ \left( \begin{matrix}
			a & b \\
			c & d
		\end{matrix} \right) \in \SL(H^1(\cE_\tau,\IR)) \; \middle| \; a=d~,~~c = -Db~,~~ad - cb = 1\right\} \; \simeq \; \mtU(1)~.
\end{align}
This shows that the Mumford--Tate group $\text{MT}(h) \defineas h(\IU(\IR))^{\text{alg}} \simeq \mtU(1)$ is Abelian, and therefore the Hodge structure is of CM-type according to the definition of appendix \ref{app:Mumford-Tate_group}. 

The reverse is also true, that is, if $\tau \notin \IQ(\sqrt{-D})$ for any $D \in \IZ_+$, then the Mumford--Tate group is not Abelian. To see this, we use the obvious fact that if $\tau$ is not quadratic imaginary, then either $\tau_1$ or $|\tau|^2$ is irrational. Then from eq.~\eqref{eq:Elliptic_curve_h(z)_general}, it is relatively easy to see that the image $ h(\IU(\IR))$ is not given by $\IQ$-algebraic equations. Therefore the Mumford--Tate group must be larger than $\IU(\IR)$. It can be shown that the Mumford--Tate group is a reductive subgroup of the automorphism group $G$ (see (I.B.6) of ref.~\cite{Green2012a}) which for elliptic curves is given by $G = \SL(H^1(\cE_\tau,\IR)) \simeq \SL(2,\IR)$. However, there are only two such groups that are non-trivial, $\mtU(1)$ and $\SL(2,\IR)$. We thus deduce that $\text{MT}(h) = \SL(2,\IR)$, and therefore it is not Abelian.

\subsubsection*{Mumford--Tate group and Hodge tensors}
\vskip-5pt
We can also find the Mumford--Tate group by using the Hodge tensors. To keep the notation compact, in this section, we denote $H^1(\cE_\tau,\IQ) = H_\IQ$, and $H^1(\cE_\tau,\IC) = H_\IC$ Consider the Hodge tensors in $T^{1,1}_\IQ \defineas H_\IQ \otimes H_\IQ^\vee$. As can be seen from eq.~\eqref{eq:Hodge_tensors_Hodge_decomposition}, this space has an induced Hodge structure of weight 0, so the Hodge classes of $T^{1,1}_\IQ$ are given by 
\begin{align} \nonumber
	\Hg^{1,1} \defineas (T^{1,1}_\IQ) \cap \left(T^{1,1}_\IC\right)^{0,0}~, \quad \text{where} \quad \left(T^{1,1}_\IC\right)^{0,0} \= \left(H_\IC^{1,0} \otimes (H_\IC^{1,0})^{\vee} \right)  \oplus \left( H_\IC^{0,1} \otimes (H_\IC^{0,1})^{\vee} \right)~.
\end{align}
As discussed at the end of appendix \ref{app:Mumford-Tate_group}, by identifying the elements $t \in H_\IQ \otimes H_\IQ^\vee$ with maps $t: H_\IQ \to H_\IQ$, we see that the Hodge tensors in $\Hg^{1,1}$ are precisely the Hodge endomorphisms \eqref{eq:Elliptic_curve_Hodge_endomorphisms}. Recall also from the end of appendix \ref{app:Mumford-Tate_group} that the Mumford--Tate group $\text{MT}(h)$ is the subgroup of $G = \SL(H^1(\cE_\tau,\IR))$ that fixes the elements of $\Hg^{*,*}$, and consequently the elements of $\Hg^{1,1}$. The Mumford--Tate group acts on $T^{1,1}_\IQ =  H_\IQ \otimes H_\IQ^\vee$ by the adjoint action, and therefore the the elements of the Mumford--Tate group have to commute with the Hodge endomorphisms. 

Since there are non-trivial Hodge endomorphisms only if $\tau \in \IQ(\sqrt{-D})$, let us consider this case, and let $\mtM$ be a matrix representing an element of $\text{End}_\text{Hdg}(H_\IQ)$, and let $\mtS$ be an arbitrary element of the Mumford--Tate group.
\begin{align}
	\mtM &\= \left( \begin{matrix}
		a & b \\
		-Db & a
	\end{matrix} \right) \in \text{Mat}(2{\times}2,\IQ)~, \qquad  \mtS \= \left( \begin{matrix}
	\alpha & \beta \\
	\gamma & \delta
\end{matrix} \right) \in \SL(2,\IR)~.
\end{align}
Then the condition $\mtS^{-1} \mtM \mtS = \mtM$ for all $a,b \in \IQ$ implies that 
\begin{align}
	\mtS \= \left( \begin{matrix}
		\alpha & \beta \\
		-D \beta & \alpha
	\end{matrix} \right)~, \qquad \text{with} \qquad \alpha^2 + D \beta^2 \= 1~.
\end{align}
By comparing to eq.~\eqref{eq:Elliptic_curve_MT_group} we that this is equivalent to requiring that $\mtS \in h(\IU(\IR))^{\text{alg.}} \simeq \mtU(1)$, so we again see that the Mumford--Tate group consisting of these matrices is Abelian if $\tau \in \IQ(\sqrt{-D})$.

\newpage
\section{\texorpdfstring{$\cN=2$}{N=2} Minimal Models with Boundary States in the NS-Sector} \label{app:Minimal_Model_NS-states}
\vskip-10pt
To illustrate a case where non-trivial stabiliser groups $\text{Stab}(\alpha)$ arise for some $\alpha \in \cR$, we consider again minimal models. Unlike in section \ref{sect:minimal_models}, however, we take the boundary states to lie in the NS-sector. This choice gives rise to slightly different values of the relevant $\mtS$-matrix elements, which in turn necessitates analysis of stabiliser groups and the associated fixed fields. As the purpose of this appendix is to illustrate these structures, we perform the analysis in two ways. The first approach is analoguous to section~\ref{sect:minimal_models}, whereas the second one follows more closely the general argument presented in section \ref{sect:CM_Hodge_Structures_from_RCFTs}.

\subsection{Proof using direct computation}
As in section \ref{sect:minimal_model_Hodge_substructures}, we wish to explicitly construct the Hodge substructures of the Hodge structure associated with the $\mathcal{N}=2$ minimal models at level $k$. To do so, we find the field extensions $\IQ(\mtS_{\Rl})$, the stabiliser group $\text{Stab}(\Rl)$ and the fixed field $\IQ(\mtS_\Rl)^{\text{Stab}(\Rl)}$ for every representation $\Rl$ defined in eq.~\eqref{eq:Rl}.\footnote{Note that since $\mathbb{Q}(\mtS)$ is Galois and $\text{Gal}(\mathbb{Q}(\mtS)/\mathbb{Q})$ is Abelian, all subfields are Galois over $\mathbb{Q}$.}

In order to construct the number field $\IQ( \mtS_\Rl)$, we observe that the modular S-matrix~\eqref{eq:S-matrix_minimal_model} for the representations $\Rl \in \cR$, c.f. eq.~\eqref{eq:Rl}, can be written as
\begin{equation} \label{eq:app_S-matrix_generators_k}
	\begin{aligned}
		\mtS_{\Rl,(L,M,S)} &\= \frac{(-1)^{S/2+1}}{2(k+2)} \ii \left( \zeta_{2(k+2)}^{(l+1)(L+1-M)} -  \zeta_{2(k+2)}^{-(l+1)(L+1+M)}\right) \\
		&\= - \frac{(-1)^\frac{L+M+S}2}{2(k+2)} (\ii \zeta_{2(k+2)}^{l+1})^{-L-M-1} 
		\left(1 + (-1)^{S} (\zeta_{2(k+2)}^{2(l+1)} )^{(L+1)} \right)	 \ .
	\end{aligned}	
\end{equation}
Here we have used that the boundary labels $(L,M,S) \in \HWR_{\text{NS}}$ obey $L+M+S \in 2\IZ$. For such states, it also follows that
\begin{equation}
	\frac{\mtS_{\Rl,(0,0,0)}}{\mtS_{\Rl,(0,2,0)}}\= \zeta_{2(k+2)}^{2(l+1)} \ , 
\end{equation}
which together with eq.~\eqref{eq:app_S-matrix_generators_k} implies that $\IQ( \mtS_\Rl)$ is generated by the root of unity $\ii \zeta_{2(k+2)}^{l+1} = \zeta_{4(k+2)}^{k+2l+4}  \in \IQ( \mtS_\Rl)$, that is $\IQ( \mtS_\Rl) = \IQ(\ii \zeta_{2(k+2)}^{l+1})$. To further simplify the expression of this cyclotomic number field, we decompose the integers $l+1$ and $k+2$ as
\begin{equation} \label{eq:app_l_k_decomposition}
	l+1 \= 2^r \delta_l~,~~k+2 \= 2^s \delta_k \qquad \text{with} \qquad \delta_l, \delta_r \in \IZ \setminus 2 \IZ \  ,
\end{equation}   
in terms of the unique non-negative integral binary exponents $r$ and $s$.

Then the number field~$\IQ( \mtS_\Rl)$ can be expressed as\footnote{Here and in the following we have used that a primitive element of a field extension of $\IQ$ by roots of unity $\zeta_{i_1}, \ldots, \zeta_{i_m}$ is given by the root of unity $\zeta_{\lcm(i_1,\ldots,i_m)}$, i.e. $\IQ(\zeta_{i_1}, \ldots, \zeta_{i_m}) = \IQ(\zeta_{\lcm(i_1,\ldots,i_m)})$.}
\begin{equation}   
	\IQ( \mtS_\Rl) \= \IQ(\ii \zeta_{2(k+2)}^{l+1}) 
	\= \begin{cases} 
		\  \IQ(\zeta_{\kappa}) & \text{ for } r = s +1  \ , \\ 
		\ \IQ(\zeta_{2\kappa}) & \text{ otherwise} \ ,
	\end{cases}
\end{equation}
with
\begin{equation}   \label{eq:app_DefSrKappa}
	\kappa \; \defineas \; \frac12 \lcm\left( 4, \frac{2(k+2)}{\gcd\left( l+1 , 2(k+2)\right)} \right) 
	\= \lcm\left( 2, \frac{k+2}{\gcd\left( l+1 , k+2\right)} \right) \ .
\end{equation}
To verify the last equal sign in the definition of $\kappa$, we note that with the decomposition of eq.~\eqref{eq:app_l_k_decomposition}
\begin{equation}
	\begin{aligned}
		&\lcm\left( 4, \frac{2(k+2)}{\gcd\left( l+1 , 2(k+2)\right)} \right)  
		=\frac{ \lcm\left( 4, 2^{s+1 - \min(r,s+1)} \right) \delta_k}{\gcd(\delta_l,\delta_k)} 
		=\frac{ \lcm\left( 4, 2^{\max(s-r+1,0)} \right) \delta_k}{\gcd(\delta_l,\delta_k)} \\
		&\qquad =\frac{ \lcm\left( 4, 2^{\max(s-r+1,1)} \right) \delta_k}{\gcd(\delta_l,\delta_k)} 
		=\frac{ \lcm\left( 4, 2^{s+1 - \min(r,s)} \right) \delta_k}{\gcd(\delta_l,\delta_k)} 
		= \lcm\left( 4, \frac{2(k+2)}{\gcd\left( l+1 , k+2\right)} \right) \ .
	\end{aligned}  
\end{equation}
For the construction of the number field $\IQ( \mtS_{\Rl})^{\text{Stab}(\Rl)}$, we first determine the action of the Galois group  $\Gal( \IQ(\mtS_\Rl)/\IQ)$ on the representations $\Rl \in \cR$. We analyse the generic case $r \ne s+1$ (resp. the non-generic case $r = s+1$). Recall from eqs.~\eqref{eq:GalCorresp} and \eqref{eq:minimal_model_rho_definition} that $\Gal(\IQ(\zeta_{n})/\IQ) \simeq (\IZ/n\IZ)^\times$, and let $\rho_a \in \Gal(\IQ(\zeta_{2\kappa})/\IQ)$ (resp. $\rho_a \in \Gal(\IQ(\zeta_{\kappa})/\IQ)$) be the Galois automorphism corresponding to $[a] \in (\IZ/ 2\kappa \IZ)^\times$ (resp. $[a] \in (\IZ/ \kappa \IZ)^\times$). It transforms the primitive root $\ii \zeta_{2(k+2)}^{l+1}$ as
\begin{equation}
\rho_a(\ii \zeta_{2(k+2)}^{l+1}) \= \ii^a \zeta_{2(k+2)}^{a(l+1)} \= \pm \ii \zeta_{2(k+2)}^{(\wt l+1)}~,
\end{equation}
for some integer $l'$ in the range $0 \leq \wt l \leq 2(k+2)$ such that $l' \equiv a(l+1) -1 \!\! \mod k+2$. In the above, we have also used that $a$ is odd. Since $\zeta_{2(k+2)}^{k+2} = -1$, we can find $l'$ in the range $0 \leq l' \leq (k+2)$ with $l' \equiv \wt l \mod (k+2)$ so that $\zeta_{2(k+2)}^{(\wt l+1)} = \pm \zeta_{2(k+2)}^{(l'+1)}$. Furthermore, due to the fact that $l$ is limited to the range $0 \leq l \leq k$, it can be shown\footnote{\label{foot:l'_neq_k+1}Assume that $l' = k+1$. This implies $\lambda (k+2) = a (l+1)$ for some integer $\lambda$. Note that $a$ does not divide $\lambda$ as otherwise $l$ is not in the required range $0 \le l \le k$. Then we get $\kappa=\lcm(2, \frac{\lambda(k+2)}{\gcd(\lambda(l+1),\lambda( k+2))}) = \lcm(2, \frac{a(l+1)}{\gcd(\lambda(l+1),a( l+1))}) =  \lcm(2, \frac{a}{\gcd(a,\lambda)})$, which contradicts $a$ being coprime to $\kappa$ (as $\gcd(a,\lambda) \ne a$). Hence $l \neq k+1$.} that $l' \neq k+1$, so we have that 
\begin{align} \label{eq:NS_minimal_model_Galois_action}
	\rho_a( \ii \zeta_{2(k+2)}^{l+1} ) \=  \pm  \ii \zeta_{2(k+2)}^{l'+1}~, \qquad \text{with} \qquad 0 \leq l' \leq k~\text{ and }~l' \equiv a(l+1) -1 \mod k+2~.
\end{align}
As a consequence the Galois automorphism $\rho_a$ acts on the modular S-matrix~\eqref{eq:app_S-matrix_generators_k} as
\begin{equation} \label{eq:app_minimal_model_S-matrix_Galois_action}
	\rho_a( \mtS_{\Rl,(L,M,S)} ) \=  \pm \mtS_{\Rla{l'},(L,M,S)} \ ,
\end{equation}
which on the set of representations $\cR$ induces the action
\begin{equation} \label{eq:app_FuncVarSigma}
	\varrho_a:\ \cR \to \cR~, \qquad \Rl \mapsto \Rla{l'} 
\end{equation}
Now we are ready to determine the stabiliser subgroup $\text{Stab}(\Rl)$ of the Galois group $\Gal(\IQ(\zeta_{2\kappa})/\IQ)$ (resp. $\Gal(\IQ(\zeta_{\kappa})/\IQ)$) and number field $\IQ( \mtS_{\Rl})^{\text{Stab}(\Rl)}$. For a given representation $\Rl \in \cR$ we need to calculate the conjugacy classes $[a] \in (\IZ/2\kappa \IZ)^\times$ (resp. $[a] \in (\IZ/ \kappa \IZ)^\times$) with $\varrho_a(\Rl) = \Rl$, which according to eq.~\eqref{eq:app_FuncVarSigma} amounts to solving the equation
\begin{equation} \label{eq:app_stabiliser_congruence}
	(a-1)(l+1) \= \lambda (k+2) \ \Longleftrightarrow \
	(a-1)\frac{(l+1)}{\gcd(l+1,k+2)} \= \lambda \frac{(k+2)}{\gcd(l+1,k+2)} 
	\quad \text{for $\lambda\in \IZ$} \ .
\end{equation}
To ensure that $a$ defines a conjugacy class $[a] \in (\IZ/\kappa \IZ)^\times$, $a$ need to be odd. Such solutions take the form
\begin{equation}
	a-1 \= m\,\lcm\left(2,\frac{k+2}{\gcd(l+1,k+2)}\right) 
	\ \Longleftrightarrow \
	a \= 1+ m\,\kappa  
	\quad \text{with $m \in \IZ$} \ .
\end{equation}
As the classes $[a]$ are represented by integers $a$ modulo $2\kappa$ (resp. $a$ modulo $\kappa$) and because $1+\kappa$ is always coprime to $2\kappa$, the stabiliser group $\text{Stab}(\Rl)$ is isomorphic to $\IZ/2\IZ$ (resp. is trivial). In summary, we find
\begin{equation} \label{eq:app_minimal_model_stabiliser}
	\text{Stab}(\Rl) \= 
	\begin{cases} 
		\left\langle \,  \rho_1 \, \right\rangle \simeq \langle \,1\, \rangle & \text{for $r = s+1$} \ , \\
		\left\langle \,  \rho_{1+\kappa} \, \right\rangle \simeq   \IZ/2\IZ  & \text{otherwise} \ .
	\end{cases}
\end{equation}
As a consequence the invariant number field  $\IQ( \mtS_{\Rl})^{\text{Stab}(\Rl)}$ is given, in all cases, by
\begin{equation}
	\IQ( \mtS_{\Rl})^{\text{Stab}(\Rl)} = \IQ(\zeta_{\kappa}) \ ,
\end{equation}
and the conveniently scaled vectors $\ket{\Rl}_\cS$ defined in eq.~\eqref{eq:R_state_generator_expansion} in terms of the states $\ket{\Rl}_\cI$ (see eq.~\eqref{eq:intermediate_scaled_states}) are given by
\begin{equation} \label{eq:app_scaledminimalmodel}
	\ket{\Rl}_\cS := 
	\begin{cases} 
	\phantom{\zeta_{2\kappa}} \ket{\Rl}_\mathcal{I} \in V_\mathbb{Q} \otimes_\mathbb{Q} \mathbb{Q}(\zeta_{\kappa}) & \text{ for } r=s+1~,\\
	\zeta_{2\kappa} \ket{\Rl}_\mathcal{I} \in V_\mathbb{Q} \otimes_\mathbb{Q}\mathbb{Q}(\zeta_{\kappa}) & \text{ otherwise}~,
	\end{cases}
\end{equation}
as can be seen by noting that $\rho_{1+\kappa}(\zeta_\kappa) = -\zeta_\kappa$. On these states, the field automorphisms $\theta_a \in \Gal(\IQ(\zeta_{\kappa})/\IQ)$ corresponding to the elements $[a] \in (\IZ/\kappa\IZ)^\times$ act as
\begin{align} \label{eq:app_minimal_model_states_Galois_transformation}
	\theta_a \ket{\Rl} \= \ket{\vartheta_a(\Rl)} = \ket{\Rla{l'}}~, \quad \text{with} \quad 0 \leq l' \leq k~\text{ and }~l' \equiv a(l+1) -1 \mod k+2~.
\end{align}
\subsection{Proof using a tower of field extensions}
There is an alternative way of establishing the Galois action of eqs.~\eqref{eq:app_minimal_model_S-matrix_Galois_action} and \eqref{eq:app_FuncVarSigma}, which follows more closely the discussion of section \ref{sect:CM_Hodge_Structures_from_RCFTs} by utilising the existence of the tower of field extension
\begin{equation} \label{eq:app_TowerExt}
	\IQ(\mtS)  \supseteq \IQ( \mtS_\Rl) \supseteq \IQ( \mtS_{\Rl})^{\text{Stab}(\Rl)} \, ,
\end{equation} 
and the fact that the Galois group $\Gal(\IQ(\mtS)/\IQ)$ projects onto $\Gal(\IQ(\mtS_\Rl)/\IQ)$. Since this way of thinking helps to illustrate that of section \ref{sect:CM_Hodge_Structures_from_RCFTs} and more readily generalises to a wider variety of theories, we discuss it briefly here. However, to keep this discussion as accessible as possible, we restrict to the case where the level $k$ is odd. The other cases can be treated with similar arguments.

To determine the number field $\IQ(\mtS)$ generated by the entries of the S-matrix~\eqref{eq:S-matrix_minimal_model}, we note that\footnote{Note that $(0,0,0)$, $(1,-1,0)$, $(0,-2,0)$, $(1,1,0)$, $(0,2,0)$ are representations contained in any minimal model for arbitrary level $k\geq1$.}
\begin{equation}
	\zeta_{2(k+2)} \= \left( 1+   \frac{\mtS_{(0,0,0),(1,-1,0)}}{\mtS_{(0,-2,0),(1,1,0)}} \right)  \left(  \frac{\mtS_{(0,-2,0),(0,0,0)}}{\mtS_{(0,0,0),(1,-1,0)}}    \right)~, \qquad	\ii \= \frac{\mtS_{(1,1,0),(0,2,0)}}{1-\frac{\mtS_{(0,-2,0),(0,-2,0)}}{\mtS_{(0,-2,0),(0,0,0)}}}~.
\end{equation}
By the expression~\eqref{eq:S-matrix_minimal_model}, every $\mtS$-matrix element can be generated by these so that
\begin{equation}
	\IQ(\mtS) \= \IQ(\zeta_{2(k+2)},\ii) \= \IQ(\zeta_{4(k+2)})~.
\end{equation}
Then, as above, we determine the stabiliser groups $\text{Stab}(\Rl)$ introduced in eq.~\eqref{eq:Stabiliser_group_definition}, but this time to do this we use the fact that the  various field extensions we consider have the structure displayed in figure~\ref{fig:field_extensions} with the corresponding projections between the Galois groups.\footnote{The inclusion $\IQ(\zeta_\kappa) \subseteq \IQ(\zeta_{k+2})$ follows from the fact that for $k$ odd $\IQ(\zeta_{\kappa}) = \IQ(\zeta_{\kappa/2})$, which is manifestly a subfield of $\IQ(\zeta_{k+2})$.}
\begin{figure}[H]
\begin{center}
	\begin{tikzcd}[row sep=large]
		& \IQ(\zeta_{4(k{+}2)})
		\ar[dl,-]\ar[dd,-]\ar[dr,-]     &                          \\
		\IQ(\zeta_{2 \kappa}) \ar[dr,-]
		&                           & \IQ(\zeta_{k{+}2})
		\ar[dl,-]               \\  
		& \IQ(\zeta_\kappa)                &
	\end{tikzcd}
	\begin{tikzcd}[row sep=large,every label/.append style = {font=\normalsize}]
		& (\IZ/4(k{+}2)\IZ)^\times
		\ar[dl,->>,"p"']\ar[dd,->>]\ar[dr,->>,"\pi"]     &                          \\
		(\IZ/2\kappa\IZ)^\times \ar[dr,->>]
		&                           & (\IZ/(k{+}2)\IZ)^\times
		\ar[dl,->>]               \\  
		& (\IZ/\kappa\IZ)^\times               &
	\end{tikzcd}
\vskip8pt
\capt{4.3in}{fig:field_extensions}{Relations between different field extensions of $\IQ$ relevant for the study of the Galois symmetry of minimal models of an odd level $k$.}	
\end{center}
\vskip-25pt
\end{figure}
Note that the short exact sequence
\begin{align}
	1 \longrightarrow \Gal(\IQ(\zeta_{4(k+2)})/\IQ(\zeta_{k{+}2}))  \longrightarrow  \Gal(\IQ(\zeta_{4(k+2)})/\IQ)  \overset{\wt\pi}{\longrightarrow}  \Gal(\IQ(\zeta_{4(k+2)})/\IQ(\ii))  \longrightarrow  1 \ ,
\end{align}
where $\wt\pi$ is induced from the map~$\pi$ in figure~\ref{fig:field_extensions}, splits so that we have
\begin{align}
\begin{split}
	\Gal(\IQ(\mtS)/\IQ) &\simeq \Gal(\IQ(\zeta_{4(k+2)})/\IQ(\zeta_{k+2})) \times \Gal(\IQ(\zeta_{4(k+2)})/\IQ(\ii)) \\
	&\simeq (\IZ/4\IZ)^\times \times (\IZ/(k{+}2)\IZ)^\times~.
\end{split}
\end{align}
Every element $\sigma_a \in \Gal(\IQ(\zeta_{4(k{+}2)})/\IQ)$ corresponds to a pair of elements
\begin{align}
(\iota_b,\theta_c) \in \Gal(\IQ(\zeta_{4(k+2)})/\IQ(\zeta_{k+2})) \times \Gal(\IQ(\zeta_{4(k+2)})/\IQ(\ii))~,	
\end{align}
with $[b] \in (\IZ/4\IZ)^\times$ and $[c] \in (\IZ/(k{+}2)\IZ)^\times$ which act as
\begin{align}
	\iota_b \; : \; \ii \longmapsto \ii^b~, \qquad \theta_c \; : \; \zeta_{k+2} \longmapsto \zeta_{k+2}^c~,
\end{align}
and we can write the action of $\sigma_a$ on any $x \in \IQ(\mtS)$ as
\begin{align}
	\sigma_a(x) \= \iota_b(\theta_c(x))~.
\end{align}
This correspondence is useful since the Galois group $\Gal(\IQ(\zeta_{k+2})/\IQ(\ii))$ manifestly acts on the set $\cR$ of representations. To see this, let $[c] \in (\IZ/(k{+}2)\IZ)^\times$, and consider the action of $\theta_c \in \Gal(\IQ(\zeta_{k+2})/\IQ(\ii))$. As above, from the form~\eqref{eq:app_S-matrix_generators_k} of the S-matrix, we can immediately read off the action of $\sigma_a$:
\begin{align} \label{eq:Gal_k+2_action_S_matrix}
	\theta_c(\mtS_{\Rl,(L,M,S)}) \= \mtS_{\Rla{l'},(L,M,S)} \qquad \text{with} \qquad 0 \leq l' \leq k~,~~	l' \; \equiv \; c (l+1) - 1 \mod k{+}2~,
\end{align}
where $l' \neq k{+}1$ follows directly from the fact that $c$ is invertible modulo $k{+}2$, so $l' = k{+}1$ would imply $l{+}1 \equiv 0 \mod k{+}2$, which is not possible for $0 \leq l \leq k$. Therefore we conclude that $\vartheta_c(\Rl) = \Rla{l'}$ gives a well-defined permutation of $\cR$.

From this it follows that the elements $\rho_d \in \Gal(\IQ(\zeta_{2\kappa})/\IQ)$ also have a well-defined action on $\cR$. Namely, we can use the fact that every such element can be written as a restriction of some field automorphism ${\sigma_a \in \Gal(\IQ(\zeta_{4(k+2)})/\IQ)}$. In terms of the multiplicative groups of units, the restriction morphism corresponds to the~map $p$ of figure \ref{fig:field_extensions} given explicitly by reduction modulo $2 \kappa$
\begin{align} \label{eq:restriction_morphism_minimal_model}
	\begin{split}
		p \; : \; [a] \longmapsto [d]~,\qquad \text{where} \qquad a \; \equiv \; b \! \mod 2\kappa~.
	\end{split}	
\end{align}
Then it follows that for some $\sigma_a \in \Gal(\IQ(\zeta_{4(k+2)})/\IQ)$
\begin{align} \label{eq:lift_restriction_S}
	\rho_{d}(\mtS_{\Rl,(L,M,S)}) \= \sigma_{a}(\mtS_{\Rl,(L,M,S)}) \= \iota_b(\theta_c(\mtS_{\Rl,(L,M,S)})) \= \pm \mtS_{\Rla{l'},(L,M,S)} ~,
\end{align} 
where we have used that fact that from the form \eqref{eq:app_S-matrix_generators_k} of the S-matrix, it is manifest that $\iota_b$ acts by an overall sing. Note also that the field automorphism $\sigma_a$ is not necessarily unique, but this does not affect the above argument.

To find the group $\text{Stab}(\Rl)$, we first find the kernel of the projection $\pi$ from  $\Gal(\IQ(\zeta_{4(k+2)})/\IQ)$ to $\Gal(\IQ(\zeta_{k+2})/\IQ(\ii))$, given explicitly by reduction modulo $k{+}2$:
\begin{align} \label{eq:restriction_morphism_S_k+2}
	\begin{split}
			\pi \; : \; [a] \longmapsto [c]~,\qquad \text{where} \qquad a \; \equiv \; b \! \mod k{+}2~.
		\end{split}
\end{align}
The kernel of this map is given by
\begin{align}
	\text{ker}(\pi) \= \{[a] \in (\IZ/4(k{+}2)\IZ) \; | \; a \equiv 1 \! \mod k{+}2\} \= \{[1 + v (k{+}2)] \; | \; v \in \{0,2\}\}~,
\end{align}
where we have noted that the even integers $1+(k{+}2)$ and $1+3(k{+}2)$ are not coprime to $4(k{+}2)$.

We can now use the restriction morphism \eqref{eq:restriction_morphism_minimal_model} again. Under it, the image of $\text{ker}(\pi)$ is $\{[1],[1+\kappa]\} \subset (\IZ/2\kappa\IZ)^\times$.\footnote{To see this, note that $[1+2(k{+}2)]$ maps to a non-trivial element of $(\IZ/2\kappa\IZ)^\times$ as otherwise we would have that $2\kappa|2(k{+}2)$ which would imply that $4|2(k{+}2)$, but $k{+}2$ is odd. We know that $[1+2(k{+}2)] = [1+n \kappa]$ for some $n \in \IZ$ since $\kappa|2(k{+}2)$. The only non-trivial possibility is $[1+n \kappa] = [1+\kappa]$.} Then, as in eq.~\eqref{eq:lift_restriction_S}, it follows that
\begin{align}
	\rho_{1+\kappa}(\mtS_{\Rl,(L,M,S)}) \= \sigma_{1+2(k{+}2)}(\mtS_{\Rl,(L,M,S)}) \= \iota_{b}(\theta_1(\mtS_{\Rl,(L,M,S)})) \= \pm \mtS_{\Rl,(L,M,S)}~,
\end{align} 
Therefore $\rho_{1+\kappa} \in \text{Stab}(\Rl)$. Either recalling from section \ref{sect:CM_Hodge_Structures_from_RCFTs} that the group $\text{Stab}(\Rl)$ is at most of order 2, or by using an argument analogous to that in eq.~\eqref{eq:app_stabiliser_congruence}, it follows that
\begin{align}
	\text{Stab}(\Rl) \= \langle\rho_{1+\kappa}\rangle \; \simeq \; \IZ/2\IZ~,
\end{align}
which is the result \eqref{eq:app_minimal_model_stabiliser} found above by direct computation. 

\newpage
\bibliographystyle{JHEP}
\bibliography{CM_Hodge_Structures_from_RCFTs}

\providecommand{\href}[2]{#2}\begingroup\raggedright\begin{thebibliography}{100}

\bibitem{Friedan:1983xq}
D.~Friedan, Z.-a. Qiu and S.~H. Shenker, \emph{{Conformal Invariance, Unitarity
  and Two-Dimensional Critical Exponents}},
  \href{https://doi.org/10.1103/PhysRevLett.52.1575}{\emph{Phys. Rev. Lett.}
  {\bfseries 52} (1984) 1575}.

\bibitem{Moore:1989vd}
G.~W. Moore and N.~Seiberg, \emph{{L}ectures on {RCFT}},  in \emph{{1989 Banff
  NATO ASI: Physics, Geometry and Topology}}, 9, 1989.

\bibitem{Ginsparg:1988ui}
P.~H. Ginsparg, \emph{{A}pplied {C}onformal {F}ield {T}heory},  in \emph{{Les
  Houches Summer School in Theoretical Physics: Fields, Strings, Critical
  Phenomena}}, 9, 1988, \href{https://arxiv.org/abs/hep-th/9108028}{{\ttfamily
  hep-th/9108028}}.

\bibitem{DiFrancesco:1997nk}
P.~Di~Francesco, P.~Mathieu and D.~Senechal, \emph{{Conformal Field Theory}},
  Graduate Texts in Contemporary Physics. Springer-Verlag, New York, 1997,
  \href{https://doi.org/10.1007/978-1-4612-2256-9}{10.1007/978-1-4612-2256-9}.

\bibitem{Anderson:1988to}
G.~Anderson and G.~Moore, \emph{Rationality in conformal field theory},
  \href{https://doi.org/10.1007/BF01223375}{\emph{Communications in
  Mathematical Physics} {\bfseries 117} (1988) 441}.

\bibitem{Belavin:1984vu}
A.~A. Belavin, A.~M. Polyakov and A.~B. Zamolodchikov, \emph{{Infinite
  Conformal Symmetry in Two-Dimensional Quantum Field Theory}},
  \href{https://doi.org/10.1016/0550-3213(84)90052-X}{\emph{Nucl. Phys. B}
  {\bfseries 241} (1984) 333}.

\bibitem{Moore:1988uz}
G.~W. Moore and N.~Seiberg, \emph{{Polynomial Equations for Rational Conformal
  Field Theories}},
  \href{https://doi.org/10.1016/0370-2693(88)91796-0}{\emph{Phys. Lett. B}
  {\bfseries 212} (1988) 451}.

\bibitem{Friedan:1980jf}
D.~Friedan, \emph{{Nonlinear Models in Two Epsilon Dimensions}},
  \href{https://doi.org/10.1103/PhysRevLett.45.1057}{\emph{Phys. Rev. Lett.}
  {\bfseries 45} (1980) 1057}.

\bibitem{Friedan:1980jm}
D.~H. Friedan, \emph{{Nonlinear Models in Two + Epsilon Dimensions}},
  \href{https://doi.org/10.1016/0003-4916(85)90384-7}{\emph{Annals Phys.}
  {\bfseries 163} (1985) 318}.

\bibitem{Zamolodchikov:1986gt}
A.~B. Zamolodchikov, \emph{{Irreversibility of the Flux of the Renormalization
  Group in a 2D Field Theory}}, {\emph{JETP Lett.} {\bfseries 43} (1986) 730}.

\bibitem{Benjamin:2020flm}
N.~Benjamin, C.~A. Keller, H.~Ooguri and I.~G. Zadeh, \emph{{On Rational Points
  in CFT Moduli Spaces}},
  \href{https://doi.org/10.1007/JHEP04(2021)067}{\emph{JHEP} {\bfseries 04}
  (2021) 067} [\href{https://arxiv.org/abs/2011.07062}{{\ttfamily
  2011.07062}}].

\bibitem{Keller:2023ssv}
C.~A. Keller, \emph{{Conformal perturbation theory on K3: the quartic Gepner
  point}}, \href{https://doi.org/10.1007/JHEP01(2024)197}{\emph{JHEP}
  {\bfseries 01} (2024) 197}
  [\href{https://arxiv.org/abs/2311.12974}{{\ttfamily 2311.12974}}].

\bibitem{Moore:1998pn}
G.~W. Moore, \emph{{Arithmetic and attractors}},
  \href{https://arxiv.org/abs/hep-th/9807087}{{\ttfamily hep-th/9807087}}.

\bibitem{Wendland:2000ye}
K.~Wendland, \emph{{Moduli spaces of unitary conformal field theories}}, {PhD
  thesis}, Universit\"at Bonn, Bonn, 2000.

\bibitem{Hosono:2002yb}
S.~Hosono, B.~H. Lian, K.~Oguiso and S.-T. Yau, \emph{{Classification of c = 2
  rational conformal field theories via the Gauss product}},
  \href{https://doi.org/10.1007/s00220-003-0927-0}{\emph{Commun. Math. Phys.}
  {\bfseries 241} (2003) 245}
  [\href{https://arxiv.org/abs/hep-th/0211230}{{\ttfamily hep-th/0211230}}].

\bibitem{Gukov:2002nw}
S.~Gukov and C.~Vafa, \emph{{Rational conformal field theories and complex
  multiplication}},
  \href{https://doi.org/10.1007/s00220-003-1032-0}{\emph{Commun. Math. Phys.}
  {\bfseries 246} (2004) 181}
  [\href{https://arxiv.org/abs/hep-th/0203213}{{\ttfamily hep-th/0203213}}].

\bibitem{Chen:2005gm}
M.~Chen, \emph{{Complex multiplication, rationality and mirror symmetry for
  Abelian varieties}},
  \href{https://doi.org/10.1016/j.geomphys.2008.01.001}{\emph{J. Geom. Phys.}
  {\bfseries 58} (2008) 633}
  [\href{https://arxiv.org/abs/math/0512470}{{\ttfamily math/0512470}}].

\bibitem{Kidambi:2024vwl}
A.~Kidambi, M.~Okada and T.~Watari, \emph{{Notes on Characterizations of 2d
  Rational SCFTs: Algebraicity, Mirror Symmetry, and Complex Multiplication}},
  \href{https://doi.org/10.1002/prop.202400161}{\emph{Fortsch. Phys.}
  {\bfseries 73} (2025) 2400161}
  [\href{https://arxiv.org/abs/2408.00861}{{\ttfamily 2408.00861}}].

\bibitem{Distler:1992gi}
J.~Distler, \emph{{Notes on N=2 sigma models}},
  \href{https://arxiv.org/abs/hep-th/9212062}{{\ttfamily hep-th/9212062}}.

\bibitem{Greene:1996cy}
B.~R. Greene, \emph{{String theory on Calabi-Yau manifolds}},  in
  \emph{{Theoretical Advanced Study Institute in Elementary Particle Physics
  (TASI 96): Fields, Strings, and Duality}}, pp.~543--726, 6, 1996,
  \href{https://arxiv.org/abs/hep-th/9702155}{{\ttfamily hep-th/9702155}}.

\bibitem{Jockers:2024ffr}
H.~Jockers, M.~Sarve and I.~G. Zadeh, \emph{{Minimally extended current
  algebras of toroidal conformal field theories}},
  \href{https://doi.org/10.1007/JHEP07(2024)187}{\emph{JHEP} {\bfseries 07}
  (2024) 187} [\href{https://arxiv.org/abs/2404.18269}{{\ttfamily
  2404.18269}}].

\bibitem{MR990016}
Y.~Andr\'{e}, \emph{{$G$}-functions and geometry}, Aspects of Mathematics, E13.
  Friedr. Vieweg \& Sohn, Braunschweig, 1989,
  \href{https://doi.org/10.1007/978-3-663-14108-2}{10.1007/978-3-663-14108-2}.

\bibitem{MR1472499}
F.~Oort, \emph{Canonical liftings and dense sets of {CM}-points},  in
  \emph{Arithmetic geometry ({C}ortona, 1994)}, Sympos. Math., XXXVII,
  pp.~228--234.
\newblock Cambridge Univ. Press, Cambridge, 1997.

\bibitem{MR1273413}
E.~Cattani, P.~Deligne and A.~Kaplan, \emph{On the locus of {H}odge classes},
  \href{https://doi.org/10.2307/2152824}{\emph{J. Amer. Math. Soc.} {\bfseries
  8} (1995) 483}.

\bibitem{MR2520786}
J.~Pila, \emph{Rational points of definable sets and results of
  {A}ndr\'{e}-{O}ort-{M}anin-{M}umford type},
  \href{https://doi.org/10.1093/imrn/rnp022}{\emph{Int. Math. Res. Not. IMRN}
  (2009) 2476}.

\bibitem{MR2800724}
J.~Pila, \emph{O-minimality and the {A}ndr\'{e}-{O}ort conjecture for {$\mathbb
  C^n$}}, \href{https://doi.org/10.4007/annals.2011.173.3.11}{\emph{Ann. of
  Math. (2)} {\bfseries 173} (2011) 1779}.

\bibitem{MR3245009}
B.~Klingler and A.~Yafaev, \emph{The {A}ndr\'{e}-{O}ort conjecture},
  \href{https://doi.org/10.4007/annals.2014.180.3.2}{\emph{Ann. of Math. (2)}
  {\bfseries 180} (2014) 867}
  [\href{https://arxiv.org/abs/1209.0936}{{\ttfamily 1209.0936}}].

\bibitem{MR4689371}
G.~Baldi, B.~Klingler and E.~Ullmo, \emph{On the distribution of the {H}odge
  locus}, \href{https://doi.org/10.1007/s00222-023-01226-0}{\emph{Invent.
  Math.} {\bfseries 235} (2024) 441}
  [\href{https://arxiv.org/abs/2107.08838}{{\ttfamily 2107.08838}}].

\bibitem{Grimm:2021vpn}
T.~W. Grimm, \emph{{Taming the landscape of effective theories}},
  \href{https://doi.org/10.1007/JHEP11(2022)003}{\emph{JHEP} {\bfseries 11}
  (2022) 003} [\href{https://arxiv.org/abs/2112.08383}{{\ttfamily
  2112.08383}}].

\bibitem{Bakker:2021uqw}
B.~Bakker, T.~W. Grimm, C.~Schnell and J.~Tsimerman, \emph{{Finiteness for
  self-dual classes in integral variations of Hodge structure}},
  \href{https://arxiv.org/abs/2112.06995}{{\ttfamily 2112.06995}}.

\bibitem{Palti:2019pca}
E.~Palti, \emph{{The Swampland: Introduction and Review}},
  \href{https://doi.org/10.1002/prop.201900037}{\emph{Fortsch. Phys.}
  {\bfseries 67} (2019) 1900037}
  [\href{https://arxiv.org/abs/1903.06239}{{\ttfamily 1903.06239}}].

\bibitem{Agmon:2022thq}
N.~B. Agmon, A.~Bedroya, M.~J. Kang and C.~Vafa, \emph{{Lectures on the string
  landscape and the Swampland}},
  \href{https://arxiv.org/abs/2212.06187}{{\ttfamily 2212.06187}}.

\bibitem{Grimm:2024fip}
T.~W. Grimm and D.~van~de Heisteeg, \emph{{Exact flux vacua, symmetries, and
  the structure of the landscape}},
  \href{https://doi.org/10.1007/JHEP01(2025)005}{\emph{JHEP} {\bfseries 01}
  (2025) 005} [\href{https://arxiv.org/abs/2404.12422}{{\ttfamily
  2404.12422}}].

\bibitem{DeWolfe:2005gy}
O.~DeWolfe, \emph{{Enhanced symmetries in multiparameter flux vacua}},
  \href{https://doi.org/10.1088/1126-6708/2005/10/066}{\emph{JHEP} {\bfseries
  10} (2005) 066} [\href{https://arxiv.org/abs/hep-th/0506245}{{\ttfamily
  hep-th/0506245}}].

\bibitem{Kanno:2017nub}
K.~Kanno and T.~Watari, \emph{{Revisiting arithmetic solutions to the $W=0$
  condition}}, \href{https://doi.org/10.1103/PhysRevD.96.106001}{\emph{Phys.
  Rev. D} {\bfseries 96} (2017) 106001}
  [\href{https://arxiv.org/abs/1705.05110}{{\ttfamily 1705.05110}}].

\bibitem{Kachru:2020sio}
S.~Kachru, R.~Nally and W.~Yang, \emph{{Supersymmetric flux compactifications
  and Calabi-Yau modularity}},
  \href{https://doi.org/10.4310/cntp.240904012658}{\emph{Commun. Num. Theor.
  Phys.} {\bfseries 18} (2024) 653}
  [\href{https://arxiv.org/abs/2001.06022}{{\ttfamily 2001.06022}}].

\bibitem{Kachru:2020abh}
S.~Kachru, R.~Nally and W.~Yang, \emph{{Flux Modularity, F-Theory, and Rational
  Models}},  \href{https://arxiv.org/abs/2010.07285}{{\ttfamily 2010.07285}}.

\bibitem{Schimmrigk:2020dfl}
R.~Schimmrigk, \emph{{On flux vacua and modularity}},
  \href{https://doi.org/10.1007/JHEP09(2020)061}{\emph{JHEP} {\bfseries 09}
  (2020) 061} [\href{https://arxiv.org/abs/2003.01056}{{\ttfamily
  2003.01056}}].

\bibitem{Candelas:2023yrg}
P.~Candelas, X.~de~la Ossa, P.~Kuusela and J.~McGovern, \emph{{Flux vacua and
  modularity for $\mathbb{Z}_2$ symmetric Calabi-Yau manifolds}},
  \href{https://doi.org/10.21468/SciPostPhys.15.4.146}{\emph{SciPost Phys.}
  {\bfseries 15} (2023) 146}
  [\href{https://arxiv.org/abs/2302.03047}{{\ttfamily 2302.03047}}].

\bibitem{Jockers:2023zzi}
H.~Jockers, S.~Kotlewski and P.~Kuusela, \emph{{Modular Calabi-Yau fourfolds
  and connections to M-theory fluxes}},
  \href{https://doi.org/10.1007/JHEP12(2024)052}{\emph{JHEP} {\bfseries 12}
  (2024) 052} [\href{https://arxiv.org/abs/2312.07611}{{\ttfamily
  2312.07611}}].

\bibitem{Gukov:1999ya}
S.~Gukov, C.~Vafa and E.~Witten, \emph{{CFT's from Calabi-Yau four folds}},
  \href{https://doi.org/10.1016/S0550-3213(00)00373-4}{\emph{Nucl. Phys. B}
  {\bfseries 584} (2000) 69}
  [\href{https://arxiv.org/abs/hep-th/9906070}{{\ttfamily hep-th/9906070}}].

\bibitem{Taylor:1999ii}
T.~R. Taylor and C.~Vafa, \emph{{R R flux on Calabi-Yau and partial
  supersymmetry breaking}},
  \href{https://doi.org/10.1016/S0370-2693(00)00005-8}{\emph{Phys. Lett. B}
  {\bfseries 474} (2000) 130}
  [\href{https://arxiv.org/abs/hep-th/9912152}{{\ttfamily hep-th/9912152}}].

\bibitem{Mayr:2000hh}
P.~Mayr, \emph{{On supersymmetry breaking in string theory and its realization
  in brane worlds}},
  \href{https://doi.org/10.1016/S0550-3213(00)00552-6}{\emph{Nucl. Phys. B}
  {\bfseries 593} (2001) 99}
  [\href{https://arxiv.org/abs/hep-th/0003198}{{\ttfamily hep-th/0003198}}].

\bibitem{Louis:2012ux}
J.~Louis, P.~Smyth and H.~Triendl, \emph{{Supersymmetric Vacua in N=2
  Supergravity}}, \href{https://doi.org/10.1007/JHEP08(2012)039}{\emph{JHEP}
  {\bfseries 08} (2012) 039} [\href{https://arxiv.org/abs/1204.3893}{{\ttfamily
  1204.3893}}].

\bibitem{Jockers:2024ocq}
H.~Jockers and S.~Kotlewski, \emph{{Geometry of N=2 Minkowski vacua of gauged
  N=2 supergravity theories in four dimensions}},
  \href{https://doi.org/10.1103/PhysRevD.110.046019}{\emph{Phys. Rev. D}
  {\bfseries 110} (2024) 046019}
  [\href{https://arxiv.org/abs/2404.11655}{{\ttfamily 2404.11655}}].

\bibitem{Lerche:1989uy}
W.~Lerche, C.~Vafa and N.~P. Warner, \emph{{Chiral Rings in N=2 Superconformal
  Theories}}, \href{https://doi.org/10.1016/0550-3213(89)90474-4}{\emph{Nucl.
  Phys. B} {\bfseries 324} (1989) 427}.

\bibitem{Verlinde:1988sn}
E.~P. Verlinde, \emph{{Fusion Rules and Modular Transformations in 2D Conformal
  Field Theory}},
  \href{https://doi.org/10.1016/0550-3213(88)90603-7}{\emph{Nucl. Phys. B}
  {\bfseries 300} (1988) 360}.

\bibitem{Gepner:1987qi}
D.~Gepner, \emph{{Space-Time Supersymmetry in Compactified String Theory and
  Superconformal Models}},
  \href{https://doi.org/10.1016/0550-3213(88)90397-5}{\emph{Nucl. Phys. B}
  {\bfseries 296} (1988) 757}.

\bibitem{Gepner:1989gr}
D.~Gepner, \emph{{L}ectures on {N=2} {S}tring {T}heory},  in \emph{{Trieste
  School and Workshop on Superstrings}}, pp.~80--144, 4, 1989.

\bibitem{Blumenhagen:2009zz}
R.~Blumenhagen and E.~Plauschinn, \emph{{Introduction to Conformal Field
  Theory}: {With Applications to String Theory}}, vol.~779 of \emph{Lecture
  Notes in Physics}. Springer-Verlag, New York, 2009,
  \href{https://doi.org/10.1007/978-3-642-00450-6}{10.1007/978-3-642-00450-6}.

\bibitem{Harvey:1987da}
J.~A. Harvey, G.~W. Moore and C.~Vafa, \emph{{Q}uasicrystalline
  {C}ompactification},
  \href{https://doi.org/10.1016/0550-3213(88)90627-X}{\emph{Nucl. Phys. B}
  {\bfseries 304} (1988) 269}.

\bibitem{Moore:1988qv}
G.~W. Moore and N.~Seiberg, \emph{{Classical and Quantum Conformal Field
  Theory}}, \href{https://doi.org/10.1007/BF01238857}{\emph{Commun. Math.
  Phys.} {\bfseries 123} (1989) 177}.

\bibitem{Schellekens:1990xy}
A.~N. Schellekens and S.~Yankielowicz, \emph{{Simple Currents, Modular
  Invariants and Fixed Points}},
  \href{https://doi.org/10.1142/S0217751X90001367}{\emph{Int. J. Mod. Phys. A}
  {\bfseries 5} (1990) 2903}.

\bibitem{Segal:2002ei}
G.~Segal, \emph{{The definition of conformal field theory}},  in
  \emph{{Symposium on Topology, Geometry and Quantum Field Theory
  (Segalfest)}}, pp.~421--575, 6, 2002.

\bibitem{MR1029428}
B.~L. Feigin and D.~B. Fuchs, \emph{Cohomology of some nilpotent subalgebras of
  the {V}irasoro and {K}ac-{M}oody {L}ie algebras},
  \href{https://doi.org/10.1016/0393-0440(88)90005-8}{\emph{J. Geom. Phys.}
  {\bfseries 5} (1988) 209}.

\bibitem{MR1186962}
D.~Kazhdan and G.~Lusztig, \emph{Tensor structures arising from affine {L}ie
  algebras. {I}, {II}}, \href{https://doi.org/10.2307/2152745}{\emph{J. Amer.
  Math. Soc.} {\bfseries 6} (1993) 905}.

\bibitem{Dijkgraaf:1988tf}
R.~Dijkgraaf and E.~P. Verlinde, \emph{{Modular Invariance and the Fusion
  Algebra}}, \href{https://doi.org/10.1016/0920-5632(88)90371-4}{\emph{Nucl.
  Phys. B Proc. Suppl.} {\bfseries 5} (1988) 87}.

\bibitem{Fuchs:1996dd}
J.~Fuchs, A.~N. Schellekens and C.~Schweigert, \emph{{A Matrix S for all simple
  current extensions}},
  \href{https://doi.org/10.1016/0550-3213(96)00247-7}{\emph{Nucl. Phys. B}
  {\bfseries 473} (1996) 323}
  [\href{https://arxiv.org/abs/hep-th/9601078}{{\ttfamily hep-th/9601078}}].

\bibitem{Brunner:2000nk}
I.~Brunner and V.~Schomerus, \emph{{D-branes at singular curves of Calabi-Yau
  compactifications}},
  \href{https://doi.org/10.1088/1126-6708/2000/04/020}{\emph{JHEP} {\bfseries
  04} (2000) 020} [\href{https://arxiv.org/abs/hep-th/0001132}{{\ttfamily
  hep-th/0001132}}].

\bibitem{DeBoer:1990em}
J.~De~Boer and J.~Goeree, \emph{{Markov traces and II(1) factors in conformal
  field theory}}, \href{https://doi.org/10.1007/BF02352496}{\emph{Commun. Math.
  Phys.} {\bfseries 139} (1991) 267}.

\bibitem{Coste:1993af}
A.~Coste and T.~Gannon, \emph{{Remarks on Galois symmetry in rational conformal
  field theories}},
  \href{https://doi.org/10.1016/0370-2693(94)91226-2}{\emph{Phys. Lett. B}
  {\bfseries 323} (1994) 316}.

\bibitem{Warner:1989dj}
N.~P. Warner, \emph{{L}ectures on {N=2} {S}uperconformal {T}heories and
  {S}ingularity {T}heory},  in \emph{{Trieste School and Workshop on
  Superstrings}}, 6, 1989.

\bibitem{Schwimmer:1986mf}
A.~Schwimmer and N.~Seiberg, \emph{{Comments on the N=2, N=3, N=4
  Superconformal Algebras in Two-Dimensions}},
  \href{https://doi.org/10.1016/0370-2693(87)90566-1}{\emph{Phys. Lett. B}
  {\bfseries 184} (1987) 191}.

\bibitem{Ishibashi:1988kg}
N.~Ishibashi, \emph{{The Boundary and Crosscap States in Conformal Field
  Theories}}, \href{https://doi.org/10.1142/S0217732389000320}{\emph{Mod. Phys.
  Lett. A} {\bfseries 4} (1989) 251}.

\bibitem{Ishibashi:1988tf}
N.~Ishibashi and T.~Onogi, \emph{{O}pen {S}tring {M}odel {B}uilding},
  \href{https://doi.org/10.1016/0550-3213(89)90054-0}{\emph{Nucl. Phys. B}
  {\bfseries 318} (1989) 239}.

\bibitem{Cardy:1989ir}
J.~L. Cardy, \emph{{Boundary Conditions, Fusion Rules and the Verlinde
  Formula}}, \href{https://doi.org/10.1016/0550-3213(89)90521-X}{\emph{Nucl.
  Phys. B} {\bfseries 324} (1989) 581}.

\bibitem{Govindarajan:2000my}
S.~Govindarajan and T.~Jayaraman, \emph{{On the Landau-Ginzburg description of
  boundary CFTs and special Lagrangian submanifolds}},
  \href{https://doi.org/10.1088/1126-6708/2000/07/016}{\emph{JHEP} {\bfseries
  07} (2000) 016} [\href{https://arxiv.org/abs/hep-th/0003242}{{\ttfamily
  hep-th/0003242}}].

\bibitem{Hori:2000ck}
K.~Hori, A.~Iqbal and C.~Vafa, \emph{{D-branes and mirror symmetry}},
  \href{https://arxiv.org/abs/hep-th/0005247}{{\ttfamily hep-th/0005247}}.

\bibitem{Douglas:1999hq}
M.~R. Douglas and B.~Fiol, \emph{{D-branes and discrete torsion. 2.}},
  \href{https://doi.org/10.1088/1126-6708/2005/09/053}{\emph{JHEP} {\bfseries
  09} (2005) 053} [\href{https://arxiv.org/abs/hep-th/9903031}{{\ttfamily
  hep-th/9903031}}].

\bibitem{Brunner:1999jq}
I.~Brunner, M.~R. Douglas, A.~E. Lawrence and C.~R{\"o}melsberger,
  \emph{{D-branes on the quintic}},
  \href{https://doi.org/10.1088/1126-6708/2000/08/015}{\emph{JHEP} {\bfseries
  08} (2000) 015} [\href{https://arxiv.org/abs/hep-th/9906200}{{\ttfamily
  hep-th/9906200}}].

\bibitem{Witten:1998cd}
E.~Witten, \emph{{D-branes and K-theory}},
  \href{https://doi.org/10.1088/1126-6708/1998/12/019}{\emph{JHEP} {\bfseries
  12} (1998) 019} [\href{https://arxiv.org/abs/hep-th/9810188}{{\ttfamily
  hep-th/9810188}}].

\bibitem{Witten:1993jg}
E.~Witten, \emph{{On the Landau-Ginzburg description of N=2 minimal models}},
  \href{https://doi.org/10.1142/S0217751X9400193X}{\emph{Int. J. Mod. Phys. A}
  {\bfseries 9} (1994) 4783}
  [\href{https://arxiv.org/abs/hep-th/9304026}{{\ttfamily hep-th/9304026}}].

\bibitem{Greene:1990ud}
B.~R. Greene and M.~R. Plesser, \emph{{Duality in {Calabi-Yau} Moduli Space}},
  \href{https://doi.org/10.1016/0550-3213(90)90622-K}{\emph{Nucl. Phys. B}
  {\bfseries 338} (1990) 15}.

\bibitem{Green2012a}
M.~Green, P.~Griffiths and M.~Kerr, \emph{Mumford-{T}ate groups and domains},
  vol.~183 of \emph{Annals of Mathematics Studies}. Princeton University Press,
  Princeton, NJ, 2012.

\bibitem{Lang1983a}
S.~Lang, \emph{Complex multiplication}, vol.~255 of \emph{Grundlehren der
  mathematischen Wissenschaften [Fundamental Principles of Mathematical
  Sciences]}. Springer-Verlag, New York, 1983,
  \href{https://doi.org/10.1007/978-1-4612-5485-0}{10.1007/978-1-4612-5485-0}.

\bibitem{Lang2002a}
S.~Lang, \emph{Algebra}, vol.~211 of \emph{Graduate Texts in Mathematics}.
  Springer-Verlag, New York, third~ed., 2002,
  \href{https://doi.org/10.1007/978-1-4613-0041-0}{10.1007/978-1-4613-0041-0}.

\bibitem{Martinec:1988zu}
E.~J. Martinec, \emph{{Algebraic Geometry and Effective Lagrangians}},
  \href{https://doi.org/10.1016/0370-2693(89)90074-9}{\emph{Phys. Lett. B}
  {\bfseries 217} (1989) 431}.

\bibitem{Greene:1988ut}
B.~R. Greene, C.~Vafa and N.~P. Warner, \emph{{Calabi-Yau Manifolds and
  Renormalization Group Flows}},
  \href{https://doi.org/10.1016/0550-3213(89)90471-9}{\emph{Nucl. Phys. B}
  {\bfseries 324} (1989) 371}.

\bibitem{Witten:1993yc}
E.~Witten, \emph{{Phases of N=2 theories in two-dimensions}},
  \href{https://doi.org/10.1016/0550-3213(93)90033-L}{\emph{Nucl. Phys. B}
  {\bfseries 403} (1993) 159}
  [\href{https://arxiv.org/abs/hep-th/9301042}{{\ttfamily hep-th/9301042}}].

\bibitem{Witten:1991zz}
E.~Witten, \emph{{Mirror manifolds and topological field theory}},
  {\emph{AMS/IP Stud. Adv. Math.} {\bfseries 9} (1998) 121}
  [\href{https://arxiv.org/abs/hep-th/9112056}{{\ttfamily hep-th/9112056}}].

\bibitem{Ooguri:1996ck}
H.~Ooguri, Y.~Oz and Z.~Yin, \emph{{D-branes on Calabi-Yau spaces and their
  mirrors}}, \href{https://doi.org/10.1016/0550-3213(96)00379-3}{\emph{Nucl.
  Phys. B} {\bfseries 477} (1996) 407}
  [\href{https://arxiv.org/abs/hep-th/9606112}{{\ttfamily hep-th/9606112}}].

\bibitem{Hori:2003ic}
K.~Hori, S.~Katz, A.~Klemm, R.~Pandharipande, R.~Thomas, C.~Vafa et~al.,
  \emph{{Mirror symmetry}}, vol.~1 of \emph{Clay mathematics monographs}. AMS,
  Providence, USA, 2003.

\bibitem{Goddard:1986ee}
P.~Goddard, A.~Kent and D.~I. Olive, \emph{{Unitary Representations of the
  Virasoro and Supervirasoro Algebras}},
  \href{https://doi.org/10.1007/BF01464283}{\emph{Commun. Math. Phys.}
  {\bfseries 103} (1986) 105}.

\bibitem{DiVecchia:1986fwg}
P.~Di~Vecchia, J.~L. Petersen, M.~Yu and H.~B. Zheng, \emph{{Explicit
  Construction of Unitary Representations of the N=2 Superconformal Algebra}},
  \href{https://doi.org/10.1016/0370-2693(86)91099-3}{\emph{Phys. Lett. B}
  {\bfseries 174} (1986) 280}.

\bibitem{Fuchs:1999zi}
J.~Fuchs and C.~Schweigert, \emph{{Symmetry breaking boundaries. 1. General
  theory}}, \href{https://doi.org/10.1016/S0550-3213(99)00406-X}{\emph{Nucl.
  Phys. B} {\bfseries 558} (1999) 419}
  [\href{https://arxiv.org/abs/hep-th/9902132}{{\ttfamily hep-th/9902132}}].

\bibitem{Fuchs:1999xn}
J.~Fuchs and C.~Schweigert, \emph{{Symmetry breaking boundaries. 2. More
  structures: Examples}},
  \href{https://doi.org/10.1016/S0550-3213(99)00669-0}{\emph{Nucl. Phys. B}
  {\bfseries 568} (2000) 543}
  [\href{https://arxiv.org/abs/hep-th/9908025}{{\ttfamily hep-th/9908025}}].

\bibitem{Fuchs:2000gv}
J.~Fuchs, C.~Schweigert and J.~Walcher, \emph{{Projections in string theory and
  boundary states for Gepner models}},
  \href{https://doi.org/10.1016/S0550-3213(00)00487-9}{\emph{Nucl. Phys. B}
  {\bfseries 588} (2000) 110}
  [\href{https://arxiv.org/abs/hep-th/0003298}{{\ttfamily hep-th/0003298}}].

\bibitem{Schellekens:1989am}
A.~N. Schellekens and S.~Yankielowicz, \emph{{Extended Chiral Algebras and
  Modular Invariant Partition Functions}},
  \href{https://doi.org/10.1016/0550-3213(89)90310-6}{\emph{Nucl. Phys. B}
  {\bfseries 327} (1989) 673}.

\bibitem{Schellekens:1989dq}
A.~N. Schellekens and S.~Yankielowicz, \emph{{Modular Invariants From Simple
  Currents: An Explicit Proof}},
  \href{https://doi.org/10.1016/0370-2693(89)90948-9}{\emph{Phys. Lett. B}
  {\bfseries 227} (1989) 387}.

\bibitem{Gato-Rivera:1991bqv}
B.~Gato-Rivera and A.~N. Schellekens, \emph{{Complete classification of simple
  current modular invariants for $(Z_p)^k$}},
  \href{https://doi.org/10.1007/BF02099282}{\emph{Commun. Math. Phys.}
  {\bfseries 145} (1992) 85}.

\bibitem{Recknagel:1997sb}
A.~Recknagel and V.~Schomerus, \emph{{D-branes in Gepner models}},
  \href{https://doi.org/10.1016/S0550-3213(98)00468-4}{\emph{Nucl. Phys. B}
  {\bfseries 531} (1998) 185}
  [\href{https://arxiv.org/abs/hep-th/9712186}{{\ttfamily hep-th/9712186}}].

\bibitem{Gepner:1987vz}
D.~Gepner, \emph{{Exactly Solvable String Compactifications on Manifolds of
  SU(N) Holonomy}},
  \href{https://doi.org/10.1016/0370-2693(87)90938-5}{\emph{Phys. Lett. B}
  {\bfseries 199} (1987) 380}.

\bibitem{Borcea}
C.~Borcea, \emph{Calabi-{Y}au threefolds and complex multiplication},  in
  \emph{Essays on mirror manifolds}, pp.~489--502.
\newblock Int. Press, Hong Kong, 1992.

\bibitem{Kidambi:2022wvh}
A.~Kidambi, M.~Okada and T.~Watari, \emph{{Towards Hodge Theoretic
  Characterizations of 2d Rational SCFTs}},
  \href{https://arxiv.org/abs/2205.10299}{{\ttfamily 2205.10299}}.

\bibitem{Okada:2022jnq}
M.~Okada and T.~Watari, \emph{{Towards Hodge Theoretic Characterizations of 2d
  Rational SCFTs: II}},  \href{https://arxiv.org/abs/2212.13028}{{\ttfamily
  2212.13028}}.

\bibitem{Kazama:1988uz}
Y.~Kazama and H.~Suzuki, \emph{{Characterization of N=2 Superconformal Models
  Generated by Coset Space Method}},
  \href{https://doi.org/10.1016/0370-2693(89)91378-6}{\emph{Phys. Lett. B}
  {\bfseries 216} (1989) 112}.

\bibitem{Narain:1986qm}
K.~S. Narain, M.~H. Sarmadi and C.~Vafa, \emph{{Asymmetric Orbifolds}},
  \href{https://doi.org/10.1016/0550-3213(87)90228-8}{\emph{Nucl. Phys. B}
  {\bfseries 288} (1987) 551}.

\bibitem{Shatashvili:1994zw}
S.~L. Shatashvili and C.~Vafa, \emph{{Superstrings and manifold of exceptional
  holonomy}}, \href{https://doi.org/10.1007/BF01671569}{\emph{Selecta Math.}
  {\bfseries 1} (1995) 347}
  [\href{https://arxiv.org/abs/hep-th/9407025}{{\ttfamily hep-th/9407025}}].

\bibitem{Roiban:2001cp}
R.~Roiban and J.~Walcher, \emph{{Rational conformal field theories with G(2)
  holonomy}}, \href{https://doi.org/10.1088/1126-6708/2001/12/008}{\emph{JHEP}
  {\bfseries 12} (2001) 008}
  [\href{https://arxiv.org/abs/hep-th/0110302}{{\ttfamily hep-th/0110302}}].

\bibitem{Bonisch:2022mgw}
K.~B\"onisch, A.~Klemm, E.~Scheidegger and D.~Zagier, \emph{{D-brane masses at
  special fibres of hypergeometric families of Calabi-Yau threefolds, modular
  forms, and periods}},  \href{https://arxiv.org/abs/2203.09426}{{\ttfamily
  2203.09426}}.

\bibitem{Candelas:2019llw}
P.~Candelas, X.~de~la Ossa, M.~Elmi and D.~Van~Straten, \emph{{A One Parameter
  Family of Calabi-Yau Manifolds with Attractor Points of Rank Two}},
  \href{https://doi.org/10.1007/JHEP10(2020)202}{\emph{JHEP} {\bfseries 10}
  (2020) 202} [\href{https://arxiv.org/abs/1912.06146}{{\ttfamily
  1912.06146}}].

\bibitem{Schimmrigk:2008mp}
R.~Schimmrigk, \emph{{Emergent spacetime from modular motives}},
  \href{https://doi.org/10.1007/s00220-010-1179-4}{\emph{Commun. Math. Phys.}
  {\bfseries 303} (2011) 1} [\href{https://arxiv.org/abs/0812.4450}{{\ttfamily
  0812.4450}}].

\bibitem{Schimmrigk:2006aa}
R.~Schimmrigk, \emph{Arithmetic spacetime geometry from string theory},
  \href{https://doi.org/10.1142/S0217751X06034343}{\emph{Int.J.Mod.Phys.}
  {\bfseries A21} (2006) 6323}
  [\href{https://arxiv.org/abs/hep-th/0510091}{{\ttfamily hep-th/0510091}}].

\bibitem{Kondo:2018mha}
S.~Kondo and T.~Watari, \emph{{String-theory Realization of Modular Forms for
  Elliptic Curves with Complex Multiplication}},
  \href{https://doi.org/10.1007/s00220-019-03302-0}{\emph{Commun. Math. Phys.}
  {\bfseries 367} (2019) 89}
  [\href{https://arxiv.org/abs/1801.07464}{{\ttfamily 1801.07464}}].

\bibitem{Kondo:2019jpi}
S.~Kondo and T.~Watari, \emph{{Modular parametrization as Polyakov path
  integral: cases with CM elliptic curves as target spaces}},
  \href{https://doi.org/10.4310/CNTP.2022.v16.n2.a3}{\emph{Commun. Num. Theor.
  Phys.} {\bfseries 16} (2022) 353}
  [\href{https://arxiv.org/abs/1912.13294}{{\ttfamily 1912.13294}}].

\bibitem{Weil1949a}
A.~Weil, \emph{Numbers of solutions of equations in finite fields},
  \href{https://doi.org/10.1090/S0002-9904-1949-09219-4}{\emph{Bull. Amer.
  Math. Soc.} {\bfseries 55} (1949) 497}.

\bibitem{Dwork1960a}
B.~Dwork, \emph{On the rationality of the zeta function of an algebraic
  variety}, \href{https://doi.org/10.2307/2372974}{\emph{Amer. J. Math.}
  {\bfseries 82} (1960) 631}.

\bibitem{Grothendieck1995a}
A.~Grothendieck, \emph{Formule de {L}efschetz et rationalit\'{e} des fonctions
  {$L$}},  in \emph{S\'{e}minaire {B}ourbaki, {V}ol. 9}, pp.~Exp. No. 279,
  41--55.
\newblock Soc. Math. France, Paris, 1995.

\bibitem{Deligne1974a}
P.~Deligne, \emph{La conjecture de {W}eil. {I}}, {\emph{Inst. Hautes \'{E}tudes
  Sci. Publ. Math.} (1974) 273}.

\bibitem{Deligne1980a}
P.~Deligne, \emph{La conjecture de {W}eil. {II}}, {\emph{Inst. Hautes
  \'{E}tudes Sci. Publ. Math.} (1980) 137}.

\bibitem{Silverman}
J.~H. Silverman, \emph{Advanced topics in the arithmetic of elliptic curves},
  vol.~151 of \emph{Graduate Texts in Mathematics}. Springer-Verlag, New York,
  1994,
  \href{https://doi.org/10.1007/978-1-4612-0851-8}{10.1007/978-1-4612-0851-8}.

\bibitem{LMFDB}
T.~{LMFDB Collaboration}, ``The {L}-functions and modular forms database.''
  \url{https://www.lmfdb.org}, 2025.

\bibitem{Candelas:2000fq}
P.~Candelas, X.~de~la Ossa and F.~Rodriguez-Villegas, \emph{{Calabi-Yau
  manifolds over finite fields. 1.}},
  \href{https://arxiv.org/abs/hep-th/0012233}{{\ttfamily hep-th/0012233}}.

\bibitem{Candelas:2004sk}
P.~Candelas, X.~de~la Ossa and F.~Rodriguez~Villegas, \emph{{Calabi-Yau
  manifolds over finite fields. 2.}}, {\emph{Fields Inst. Commun.} {\bfseries
  38} (2013) 121} [\href{https://arxiv.org/abs/hep-th/0402133}{{\ttfamily
  hep-th/0402133}}].

\bibitem{Watkins:2011}
M.~Watkins, \emph{{Computing with Hecke Gr\"ossencharacters}},
  {\emph{Publications math\'ematiques de Besan\c con. Alg\`ebre et th\'eorie
  des nombres. Actes de la conf\'erence ``Th\'eorie des nombres et
  applications'', CIRM, Luminy, du 30/11 au 04/12 2009} (2011) 119}.

\bibitem{Watkins:2018}
M.~Watkins, \emph{{Jacobi sums and Gr\"ossencharacters}}, {\emph{Publications
  math\'ematiques de Besan\c con. Alg\`ebre et th\'eorie des nombres} (2018)
  111}.

\bibitem{vanderPut1986a}
M.~van~der Put, \emph{The cohomology of {M}onsky and {W}ashnitzer},
  {\emph{M\'em. Soc. Math. France (N.S.)} (1986) 4, 33}.

\bibitem{Okada:2023udq}
M.~Okada and T.~Watari, \emph{{A note on varieties of weak CM-type}},
  \href{https://doi.org/10.1016/j.geomphys.2023.105084}{\emph{J. Geom. Phys.}
  {\bfseries 197} (2024) 105084}
  [\href{https://arxiv.org/abs/2306.10282}{{\ttfamily 2306.10282}}].

\bibitem{Koblitz}
N.~Koblitz, \emph{Introduction to elliptic curves and modular forms}, vol.~97
  of \emph{Graduate Texts in Mathematics}. Springer-Verlag, New York,
  second~ed., 1993,
  \href{https://doi.org/10.1007/978-1-4612-0909-6}{10.1007/978-1-4612-0909-6}.

\end{thebibliography}\endgroup

\end{document}